\def\({ \left( }
\def\){ \right) }
\def\b{\begin{equation}}
\def\e{\end{equation}}
\def\={\ =\ }

\def\+{\ +\ }
\def\-{\ -\ }

\def\mumu{$\mu^+\mu^-$}
\def\ee{$e^+e^-$}
\def\rr{}

\def\dofigs#1#2#3#4{\centerline{\epsfxsize=#1\epsfbox{#2}\hfill%
   \epsfxsize=#3\epsfbox{#4}}}

\def\etal{{\it et al.}}
\def\ie{{\it i.e.}}
\def\vs{{\it vs.}}

\def\~{{$\tilde{\phantom{a}}$}}

\def\lsim{\mathrel {\vcenter {\baselineskip 0pt \kern 0pt
    \hbox{$<$} \kern 0pt \hbox{$\sim$} }}}
\def\gsim{\mathrel {\vcenter {\baselineskip 0pt \kern 0pt
    \hbox{$>$} \kern 0pt \hbox{$\sim$} }}}
\def\lambdabar{\protect\@lambdabar}
\def\@lambdabar{%
  \relax
  \bgroup
  \def\@tempa{\hbox{\raise.73\ht0
  \hbox to0pt{\kern.25\wd0\vrule width.5\wd0
  height.1pt depth.1pt\hss}\box0}}%
  \mathchoice{\setbox0\hbox{$\displaystyle\lambda$}\@tempa}%
  {\setbox0\hbox{$\textstyle\lambda$}\@tempa}%
  {\setbox0\hbox{$\scriptstyle\lambda$}\@tempa}%
  {\setbox0\hbox{$\scriptscriptstyle\lambda$}\@tempa}%
  \egroup}
\def\E{\,\rlap/\!E_T}
\def\srts{\sigma_{\!\!\!\sqrt s}^{\vphantom y}}
\def\mumu{$\mu^+\mu^-$}
\documentstyle[floats,epsfig,color,pra,aps]{revtex}
\begin{document}
\draft
\title{\hfill BNL-65623\phantom{Laaaaaaaaa}\\
\hfill Fermilab-PUB-98/179\\
\hfill LBNL-41935\phantom{Laaaaaaa}\\
Status of Muon Collider Research 
and Development and Future Plans}
\author{
Charles~M.~Ankenbrandt,$^1$ 
 Muzaffer~Atac,$^1$            
 Bruno~Autin,$^2$             
 Valeri~I.~Balbekov,$^1$
 Vernon~D.~Barger,$^3$           
 Odette~Benary,$^4$
 J.~Scott~Berg,$^5$
 Michael~S.~Berger,$^5$
 Edgar~L.~Black,$^6$
 Alain~Blondel,$^7$
 S.~Alex~Bogacz,$^8$
 T.~Bolton,$^9$            
 Shlomo~Caspi,$^{10}$            
 Christine~Celata,$^{10}$    
 Weiren~Chou,$^1$                          
 David~B.~Cline,$^{11}$
 John~Corlett,$^{10}$
 Lucien~Cremaldi,$^{12}$                        
 H.~Thomas~Diehl,$^1$          
 Alexandr~Drozhdin,$^1$
 Richard~C.~Fernow,$^{13}$
 David~A.~Finley,$^1$
 Yasuo~Fukui,$^{14}$              
 Miguel~A.~Furman,$^{10}$
 Tony~Gabriel,$^{15}$          
 Juan~C.~Gallardo,$^{13}$\thanks{Corresponding author. 
                                 Email:gallardo@bnl.gov}
 Alper~A.~Garren,$^{11}$
 Stephen~H.~Geer,$^1$
 Ilya~F.~Ginzburg,$^{16}$             
 Michael~A.~Green,$^{10}$
 Hulya~Guler,$^{17}$         
 John~F.~Gunion,$^{18}$             
 Ramesh~Gupta,$^{10}$
 Tao~Han,$^3$
 Gail~G.~Hanson,$^5$
 Ahmed~Hassanein,$^{19}$  
 Norbert~Holtkamp,$^1$           
 Colin~Johnson,$^2$             
 Carol~Johnstone,$^1$           
 Stephen~A.~Kahn,$^{13}$
 Daniel~M.~Kaplan,$^6$ 
 Eun~San~Kim,$^{10}$            
 Bruce~J.~King,$^{13}$           
 Harold~G.~Kirk,$^{13}$             
 Yoshitaka~Kuno,$^{14}$             
 Paul~Lebrun,$^1$       
 Kevin~Lee,$^{11}$
 Peter~Lee,$^{10}$             
 Derun~Li,$^{10}$         
 David~Lissauer,$^{13}$
 Laurence~S.~Littenberg,$^{13}$             
 Changguo~Lu,$^{17}$
 Alfredo~Luccio,$^{13}$  
 Joseph~D.~Lykken,$^1$ 
 Kirk~T.~McDonald,$^{17}$             
 Alfred~D.~McInturff,$^{10}$
 John~R.~Miller,$^{20}$           
 Frederick~E.~Mills,$^1$          
 Nikolai~V.~Mokhov,$^1$            
 Alfred~Moretti,$^1$
 Yoshiharu~Mori,$^{14}$             
 David~V.~Neuffer,$^1$             
 King-Yuen~Ng,$^1$         
 Robert~J.~Noble,$^1$         
 James~H.~Norem,$^{19,1}$                       
 Yasar~Onel,$^{21}$             
 Robert~B.~Palmer,$^{13}$ 
 Zohreh~Parsa,$^{13}$       
 Yuriy~Pischalnikov,$^{11}$              
 Milorad~Popovic,$^1$             
 Eric~J.~Prebys,$^{17}$             
 Zubao~Qian,$^1$
 Rajendran~Raja,$^1$
 Claude~B.~Reed,$^{19}$             
 Pavel~Rehak,$^{13}$            
 Thomas~Roser,$^{13}$           
 Robert~Rossmanith,$^{22}$
 Ronald~M.~Scanlan,$^{10}$                        
 Andrew~M.~Sessler,$^{10}$ 
 Brad~Schadwick,$^{10}$             
 Quan-Sheng~Shu,$^8$         
 Gregory~I.~Silvestrov,$^{23}$             
 Alexandr~N.~Skrinsky,$^{23}$
 Dale~Smith,$^{19}$
 Panagiotis~Spentzouris,$^1$  
 Ray~Stefanski,$^1$     
 Sergei~Striganov,$^1$             
 Iuliu~Stumer,$^{13}$          
 Don~Summers,$^{12}$             
 Valeri~Tcherniatine,$^{13}$
 Lee~C.~Teng,$^{19}$   
 Alvin~V.~Tollestrup,$^1$             
 Ya\u{g}mur~Torun,$^{13,24}$             
 Dejan~Trbojevic,$^{13}$            
 William~C.~Turner,$^{10}$
 Sven~E.~Vahsen,$^{17}$            
 Andy~Van~Ginneken,$^1$            
 Tatiana~A.~Vsevolozhskaya,$^{23}$            
 Weishi~Wan,$^1$
 Haipeng~Wang,$^{13}$             
 Robert~Weggel,$^{13}$             
 Erich~H.~Willen,$^{13}$          
 Edmund~J.~N.~Wilson,$^2$             
 David~R.~Winn,$^{25}$           
 Jonathan~S.~Wurtele,$^{26}$
 Takeichiro~Yokoi,$^{14}$             
 Yongxiang~Zhao,$^{13}$        
 Max~Zolotorev$^{10}$   % no comma for last author
}   
\address{
$^1$Fermi National Laboratory, P. O. Box 500, Batavia, IL 60510\protect\\
%$^2$Rockefeller University, New York, NY 10021\protect\\
$^2$CERN, 1211 Geneva 23, Switzerland\protect\\
$^3$Department of Physics, University of Wisconsin, Madison, WI 53706\protect\\
$^4$Tel-Aviv University, Ramat-Aviv, Tel-Aviv 69978, Israel\protect\\
$^5$Physics Department, Indiana University, Bloomington, IN 47405\protect\\
$^6$Illinois Institute of Technology, Physics Div., Chicago IL 60616\protect\\
$^7$\'{E}cole Polytechnique, Lab. de Physique Nucl\'{e}aire et de Hautes Energies, Palaiseau, F-91128, France\protect\\
$^8$Jefferson Laboratory, 12000 Jefferson Ave.,
Newport News, VA 23606\protect\\
$^9$Kansas State University, Manhattan, KS 66502-2601\protect\\
$^{10}$Lawrence Berkeley National Laboratory,
1 Cyclotron Rd., Berkeley, CA 94720\protect\\
$^{11}$University of California Los Angeles, Los Angeles, CA 90095\protect\\
$^{12}$University of Mississippi, Oxford, MS 38677\protect\\
$^{13}$Brookhaven National Laboratory,
Upton, NY 11973\protect\\
$^{14}$KEK High Energy Accelerator Research Organization,
1-1 Oho, Tsukuba 305, Japan\protect\\
$^{15}$Oak Ridge National Laboratory, Oak Ridge, TN 37831\protect\\
$^{16}$Institute of Mathematics, Prosp.\ ac. Koptyug 4, 630090 Novosibirsk, 
Russia\protect\\
$^{17}$Joseph Henry Laboratories, Princeton University, Princeton, NJ 
08544\protect\\
$^{18}$Physics Department, University of California,
Davis, CA 95616\protect\\
$^{19}$Argonne National Laboratory, 
Argonne, IL 60439\protect\\
$^{20}$Magnet Science \& Technology, National High Magnetic Field Laboratory, FL 32310\protect\\
$^{21}$Physics Department, Van Allen Hall,
University of Iowa, Iowa City, IA 52242\protect\\
$^{22}$Research Center Karlsruhe, D-76021 Karlsruhe, Germany\protect\\
$^{23}$ Budker Institute of Nuclear Physics, 
630090 Novosibirsk, Russia\protect\\
$^{24}$Department of Physics and Astronomy, SUNY, Stony Brook, NY 11790\protect\\
$^{25}$Fairfield University, Fairfield, CT 06430\protect\\
$^{26}$University of California Berkeley,
Berkeley, CA 94720
}

\date{\today}
\maketitle
\begin{abstract}

  The status of the research on muon colliders is discussed and plans
are outlined for future theoretical and experimental studies.  Besides
continued work on the parameters of a 3-4 and 0.5~TeV center-of-mass (CoM)
energy
collider, many studies are now concentrating on a machine near 0.1~TeV
(CoM) that could be a factory for the $s$-channel production of Higgs
particles.  We discuss the research on the various components in such
muon colliders, starting from the proton accelerator needed to
generate pions from a heavy-$Z$ target and proceeding through the 
phase rotation and decay 
($\pi \rightarrow \mu\,\nu_{\mu} $) channel, muon cooling,
acceleration, storage in a collider ring and the collider
detector. We also present theoretical and experimental R \& D
plans for the next several years that should lead to a better 
understanding of the design and feasibility issues for all of the
components. This report is an update of the progress on the R \& D
since the Feasibility Study of Muon Colliders presented at the
Snowmass'96 Workshop  [R.~B.~Palmer, A.~Sessler and A.~Tollestrup, 
{\em Proceedings of the 1996 DPF/DPB Summer Study on High-Energy Physics} 
(Stanford Linear Accelerator Center, Menlo Park, CA, 1997)].

\end{abstract}
%\end{frontmatter}

\tableofcontents
\listoffigures
\listoftables
\section{INTRODUCTION}
The Standard Model of electroweak and strong interactions
has passed precision experimental tests at the highest 
energy scale accessible today.
Theoretical arguments indicate that new physics 
\textit{beyond  the Standard Model} associated with the 
electroweak gauge symmetry breaking and fermion mass
generation will emerge in 
% particle-antiparticle
% quark-antiquark and lepton-antilepton    glue-glue dominates in p-pbar!
parton collisions at or
approaching the TeV energy scale. It is likely that both hadron-hadron  and lepton-antilepton 
colliders will be required to discover and make precision measurements of the
new phenomena.  The next big step  forward in advancing the hadron-hadron
collider energy frontier will be  provided by the CERN Large Hadron Collider
(LHC), a proton-proton collider with a center-of-mass (CoM) energy of 14~TeV 
 which is due to come into operation in 2005. Note that in a high energy hadron beam, valence quarks carry momenta which are, approximately, between ${1\over 6}$ to ${1\over 9}$ of the hadron momentum. The LHC will therefore provide hard parton-parton collisions with typical center of mass energies of $2.3 - 1.5 $~TeV.

The route towards TeV-scale lepton-antilepton colliders is less clear. The lepton-antilepton colliders built so far have been $e^+e^-$ colliders, such as the
Large Electron Positron collider (LEP) at CERN and the Stanford Linear  Collider (SLC) at SLAC. In a circular ring such as LEP the  energy lost per revolution in keV is
$88.5\times E^4 / \rho,$ where the electron energy $E$ is in GeV, and the 
radius of
the orbit $\rho$ is in meters. Hence, the energy loss grows rapidly as $E$
increases. This limits the center-of-mass energy that  would be achievable in a
LEP-like collider. The problem can be  avoided by building a linear machine
(the SLC is partially linear), but with current technologies, such a        
machine must be very long (30-40~km) to attain the TeV energy scale.
Even so, radiation during the beam-beam interaction (beamstrahlung) limits
the precision of the CoM energy \cite{Tigner92}.

\begin{figure*}[bth!]
\centerline{\epsfig{file=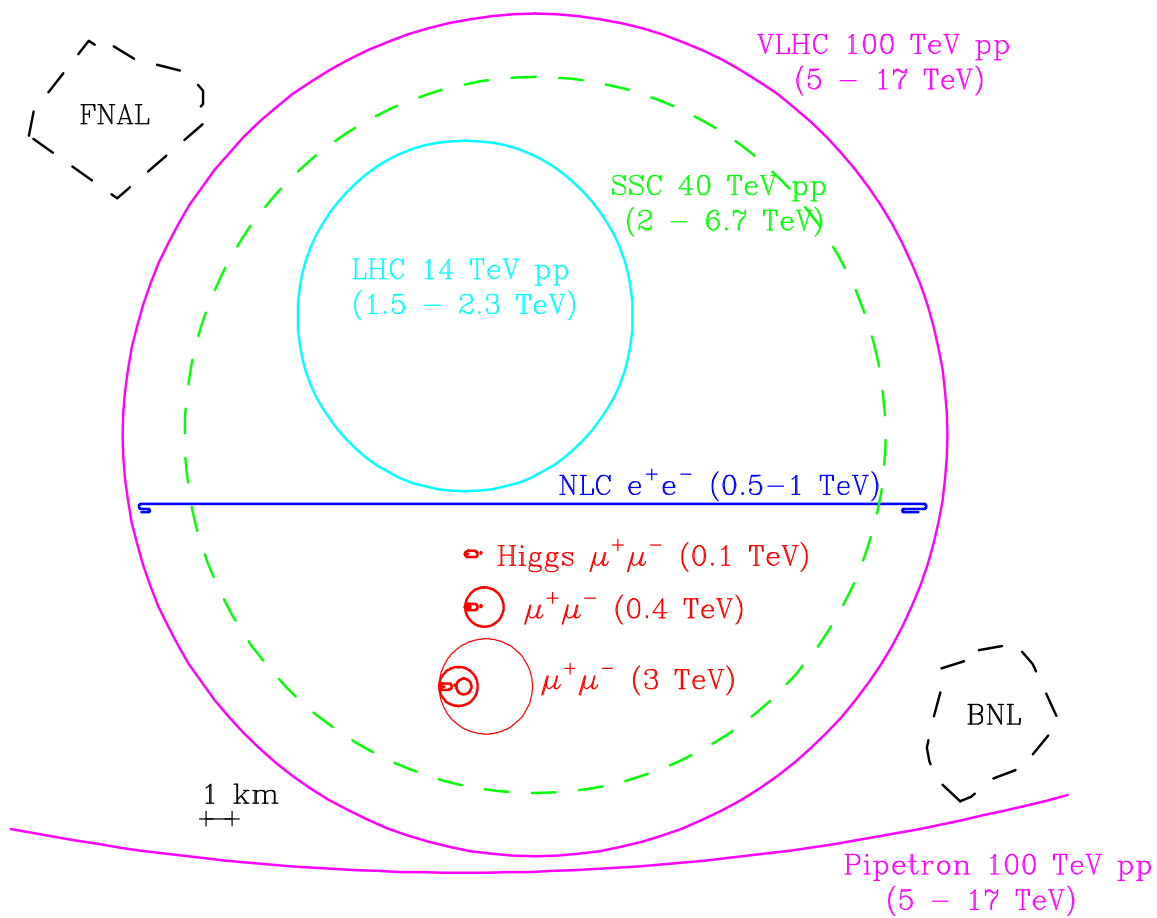,height=4.75in,width=6.in}}
\vspace{0.5cm}
\caption[Comparative sizes of various proposed high energy colliders compared with the FNAL and BNL sites.]
{Comparative sizes of various proposed high energy colliders compared with the FNAL and BNL sites. The energies in parentheses give for lepton colliders their CoM energies and for hadron colliders the approximate range of CoM energies attainable for hard parton-parton collisions.}
\label{compare}
\end{figure*}

For a lepton with mass $m$ the radiative energy losses 
are inversely proportional to $m^4$. Hence, the energy-loss  problem can be
solved by using heavy leptons. In practice this means using  muons, which have
a mass $\approx 207$ times that of an electron. The resulting % significant 
reduction in
radiative losses enables higher energies to be reached and  smaller collider
rings to be used \cite{ref01,ref01a}. Parameters for 10 to 100~TeV collider have been discussed\cite{skrinskiultimate,king_highe}. 
Estimated sizes of the accelerator  complexes required for
0.1-TeV, 0.5-TeV and 4-TeV muon colliders are compared with  the sizes of other
possible future colliders,  and with the FNAL and BNL sites in 
Fig.~\ref{compare}. Note that muon colliders with CoM energies up to $\approx 4$~TeV would fit on these existing
laboratory sites. The cost  of building a muon collider is not yet known. However, since muon colliders are relatively  small, they may be significantly less expensive than alternative machines. 

Since muons decay quickly, large numbers of them must be produced to operate
a muon collider at high luminosity.  Collection of muons from the decay of
pions produced in proton-nucleus interactions results in a large initial
phase volume for the muons, which must be reduced (cooled) by a factor of $10^6$ for a practical collider. This may be compared with the antiproton stochastic cooling achieved in the Tevatron. In this case the 6-dimensional (6-D) phase space is reduced by approximately a factor of $10^6,$ while with stacking the phase space density \cite{tevcooling,reviewtev} is increased by a factor of $10^{10}.$  The technique of ionization cooling
is proposed for the $\mu^+\mu^-$ collider \cite{Kolomensky,Budker67,Ado,Balbekov96}. This technique is uniquely applicable to muons because of
their minimal interaction with matter.

Muon  colliders also offer some significant physics advantages. The small
radiative losses  permit very small beam-energy spreads to be achieved.  For
example, momentum spreads as low as $\Delta P/P = 0.003\%$ are believed  to be
possible for a low-energy collider. By measuring the time-dependent decay
asymmetry resulting from the naturally polarized  muons, it has been
shown \cite{ref7} that the beam energy could be determined with a precision of 
$\Delta E/E = 10^{-6}$. The small beam-energy spread, together with the precise
energy determination, would facilitate measurements of the masses
and widths of any new resonant states scanned by the collider. In addition,
since the cross-section for producing a Higgs-like scalar particle in the
s-channel (direct lepton-antilepton annihilation) is proportional to $m^2$,
this extremely important process could be studied only at a muon collider and not at an $e^+e^-$ collider \cite{ref2b}. Finally, 
the decaying muons will produce copious quantities of neutrinos.  Even 
short straight sections in a muon-collider ring will result in neutrino
beams several orders of magnitude higher in intensity than presently
available, permitting greatly extended studies of neutrino
oscillations,  nucleon structure functions, the CKM matrix, and
 precise indirect measurements of the $W$-boson mass\cite{stevealw} (see section~\textbf{II.I}).

The concept of muon colliders was introduced by 
G.~I.~Budker \cite{ref01,ref01a}, and developed further by A.~N.~Skrinsky \etal \cite{Skrinsky71,Budker78,Skrinsky80,ref1a,Skrinsky82,ref1,Skrinsky96a,%
Skrinsky96b} and D.~Neuffer \cite{ref2b,ref2,ref2a,ref2c}. 
They pointed out the significant challenges in
designing an accelerator complex that can  make, accelerate, and collide
$\mu^+$ and $\mu^-$ bunches all within the muon lifetime of $2.2\,\mu$s 
($c\tau=659$~m). A concerted study of a muon collider design has been underway in the U.S. since 1992 \cite{PortJeff,Napa,ref3a,Fontana,ref3b,Tamura94,ref3c,Snowmass96,SantaB96,LBL97,Vancouver97,BNLlattice97,FNALcool97,mup,sanfrisco,Alabama98,BNLcool98}.
By the Sausalito
workshop \cite{ref3b} in 1995 it was  realized that with new ideas and modern
technology, it may be feasible to make muon bunches containing a few  times
$10^{12}$ muons, compress their phase space and  accelerate them 
%the resulting still intense bunches 
up to the multi-TeV energy scale before more than about 3/4 of them have
decayed. With careful design of the collider ring and  shielding it  appears possible to
reduce to acceptable levels the backgrounds  within the detector that arise
from the very large flux of electrons produced  in muon decays. These
realizations led to  an intense activity,  which resulted in the muon-collider 
feasibility study report \cite{status96,ref6a} prepared for the 
1996 DPF/DPB Summer Study on High-Energy Physics (the Snowmass'96  
workshop).   Since then, the physics 
prospects at a muon collider have been studied extensively \cite{Barger,gunion,ref100}, and the potential physics program at a muon 
collider facility has been explored in workshops \cite{mup} and 
conferences \cite{sanfrisco}.
%Studies \cite{ref4a,ref4,ref4b,ref5,ref6,ref0,ref0a,ref0b}  have been made for 
%muon colliders at 3-4~TeV, 0.4-0.5~TeV and $\approx 100$~GeV.

 Encouraged by further progress in developing the 
muon-collider concept, together with the growing interest and involvement of 
the high-energy-physics community, the {\it Muon Collider Collaboration} 
became a formal
entity in May of 1997. The collaboration is led
by an executive board with members from Brookhaven National Laboratory (BNL), 
Fermi National Accelerator Laboratory (FNAL), 
Lawrence Berkeley National Laboratory (LBNL), 
Budker Institute for Nuclear Physics (BINP), 
University of California at Los Angeles (UCLA),
University of Mississippi and Princeton University.
The goal of the collaboration is to complete  within a
few years the R\&D needed to determine whether a Muon Collider is technically
feasible, and if it is, to design the First Muon Collider.

\begin{table*}[thb!]
\centering  
\caption[Baseline parameters for high- and low-energy muon colliders. ]
{Baseline parameters for high- and low-energy muon colliders.
Higgs/year assumes a cross section $\sigma=5\times 10^4$~fb; a Higgs width 
$\Gamma=2.7$~MeV; 1~year = $10^7$~s.}
\label{sum}
\begin{tabular}{llccccc}
%\hline
\rr CoM energy         &\rr TeV   &\rr 3 &\rr 0.4 &
\multicolumn{3}{c}{0.1 }  \\
% & & & & & & \\
$p$ energy       & GeV        &  16  & 16 & \multicolumn{3}{c}{16}\\
$p$'s/bunch      &    &  $2.5\times 10^{13}$  & $2.5\times 10^{13}$  &
\multicolumn{3}{c}{$5\times 10^{13}$  }  \\  
Bunches/fill   &           & 4 & 4 & \multicolumn{3}{c}{2 }  \\
Rep.~rate  & Hz     &  15 & 15 & \multicolumn{3}{c}{15 }  \\
$p$ power        & MW         &  4   & 4 & \multicolumn{3}{c}{4}  \\ 
$\mu$/bunch  &    & $2\times 10^{12}$ & $2\times 10^{12}$ &
\multicolumn{3}{c}{$4\times 10^{12}$ }  \\
\rr $\mu$ power  &\rr MW     & \rr 28 &\rr 4 & \multicolumn{3}{c}{\rr 1 }  \\
\rr Wall power    &\rr MW    &  \rr  204 &\rr 120  & \multicolumn{3}{c}{\rr
81 }  \\
Collider circum.   & m          &  6000 & 1000 & \multicolumn{3}{c}{350 }  \\
Ave bending field & T       & 5.2 & 4.7 &\multicolumn{3}{c}{3 }  \\
%Depth   & m          &  500 & 100 & \multicolumn{3}{c}{10 }  \\
%\hline
\rr Rms ${\Delta p/p}$       &\rr  \%          &\rr 0.16 &\rr 0.14 &\rr
0.12 &\rr 0.01&\rr
0.003 \\
%\hline
6-D $\epsilon_{6,N}$    &  $(\pi \textrm{m})^3$&$1.7\times 10^{-10}$&$1.7\times
10^{-10}$&$1.7\times 10^{-10}$&$1.7\times 10^{-10}$&$1.7\times 10^{-10}$\\
Rms $\epsilon_n$     &$\pi$ mm-mrad     &  50 & 50 & 85 & 195 & 290\\
$\beta^*$         & cm          & 0.3 & 2.6 & 4.1 &  9.4 & 14.1\\
$\sigma_z$         & cm          & 0.3 & 2.6 & 4.1 &  9.4 & 14.1 \\
$\sigma_r$ spot    &$\mu$m     & 3.2 & 26 & 86 & 196 & 294\\
$\sigma_{\theta}$ IP    &mrad     & 1.1 & 1.0 & 2.1 & 2.1 & 2.1\\
Tune shift     &             &0.044 &0.044 & 0.051 &0.022 & 0.015\\
$n_{\rm turns}$ (effective) &     &  785 & 700 & 450 & 450 & 450 \\
%\hline
\rr Luminosity     &\rr cm$^{-2}$s$^{-1}$&\rr $7\times 10^{34}$ & $10^{33}$ &\rr
$1.2\times 10^{32}$ &\rr $2.2\times 10^{31}$&\rr $10^{31}$ \\
 & & & & & & \\
Higgs/year    &  &  & & $1.9\times 10^3$ & $4\times 10^3$ & $3.9\times 10^3$ \\
%\hline
\end{tabular}
\end{table*}

\begin{figure*}[tbh!]
\centerline{\epsfig{file=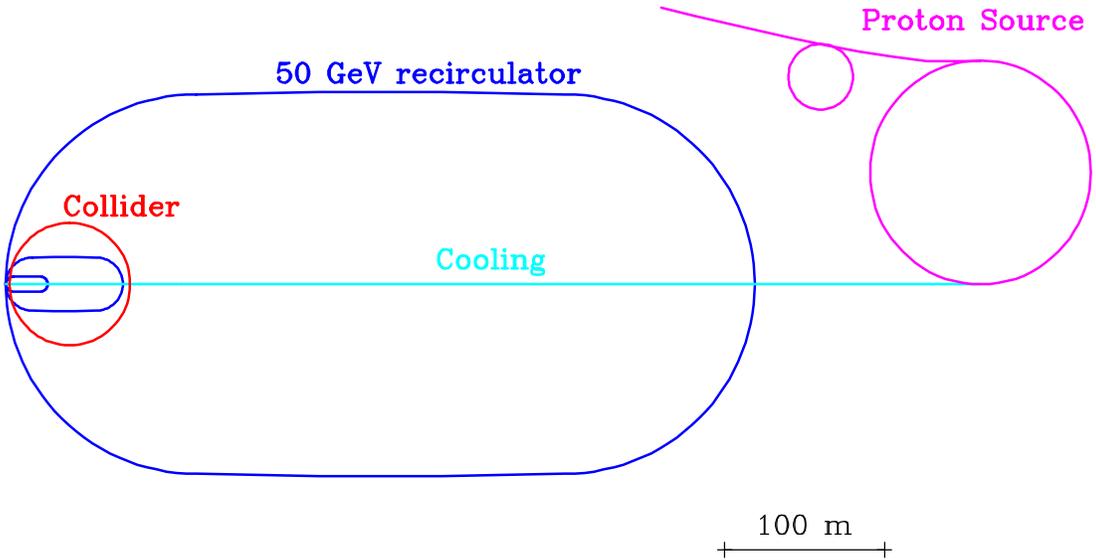,height=2.9in,width=5.7in}}
\caption{Plan of a 0.1-TeV-CoM muon collider.}
\label{plan1}
\end{figure*}
 
\begin{figure*}[bth!]
\centerline{\epsfig{file=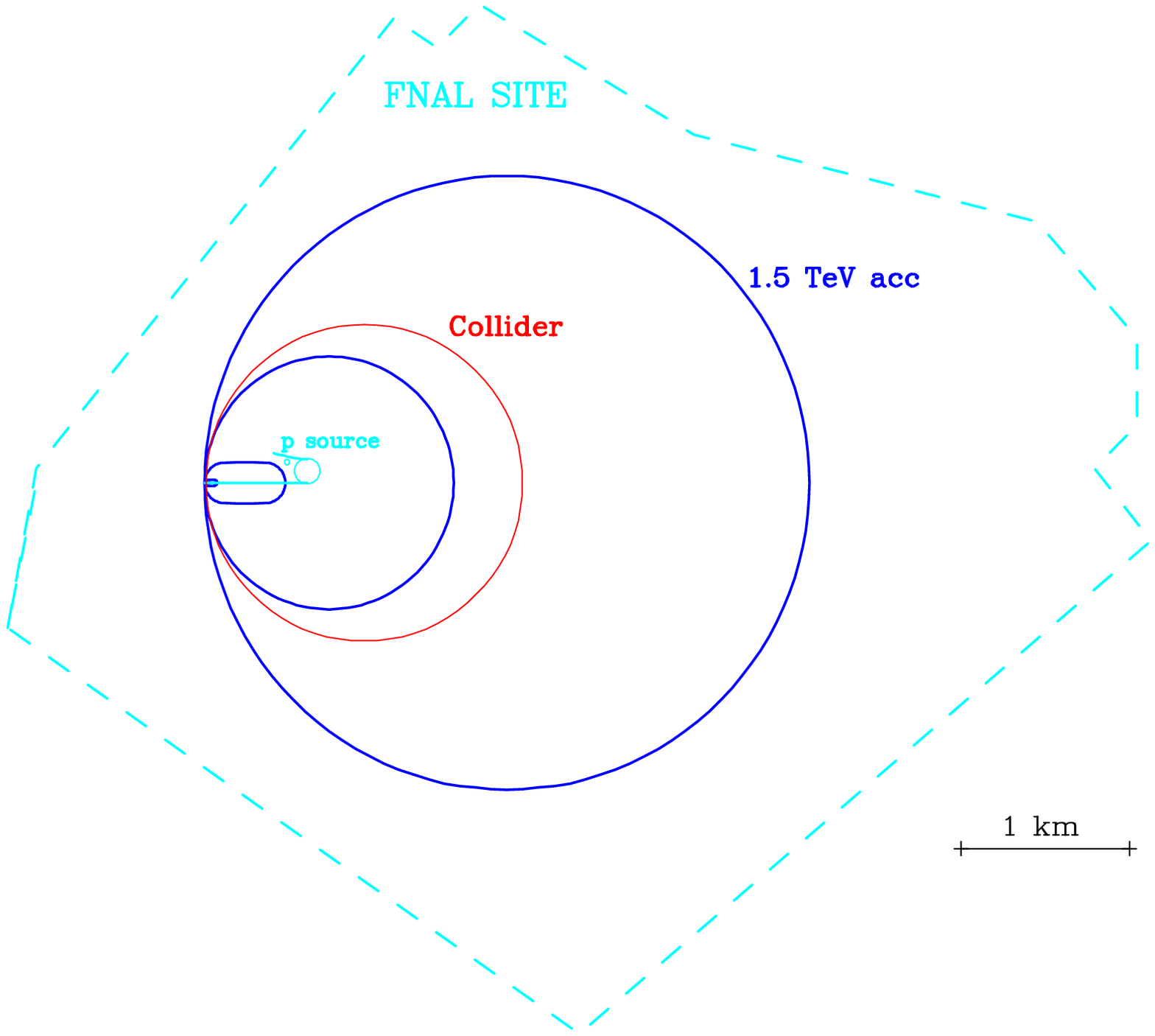,height=5.0in,width=5.45in}}
\caption{Plan of a 3-TeV-CoM muon collider shown on the Fermi National
Laboratory site as an example.}
\label{plan2}
\end{figure*}

Table~\ref{sum} gives the parameters of the muon colliders under study \cite{ref4a,ref4,ref4b,ref5,ref6,ref0,ref0a,ref0b}, which have CoM energies of
0.1~TeV, 0.4~TeV and 3~TeV and
Figs.~\ref{plan1} and \ref{plan2} show possible outlines of the  0.1~TeV and 
3~TeV  machines. In the former case, parameters are given in the table  for
operation with three  different beam-energy spreads: ${\Delta p/p} = 0.12$,
0.01, and 0.003\%.  In all cases, proton bunches containing 2.5-$5\times
10^{13}$  particles are accelerated to energies of 16~GeV. The protons interact
in a target to  produce  ${\cal O}(10^{13})$ charged pions of each sign.   A
large fraction of these pions can be captured in a high-field solenoid.  Muons
are produced by allowing the pions to decay into a lower-field solenoidal
channel. To  collect as many particles as possible within a useful energy
interval, rf  cavities are used to accelerate the lower-energy particles and
decelerate the higher-energy particles (so-called phase rotation).  With two
proton bunches every  accelerator cycle, the first used to make and collect
positive muons and the  second to make and collect negative muons, there are
about $10^{13}$ muons of  each charge available at the end of the decay channel
per accelerator cycle.  If the proton accelerator is cycling at 15~Hz, then in
an operational year  ($10^7$~s), about $10^{21}$ positive and negative muons
would be produced and collected.

As stated before, the muons exiting the decay channel populate a very diffuse phase space. The
next step in the muon-collider complex is to \textit{cool} the muon bunch, \ie,
to turn the diffuse muon cloud into a very \textit{bright} bunch with small  
longitudinal
and transverse dimensions, suitable for accelerating and  injecting into a
collider. The cooling must be done within a time that is short compared to
the muon lifetime. Conventional cooling techniques  (stochastic cooling \cite{Ruggiero92} and electron cooling \cite{Budker78}) take too long. 
The technique proposed for cooling muons is called ionization cooling \cite{Kolomensky,Ado,Balbekov96}, and 
will be discussed in detail in sec.~V.  Briefly, the muons traverse 
some material in which they  lose
both longitudinal and transverse momentum by ionization losses ($dE/dx$).  The
longitudinal momentum is then replaced using an rf accelerating cavity,  and
the process is repeated many times until there is a large reduction in  the
transverse phase space occupied by the muons. The energy spread within  the
muon beam can also be reduced by using a wedge-shaped absorber in a region of 
dispersion (where the transverse position is momentum dependent). 
The wedge is arranged so that the higher-energy particles
pass through more material than lower-energy particles.  Initial calculations
suggest that the 6-D phase space occupied by the initial muon
bunches can be reduced by a factor of $10^5$-$10^6$ before multiple Coulomb
scattering and energy straggling limit further reduction. We reiterate that ionization
cooling is uniquely suited to muons because of the absence of strong nuclear
interactions and electromagnetic shower production for these particles at energies around 200~MeV/$c.$

Rapid acceleration to the collider beam energy is needed to avoid % further
excessive particle loss from decay. It can be achieved, initially in a linear
accelerator, and later in  recirculating linear accelerators, rapid-cycling
synchrotron, or fixed-field-alternating-gradient (FFAG) accelerators. Positive 
and negative muon bunches are then injected in opposite directions into a 
collider storage ring and brought into collision at the interaction point.  The
bunches circulate and collide for many revolutions before decay has  depleted
the beam intensities to an uninteresting level. Useful luminosity can be delivered for about 800 revolutions for the high-energy collider and 450 revolutions for the low-energy one.

There are many interesting and challenging problems that need to be  resolved
before the feasibility of building a muon collider can be  demonstrated. For
example, (i) heating from the very intense proton bunches may  require the
use of  of a liquid-jet target,  and (ii) attaining the desired cooling factor
in the ionization-cooling channel may require the development of
rf cavities with thin beryllium windows operating at liquid-nitrogen
temperatures in high solenoidal fields. In addition, the development of long
liquid-lithium lenses  may be desirable to provide  stronger radial focusing 
for the final cooling stages. 

This article describes  the status of our muon-collider feasibility
studies, and is organized as follows. Section~\textbf{II} gives a brief summary of the
physics potential of muon colliders, including physics at the accelerator
complex required for a muon collider. Section~\textbf{III} describes the proton-driver
specifications for a muon collider, and two site-dependent
examples that have been studied in some detail. 
Section~\textbf{IV} presents pion production,
capture, and the pion-decay channel, and section~\textbf{V} discusses the design of the
ionization-cooling channel needed to produce an intense muon beam suitable
for acceleration and injection into the final collider. Sections~\textbf{VI} and \textbf{VII} 
describe the acceleration scenario and collider ring, respectively. 
Section~\textbf{VIII}
discusses backgrounds at the collider interaction point and section~\textbf{IX} deals with possible detector scenarios.  A summary of the conclusions is given in section~\textbf{X}.
\section{THE PHYSICS POTENTIAL OF MUON COLLIDERS}
\subsection{Brief overview}
The physics agenda at a muon collider falls into three categories: 
First Muon Collider (FMC) 
physics at a machine with center-of-mass energies of 100 to 500~GeV;
Next Muon Collider (NMC) physics at 3--4~TeV 
center-of-mass energies; and
front-end physics with a high-intensity muon source.

\begin{figure*}[tbh!]
\centering\leavevmode
\epsfxsize=5in\epsffile{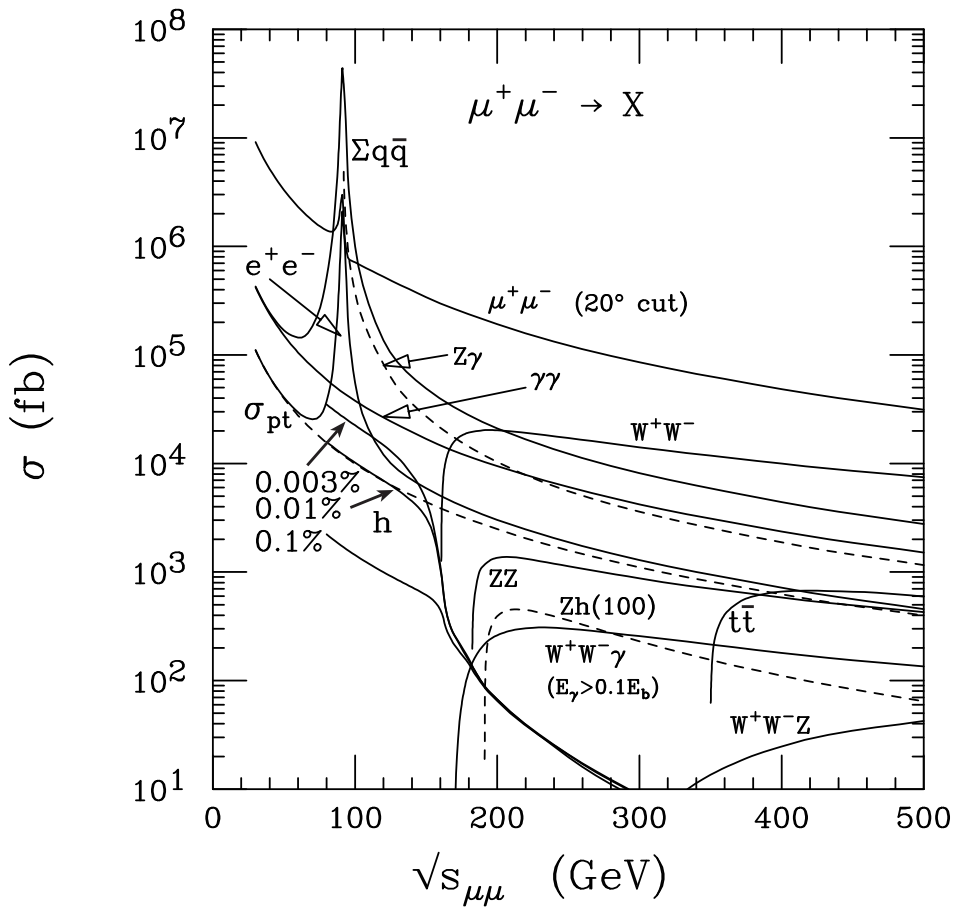}
\caption[Cross sections for SM processes versus the 
CoM energy at the FMC]{Cross sections for SM processes versus the 
CoM energy at the FMC. 
$\sigma_{pt}\equiv \sigma(\mu^+\mu^-\to \gamma^*\to e^+e^-)$.
For the $s$-channel Higgs boson production, 
three different beam energy
resolutions of 0.003\%, 0.01\% and 0.1\% are presented.
}
\label{fmc_han}
\end{figure*}

The FMC will be a unique facility for neutral Higgs boson (or  
techni-resonance) studies through $s$-channel resonance production.  
Measurements can also be made of the threshold cross sections for 
production of $W^+W^-$,  
$t\bar t$, $Zh$, and pairs of supersymmetry particles ---
$\chi_1^+\chi_1^-$, $\chi_2^0\chi_1^0$,  
$\tilde\ell^+\tilde\ell^-$ and $\tilde\nu\bar{\tilde\nu}$ --- that will  
determine the corresponding masses to high precision.
A $\mu^+\mu^-\to Z^0$ factory, utilizing the partial  
polarization of the muons, could allow significant improvements in  
$\sin^2\theta_{\rm w}$ precision and in $B$-mixing and CP-violating studies.
In Fig.~\ref{fmc_han}, we show the cross sections 
for SM processes versus the CoM energy at the FMC. For the 
unique $s$-channel Higgs boson production, where $\sqrt s_{\mu\mu} =m_H$,
results  for three different beam energy resolutions are presented.

The NMC will be particularly valuable for reconstructing supersymmetric  
particles of high mass from 
their complex cascade decay chains. Also, any $Z'$  
resonances within the kinematic reach of the machine would give enormous  
event rates. The effects of virtual $Z'$ states would be detectable to high  
mass. If no Higgs bosons exist below $\sim$1~TeV, then the NMC would be the  
ideal machine for the study of strong $WW$ scattering at TeV energies.

At the front end, a high-intensity muon source will permit searches for rare  
muon processes sensitive to branching ratios 
that are orders of magnitude below  
present upper limits. Also, a high-energy muon-proton collider can be  
constructed to probe high$-Q^2$ phenomena beyond the reach of the HERA $ep$  
collider. In addition, 
the decaying muons will provide high-intensity neutrino  
beams for precision neutrino cross-section measurements and for long-baseline  
experiments \cite{sgeer,sgeerjhf,sgeerphyrevd,moha1,harris1,bjknu1,bjknu2,ref8b,yu1,cline80}.
Plus, there are numerous other new physics possibilities 
for muon facilities \cite{ref6a,mup} that we will not
discuss in detail in this document.

\subsection{Higgs boson physics}

The expectation that there will be a light (mass below $2M_W$) 
SM-like Higgs boson
provides a major motivation for the FMC, since such a Higgs boson
can be produced with a very high rate directly in the $s$-channel.
Theoretically, the lightest Higgs boson $h^0$ of the most
general supersymmetric model is predicted to have mass below $150$~GeV
and to be very SM-like in the usual decoupling limit. Indeed,
in the minimal supersymmetric model, which contains
the five Higgs bosons $h^0,\, H^0,\, A^0,\, H^\pm$, one finds
$m_{h^0}\alt 130$~GeV and the $h^0$ is SM-like if $m_{A^0}\agt 130$~GeV. 
The light SUSY $h^0$ is regarded as  
the jewel in the SUSY crown. Experimentally, 
global analyses of precision electroweak data now indicate a strong 
preference for a light SM-like Higgs boson; this could well be
the smoking gun for the SUSY Higgs boson.
The goals of the FMC for studying the SUSY
Higgs sector via $s$-channel resonance production are: to measure
the light Higgs mass, width, and branching fractions with
high precision, in particular sufficient to differentiate the MSSM
$h^0$ from the SM $h_{\rm SM}$; and, to find and study 
the heavier neutral Higgs bosons $H^0$ and $A^0$.

\begin{figure*}[htb!]
\centering\leavevmode  % we may want to uncomment this line^^^^^^^^^^^^^
\epsfxsize=5in\epsffile{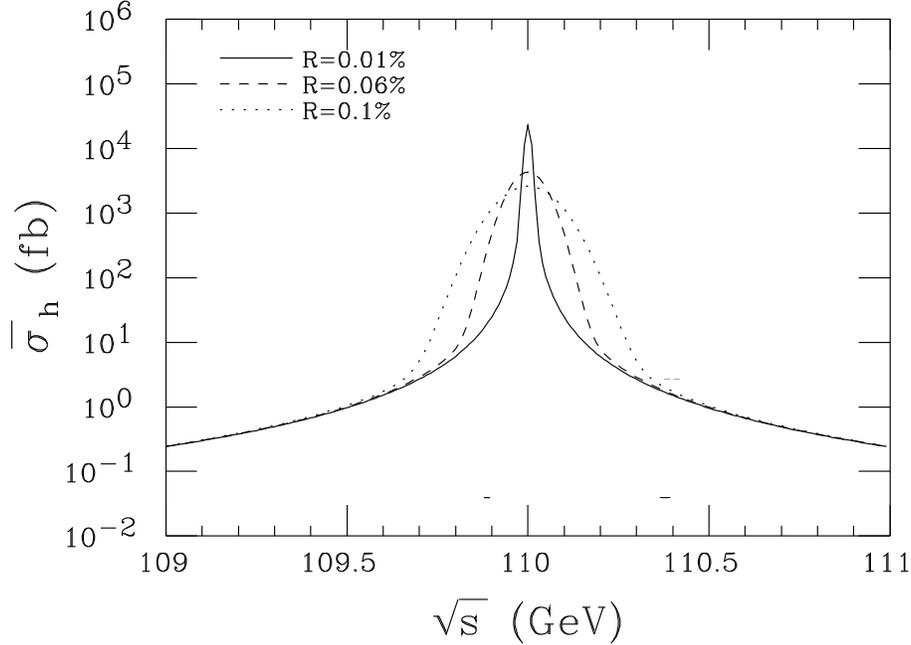}

\caption[Effective $s$-channel Higgs cross section $\bar\sigma_h$]{Effective $s$-channel Higgs cross section $\bar\sigma_h$ obtained  
by convoluting the Breit-Wigner resonance formula with a Gaussian 
distribution  
for resolution $R$. From Ref.~\cite{Barger}.\label{s-chan-higgs}}
\end{figure*}

The production of Higgs bosons in the $s$-channel with interesting rates is  
a unique feature of a muon collider \cite{Barger,bbgh}. The resonance cross  
section is
\begin{equation}
\sigma_h(\sqrt s) = {4\pi \Gamma(h\to\mu\bar\mu) \, \Gamma(h\to X)\over
\left( s - m_h^2\right)^2 + m_h^2 \left(\Gamma_{\rm tot}^h \right)^2}\,.
\end{equation}
Gaussian beams with root-mean-square (rms) 
energy resolution  down to $R=0.003\%$ are  
realizable. The corresponding rms spread $\srts$ in CoM energy is
\begin{equation}
\srts = (2{\rm~MeV}) \left( R\over 0.003\%\right) \left(\sqrt s\over  
100\rm~GeV\right) \,.
\end{equation}
The effective $s$-channel Higgs cross section convolved with a Gaussian spread,
\begin{equation}
\bar\sigma_h(\sqrt s) = {1\over \sqrt{2\pi}\,\srts} \; \int \sigma_h  
(\sqrt{\hat s}) \; \exp\left[ -\left( \sqrt{\hat s} - \sqrt s\right)^2 \over  
2\sigma_{\sqrt s}^2 \right] d \sqrt{\hat s},
\end{equation}
is illustrated in Fig.~\ref{s-chan-higgs} for 
$m_h = 110$~GeV, $\Gamma_h = 2.5$~MeV, and  
resolutions $R=0.01\%$, 0.06\% 
and 0.1\%. 
A resolution $\srts \sim \Gamma_h$ is needed to be 
sensitive to the Higgs width. 
The light Higgs  width is predicted to be
\begin{equation}
\begin{array}{lll}
\Gamma \approx 2\mbox{ to 3 MeV}& \rm if& \tan\beta\sim1.8,\\
\Gamma \approx 2\mbox{ to 800 MeV}& \rm if& \tan\beta\sim20,
\end{array}
\end{equation}
for $80{\rm~GeV}\alt m_h\alt120$~GeV, where the smaller values
apply in the decoupling limit of large $m_{A^0}$. We note
that, in the MSSM, $m_{A^0}$ is required to be in the decoupling regime 
in the context of mSUGRA boundary conditions in order that
correct electroweak symmetry breaking arises after evolution of parameters
from the unification scale. In particular, decoupling applies
in mSUGRA at $\tan\beta\sim1.8$, corresponding to the 
infrared fixed point of the top quark Yukawa coupling.

\begin{figure*}[hbt!]
\centering\leavevmode
\epsfxsize=5.5in\epsffile{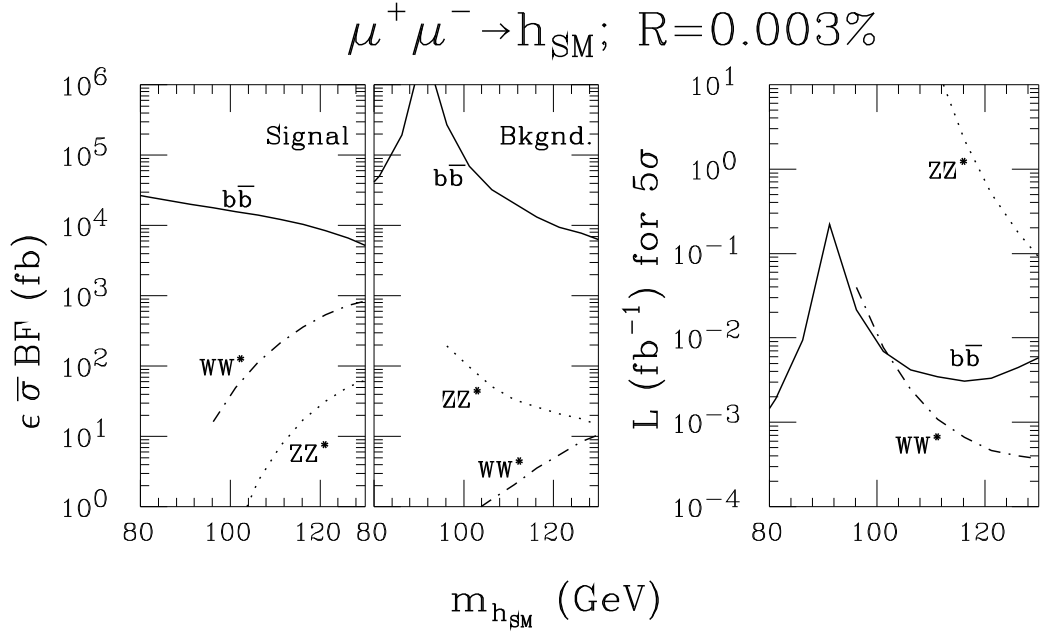}

\caption[The SM Higgs cross sections and backgrounds in $b\bar b,\ WW^*$  
and $ZZ^*$. ]{The SM Higgs cross sections and backgrounds in $b\bar b,\ WW^*$  
and $ZZ^*$. Also shown is the luminosity needed for a 5~standard deviation  
detection in $b\bar b$. From Ref.~\cite{Barger}.\label{sm-higgs}}
\end{figure*}

At $\sqrt s = m_h$, the effective $s$-channel Higgs cross section is
\begin{equation}
\bar\sigma_h \simeq {4\pi\over m_h^2} \; {{\rm BF}(h\to\mu\bar\mu) \,
{\rm BF}(h\to X) \over \left[ 1 + {8\over\pi} \left(\srts\over\Gamma_{\rm  
tot}^h\right)^2 \right]^{1/2}} \,.
\end{equation}
BF denotes the branching fraction for $h$ decay; also,
note that $\bar\sigma_h\propto 1/\srts$ for $\srts>\Gamma_{\rm tot}^h$. At  
$\sqrt s = m_h \approx 110$~GeV, the $b\bar b$ rates are
\begin{eqnarray}
\rm signal &\approx& 10^4\rm\ events\times L(fb^{-1})\,,\\
\rm background &\approx& 10^4\rm\ events\times L(fb^{-1})\,,
\end{eqnarray}
assuming a $b$-tagging efficiency $\epsilon \sim 0.5$. The effective  
on-resonance cross sections for other $m_h$ values and other channels ($ZZ^*,  
WW^*$) are shown in Fig.~\ref{sm-higgs} for the SM Higgs. 
The rates for the MSSM Higgs are nearly the same as the SM rates 
in the decoupling regime of large $m_{A^0}$.

\begin{figure*}[tbh!]
\centering\leavevmode
\epsfxsize=4in\epsffile{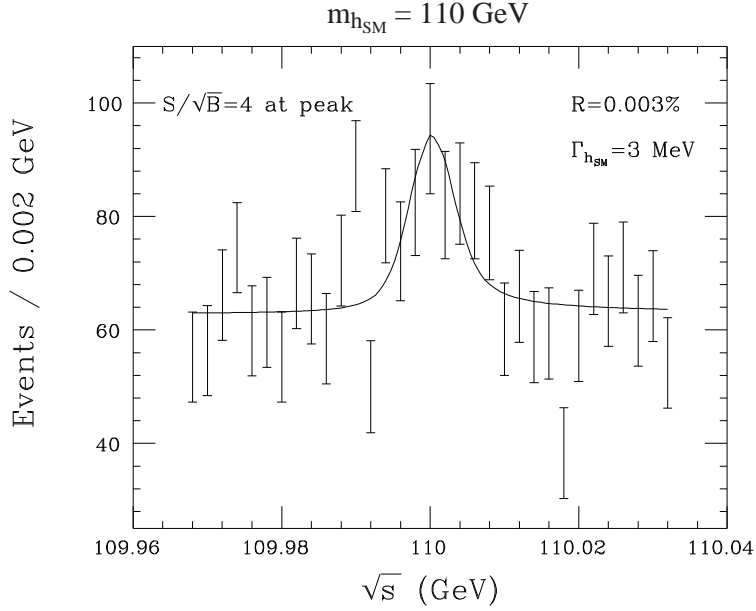}

\caption[Number of events and statistical errors in the $b\bar b$ final  
states ]{Number of events and statistical errors in the $b\bar b$ final  
states as a function of $\sqrt s$ in the vicinity of $m_{h_{\rm SM}}=110$~GeV,  
assuming $R=0.003\%$. From Ref.~\cite{Barger}.\label{bbar-events}}
\end{figure*}

The important factors that make $s$-channel Higgs physics studies possible at  
a muon collider are energy resolutions $\srts$ of order a few MeV, little  
bremsstrahlung and no beamstrahlung smearing, and precise tuning of the beam  
energy to an accuracy $\Delta E\sim10^{-6}E$ through continuous spin-rotation  
measurements \cite{ref7}. As a case study, we consider a SM-like
Higgs boson with $m_h \approx 110$~GeV.  
Prior Higgs discovery is assumed at the Tevatron 
(in $Wh, t\bar th$ production  
with $h\to b\bar b$ decay) or at the LHC (in $gg\to h$ production with $h\to  
\gamma\gamma, 4\ell$ decays with a mass measurement of $\Delta m_h \sim  
100$~MeV for an integrated luminosity of $L=300\rm~fb^{-1}$) or possibly at a  
NLC (in $Z^*\to Zh, h\to b\bar b$ giving $\Delta m_h \sim 50$~MeV for  
$L=200\rm~fb^{-1}$). A muon collider ring design would be optimized to run at  
energy $\sqrt s= m_h$. For an initial Higgs-mass uncertainty of $\Delta  
m_h\sim 100$~MeV, the maximum number of scan points required to locate the  
$s$-channel resonance peak at the muon collider is
\begin{equation}
n = {2\Delta m_h\over \srts} \approx 100
\end{equation}
for a $R=0.003\%$ resolution of $\srts \approx 2$~MeV. 
The necessary luminosity per scan  
point ($L_{\rm s.p.}$) to observe or eliminate the $h$-resonance at a  
significance level of $S/\sqrt B = 3$ is $L_{\rm s.p.} \sim
1.5\times10^{-3}\,\rm   fb^{-1}$. (The scan luminosity requirements increase
for $m_h$ closer to   $M_Z$; at $m_h\sim M_Z$ the $L_{\rm s.p.}$ needed is a
factor of 50 higher.)   The total luminosity then needed to tune to a Higgs
boson with $m_h = 110$~GeV   is $L_{\rm tot} = 0.15\rm~fb^{-1}$. If the machine
delivers    $1.5\times10^{31}\rm\, cm^{-2}\, s^{-1}$ (0.15~fb$^{-1}$/year),
then one year   of running  would suffice to complete the scan and measure the
Higgs mass to   an accuracy  $\Delta m_h \sim 1$~MeV.  Figure~\ref{bbar-events}
illustrates a simulation of such a scan.

Once the $h$-mass is determined to $\sim1$~MeV, a 3-point fine  
scan \cite{Barger} can be made across the peak with higher luminosity,  
distributed with $L_1$ at the observed peak position in $\sqrt s$ and $2.5L_1$  
at the wings ($\sqrt s = {\rm peak} \pm 2\srts$). Then, with $L_{\rm tot}=  
0.4\rm~fb^{-1}$ the following accuracies would be achievable: 16\% for  
$\Gamma_{\rm tot}^h$, 1\% for $\sigma\rm BF(b \bar b)$ and 5\% 
for $\sigma\rm  
BF(WW^*)$. The ratio $r = {\rm BF}(WW^*)/ {\rm BF} (b\bar b)$ is sensitive to  
$m_{A^0}$ for $m_{A^0}$ values below 500~GeV.
For example, $r_{\rm MSSM}/r_{\rm SM} =  
0.3, 0.5, 0.8$ for $m_{A^0} = 200, 250, 400$~GeV \cite{Barger}. Thus, 
using $s$-channel measurements of the $h$, 
it may be possible not only to distinguish the $h^0$ from the
SM $h_{SM}$ but also to infer $m_{A^0}$. 

The study of the other neutral MSSM Higgs bosons at a muon collider via the  
$s$-channel is also of major interest. Finding the $H^0$ and $A^0$ may not be  
easy at other colliders. At the LHC the region $m_{A^0}>200$~GeV is deemed to be  
inaccessible for $3\alt\tan\beta\alt5$--10 \cite{froid}. At an NLC the  
$e^+e^-\to H^0 A^0$ production process may be kinematically inaccessible if  
$H^0$ and $A^0$ are heavy
(mass $>230$~GeV for $\sqrt s=500$~GeV). 
At a $\gamma\gamma$ collider, very high luminosity  
(${\sim}200\rm\ fb^{-1}$) would be needed for $\gamma\gamma\to H^0, A^0$  
studies.

At a muon collider the resolution requirements for $s$-channel $H^0$ and  
$A^0$ studies are not as demanding as for the $h^0$ because the $H^0, A^0$  
widths are broader; typically $\Gamma\sim30$~MeV for $m_{A^0}<2m_t$ and  
$\Gamma\sim3$~GeV for $m_{A^0}>2m_t$. Consequently $R\sim0.1\%$ ($\srts \sim  
70$~MeV) is adequate for a scan. This is important, since higher
instantaneous luminosities (corresponding to $L\sim 2-10~{\rm fb}^{-1}
/{\rm yr}$)
are possible for $R\sim 0.1\%$ (as contrasted with the $L\sim 0.15~{\rm
fb}^{-1}/{\rm yr}$ for the much smaller $R\sim 0.003\%$ preferred for studies
of the $h^0$). A luminosity per scan point $L_{\rm s.p.}\sim  
0.1\rm~fb^{-1}$ probes the parameter space with $\tan\beta>2$.
The $\sqrt s$-range over which the scan should be made depends on other  
information available to indicate 
the $A^0$ and $H^0$ mass range of interest. A wide
scan would not be necessary if $r$ is measured
with the above-described precision 
to obtain an approximate value of $m_{A^0}$.

\begin{figure*}[hbt!]
\centering\leavevmode
\epsfxsize=4in\epsffile{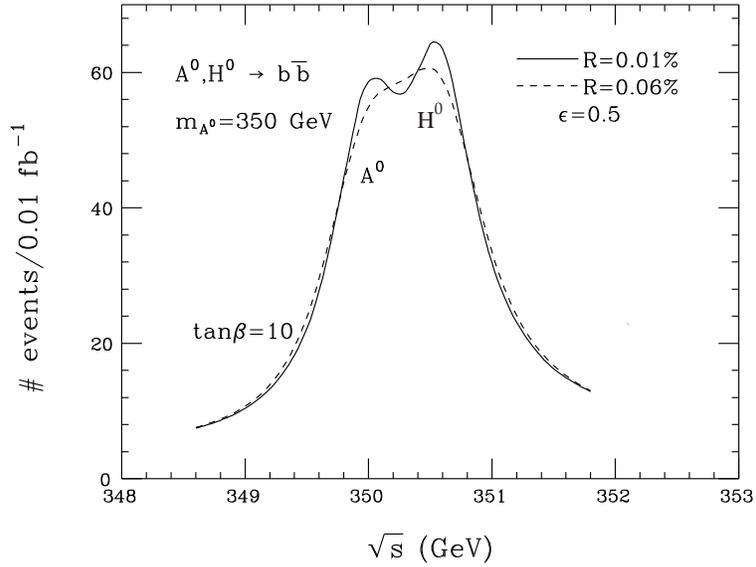}

\caption[Separation of $A^0$ and $H^0$ signals for $\tan\beta=10$]{Separation of $A^0$ and $H^0$ signals for $\tan\beta=10$. From  
Ref.~\cite{Barger}. \label{H0-A0-sep}}
\end{figure*}

In the MSSM, $m_{A^0}\approx m_{H^0}\approx m_{H^\pm}$ at large $m_{A^0}$
(as expected for mSUGRA boundary conditions), with a  
very close degeneracy in these masses for large $\tan\beta$. In such a  
circumstance, only an $s$-channel scan 
with the good resolution possible at a muon collider may allow  
separation of the $A^0$ and $H^0$ states; see Fig.~\ref{H0-A0-sep}.

\subsection{Light particles in technicolor models}

In most technicolor models, there will be light neutral and colorless
technipion resonances, $\pi_T^0$ and $\pi_T^{0\prime}$,
with masses below 500~GeV. Sample models include the recent 
top-assisted technicolor models \cite{techni}, in which
the technipion masses are typically above 100~GeV,
and models \cite{dominici} in which the masses of the 
neutral colorless resonances come
primarily from the one-loop effective potential
and the lightest state typically has mass as low as 10 to 100~GeV.
The widths of these light neutral and colorless
states in the top-assisted models
will be of order 0.1 to 50~GeV \cite{bhat}. In the one-loop models,
the width of the lightest technipion is  typically 
in the range from 3 to 50~MeV. Neutral technirho and techniomega
resonances are also a typical feature of technicolor models.
In all models, these resonances 
are predicted to have substantial Yukawa-like couplings to muons
and would be produced in the $s$-channel at a muon collider,
\begin{equation}
\mu^+\mu^-\to\pi^0_T, \, \pi^{0\prime}_T,\, \rho^0_T,\, \omega^0_T,
\end{equation}
with high event rates.
The peak cross sections for these processes are estimated to be  
$\approx 10^4$--$10^7$~fb \cite{bhat}. The dominant decay modes 
depend on eigenstate composition and other details but
typically are \cite{bhat}
\begin{eqnarray}
\pi^0_T &\to& gg,\, b\bar b, \, \tau\bar\tau,\, c\bar c,\, t\bar t\,,\\ 
\pi_T^{0\prime} &\to& gg,\, b\bar b,\, c\bar c,\, t\bar t,\, \tau^+\tau^-\,,\\
\rho^0_T &\to& \pi_T\pi_T,\, W\pi_T,\, WW \,,\\
\omega^0_T &\to& c\bar c,\, b\bar b,\, \tau\bar\tau,\, t\bar t,\,  
\gamma\pi^0_T,\, Z\pi^0_T \,.
\end{eqnarray}
Such resonances would be easy to find and study at a muon collider.

\subsection{Exotic narrow resonance possibilities}

There are important types of exotic physics that would be best
probed in $s$-channel production of a narrow resonance at a muon
collider. Many extended Higgs sector models
contain a doubly-charged Higgs boson $\Delta^{--}$ (and its $\Delta^{++}$
partner) that couples to $\mu^-\mu^-$ via a Majorana coupling.
The $s$-channel process
$\mu^-\mu^- \rightarrow \Delta^{--}$ has been shown \cite{gunmummum}
to probe extremely small values of this Majorana coupling, in particular
values naturally expected in models where such couplings
are responsible for neutrino mass generation. 
In supersymmetry, it is possible that there is R-parity violation.
If R-parity violation is of
the purely leptonic type, the coupling $\lambda_{\mu\tau\mu}$ for 
$\mu^-\mu^+\to\tilde\nu_\tau$ is very possibly the largest
such coupling and could be related to neutrino mass
generation.  This coupling can be probed down to
quite small values via $s$-channel
$\tilde\nu_\tau$ production at the muon collider \cite{fgh}.
\subsection{$Z$-factory}

A muon collider operating at the $Z$-boson resonance energy is an interesting  
option for measurement of polarization asymmetries, $B_s^0$--$\bar B_s^0$  
mixing, and of CP violation in the $B$-meson system \cite{demarteau}. 
The muon collider advantages are the partial muon beam polarization, 
%the separation of  $b$ and $\bar b$ in $Z\to b\bar b$ events,
and the long $B$-decay length for  
$B$-mesons produced at this $\sqrt s$. The left-right asymmetry $A_{LR}$ is  
the most accurate measure of $\sin^2\theta_{\rm w}$, since the uncertainty is  
statistics dominated. The present LEP and SLD polarization measurements show  
standard deviations of 2.4 in $A_{LR}^0$, 1.9 in $A_{FB}^{0,b}$ and 1.7 in  
$A_{FB}^{0,\tau}$ \cite{e-l}. The CP angle $\beta$
%(see Fig.~\ref{uni-triangle})
could be  measured from $B^0\to K_s J/\psi$ decays. 
To achieve significant improvements over existing measurements and 
those at future $B$-facilities, a data sample of $10^8\,Z$-boson events/year  would be needed. 
This corresponds to a luminosity $>0.15\rm~fb^{-1}$ /year, 
which is well within the domain of muon collider expectations;
 $R\sim 0.1\%$ would be more than adequate, given the
substantial $\sim 2.4$~GeV width of the $Z$.

\subsection{Threshold measurements at a muon collider}

With 10~fb$^{-1}$ integrated luminosity devoted to a measurement of a  
threshold cross-section, the following precisions on particle masses may be  
achievable \cite{bbgh2}:
\begin{equation}
\begin{array}{ll}
\mu^+\mu^-\to W^+W^-& \Delta M_W = 20\rm\ MeV\,,\\
\mu^+\mu^-\to t\bar t & \Delta m_t = 0.2\rm\ GeV\,,\\
\mu^+\mu^-\to Zh& \Delta m_h = 140 \rm\ MeV\ \ 
\end{array}
\end{equation}
(if\  $m_h = 100\rm\ GeV)\,.$ 
Precision $M_W$ and $m_t$ measurements allow important tests of electroweak  
radiative corrections through the relation
\begin{equation}
M_W = M_Z \left[ 1 - {\pi\alpha\over \sqrt 2 \, G_\mu \, M_W^2 (1-\delta r)}  
\right]^{1/2} \,,
\end{equation}
where $\delta r$ represents loop corrections. In the SM, $\delta r$ depends  
on $m_t^2$ and $\log m_h$. The optimal precision for tests of this relation is  
$\Delta M_W \approx {1\over 140}\Delta m_t$, so the uncertainty on $M_W$ is  
the most critical. With $\Delta M_W=20$~MeV the SM Higgs mass could be  
inferred to an accuracy
\begin{equation}
\Delta m_{h_{\rm SM}} =  30{\rm\ GeV} \left(m_h\over 100\rm\ GeV\right)\,.
\end{equation}
Alternatively, once $m_h$ is known from direct measurements, SUSY loop  
contributions can be tested.

In top-quark production at a muon collider above the threshold region, modest  
muon polarization would allow sensitive tests of anomalous top quark  
\mbox{couplings \cite{parke}}.

One of the important physics opportunities for the First Muon Collider is the  
production of the lighter chargino, $\tilde\chi_1^+$ \cite{carprot}. Fine-tuning arguments in mSUGRA suggest that it should be lighter than  
200~GeV. A search at the upgraded Tevatron for the process $q\bar  
q\to\tilde\chi_1^+\tilde\chi_2^0$ with $\tilde\chi_1^+\to  
\tilde\chi_1^0\ell^+\nu$ and $\tilde\chi_2^0\to\tilde\chi_1^0\ell^+\ell^-$  
decays can potentially reach masses $m_{\tilde\chi_1^+}\simeq  
m_{\tilde\chi_2^0}\sim 170$~GeV with 2~fb$^{-1}$ luminosity and $\sim230$~GeV  
with 10~fb$^{-1}$ \cite{teva2000}. The mass difference $M(\tilde\chi_2^0) -  
M(\tilde\chi_1^0)$ can be determined from the $\ell^+\ell^-$ mass  
distribution.
\begin{figure*}[tbh!]
\centering\leavevmode
\epsfxsize=4in\epsffile{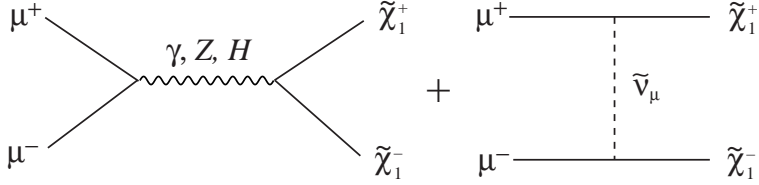}
\caption[Diagrams for production of the lighter chargino]{Diagrams for production of the lighter chargino.\label{light-chargino}}
\end{figure*}

The two contributing diagrams in the chargino pair production process are  
shown in Fig.~\ref{light-chargino}; 
the two amplitudes interfere destructively. The  
$\tilde\chi_1^+$  and $\tilde\nu_\mu$ 
masses can be inferred from the shape of  
the cross section in the threshold region \cite{bbh-new}. The chargino decay  
is $\tilde\chi_1^+\to f\bar f' \tilde\chi_1^0$. Selective cuts suppress the  
background from $W^+W^-$ production and leave $\sim 5\%$ signal efficiency for  
4\,jets${}+\E$ events. Measurements at two energies in the threshold region  
with total luminosity $L=50\rm~fb^{-1}$ and resolution $R=0.1\%$ can give the  
accuracies listed in table~\ref{chargino-table} on the chargino mass for the specified values of  
$m_{\tilde\chi_1^+}$ and $m_{\tilde\nu_\mu}$.

\begin{table*}[bht!]
\caption[Achievable uncertainties $\Delta m$ with 50~fb$^{-1}$ luminosity]{Achievable uncertainties with 50~fb$^{-1}$ luminosity on the mass  
of the lighter chargino for representative $m_{\tilde\chi_1^+}$ and  
$m_{\tilde\nu_\mu}$ masses. From Ref.~\cite{bbh-new}. \label{chargino-table}}
\centering\leavevmode
%\begin{tabular}{c}
\begin{tabular}{ccc}
$\Delta m_{\tilde\chi_1^+}$ (MeV) & $m_{\tilde\chi_1^+}$ (GeV) & $m_{\tilde\nu_\mu}$ (GeV)\\
35 & 100 & 500\\
45 & 100 & 300\\
150 & 200 & 500\\
300 & 200 & 300\\
\end{tabular}
\end{table*}

\subsection{Heavy particles of supersymmetry}

\begin{figure}[bht!]
\centering\leavevmode
\epsfxsize=3.25in\epsffile{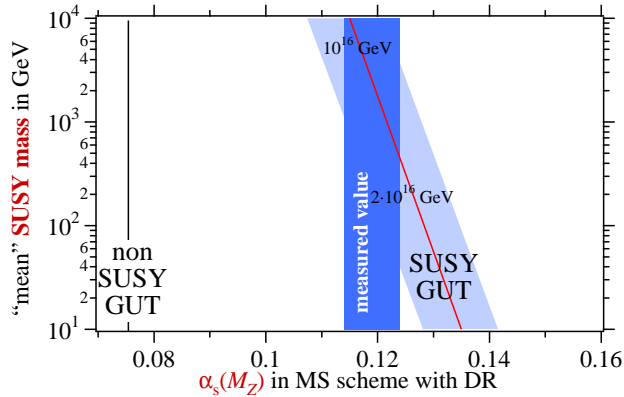}
\bigskip
\caption[$\alpha_s$ prediction in supersymmetric GUT]{$\alpha_s$ prediction in supersymmetric GUT with minimal particle content in the Dimensional Regularization scheme. \label{alpha_s}}
\end{figure}

The requirements of gauge coupling unification can be used to predict the  
mean SUSY mass scale, given the value of the strong coupling constant 
at the $Z$-mass  
scale. Figure~\ref{alpha_s} shows the SUSY GUT 
predictions versus $\alpha_s(M_Z)$. For the  
value $\alpha_s(M_Z) = 0.1214\pm0.0031$ from a new global fit to precision  
electroweak data \cite{e-l}, a mean SUSY mass of order 1~TeV is expected. 
Thus, it is likely that some SUSY particles will have masses at the TeV scale.
Large masses for the squarks of the first family are perhaps the
most likely in that this would provide a simple cure for possible flavor
changing neutral current difficulties. 

At the LHC, mainly squarks and gluinos will be produced; 
these decay to lighter  
SUSY particles. The LHC will be  a great SUSY machine, but some sparticle  
measurements will be very difficult or impossible there \cite{hinch,paige},  
namely: (i)~the determination of the LSP mass (LHC measurements give SUSY mass  
differences); (ii)~study of sleptons of mass $\agt200$~GeV because Drell-Yan  
production becomes too small at these masses; (iii)~study of heavy gauginos  
$\tilde\chi_2^\pm$ and $\tilde\chi_{3,4}^0$, which are mainly Higgsino and  
have small direct production rates and small branching fractions to channels  
usable for detection; (iv)~study of heavy Higgs bosons $H^\pm,\ H^0,\ A^0$  
when the MSSM $\tan\beta$ parameter is not large and their
masses are larger than $2m_t$, so that cross sections
are small and decays to $t\bar t$ are likely to be
dominant (their detection is deemed impossible if SUSY decays dominate).

\begin{figure*}[tbh!]
\centering\leavevmode
\epsfxsize=5.5in\epsffile{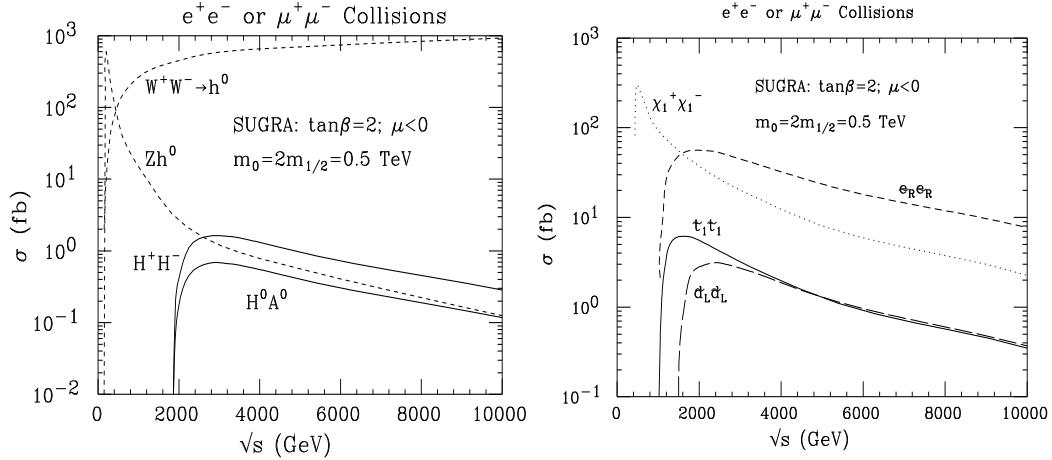}
\caption[Cross sections for pair production of Higgs bosons and scalar  
particles]{Cross sections for pair production of Higgs bosons and scalar  
particles at a high-energy muon collider. 
From Ref.~\cite{sanfran95}. \label{pair-product}}
\end{figure*}

Detection and study of the many scalar particles predicted in
supersymmetric models could be a particularly valuable contribution
of a high energy lepton collider. However, since pair production of  
scalar particles at a lepton collider is $P$-wave suppressed,
energies well above threshold are needed for sufficient production rates; see  
Fig.~\ref{pair-product}. For scalar particle masses of order 1 TeV a
collider energy of 3 to 4 TeV is needed to get past the threshold
suppression. A muon collider operating in this energy range
with high luminosity ($L\sim10^2$ to  
$10^3\rm~fb^{-1}/year$) would provide sufficient event rates to reconstruct  
heavy sparticles from their complex cascade decay chains \cite{paige,lykken}.

In string models, it is very natural to have
extra $Z$ bosons in addition to low-energy supersymmetry.
The $s$-channel production of a $Z'$ boson at the  
resonance energy would give enormous event rates at the NMC.  Moreover, the  
$s$-channel contributions of $Z'$ bosons with mass far above the kinematic  
reach of the collider could be revealed as contact interactions \cite{god}.
\subsection{Strong scattering of weak bosons}

\begin{figure}[bth!]
\centering\leavevmode
\epsfxsize=3in\epsffile{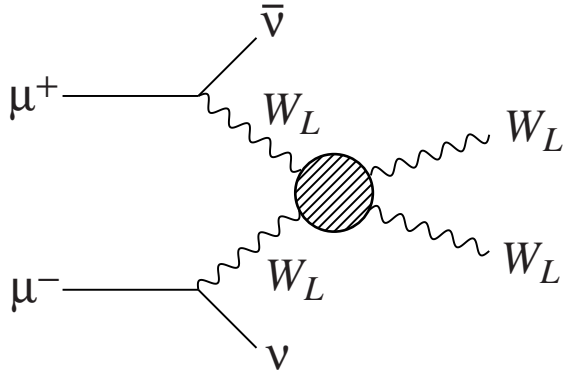}
\caption{Symbolic diagram for strong $WW$ scattering. \label{strong-ww}}
\end{figure}

The scattering of weak bosons can be studied at a high-energy muon collider  
through the process in Fig.~\ref{strong-ww}. 
The amplitude for the scattering of  
longitudinally polarized $W$-bosons behaves like
\begin{equation}
A(W_LW_L\to W_LW_L) \sim m_H^2/v^2
\end{equation}
if there is a light Higgs boson, and like 
\begin{equation}
A(W_LW_L\to W_LW_L) \sim s_{WW}^{\vphantom y} /v^2
\end{equation}
if no light Higgs boson exists; here $s_{WW}^{\vphantom y}$ is the square of  
the $WW$ CoM energy and $v=246$~GeV. In the latter scenario, partial-wave  
unitarity of $W_LW_L\to W_LW_L$ requires that the scattering of 
weak bosons  becomes strong at energy scales of order 1 to 2 TeV. 
Thus, subprocess energies  
$\sqrt{\smash{s_{WW}^{\vphantom y}}}\agt 1.5$~TeV are needed to probe strong  
$WW$ scattering effects.

The nature of the dynamics in the $WW$ sector is unknown. Models for this
scattering assume heavy resonant particles (isospin scalar and vector) or a  
non-resonant amplitude based on
a unitarized extrapolation of the low-energy theorem behavior  
$A\sim s_{WW}^{\vphantom y}/v^2$. In all models, impressive signals of strong  
$WW$ scattering are obtained at the NMC, with cross sections typically of  
order 50~fb$^{-1}$ \cite{W-mass}. Event rates are such that the
various weak-isospin channels ($I=0,1,2$) could be studied in detail
as a function of $s_{WW}$. After several years of operation,
it would even be possible to perform such a study after projecting
out the different final polarization states ($W_LW_L$, $W_LW_T$
and $W_TW_T$), thereby enabling one to verify that it is the
$W_LW_L$ channel in which the strong scattering is taking place.

\subsection{Front end physics}

New physics is likely to have important lepton flavor dependence and
may be most apparent for heavier flavors. The intense muon source
available at the front end of the muon collider will provide many
opportunities for uncovering such physics.

\subsubsection{Rare muon decays}

The planned muon flux of ${\sim}10^{14}$ muons/sec for a muon collider  
dramatically eclipses the flux, ${\sim}10^8$ muons/sec, of present sources. 
With an intense source, 
the rare muon processes $\mu\to e\gamma$ (for which the current branching  
fraction limit is $0.49\times10^{-12}$), 
$\mu N\to eN$ conversion, and the muon  
electric dipole moment can be probed at very interesting levels. A generic  
prediction of supersymmetric grand unified theories is that these lepton  
flavor violating or CP-violating processes should occur via loops at  
significant rates, e.g.\ BF$(\mu\to e\gamma)\sim 10^{-13}$. 
Lepton-flavor violation can also occur via $Z'$ bosons, leptoquarks, 
and heavy neutrinos \cite{marciano}.

\subsubsection{Neutrino flux}

The decay of a muon beam leads to neutrino beams of 
well defined flavors. 
A muon collider would yield a neutrino flux 1000  
times that presently available \cite{chuckandsteve}. This would result
in ${\sim}10^6$ $\nu N$  
and $\bar\nu N$ events per year, which could be used
to measure charm production  
($\sim$6\% of the total cross section) and measure $\sin^2\theta_{\rm w}$  
(and infer the $W$-mass to an accuracy $\Delta M_W \simeq 30$--50~MeV in one  
year) \cite{sgeer,sgeerjhf,sgeerphyrevd,moha1,harris1,bjknu1,bjknu2,ref8b,yu1}.

\subsubsection{Neutrino oscillations}

A special purpose muon ring has been proposed \cite{sgeerjhf} to store  
${\sim}10^{21}$ $\mu^+$ or $\mu^-$ per year and obtain ${\sim}10^{20}$  
neutrinos per year from muon decays along $\sim$75-m straight sections of 
the  
ring, which would be pointed towards a distant neutrino detector. 
The  neutrino fluxes from $\mu^-\to\nu_\mu\bar\nu_e e^-$ or from  
$\mu^+\to\bar\nu_\mu\nu_e e^+$ decays can be calculated with little systematic
error. Then, for example, from the decays of  
stored $\mu^-$'s, the following neutrino oscillation channels could be studied  
by detection of the charged leptons from the interactions of neutrinos in the  
detector:
$$ \begin{array}{cc}
\underline{\rm\ oscillation\ }& \underline{\rm\ detect\ }\\
\nu_\mu\to\nu_e& e^-\\
\nu_\mu\to\nu_\tau& \tau^-\\
\bar\nu_e\to\bar\nu_\mu& \mu^+\\
\bar\nu_e\to\bar\nu_\tau& \tau^+
\end{array}
$$
The detected $e^-$ or $\mu^+$ have the ``wrong sign" from the leptons  
produced by the interactions of the $\bar\nu_e$ and $\nu_\mu$ flux. The known  
neutrino fluxes from muon decays could be used for long-baseline oscillation  
experiments at any detector on Earth. The probabilities for vacuum  
oscillations between two neutrino flavors are given by
\begin{equation}
P(\nu_a\to\nu_b) = \sin^2 2\theta \, \sin^2(1.27\delta m^2 L/E)
\end{equation}
with $\delta m^2$ in eV$^2$ and $L/E$ in km/GeV.  In a very long baseline  
experiment from Fermilab to the Gran Sasso laboratory 
or the Kamioka mine
($L={\cal O}(10^4)$~km) with  
$\nu$-energies $E_\nu=20$ to 50~GeV ($L/E = 500$--200~km/GeV),  neutrino  
charged-current interaction rates of ${\sim}10^3$/year would result. 
In a long baseline experiment from Fermilab to the Soudan mine 
(L=732 km), the corresponding interaction rate is ${\sim}10^4$/year.
Such an  experiment would have sensitivity to oscillations down to $\delta  
m^2\sim 10^{-4}\,-\,10^{-5}\rm\,eV^2$ for $\sin^22\theta=1$ \cite{sgeerjhf}.

\subsubsection{$\mu p$ collider}

The possibility of colliding 200-GeV muons with 1000-GeV protons at Fermilab  
is under study. This collider would reach a maximum  
$Q^2\sim8\times10^4$~GeV$^2$, which is $\sim$90~times the reach of the HERA  
$ep$ collider, and deliver a luminosity ${\sim}10^{33}\rm\,cm^{-2}\,s^{-1}$,  
which is $\sim$300 times the HERA luminosity. The $\mu  p $ collider would  
produce ${\sim}10^6$ neutral-current deep-inelastic-scattering events per year  
at $Q^2>5000\rm~GeV^2$, which is more than a factor of
$10^3$ higher than at HERA.  In the new physics realm,
leptoquark couplings and contact interactions,
if present, are likely to be larger for muons than for electrons.
This $\mu p$ collider would have sufficient sensitivity 
to probe leptoquarks up to a  
mass $M_{LQ} \sim 800$~GeV and contact interactions to a scale  
$\Lambda\sim6$--9~TeV \cite{cheung}. 

\subsection{Summary of the physics potential}

The First Muon Collider offers unique probes of supersymmetry (particularly  
$s$-channel Higgs boson resonances) and technicolor models (via
$s$-channel production of techni-resonances),
high-precision threshold measurements of  
$W,\ t$ and SUSY particle masses, tests of SUSY radiative corrections that  
indirectly probe the existence of high-mass squarks, and a possible $Z^0$  
factory for improved precision in polarization measurements and for  
$B$-physics studies of CP violation and mixing.

The Next Muon Collider guarantees access to heavy SUSY scalar particles and  
$Z'$ states or to strong $WW$ scattering if there are no Higgs bosons and no  
supersymmetry.

The Front End of a muon collider offers dramatic improvements in sensitivity  
for flavor-violating transitions (e.g., \ $\mu\to e\gamma$), 
access to high-$Q^2$  
phenomena in deep-inelastic muon-proton and neutrino-proton interactions, and  
the ability to probe very small $\delta m^2$
via neutrino-oscillation studies in long-baseline experiments.

The muon collider would be crucial to unraveling the flavor dependence
of any type of new physics that is found at the next generation
of colliders.

Thus, muon colliders are robust options for probing new  
physics that may not be accessible at other colliders.
\section{PROTON DRIVER}  
 The overview of the required parameters is followed by a description of  
designs that have been studied in some detail.  The section concludes
with a discussion of the outstanding open issues.
\subsection{Specifications}
The proton driver requirements are determined by the design 
luminosity of the collider, and the efficiencies of muon 
collection, cooling, transport and acceleration.  The baseline 
specification is for a 4-MW, 16-GeV or a 7-MW, 30-GeV proton driver, with a repetition 
rate of 
15~Hz and $10^{14}$ protons per cycle in 2~bunches (for the 100-GeV machine) 
or 4~bunches (for the higher energies) of 
$5\times 10^{13}$ or $2.5\times 10^{13}$ protons, respectively.
Half the bunches are used to make $\mu^{-}$ and the rest for $\mu^{+}$ \cite{Norem97b}.  

The total beam 
power is several MW, which is larger than that of existing synchrotrons.   
However, except for bunch length, these 
parameters are similar to those of Kaon factories \cite{kaon} and
spallation neutron sources \cite{sns}. 
 As in those cases, the proton driver must have very low losses to permit 
inexpensive maintenance of components.

The rms bunch length for the protons on target has to be about 1~ns 
to: 1) reduce the initial longitudinal emittance of muons entering the 
cooling system, and 2) optimize the production of polarized muons.
Although bunches of up to $6\times 10^{13}$ protons per cycle have been 
accelerated, the required peak current  %at extraction for 1~ns and 4~bunches, 
is 2000~A, which is unprecedented.   

Since the collection of highly polarized $\mu$'s is inefficient (see section~\textbf{IV.G}), 
the proton driver should 
eventually provide an additional factor of two or more in proton intensity to 
permit the luminosity to be maintained for polarized muon beams.

\subsection{Possible options}

Accelerator designs are site, and to some extent, time dependent, and there
have been three studies at three different energies (30~GeV \cite{Snowmass3}, 
16~GeV \cite{fnal_study} and 24~GeV \cite{agsint95,Roser96,Ahrens97,Roser98}; see also \cite{Cho96}).  In general, if the final energy is higher, 
the required currents are lower, bunch manipulation and apertures are 
easier, and the final momentum spread and space-charge tune shifts are less.  
Lowering the final energy gives somewhat more $\pi$'s/Watt, a lower rf 
requirement ($V_{rf}\sim E^{2}$) and perhaps a lower facility cost.

In the low-energy muon collider, where two bunches of protons of $5\times 10^{13}$ are required on target,  
two bunches can be merged outside the driver. These two bunches would be extracted 
simultaneously from two different extraction ports, and fed by different 
transmission lines to the same target. By arranging the path lengths of the 
two lines appropriately, % the time of arrival of 
the two bunches can be exactly merged. 
%This is explicitly assumed in the FNAL study quoted below
%\cite{fnal_study}.

\subsubsection{A generic design}

A 7-MW collider-driver design based on parameters originally 
proposed in the Snowmass Feasibility study \cite{Snowmass3}
consists of a 600-MeV linac, a 
3.6-GeV booster and a 30-GeV driver.   Both linac and 
booster are based on the BNL Spallation Neutron Source design \cite{sns}, 
using a lower 
repetition rate and a lower total number of protons per pulse.  For the 4-bunch
 case ($2.5\times 10^{13}$ protons per bunch), the 
(95\%) bunch area is assumed to be 2~eV-s at injection and $<4.5$~eV-s at 
extraction.  The driver lattice is derived from the lattice of the JHF 
driver using 90$^\circ$ FODO cells with missing dipoles in every third 
FODO cell, allowing a transition energy that is higher than the maximum 
energy or, perhaps, imaginary.

\subsubsection{FNAL study}

If a muon collider is built at an existing laboratory, then possibilities 
abound for symbiotic relationships with the other facilities and programs of 
that laboratory.  For example, the proton driver for a muon collider might 
result from an upgrade of existing proton-source capabilities, and such an 
upgrade could then also enhance other future programs that use the proton 
beams.  

Fermilab has conceived such a proton-source development 
plan \cite{HolmesED}  %  The plan is modular, 
with three major components: an upgraded linac and two rapid-cycling 
(15 Hz) synchrotrons: a prebooster and a new booster.  The two 
synchrotrons operate in series; the four proton bunches for the muon collider 
are formed in the prebooster and then accelerated sequentially in the 
prebooster and the booster.  The plan could be implemented in stages, and other 
programs would benefit from each stage, but all three components are required 
to meet the luminosity goals of the muon colliders that have been considered so 
far.

\begin{table*}[tbh!] 
\centering
\caption{Baseline proton-driver parameters of the FNAL study. %\cite{HolmesED}.
\label{key}}
\begin{tabular}
[c]{llll}
              & \textbf{Linac} & \textbf{Booster} & \textbf{Driver} \\
\hline
Energy range (GeV)          & 1    & 3     & 16  \\
Rep.\ rate (Hz)            & 15    & 15     & 15 \\
RF voltage per turn (MV)   &         & 0.15  & 1.5  \\
Circumference  (m)        &         & 158    & 474 \\
Protons per bunch  ($\times 10^{13}$)   & &$2.5 $&$2.5 $ \\
Beam emittance [95\%] ($\pi$ mm-mrad )  &         & 200& 240 \\
Bunch area [95\%] (eV-s)     &         & 1.5 & $< 2.0$  \\
Incoherent tune shift @ Inj. &   & 0.39     & 0.39 \\
\end{tabular}
\end{table*}%

Table~\ref{key} presents the major parameters of the two rings.  
%The rationale for the major design choices is presented here.  
Whenever the needs of the muon collider itself allow some flexibility, 
the parameters have been chosen to optimize the resulting facility as a proton 
source for the rest of the future program at Fermilab.  For example, the 
machine circumferences and rf-harmonic numbers result in bunch trains that are 
compatible with the existing downstream proton machines.

A muon collider requires proton bunches that are both very intense and, at the 
pion-production target, very short. Strong transverse and longitudinal 
space-charge 
forces might disrupt such bunches in the synchrotrons unless measures to 
alleviate those effects are incorporated in the design.  The Laslett 
incoherent-space-charge tune shift quantifies the severity of the transverse 
effects.  A 
useful approximation for the space-charge tune shift $\Delta \nu_{sc}$ at the 
center of a round Gaussian beam is
\begin{equation}
\Delta \nu_{sc} = - {{3 r_{p} N_{\rm tot}}\over {2 \epsilon_{n}\beta 
\gamma^{2} b}}
\label{tuneshift}
\end{equation}
In this expression $r_{p} = 1.535\times10^{-18}$~m is the so-called 
electromagnetic radius 
of the proton, $N_{\rm tot}$ is the total number of protons in the ring, 
$\epsilon_{n}$ is the 95\% 
normalized transverse emittance, $\beta$ and $\gamma$ are the usual Lorentz 
kinematical 
factors, and $b \leq 1$ is the bunching factor, defined as the ratio of the 
average beam current to the peak current.  

The approximation (\ref{tuneshift}) implies that for a given total number of 
protons, here $10^{14}$, the 
factors in the denominator are the only ways to reduce the tune shift to a 
specified maximum tolerable value, taken as 0.4.  The bunching factor can 
be raised somewhat by careful tailoring of beam distributions, but here a 
typical value of 0.25 is conservatively assumed.  Achieving the desired beam 
intensity then requires a combination of high injection energy, here taken as 
1~GeV into the first ring, and large transverse normalized emittances, here 
assumed to be about $200\pi$~mm-mrad.  The corresponding required aperture is 
about 
13~cm in the first ring and about 10~cm in the second ring.  With such large 
apertures in rapid-cycling synchrotrons, careful design of the beam pipes for 
both rings is required to manage eddy-current effects.  Two approaches 
are under consideration.  One is a thin Inconel pipe with water cooling and 
eddy-current coil corrections integrated on the pipe, as in the AGS Booster.  
The other is a ceramic beam pipe with a conductor inside to carry beam-image 
currents, as in ISIS.

The Fermilab linac presently delivers a 400-MeV beam, and is capable with 
modest modifications of accelerating as many as $3\times 10^{13}$ protons per 
cycle at 
15 Hz \cite{Popovic}.  A significant upgrade is required in order to 
deliver $10^{14}$ protons at 1~GeV.  The energy can be raised by appending 
additional side-coupled modules to the downstream end of the linac.
Increasing the linac beam intensity probably means increasing both the 
beam current and the duration of the beam pulse.  Injection into the first ring 
is by charge stripping of the $H^{-}$ beam; the incoming beam will be chopped 
and injected into pre-existing buckets to achieve high capture efficiency.

The circumference of the second ring is set equal to that of the existing 
Fermilab Booster.  This choice provides several advantages.  First, the new 
booster could occupy the same tunnel as a relocated Booster; secondly, the 
beam-batch 
length from a full second ring matches that of the present Booster, which 
simplifies matching to downstream machines for other programs.  The output 
energy of 16~GeV then results from an assumed dipole packing fraction of 0.575 
and a peak dipole field of 1.3~T, which is the highest dipole field that 
is consistent with straightforward, nonsaturating design of magnets having 
thin silicon-steel laminations.  Driving such magnets into saturation would 
cause significant heating of the magnet yoke as well as potential problems with 
tracking between the dipoles and quadrupoles.  
%The energy of 16~GeV is a good 
%choice for pion production, because it is near the value where the linear rise 
%of pion/proton ratio with energy begins to saturate.

The prebooster also has 1.3-T dipole fields, and its circumference is one third 
that of the new booster; it operates at an rf harmonic number $h=4$.  The 
strategy for achieving the required short bunches at the target while 
alleviating space-charge effects in the rings is to start with four bunches 
occupying most of the circumference of the first small ring in order to keep 
the bunching factor large, and to do a bunch-shortening rotation in 
longitudinal phase space just before extraction from the second synchrotron.  
The four bunches are accelerated in the first ring to 3~GeV, then transferred 
bunch-to-bucket into the second ring with its harmonic number $h=12$.  At that 
energy, the kinematic factor in the tune-shift formula (\ref{tuneshift})
is large enough to 
compensate for the smaller bunching factor in the second ring. The transfer 
energy of 3~GeV between the two rings roughly equalizes their space-charge tune 
shifts.

Both rings employ separated-function lattices with flexible momentum compaction in order to raise their transition energies above their respective extraction 
energies.  This not only avoids having to accelerate beam through transition 
but also provides other advantages.  Intense beams are not subject to certain 
instabilities such as the negative-mass instability below transition and 
empirically seem less susceptible to other instabilities such as the microwave 
instability.  Also, the negative natural chromaticity is beneficial for 
stabilizing the beam below transition, thereby perhaps obviating the need for 
sextupole correctors, especially in the first ring.  Having transition not too 
far above extraction also provides substantial bucket area in which to 
accomplish beam-shortening rf manipulations.

Several potential sources of instabilities in the rings have been
examined \cite{Ng}, as well as the 
possibility of compensating the latter effect 
by inductive inserts in the rings.  Space charge is the main factor 
affecting the stability of the beams;  the rings appear to be 
safe from longitudinal- and transverse-microwave 
instabilities.  Of course, standard stabilizing methods such 
as active dampers are necessary to counteract some of the 
instabilities.  Flexible momentum-compaction lattices would 
be useful not only to raise the transition energy above the 
extraction energy, but also to allow fast changes in the slip 
factor to facilitate bunch-narrowing manipulations at 
extraction time.  Compensation of longitudinal space-charge 
effects by means of ferrite-loaded inductive inserts would be 
useful, especially for the first ring.

The magnet-power-supply circuit for each ring is a 15-Hz resonant system like 
that of the existing Booster, with dipoles and quadrupoles electrically in 
series.  This implies that the second ring will accelerate only one batch at a 
time from the first ring, which is all that the muon collider needs.  Adding 
about 15\% of second harmonic to the magnet ramp reduces the required peak 
accelerating voltage by about 25\%, which is probably worth doing, especially 
for the second ring with its large voltage requirement.  

One of the advantages 
of a two-ring system is that the two rings divide the work of accelerating the 
beam.  The rf system of the first ring is relatively modest because of its 
small circumference and small energy gain; that of the second ring is 
simplified because its high injection energy means a small rf-frequency swing 
\cite{GriffinQKerns}.

ESME simulations of longitudinal motion show that the rms bunch length is 2~nsec
as desired after the bunch rotation that occurs just before extraction 
from the second synchrotron.%~\cite{Kourbanis}.  
The bunch rotation creates momentum spreads of about 2\% with longitudinal 
emittances of about 2~eV-s per bunch.  Such spreads would contribute a few 
cm in quadrature to the beam size for a short period before 
extraction.  This is thought to be tolerable, given the large apertures that 
are required in any case.  High injection 
energies help to alleviate these longitudinal effects, which result from 
space-charge voltages having the same $1/\beta\gamma^{2}$ kinematic dependence 
as the transverse tune shifts.  

\subsubsection{AGS upgrade}

\begin{figure*}
\begin{center}
\includegraphics[width=5.5in]{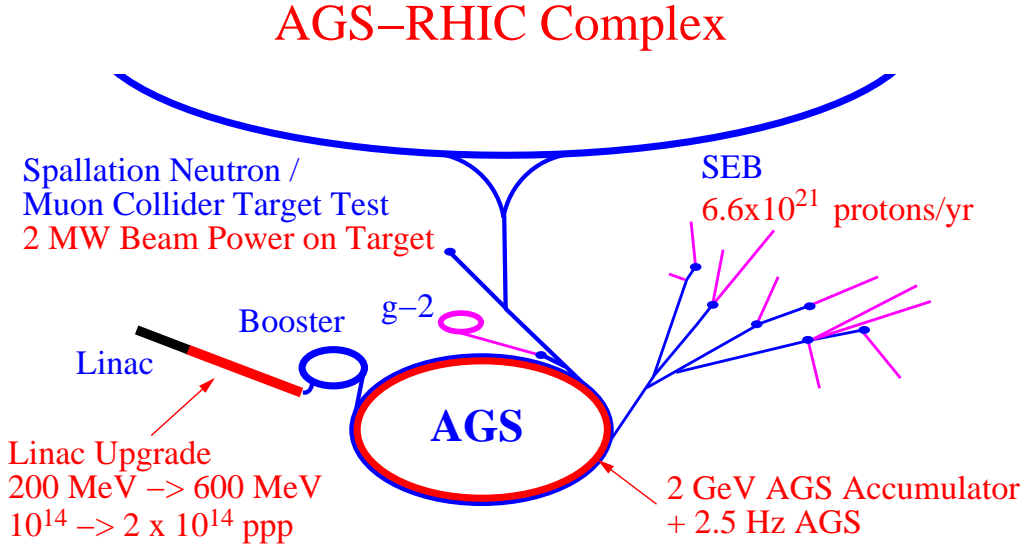} 
\end{center}
\caption[The AGS--RHIC accelerator complex]{The AGS--RHIC accelerator complex and a summary of possible intensity upgrades for the AGS.  }
\label{ags-rhic}
\end{figure*}

The third study \cite{agsint95,Roser96,Ahrens97,Roser98} is of an upgrade to the BNL AGS, which
should produce 
bunches larger than those required for the muon collider, but
at a lower repetition rate.   The AGS presently produces $6\times 10^{13}$ 
protons 
in eight bunches at 25~GeV and 0.6~Hz.  A 2.5-GeV accumulator ring in 
the AGS tunnel and AGS power-supply upgrade to 2.5-Hz operation would 
match the repetition rate to the 10-Hz  repetition rate of the 
booster. This would generate 1~MW beam power.  
With an additional upgrade of the linac energy to 600~MeV, an 
intensity of $2\times 10^{14}$ protons per pulse in four bunches of 
$5\times 10^{13}$ at 25~GeV and 2.5~Hz could be reached, raising the power to 
2~MW. The upgrades to the AGS accelerator complex are summarized in Fig.\ref{ags-rhic}. Other options are also under consideration, such as the addition of a 
second booster and 5-Hz operation, that would reach the baseline specification 
of 4~MW. 

The AGS momentum acceptance of $\pm3$\% requires that the 
longitudinal phase space occupied by one bunch be less than 4.5~eV-s.  
This high bunch density in turn generates stringent demands on the 
earlier parts of the accelerator cycle.  In particular, Landau damping
from the beam momentum spread may guard against resistive wall 
instabilities during injection and longitudinal microwave 
instabilities after transition.   Beam stability can be restored with 
a more-powerful transverse-damping system and possibly a new low-impedance 
vacuum chamber.  The transverse microwave instability is 
predicted to occur after transition crossing unless damped by 
Landau damping from incoherent tune spread or possibly high-frequency 
quadrupoles.

\subsection{Progress and open issues}

Conventional rf manipulations appear able to produce 1- to 2-ns proton 
bunches if enough rf voltage to overcome the space-charge forces is 
used, and the beam energy is far enough from transition so the final 
bunch rotation is fast. 
Both simulations and experimental work have been directed at 
demonstrating that a short pulse can be produced easily.  

An experiment at the AGS has shown that bunches with $\sigma_z=2$~ns can
be produced near transition from bunches with $\sigma_z \sim 8$~ns by
bunch rotation \cite{ref10,agsbunch}.  In this experiment, the AGS was
flattoped
near transition ($\sim 7$~GeV) while the $\gamma_t$-jump system
was used to bring the transition energy suddenly to the beam energy,
letting the bunch-energy spread expand and bunch length contract.  The
experiment also demonstrated that bunches are stable over periods of
0.1~s.  In addition, the data were used to measure the lowest two orders
of the momentum compaction factor.

The AGS bunch area, 1.5 eV-s, was comparable to that expected in the
proton driver, but the bunch charge (though as large as 3-$5 \times
10^{12}$ protons) was only about one tenth of that required by the muon
target.  The proton driver would use a flexible momentum compaction lattice which would give
much better control of the transition energy, permitting a very fast
final bunch rotation \cite{Trbojevic97b}. In addition, the rf frequency would be higher than
that of the AGS so the buckets (and bunches) would initially be only
half as long.  Thus bunch rotation could be expected to be easier with
the new machine, which should compensate for the larger charge.

Simulations with the ESME code have also 
shown that 1-2~ns bunches of $5\times 10^{13}$ can be produced at 
extraction in a 16-GeV ring with the rf and emittance shown in Table~\ref{key}.

The efficiency of capturing and accelerating beam may be increased by 
compensation of the space-charge forces in the proton driver.  The use 
of tunable inductive inserts in the ring vacuum chamber may permit 
active control and compensation of the longitudinal space charge below 
transition (since the inductive impedance is of the opposite sign to 
the capacitive space charge).  Initial experiments at the 
KEK proton synchrotron and Los Alamos PSR \cite{PSR}
with short ferrite inserts appear to 
show a reduction in the synchrotron oscillation frequency shift 
caused by space charge and a decrease in the necessary rf voltage to 
maintain a given bunch intensity. Further experiments are 
needed to demonstrate this technique fully.

The high rf voltage and beam power and the relatively small
size of the machine require high-gradient, high-power rf cavities.
Fermilab, BNL and KEK are collaborating
on research and development of such type of cavities.%\cite{give_reference1}. 
This
work includes the study of magnet alloys and hybrid cavities using 
both ferrite and new magnet alloys, high-power amplifiers and beam-loading
compensation.

The employment of barrier-bucket \cite{give_reference2} rf cavities can effectively
generate and manipulate a gap in the beam and reduce the space-charge effect. A successful test of this scheme has recently been completed \cite{give_reference3}, and two $40-$kV barrier cavities have been built by BNL and KEK
and are being installed on the AGS. Another high-gradient
barrier cavity using magnet alloys is under study at Fermilab.
%\cite{give_reference3}.
%\input{prst4_new.tex}
\section[PION PRODUCTION AND CAPTURE]{PION PRODUCTION, CAPTURE AND PHASE ROTATION CHANNEL} 
\label{subsec-comppion}

This section first discusses the choice of target technology and optimization of the target geometry, and then describes design studies for the
pion capture and phase rotation channel.  Prospects for polarized muon beams
are discussed in detail.  The section concludes with an outline of
an R\&D program for target and phase rotation issues.

\begin{figure*}[thb!]
\begin{center}
\includegraphics[width=5.5in]{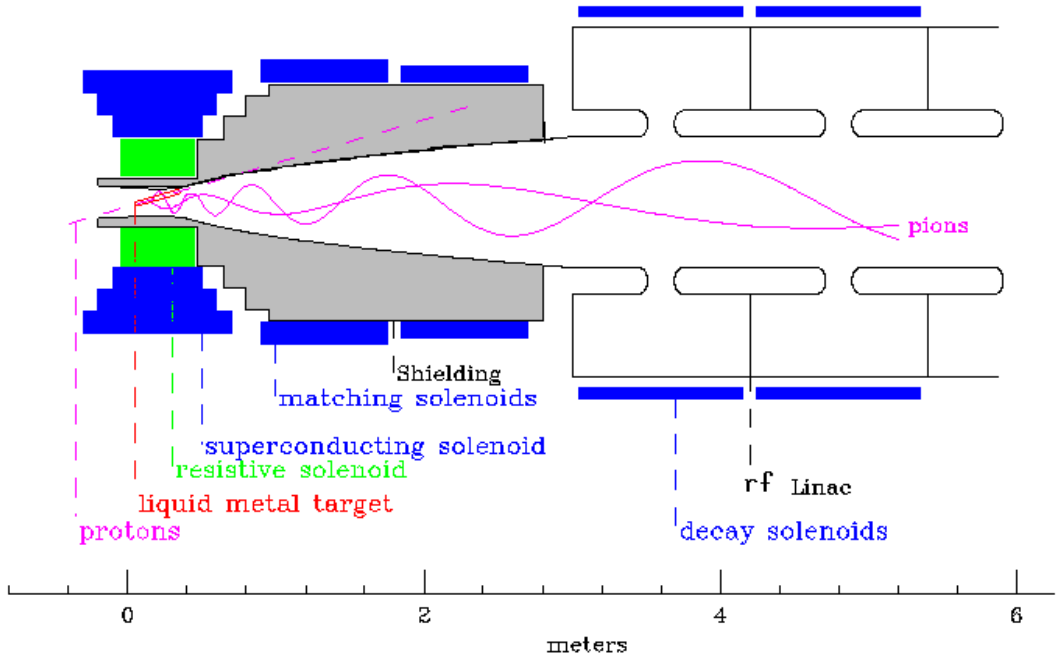} 
\end{center}
\caption[Schematic view of pion production, capture and initial phase
rotation  ]{Schematic view of pion production, capture and initial phase
rotation. A pulse of 16-30~GeV
protons is incident on a skewed target inside a high-field solenoid magnet
followed by a decay and phase rotation channel. }
 \label{capture}
 \end{figure*}

Figure~\ref{capture} gives an overview of the
configuration for production of pions by a 
proton beam impinging on a long, transversely thin target, followed by capture of
low-momentum, forward pions in a channel of solenoid magnets with rf
cavities to compress the bunch energy while letting the bunch length grow. This arrangement performs the desired rotation of the beam. 

\subsection{Pion production}

To achieve the luminosities for muon colliders presented in Table~\ref{sum},
$2\times 10^{12}$ (or $4\times 10^{12}$ in the 100~GeV CoM case) muons of each 
sign must be delivered to the collider ring in  
each pulse.  We estimate that a muon has a probability of only 1/4 of 
surviving the processes of cooling and acceleration, due to losses in
beam apertures or by decay.  
Thus, $0.8\times 10^{13}$ muons (1.6 $\times 10^{13}$ at 100~GeV) must exit the
 phase rotation channel each pulse.  For
pulses of $2.5\times 10^{13}$ protons ($5\times 10^{13}$ for 100~GeV), this 
requires 0.3~muons per initial proton.  Since
the efficiency of the phase rotation channel is about 1/2, this is equivalent 
to a capture of about 0.6~pions per proton: a very high efficiency. 

\begin{figure*}[thb!]
\begin{center}
\includegraphics[width=5in,clip=]{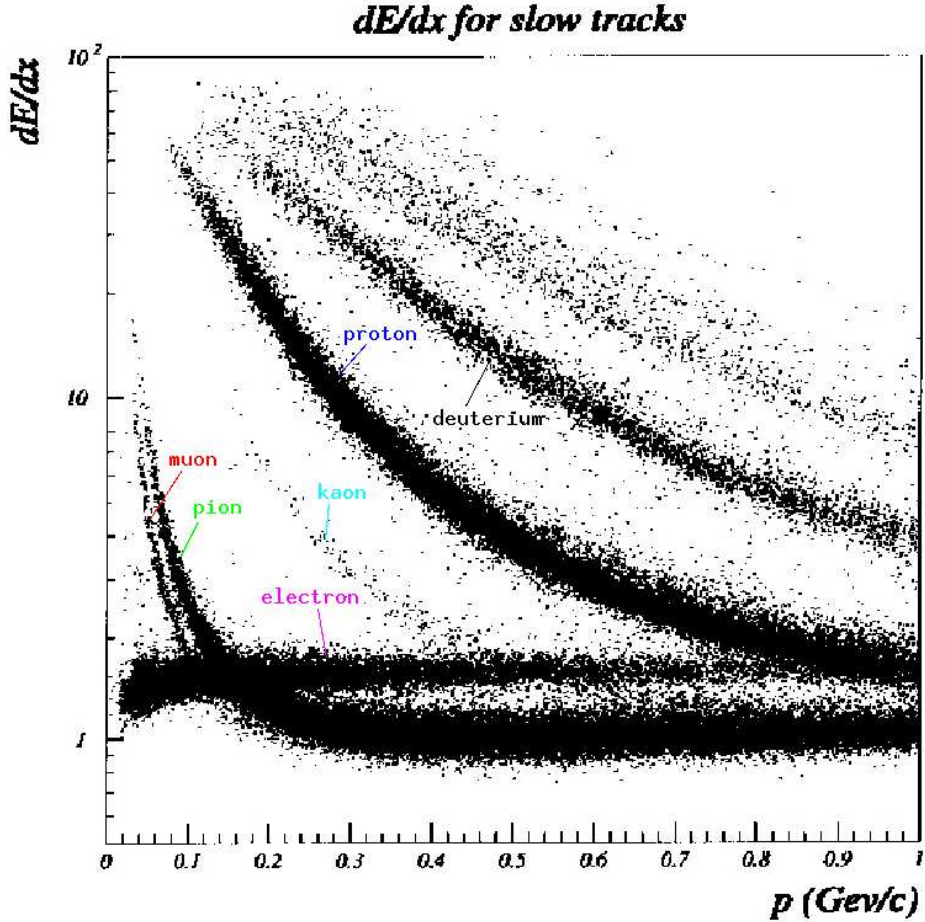} 
\end{center}
\caption[Ionization energy loss \vs~momentum. $dE/dx$ curve  ]{$dE/dx$ curve in arbitrary units for low momentum tracks; the ionization energy loss is for tracks with 30 or more hits in the TPC. The incident beam momentum is 18~GeV/$c.$ From left to right the bands correspond to muons, pions, kaons, protons and deuterium, respectively. Note the overlap of the (nearly horizontal) electron band with other species.}
 \label{figdedx_exp}
 \end{figure*}

The pions are produced in the interaction of the
proton beam with the primary target.  
Extensive simulations have been performed for pion production from 
8-30~GeV proton beams on different target materials in a high-field 
solenoid \cite{ref6a,tar-snake,ref15,Ehst97,Takahashi97}.  %tarPAC97,ref15}. %,tarmok98}. 
Three different Monte Carlo codes \cite{arc,mars,ref16,dpmjet} predict 
similar
pion yields despite significant differences in their physics models.
Some members of the Collaboration are involved in an AGS 
experiment BNL E-910 \cite{exp910} to measure 
the yield of very low momentum pions, which will validate the 
codes in the critical kinematic region. This experiment ran for 14 weeks during the Spring of 1996 and has collected over 20 million events, of which about a quarter are minimum bias triggers for inclusive cross section measurements. The targets were varied in material (Be, Cu, Au, U) and thickness (2--100\% interaction length ($\lambda_{I}$)) and three different beam momenta were used (6, 12.5, 18~GeV/$c).$ Presently, the E910 collaboration is doing a careful analysis of the large data sample obtained. Figure~\ref{figdedx_exp} shows the dE/dx energy \vs~momentum for reconstructed tracks in the TPC; there is clear particle species separation \cite{refdedx_exp}. 

\begin{figure*}[thb!]
%\noindent
\begin{minipage}{.50\linewidth} % fig 4a
\centering\epsfig{figure=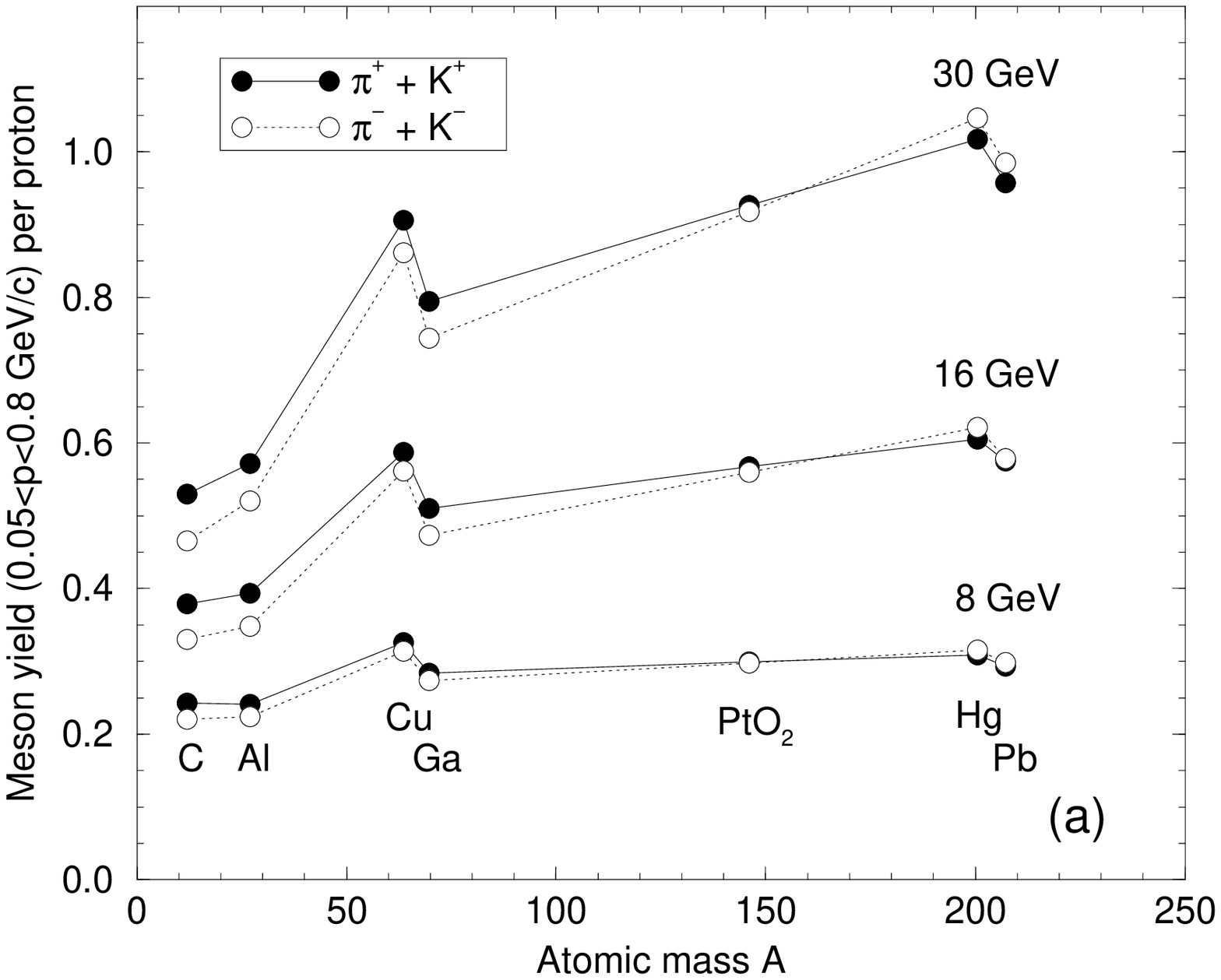,width=\linewidth}
\end{minipage}\hfill
\begin{minipage}{.50\linewidth} % fig 4b
\centering\epsfig{figure=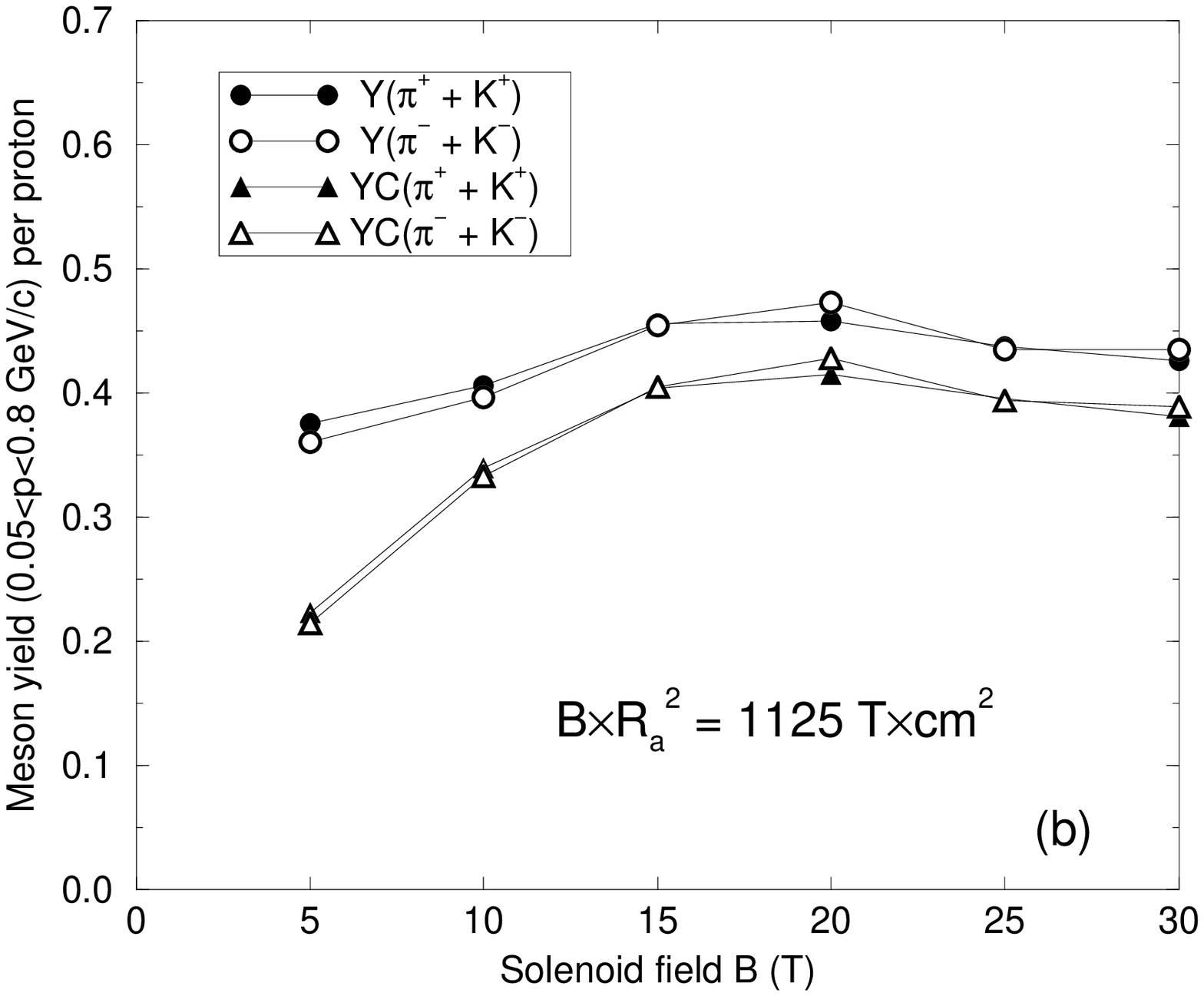,width=\linewidth}
\end{minipage}
\begin{minipage}{.50\linewidth} % fig 4c
\centering\epsfig{figure=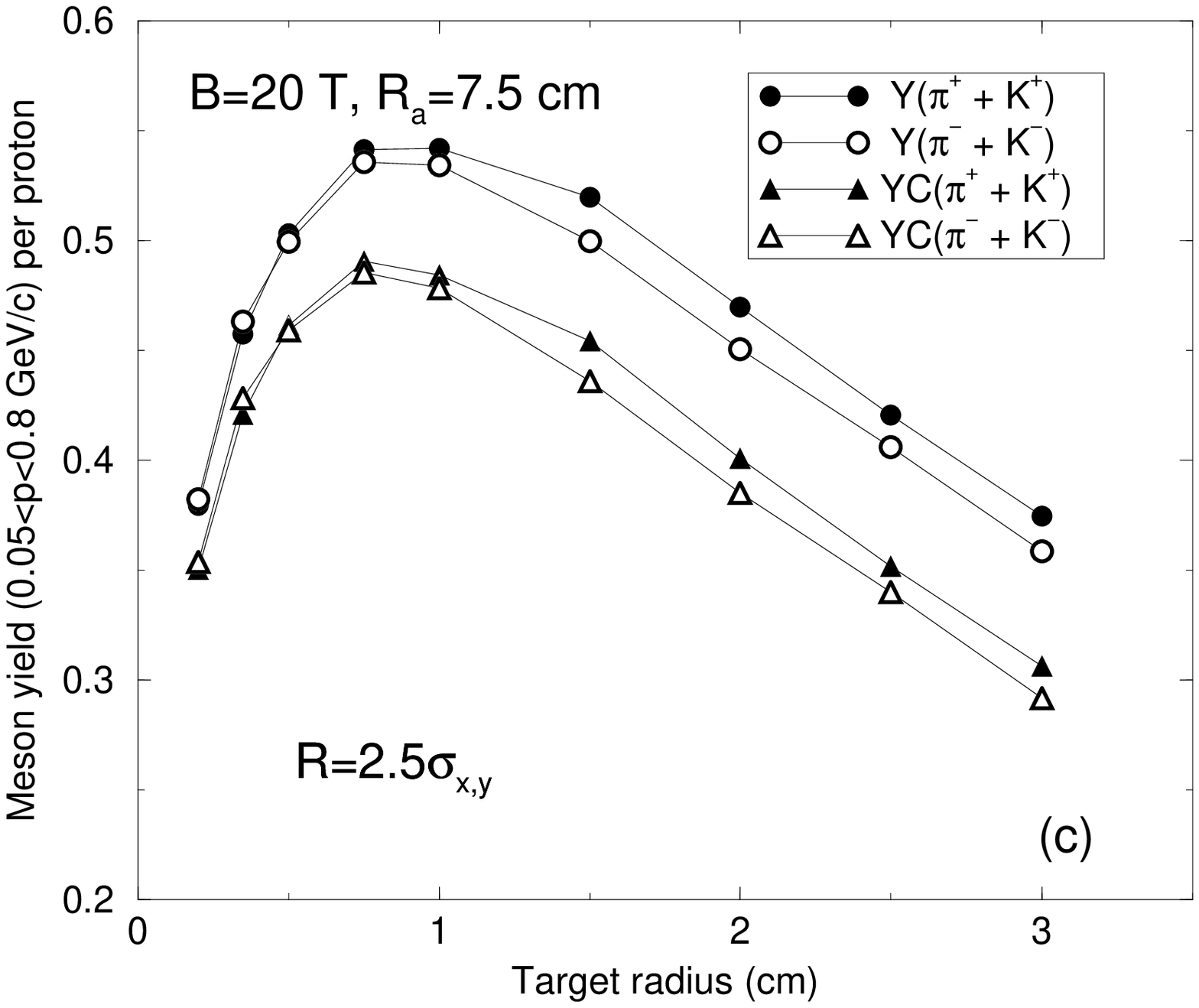,width=\linewidth}
\end{minipage}\hfill
\begin{minipage}{.50\linewidth} % fig 4d
\centering\epsfig{figure=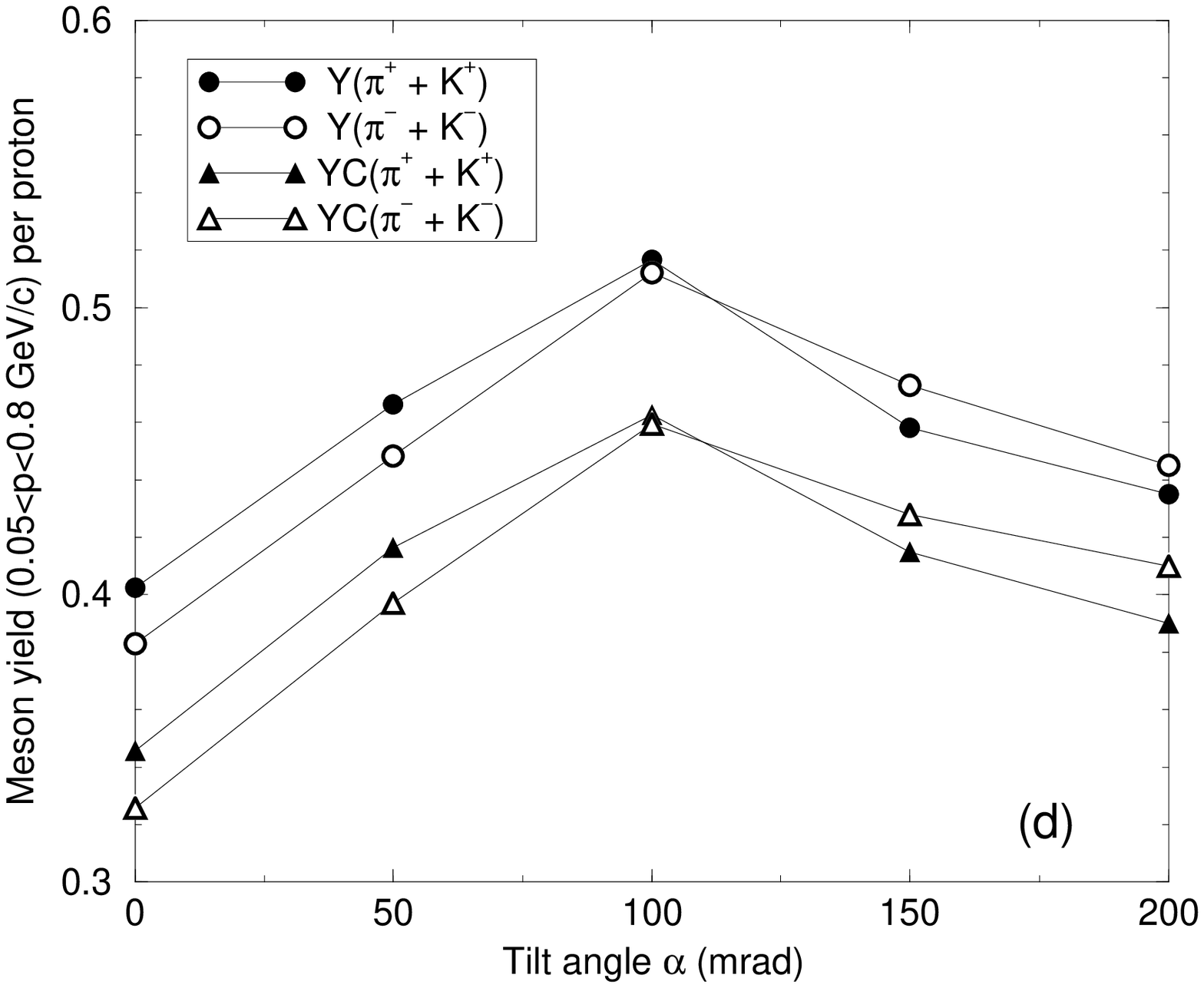,width=\linewidth}
\end{minipage}
%\vspace{10pt}
\caption[Meson yield ($\pi + K$) from different targets tilted by angle 
$\alpha$.]
{Meson yield ($\pi + K$) from different targets tilted by angle 
$\alpha$ in a solenoidal field $B$ of aperture $R_a$ as calculated with the 
MARS13(98) code. The target is aligned along the beam.
The curves labeled YC show mesons that are transported into the decay channel.
(a) Yield from a $1.5~\lambda_{I}$, 1 cm radius target irradiated
with 8, 16 and 30~GeV proton beams ($\sigma_x = \sigma_y =4$~mm) 
as a function of target atomic mass ($B =20$~T, $R_a = 7.5$~cm, $\alpha=0$);
(b) Yield from a $3-\lambda_{I}$, 1 cm radius gallium target 
tilted at $\alpha=150$~mrad in a 16~GeV proton beam 
($\sigma_x = \sigma_y = 4$~mm)
{\it vs.}\ solenoid field for a fixed adiabatic invariant $B R_a^2$;
(c) Yield as a function of radius of  a $3~\lambda_{I}$ gallium target in 
a 16~GeV proton beam ($\sigma_x =  \sigma_y = 4$~mm, $B = 20$~T, 
$R_a = 7.5$~cm, $\alpha = 100$~mrad);
(d) Yield from a $3~\lambda_{I}$, 1 cm radius gallium target {\it vs.}\
tilt angle 
between the axis of the capture solenoid and the proton beam for a 16~GeV 
proton beam ($\sigma_x = \sigma_y = 4$~mm, $B = 20$~T, $R_a = 7.5$~cm).}
\label{yield-E-A}
\end{figure*}

The pion yield is greater for relatively high $Z$ materials, and for these,
the pion yield is maximal for
longitudinal momenta of the same order as the average transverse momentum
($\approx 200$ MeV/$c$).  
Targets of varying composition ($6<Z<82$), radii (0.2-3\,cm) and thicknesses 
(0.5-3~nuclear interaction lengths) have been explored using a Monte Carlo simulation \cite{ref15}.
For a fixed number of interaction lengths, the pion yield per proton rises 
almost linearly with proton energy, and hence almost proportional 
to the energy deposited in the target.
The yield is higher for medium- and high-$Z$ target materials,
with a noticeable gain at $Z>26$ for 30~GeV proton beams, but
with only a minor effect for $E \leq 16$~GeV. This is  shown in
Fig.~\ref{yield-E-A} where results of detailed MARS13(98) \cite{mars} 
simulations are presented.
The curves show the meson yield ($\pi + K$) from the targets in the
momentum interval $0.05 \leq P \leq 0.8$~GeV/$c$  (labeled Y) and 
the number of mesons that are both
captured in the high field solenoid and transported into the decay channel 
(labeled YC). The typical statistical error is a few percent.

\subsection{Target}

The target should be 2-3 interaction lengths long to maximize
pion production.  A high-density material is favored to minimize the
size and cost of the capture solenoid magnet.  Target radii larger than about 
1~cm lead to
lower pion rates due to reabsorption, while smaller diameter targets have less production from secondary interactions.  Tilting the target by 
 100-150~mrad minimizes loss of pions by absorption in the target after one 
 turn on their helical trajectory \cite{ref4b,mcprod}. 
Another advantage of the tilted target 
 geometry is that the high energy and neutral components of the shower can be 
 absorbed in a water-cooled beam dump to the side of the focused beam.

About 30~kJ of energy is deposited in the target by each proton pulse (10\% of
the beam energy). Hence, the target absorbs 400~kW of power at the 15-Hz pulse 
rate.  Cooling of the target via 
contact with a thermal bath would lead to unacceptable
absorption of pions, and radiative cooling is inadequate for such high
power in a compact target.  Therefore,
the target must move so as to carry the energy
deposited by the proton beam to a heat exchanger 
outside the solenoid channel. 
 
Both moving solid metal and flowing liquid targets have been considered, with
the latter as the currently preferred solution.  A liquid is relatively
easy to move, easy to cool, can be readily removed and replaced, and is the
preferred target material for most spallation neutron sources under study. A
liquid flowing in a pipe was considered, but experience at ISOLDE
with short proton pulses \cite{Lettry} as well as simulations \cite{ref11,Bauer} 
suggest serious problems in shock damage to the pipe.
An open liquid jet is thus proposed.

A jet of liquid mercury has been demonstrated \cite{ref11} 
but not exposed to a beam. For our application, safety and other
considerations favor the use of a low melting point lead alloy rather than 
mercury. Gallium alloys, though with lower density, are also being considered.
Experimental and theoretical studies are underway to determine the 
consequences of beam shock heating of the liquid. 
%, and an experimental program is being planned (see section ***).
It is expected that the jet will disperse after being
exposed to the beam.  The target station must survive
damage resulting from the violence in this dispersion. This consideration will
determine the minimum beam, and thus jet, radius.
% (1 cm radius appears conservative, but 5~mm would be preferred - see below).

For a conducting liquid jet in a 
strong magnetic field, as proposed, strong eddy currents will be
induced in the jet, causing reaction forces that may disrupt its flow \cite{ref12,ref13}. The forces induced
are proportional to the square of the jet radius, and set
a maximum for this radius of order 5-10~mm.
If this maximum
is smaller than the minimum radius set by shock considerations, then multiple
smaller beams and jets could be used; {\it e.g.}, four jets of 5~mm radius with 
four beams with $2.5 \times 10^{13}$ protons per bunch.
Other alternatives include targets made from insulating materials such as 
liquid PtO$_2$ or Re$_2$O$_3$, slurries 
({\it e.g.},  Pt in water), or powders \cite{mumu98-10}.

\begin{figure*}[thb!]
%\centering
\begin{center}
\includegraphics[width=5in,clip=]{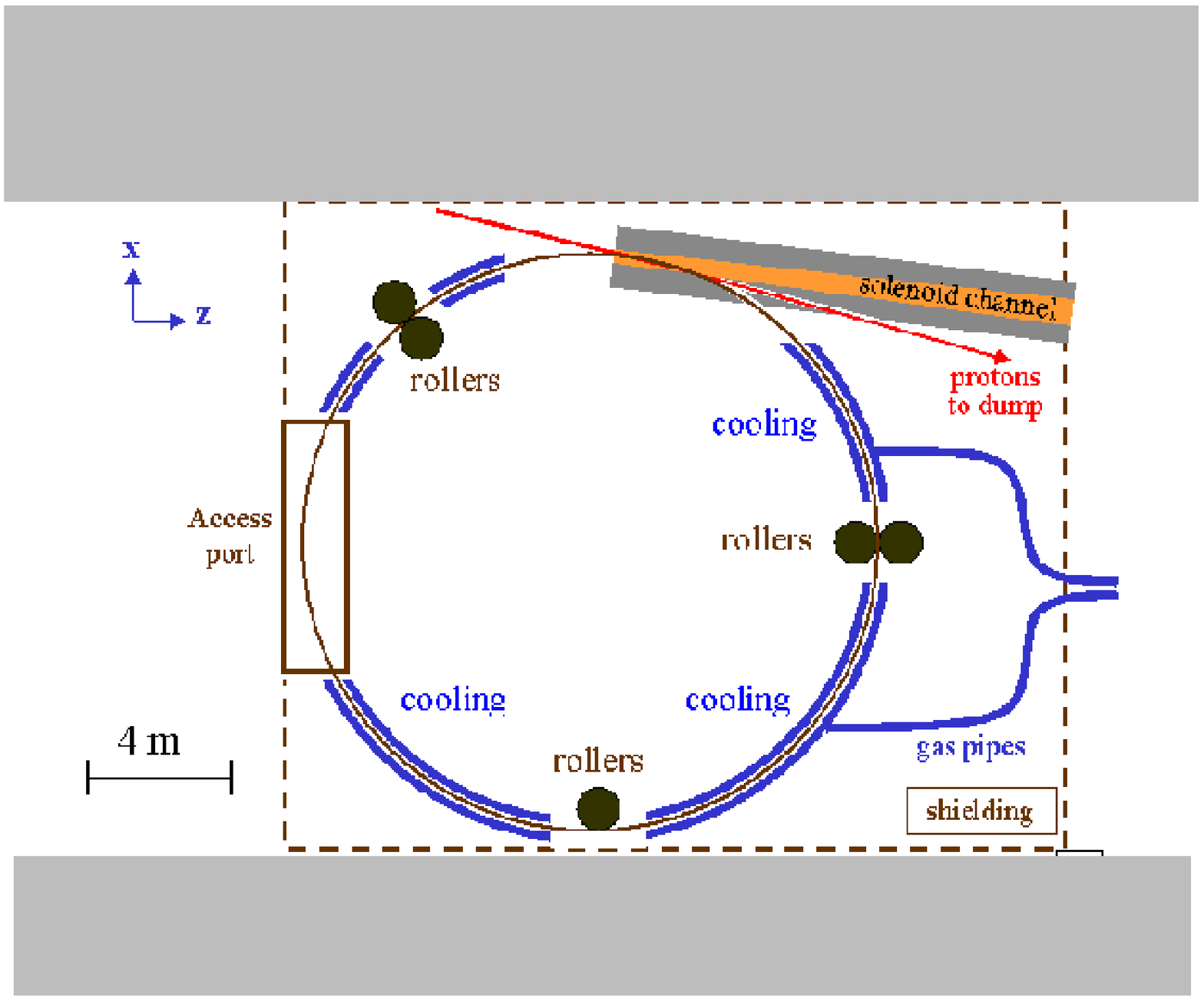} 
\end{center}
\caption{Alternative concept of a solid metal target in the form of a
rotating Cu-Ni band.}
 \label{bandsaw}
 \end{figure*}

A moving solid metal target is not the current baseline solution, but is a 
serious possibility. In this case, the target could consist of a long flat band
or hoop of copper-nickel that moves along its length (as in a
band saw) \cite{cunitarget}. The band  would be many meters in length, would be
cooled by gas jets away from the target area, and would be supported and
moved by rollers, as shown in Fig.~\ref{bandsaw}.  

The choice and parameters of the target are critical issues that need resolution. These can be resolved by experiments with a strong magnetic field and a beam, as discussed in  section~IV.H.

\subsection{Capture}

To capture all pions with transverse momenta $p_T$ less than their
typical value of 200~MeV/$c$, the product of the capture solenoid field $B$ 
and its radius $R_a$
must be greater than 1.33~T~m. The use of a high field and small radius
is preferred to minimize the corresponding transverse
emittance, which is proportional to $BR^2$: for a fixed transverse momentum 
capture, this emittance is thus proportional to $R$. A field of 20~T
and 7.5~cm radius was chosen on the basis of simulations described below. 
This gives $BR$ = 1.5~T~m, $BR^2 = 0.1125$~T~m$^2$ and 
a maximum transverse momentum capture of $p_T = 225$ MeV/$c$.

A preliminary design \cite{ref18} of the capture solenoid 
has an inner 6~T, 4~MW, water cooled, hollow conductor magnet with an inside 
diameter of 24~cm and an outside diameter of 60~cm.   
There is space for a 4~cm thick, water cooled, heavy metal shield 
inside the coil.  The outer superconducting magnet has
three coils, with inside diameters of 60 to 80~cm. It generates
an additional 14~T of field  at the target and provides the required tapered
field to match into the decay channel. Such a hybrid
solenoid has parameters compatible with those of existing 
magnets \cite{ref17}.       

The 20~T capture solenoid is matched via a transfer solenoid \cite{tar-snake} into a decay channel consisting 
of a system of superconducting solenoids 
with the same adiabatic invariant $BR^2 \propto R p_T$.
Thus, for a 1.25~T decay channel, $B$ drops by a factor of 16
between the target and decay channel, 
$R$ and $p_T$ change by factors of
4 and 1/4, respectively.  This permits improved acceptance
of transverse
momentum within the decay channel, at the cost of an increased spread
in longitudinal momentum.
Figure~\ref{yield-E-A}(b) shows the meson yield as a
function of field in the capture solenoid, with the radius of the capture
solenoid adjusted to maintain the same $BR_a^2$ as in the decay channel.
The optimum field is 20~T in the capture solenoid.  

If the axis of the target is coincident with that of the solenoid field, then
there is a relatively high probability that pions re-enter the target after
one cycle on their helical trajectory and are lost due to nuclear interactions.
When the target and proton beam 
are set at an angle of 100-150~mrad with respect to the field 
axis \cite{ref15}, %\cite{ref14,ref15}, 
the probability for such
pion interactions at the target is reduced, and the overall production rate 
is increased by 60\%, as shown in Fig.~\ref{yield-E-A}(d). 

In summary, the simulations indicate that a 20~T solenoid of 16~cm inside diameter 
surrounding a tilted target will capture about
half of all produced pions.  With target efficiency included, about 0.6
pions per proton will enter the pion decay channel \cite{ref15}. %tarmok98}. 

\subsection{Phase rotation linac}

The pions, and the muons into which they decay, have a momentum distribution with an rms spread of approximately 100\% and a peak at about 200~MeV/$c$. 
It would be difficult to handle such a wide
spread in any subsequent system. A linac is thus introduced along the decay
channel, with frequencies and phases chosen to decelerate the fast particles
and accelerate the slow ones; {\it i.e.}, to phase rotate the muon bunch. 
Several studies have been made of the design of this system, using differing 
ranges of rf frequency, delivering different final muon momenta, and differing 
final bunch lengths. In all cases, muon capture efficiencies of close to 
0.3 muons per proton are obtained. Until the early stages of the ionization 
cooling have been designed, it is not yet possible to choose between them.
Independent of the above choices is a question of the location of the focusing 
solenoid coils and rf cavity design, as discussed below in the 
section~IV.F.
\begin{table*}[htb!]
\begin{center}
{\caption{Parameters of the Lower-Energy Phase Rotation Linacs}
\label{rot}}
\begin{tabular}{cccc}
%\hline\hline
Linac     & Length & Frequency & Gradient  \\
          &  (m)     &   (MHz)     &  (MeV/m)    \\
\hline
1         &  3    &   60     &   5      \\
2         &  29   &   30     &   4       \\
3         &  5    &   60     &   4      \\
4         &  5    &   37     &   4       \\
%\hline\hline
\end{tabular}
\end{center}
\end{table*}
\subsubsection{Lower energy, longer bunch example}

This example captures muons at a mean kinetic energy of 130~MeV.
Table~\ref{rot} gives parameters of the linacs used. 
%It is seen that the 
%frequencies vary between 30 and 60~MHz, and that the overall length is 42~m.
Monte Carlo simulations \cite{ref6}, with the program MUONMC \cite{mcbob},  
were done using pion production calculated by ARC \cite{arc} for a copper target
of 1-cm radius at an angle of 150~mrad. 
A uniform solenoidal field was assumed in
  the phase rotation, and the rf was approximated by a series of kicks. 

\begin{figure*}[hbt!]
\begin{center}
\includegraphics[width=5in]{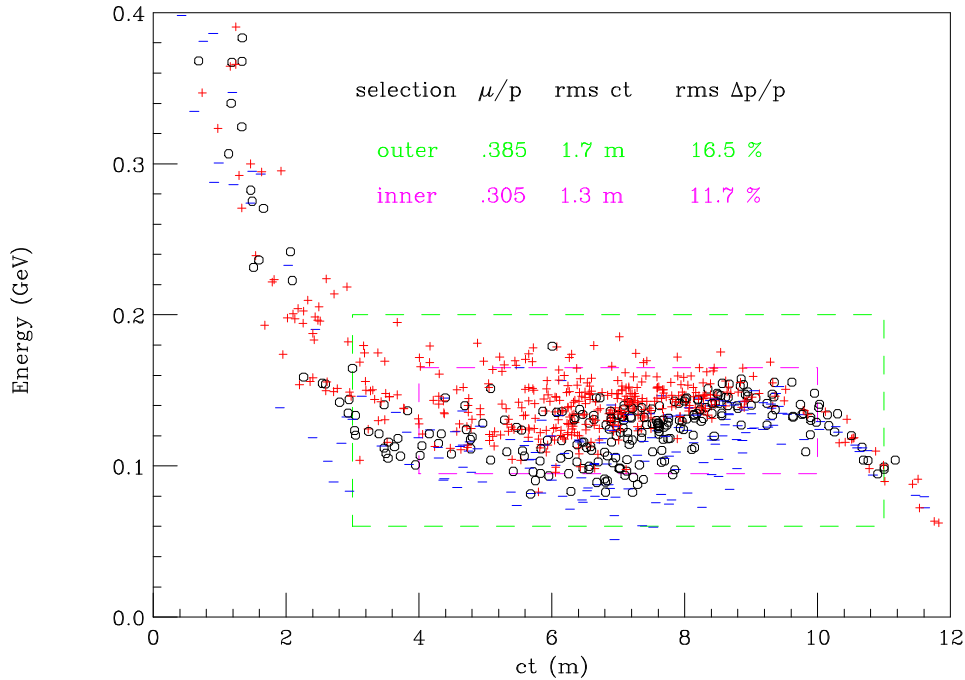} 
\end{center}
\caption[Energy {\it vs.}\ $ct$ of $\mu$'s at end of the lower-energy 
phase rotation channel.  ]
{Energy {\it vs.}\ $ct$ of $\mu$'s at end of the lower-energy
phase rotation channel.  The symbols +, o and $-$ denote muons with 
polarization $P>{1\over 3},\ -{1\over 3} < P < {1\over 3}$ and 
$P <-{1\over 3}$, respectively.}
 \label{Evsctpol2}
 \end{figure*}

 Figure~\ref{Evsctpol2} shows the energy {\it vs.}\ $ct$ at the end 
of the decay and phase rotation channel. 
The abscissa $ct$ is a measure of bunch length at the end of the
channel: the total transit time of each $\pi/\mu$ is multiplied by the velocity
of light and the total length of the channel is subtracted. Thus a ficticious reference particle at 
the center of the incident bunch at the target arrives at $ct= 0$~m.
A loose final bunch selection was defined with an energy $130 \pm 70$~MeV and 
bunch $ct$ from 3 to 
11~m. With this selection, the rms energy spread is 16.5\%, the rms $ct$
 is 1.7~m, and there are 0.39~muons per incident proton.
A tighter selection with an energy $130 \pm 35$~MeV and bunch $ct$ from 4 to 
10~m gave an rms energy spread of 11.7\%, rms $ct$ of 1.3~m, and contained 
0.31~muons per incident proton.

\subsubsection{Higher energy, shorter bunch example}

\begin{figure*}[thb!]
\centerline{\epsfig{file=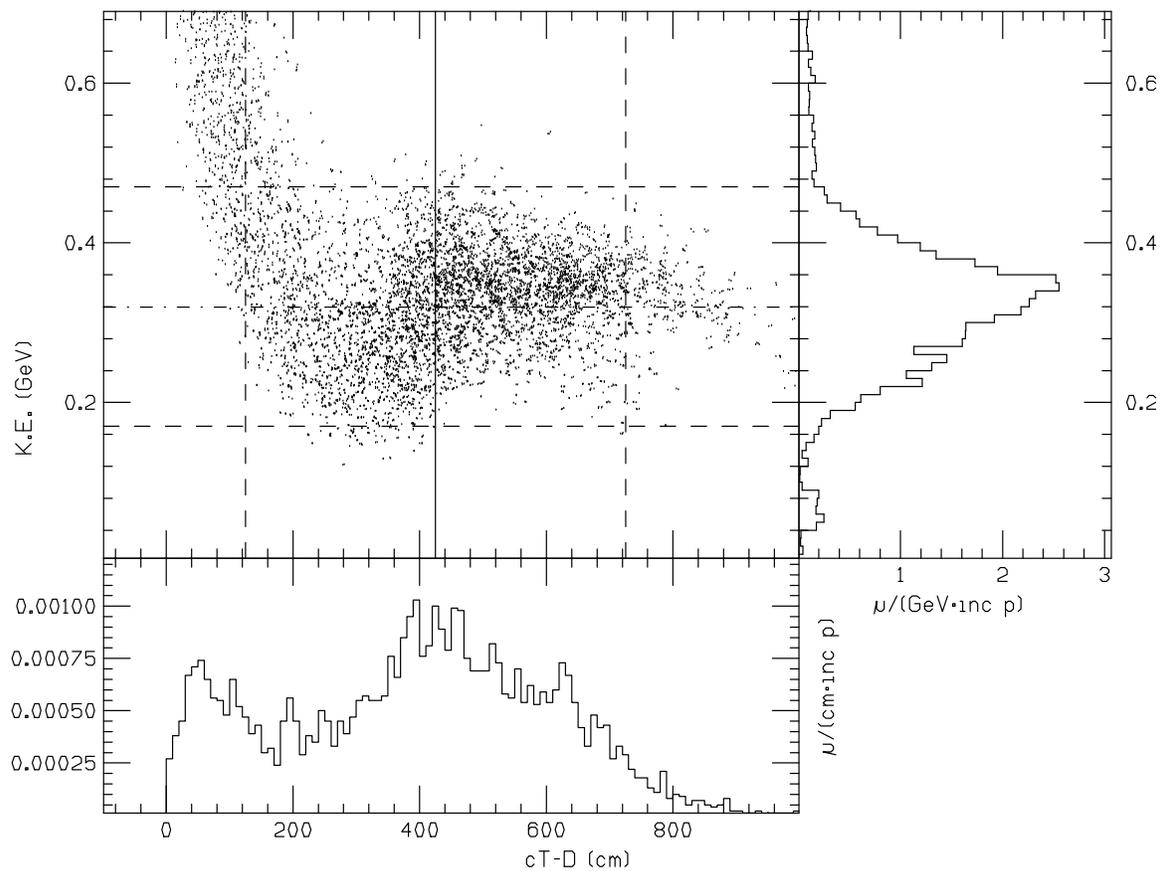,height=6.in,width=4.5in, angle=90}}
\vspace{0.5cm}
\caption[Longitudinal phase space at the end of decay channel with projections 
onto time and energy axes ]
{Longitudinal phase space at the end of decay channel with projections 
onto time and energy axes.  The four dashed lines delineate the
region deemed acceptable for the cooling channel. }
 \label{neu_van}
 \end{figure*}

  In this example the captured muons have a mean kinetic energy close to
  320~MeV.  It is based on a Monte Carlo study which uses 
the updated
  MARS pion production model \cite{ref16} to generate pions created by 16~GeV 
protons
  on a 36~cm long, 1~cm radius coaxial gallium target. Figure~\ref{neu_van} 
shows the longitudinal phase space of the muons at the end of an 80~m long, 5~T solenoidal decay channel with
  cavities of frequency in the 30-90~MHz range and acceleration gradients of
  4-18~MeV/m.  A total of 0.33~muons per proton fall within the indicated
  cut (6~m$ \times $300~MeV).  The rms bunch length inside the cut is 148~cm
  and rms energy spread is 62~MeV.  The normalized six dimensional (6-D) emittance is 217~cm$^3$
  and the transverse part is 1.86~cm (the normalized 6-D emittance $\epsilon_{6,N}$ is defined in section \textbf{V}).

 A sample simulation with lithium hydride absorbers regularly spaced in the 
  last 60~m of a 120~m decay channel and with compensating acceleration 
  captures 0.3~muons with mean kinetic energy of about 380~MeV in a
  (6~m$\times $300~MeV) window. 
The longitudinal phase space is about the same as in the
  previous example but the transverse part shrinks to 0.95~cm due to ionization
  cooling which reduces the 6-D phase space to 73.5~cm$^3.$

\subsection{Use of both signs}

Protons on the target produce pions of both signs, and a solenoid will
capture both, but the subsequent rf systems will have opposite effects
on each sign. The proposed baseline approach uses two separate proton
bunches to create separate positive and negative pion bunches and accepts
the loss of half the pions/muons during phase rotation. 

If the pions can be
charge separated with limited loss before the phase rotation cavities
are reached, then higher luminosity may be obtained. 
The separation of charged pions in a curved solenoid decay line was
studied in \cite{tar-snake}. Because
of the resulting dispersion in a bent solenoid, an initial beam
of radius $R$  with maximum-to-minimum momentum ratio $F$ will
require a large beampipe of radius $(1+F)R$ downstream to accommodate the
separated beams. A septum can then be used to capture the two beams into
separate channels. Typically the reduction in yield for a curved solenoid
compared  to a straight solenoid is about 25\% (due to the loss of
very low and very high momentum pions to the walls or septum), but this
must be weighed against the fact that both charge signs are captured for
an overall net gain. A disadvantage is that this charge separation takes
place over several meters of length during which time the beam spreads
longitudinally. This makes capture in an rf phase rotation system
difficult, although a large aperture cavity system could be incorporated
in the bent solenoid region to  alleviate this. The technique deserves
further study and may be useful to consider as an intensity upgrade to
a muon collection system.
  
\subsection{Solenoids and rf}

\begin{figure*}[hbt!]
\begin{center}
\includegraphics[height=5in,angle=-90]{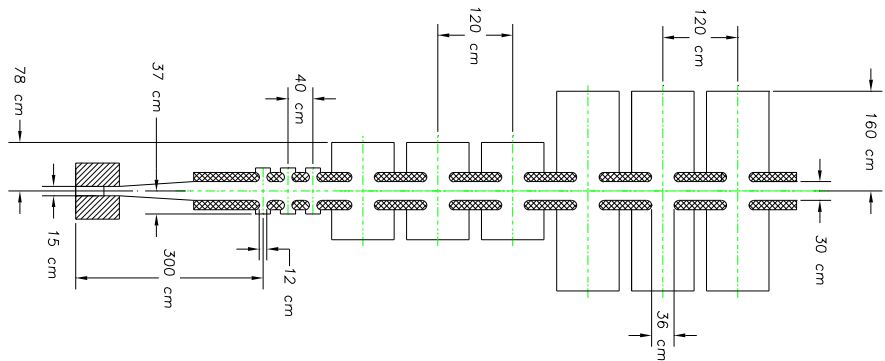}
\end{center}
\caption[Schematic of capture and phase rotation using rf cavities ]
{Schematic of capture and phase rotation using rf cavities with 
superconducting solenoids (hatched) inside the irises. Three groups of three cavities operating at 90, 50, and 30~MHz are shown from left to right, respectively.}
 \label{reentrant}
 \end{figure*}

As noted above, capture using higher frequencies appears to be less efficient, 
and most studies now use frequencies down to 30~MHz. Such cavities, when 
conventionally designed, are very large (about 6.6~m diameter). 
In the Snowmass study \cite{snowp220} a reentrant design reduced this diameter 
to 2.52~m, but this is still large, and it was first assumed that the 
5~T focusing solenoids would, for economic reasons, be placed within the 
irises of the cavities (see Fig.~\ref{reentrant}).

A study of transmission down a realistic system of iris located 
coils revealed betatron resonant excitation from the magnetic field 
periodicities, leading to significant particle loss. This was reduced by the 
use of more complicated coil shapes \cite{snowp220}, smaller gaps, and 
shorter cavities, but remained a problem. 

An alternative is to place continuous focusing coils outside the cavities 
 as shown in Fig.~\ref{capture}. In this case, cost will be minimized with 
lower magnetic fields (1.25-2.5~T) and correspondingly larger decay channel 
radii (21-30~cm). Studies are underway to determine the optimal solution.

\subsection{Polarization}

Polarization of the muon beams presents a significant physics
advantage over the 
unpolarized case, since signal and background of electroweak processes usually 
come predominantly from different polarization states. 

\subsubsection{Polarized muon production}

 In the center of mass of a decaying pion, the outgoing muon is fully 
polarized ($P = -1$ for $\mu^+$ and +1 for $\mu^-$). In the lab system the 
polarization depends on the decay angle $\theta_d$ and initial pion energy \cite{ref19,ref19a,ref19b}.  
For pion kinetic energy larger than the pion mass, the 
average polarization is about 20\%, and if nothing else is done, the 
polarization of the captured muons after the phase rotation system is 
approximately this value.

\begin{figure}[bht!] 
\centerline{
\epsfig{file=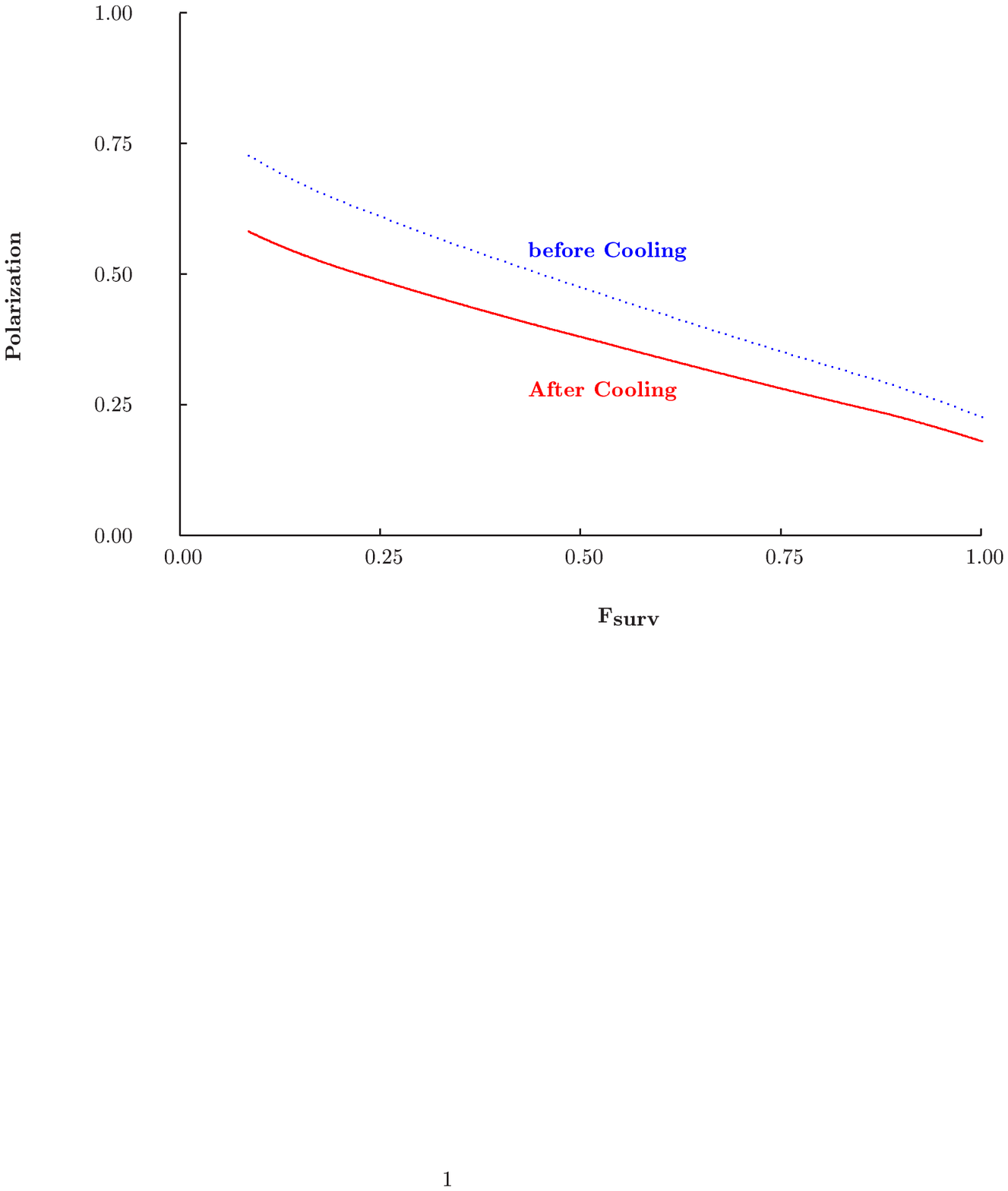,height=4.in,width=4.in}
}
\caption{Polarization {\it vs.}\ fraction $F_{\rm surv}$ of $\mu$'s accepted. 
\label{polvscutnew}}
\end{figure}

If higher polarization is required, some selection of muons from forward pion
decays  $(\cos{\theta_d} \rightarrow 1)$ is required. Figure~\ref{Evsctpol2}, 
above, showed the polarization of the phase rotated muons. The polarization 
\{P$>{1\over 3}$,
$-{1\over 3}< P<{1\over 3}$, and P$<-{1\over 3}$\} is marked by the symbols
$\mathbf{+,\,o\,}$ and $\mathbf{-}$ respectively. 
If a selection is made on the minimum energy of the muons, then greater 
polarization is obtained. The tighter the cut, the higher the 
polarization, but the less the fraction $F_{\rm surv}$  of muons that survive. 
 Figure~\ref{polvscutnew} gives the results of a Monte Carlo study.

If this selection is made on both beams, and if the proton bunch intensity is 
maintained, then each muon bunch is reduced by the factor $F_{\rm surv}$ 
and the 
luminosity would fall by $F_{\rm surv}^2$. But if, instead, proton bunches are 
merged so as to obtain half as many bunches with twice the intensity, then the 
muon bunch intensity is maintained and the luminosity (and repetition rate)  
falls only as $F_{\rm surv}$.

The luminosity could be maintained at the full 
unpolarized value if the proton source intensity could be increased. Such an 
increase in proton source intensity in the unpolarized case might be 
impractical because of the resultant excessive high energy muon beam power, 
but this restriction does not apply if the increase is used to offset losses 
in generating polarization. 

Thus, the goal of high muon beam polarization may shift the parameters of
the muon collider towards lower repetition rate and higher peak currents
in the proton driver.

\subsubsection{Polarization preservation}

The preservation of muon polarization has been discussed in some detail in \cite{ref20}.
During the ionization cooling process the muons lose energy in material and 
have a spin-flip probability $\mathcal{P},$
\begin{equation}
{\mathcal {P} }\approx \int {m_e\over m_{\mu}}\beta_v^2 {\delta E \over E}\label{pol}, 
\end{equation} 
where $\delta E/E$ is the fractional 
loss of energy due to ionization. In our case, the integrated energy loss 
is approximately 3~GeV and the typical energy is 150~MeV, so the integrated 
spin-flip probability is close to 10\%. The change in polarization 
$\delta \mathcal {P}/\mathcal {P}$ is 
twice the spin-flip probability, so the reduction in polarization is 
approximately 20\%.  This loss is included in Fig.~\ref{polvscutnew}.

   During circulation in any ring, the muon spin, if initially longitudinal, 
will precess by $\gamma(g-2)/2$ turns per revolution, where
$(g-2)/2$ is $1.166\times 10^{-3}$. A given energy spread 
${\Delta \gamma/\gamma}$ will 
introduce variations in these precessions and cause dilution of the 
polarization. 
But if the particles remain in the ring for an exact integer number of 
synchrotron 
oscillations, then their individual average $\gamma$'s will be the same and no 
dilution will occur. 

   In the collider,
 bending can be performed with the spin orientation in the vertical 
direction, and the spin rotated into 
the longitudinal direction only for the interaction region. The design of such 
spin rotators appears relatively straightforward, but long. This might be a 
preferred solution at high energies but is not practical in the 100~GeV 
machine. An alternative is to use such a small energy spread, as in the 
Higgs factory, that although the polarization vector precesses, the beam 
polarization  does not become significantly diluted. In addition, calibration of the Higgs factory collider energy to 1 part in a million \cite{ref7} requires the spins to precess continuously from turn to turn.

\subsection{R\&D program}

\begin{figure*}[tbh!]
\begin{center}
\includegraphics[ width=5in]
      {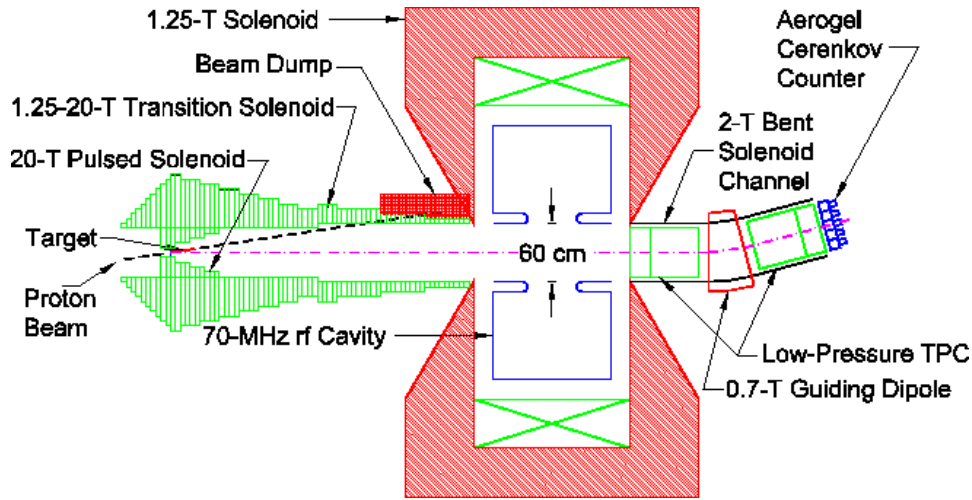}
\caption{Plan view of the full configuration of the targetry experiment.
\label{targetexpt}}
\end{center}
\end{figure*}

An R\&D program is underway to continue theoretical studies (optimization
of pion production and capture) and to clarify 
several critical issues related to targetry and phase rotation
\cite{targetprop}.  A jet of the room temperature 
eutectic liquid alloy of Ga-Sn will be exposed to nanosecond pulses of 
$1.5 \times 10^{13}$ 24~GeV protons at the Brookhaven AGS 
to study the
effect of the resulting pressure wave on the liquid.  The same jet will
also be used in conjunction with a 20~T, 20~cm bore resistive magnet at the
National High Magnetic Field Laboratory (Tallahassee, FL) to study the
effect of eddy currents on jet propagation.  Then,
a pulsed, 20~T magnet will be added to the
BNL test station to explore the full configuration of jet, magnet and
pulsed proton beam.  Also, a 70~MHz rf cavity will be exposed to the
intense flux of secondary particles downstream of the target and 20~T magnet
to determine viable operating parameters for the first phase rotation cavity.
The complete configuration of the targetry experiment is sketched in
Fig.~\ref{targetexpt}.

The first two studies should be accomplished during 1999, and the third and
fourth in the years 2000/01.
\section{IONIZATION COOLING}
\label{subsec-compcool}
\subsection{Introduction}

The design of an efficient and practical cooling system is one of the major challenges for the muon
collider project.

For a high luminosity collider, the 6-D phase space volume occupied by the muon
beam  must be reduced by a factor of $10^5 - 10^6.$ Furthermore, this phase space reduction must
be done within a time that is not long compared to the muon lifetime ($\mu$ lifetime $\approx 2~\mu
s$). Cooling by synchrotron radiation, conventional stochastic
cooling and conventional electron cooling are all too slow. Optical stochastic
cooling \cite{ref21}, electron cooling in a plasma discharge \cite{ref22}, and
cooling in a crystal lattice \cite{ref23,ref23a} are being studied, but appear
technologically difficult. The new method proposed for cooling muons is ionization cooling. This
technique \cite{Budker78,ref1a,ref1,ref24} is uniquely applicable to muons because
of their minimal interaction with matter. It is a method that seems relatively straightforward in
principle, but has proven quite challenging to implement in practice

Ionization cooling involves passing the beam through some 
material in which the muons lose both transverse and longitudinal momentum 
by ionization energy loss, commonly referred to as $dE/dx.$ The longitudinal muon 
momentum is then restored by  reacceleration, leaving a net loss of
transverse momentum (transverse cooling). The process is repeated 
many times to achieve a large cooling factor. 

The energy spread can be reduced by introducing a transverse variation in the absorber density or
thickness (e.g.\ a wedge) 
at a location where there is dispersion (the transverse position is energy dependent). This method
results in a corresponding increase of transverse phase space and is thus an exchange of longitudinal
and transverse emittances. With transverse cooling, this allows cooling in all dimensions.
 
We define the root mean square rms normalized emittance as
\begin{equation}
\epsilon_{i,N}=\sqrt{\langle\delta r_i^2\rangle \langle\delta p_i^2\rangle-\langle\delta r_i\delta p_i\rangle^2}/m_{\mu}c
\end{equation}
where $r_i$ and $p_i$ are the beam canonical conjugate variables with $i=1,2,3$ denoting the x, y
and z directions, and $\langle...\rangle$ indicates statistical averaging over the particles. The operator $\delta$
denotes the deviation from the average, so that $\delta r_i=r_i-\langle r_i\rangle$ and likewise for $\delta p_i.$ 
  The appropriate figure of merit for cooling is the final value of the 6-D relativistically invariant
emittance $\epsilon_{6,N},$ which is proportional to the area in the 6-D phase space $(x,y,z,p_x,p_y,p_z)$ since,
to a fairly good approximation, it is preserved during 
acceleration and storage in the collider ring. This quantity is the square root of the determinant of
a general quadratic moment matrix containing all possible correlations. However, until the nature
and practical implications of these correlations are understood, it is more conservative to ignore the
correlations and use the following simplified expression for 6-D normalized emittance,
\begin{equation}
\epsilon_{6,N}\approx  \epsilon_{x,N}\times \epsilon_{y,N}\times \epsilon_{z,N} %<<<
%\left(\sigma_z^{rms}\sigma_E^{rms}\gamma \beta\right) >>>
\end{equation}
Theoretical studies have shown that, assuming realistic parameters for the 
cooling hardware, ionization cooling can be expected to 
reduce the phase space volume 
occupied by the initial muon beam by a factor of $10^5$ -- $10^6$. 
A complete
cooling channel would
consist of 20 -- 30 cooling stages, each stage yielding about a factor of 
two in 6-D phase space reduction.

 It is
recognized that the feasibility of constructing a muon
ionization cooling channel is on the critical path to understanding the
viability of the whole muon collider concept.
The muon cooling channel is the most novel part of a muon collider complex.
Steady progress has been made both in improving the design of sections of the channel
and in adding detail to the computer simulations. A vigorous experimental program is needed to
verify and benchmark the computer simulations.

The following parts of this section briefly describe the physics
underling the process of ionization cooling. We will show results of simulations for some chosen
examples, and outline a six year R\&D program to demonstrate the feasibility of using ionization
cooling techniques. 

\subsection{Cooling theory}

In ionization cooling, the beam loses both transverse and longitudinal momentum
as it passes through a material. At the same time its emittance is increased due to stochastic multiple
scattering and Landau straggling. The longitudinal
momentum can be restored by reacceleration, leaving a net loss of
transverse momentum. 

The approximate equation for transverse cooling in a step $ds$ along the particle's orbit  is \cite{ref2b,ref1a,ref1,ref2a,ref24b,ref24c}
  \begin{equation}
\frac{d\epsilon_N}{ds} = -{1\over \beta^2}\frac{dE_{\mu}}{ds}\ \frac{\epsilon_N}{E_{\mu}} +
 \frac{\beta_{\perp} (0.014 GeV)^2}{2\beta^3 E_{\mu}m_{\mu}\ L_R}, \label{eq1}
  \end{equation}
where $\beta$ is the normalized velocity, $E_{\mu}$ is the total energy,
$m_{\mu}$ is the muon mass,   $\epsilon_N$ is the normalized transverse emittance,
$\beta_{\perp}$ is the betatron
function at the absorber, $dE_{\mu}/ds$ is the energy loss per unit length, and $L_R$  is the
radiation length of the material.  The betatron function is determined by the strengths of the elements
in the focusing lattice \cite{pdgaccel}. Together with the beam emittance this function
determines the local size and divergence of the beam. Note that the energy loss $dE_{\mu}/ds$ is defined here
as a positive quantity, unlike the convention often used in particle physics. The first term in this
equation is the 
cooling term, and the second describes the heating due to multiple scattering.
The heating term is minimized if $\beta_{\perp}$ is small (strong-focusing)
and $L_R$ is large (a low-Z absorber). 

The minimum, normalized transverse emittance that can be achieved for a given absorber in a given
focusing field is reached when the cooling rate equals the heating rate in 
Eq.~\ref{eq1}
\begin{equation}\epsilon_{N,min} = {\beta_{\perp} (14 MeV)^2 \over
 2 \beta m_{\mu} {dE_{\mu} \over ds} L_R }\label{equi}
\end{equation}
For a relativistic muon in liquid hydrogen with a betatron focusing value of 8 cm, which corresponds
roughly to confinement in a 15~T solenoidal field, the minimum achievable emittance is about 340~mm-mrad.

The equation for energy spread  is \cite{ref1a,ref2a,ref2c}
 \begin{equation}
{\frac{d(\Delta E_\mu)^2}{ds}}\ =
-2\ {\frac{d\left( {\frac{dE_\mu}{ds}} \right)} {dE_\mu}}
 \langle(\Delta E_{\mu})^2 \rangle\ +
{\frac{d(\Delta E_{\mu})^2_{{\rm stragg.}}}{ds}}\label{eq2}
 \end{equation}
where the first term describes the cooling (or heating) due to energy loss, 
and the second term describes the heating due to straggling. $\Delta E_{\mu}$ is the rms spread
in the energy of the beam. 

Ionization cooling of muons seems relatively straightforward in theory, but
will require extensive simulation studies and hardware development for its
optimization. There are practical problems in designing lattices that can
transport and focus the large emittance  beam. There will also be effects
from space charge and wake fields.

We have developed a number of tools for studying the ionization cooling
process. First, the basic theory was used to identify the most promising
beam properties, material type and focusing arrangements for cooling. Given
the practical limits on magnetic field strengths, this gives an estimate of
the minimum achievable emittance for a given configuration.
 Next, the differential equations for cooling and heating described above were incorporated into a computer code. Allowance for the shifts in the
betatron phase advance due to space charge and aberrations was included.
This code was used to develop an overall cooling scenario, which broke the
cooling system into a number of stages, and determined the properties of the
beam, radio frequency (rf) cavities, and focusing lattice at each stage.

Finally, several tracking codes were either written or modified to study
the cooling process in detail. Two new codes (SIMUCOOL \cite{van}, and ICOOL \cite{ref25})
use Monte Carlo techniques to track particles one at a time through the
cooling system. All the codes attempt to include all relevant physical
processes to some degree, (e.g. energy loss, straggling, multiple scattering)
and use Maxwellian models of the focusing fields. They do not yet  
take into account any space charge or wake field effects.  In addition, we
have also used a modified version of PARMELA \cite{parmela} for tracking, which does
include space charge effects, and a double precision version of GEANT \cite{ref40,paul}.

We have recently developed \cite{envelope} a model of beam cooling based on a second
order moment expansion. A computer code solving the equations for
transverse cooling gives results that agree with tracking codes. 
 The code is being extended to include energy spread and bends. It
is very fast and is appropriate for preliminary design and optimization of
the cooling channel. 
 All of these codes are actively being
updated and optimized for studying the cooling problem.

\subsection{Cooling system}

The cooling is obtained in a series of cooling stages. Each stage
consists of a succession of the following components:
\begin{enumerate}
 \item Transverse cooling sections using materials in a strong focusing (low $\beta_\perp$)
environment alternated with linear accelerators. 
 \item Emittance exchange in lattices that generate dispersion, with absorbing  wedges to reduce
momentum spread.
\item Matching sections to optimize the transmission and cooling parameters of the following
section.
 \end{enumerate}

In the examples that follow it is seen that each such stage lowers the 6-D emittance by a
factor of about 2.
Since the required total 6-D cooling is $O(10^6)$, about 20 such stages are required.
The total length of the system would be of the order of 600 m, and the total acceleration required
would be approximately
6 GeV. The fraction of muons remaining at the end of the cooling system
is estimated to be $\approx 60\%$.

The baseline solution for transverse cooling involves the use of liquid hydrogen absorbers in strong
solenoid focusing fields,  interleaved with short linac sections. The solenoidal fields in successive
absorbers are reversed to avoid build up of the canonical angular momentum \cite{angmom}. 
The focusing magnetic fields are small ($\approx$ 1~T) in the early stages where the emittances are
large, but must increase as the emittance falls.

Three transverse cooling examples have been designed and simulated. The first uses 1.25~T
solenoids to cool the very large emittance beam coming from the phase rotation channel. The muon beam at the end of the decay channel is very intense,
with approximately $7.5\times 10^{12}$ muons/bunch, but with a large normalized  transverse
emittance ($\epsilon_{x,N}(\textrm{rms})\approx 15\times 10^3$~mm-mrad) and a large normalized
longitudinal emittance ($\epsilon_{z,N}(\textrm{rms})\approx 612$~mm).
  The second
would lie toward the end of a full cooling sequence and uses 15~T solenoids. The third, using 31~T
solenoids, meets the requirements for the Higgs factory and could be the final 
cooling stage for this machine.

The baseline solution for emittance exchange involves the use of bent solenoids to generate
dispersion and wedges of hydrogen or LiH to reduce the energy spread. A simulated example is
given for exchange that would be needed after the 15~T transverse cooling case.

A lithium lens solution may prove more economical for the final stages, and might allow even lower
emittances to be obtained.
In this case, the lithium lens serves simultaneously to maintain the low
$\beta_{\perp}$, and provide $dE/dx$ for cooling. Similar lenses, with
surface fields of $10\,$T, were developed at Novosibirsk (BINP) and have been used, at low
repetition rates, as focusing elements at FNAL and CERN
\cite{reviewtev,ref26,ref26a,ref26b,ref27}. Lenses for the cooling application, which would operate
at 15~Hz, would need to employ flowing liquid lithium to provide adequate thermal cooling. Higher
surface fields would also be desirable.

Studies have simulated cooling in multiple lithium lenses, and have shown cooling through several
orders of magnitude \cite{Balbekov96}. But these studies have, so far, used
ideal matching and acceleration.
Cooling is also being studied in beam recirculators, which could lead to reduction of costs of the
cooling section \cite{balbekov,balbekov1}, but  full simulations with
all higher order effects have not yet been successfully demonstrated.

\subsection{15~T solenoid transverse cooling example}
The lattice consists of 11 identical 2~m long \textit{cells}. In each cell there is a liquid hydrogen
absorber (64~cm long, 10~cm diameter) in the 15~T solenoid focusing magnet (64~cm long, 12~cm
diameter). The direction of the fields in the magnets alternates 
from one cell to the next. 
Between the 15~T solenoids there are magnetic matching sections (1.3~m long, 32~cm inside
diameter) where the field is lowered and then reversed. Inside the matching sections are short,
805~MHz, high gradient (36~MeV/m) linacs.
\begin{figure*}[thb!]
\centerline{\epsfig{file=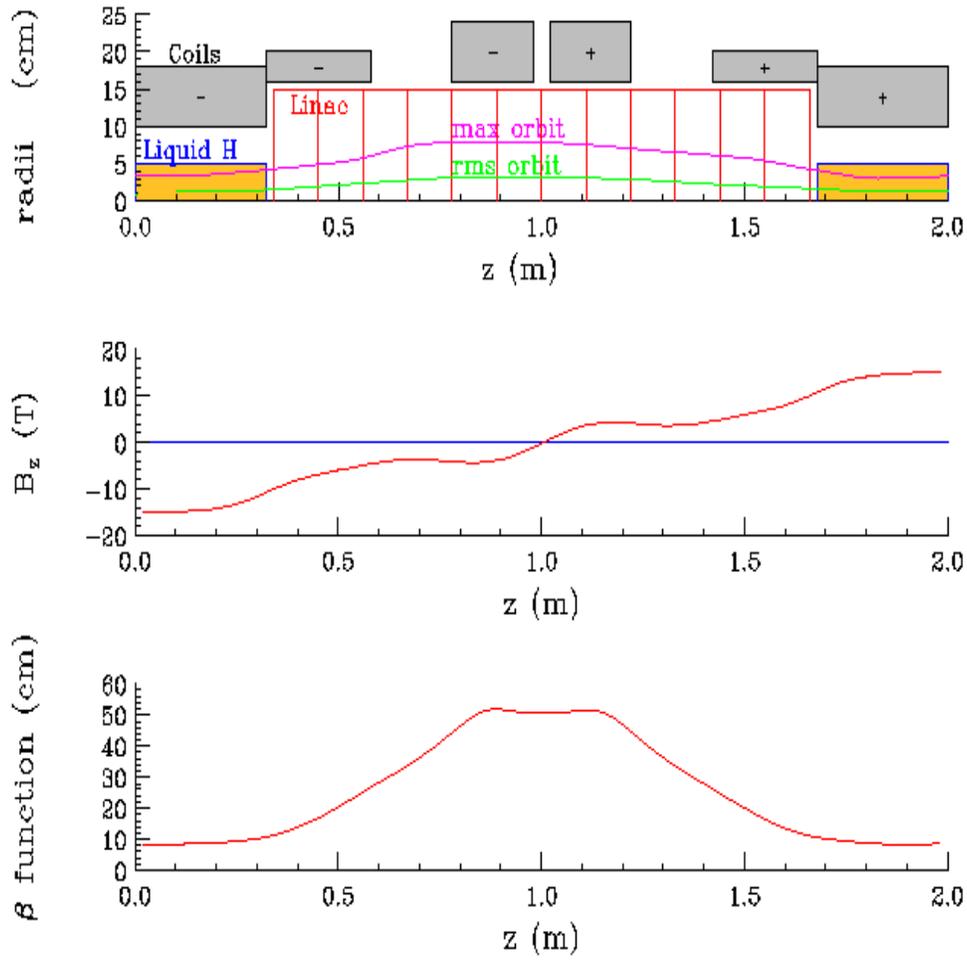,height=5.0in,width=5.0in}}
\caption[Cross section of one half period of an alternating solenoid cooling lattice ]{(a) Cross section of one half period of an alternating solenoid cooling lattice; (b) axial magnetic field \vs~z;
(c) $\beta_{\perp} $~function \vs~z.  }
\label{altsol}
 \end{figure*} 
Figure~\ref{altsol} shows the cross section of one cell of such a system, together with the betatron
function, and the magnetic field along the axis. For
convenience in modeling, the section shown in Fig.~\ref{altsol}(a) starts and ends symmetrically
in the middle of hydrogen absorber regions at the location of
the peaks in the axial magnetic field. In practice each cell would start at
the beginning of the hydrogen region and extend to the end of the rf module.

\begin{figure*}[thb!]
\centerline{\epsfig{file=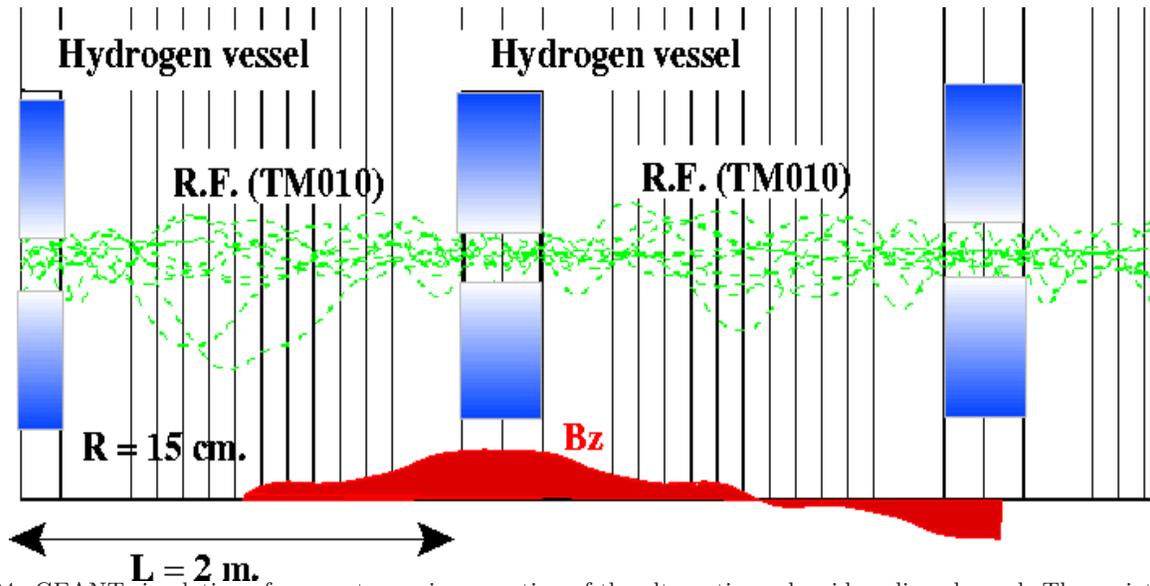,height=3.0in,width=6.0in,clip=}}
\caption[GEANT simulation of muons traversing a section of cooling channel]{GEANT simulation
of muons traversing a section of the alternating solenoid cooling channel. The variation of the
magnetic field $B_z$ is shown for $1{1\over 2}$ cells of the figure.}
 \label{dpgeant}
 \end{figure*}

A GEANT simulation of muons traversing a section of the cooling channel is shown in Fig.~\ref{dpgeant}.

Additional simulations were performed \cite{angmom,aac_paper} using the program ICOOL. The only likely significant effects which are not yet included 
are space charge and wakefields. Analytic calculations for particle bunches in free space indicate that these effects should, for the later stages, be significant but not overwhelming.
A full simulation must
be done before we are assured that no problems exist. Particles are introduced with transverse and
longitudinal emittance (186~MeV/$c,$ 1400~$\pi$~mm-mrad transverse, and
1100~$\pi$~mm~longitudinal), together with a number of naturally occurring correlations. Firstly,
the particles are given the angular momentum appropriate for the starting axial magnetic field.
Secondly, particles with large initial radius $r_o$ and/or divergence $\theta_o$ have longer
pathlengths in a solenoidal field and tend to spread out with time. This can be parameterized by
defining an initial transverse amplitude
\begin{equation}
A^2={r_o^2\over \beta_{\perp}^2 }+\theta_o^2.
\end{equation}
The temporal spreading can be minimized by introducing an initial correlation between $p_z$ and
$A^2$ that equalizes the forward velocity of the initial particles. This correlation causes the average
momentum of the beam to grow from the reference value of 186~MeV/$c$ to $\approx 195$~MeV/$c.$
Lastly, a distortion of the longitudinal bunch distribution can be introduced to reflect the asymmetric
nature of the ``alpha"-shaped rf bucket.

\begin{figure*}[htb!]
\centerline{\epsfig{file=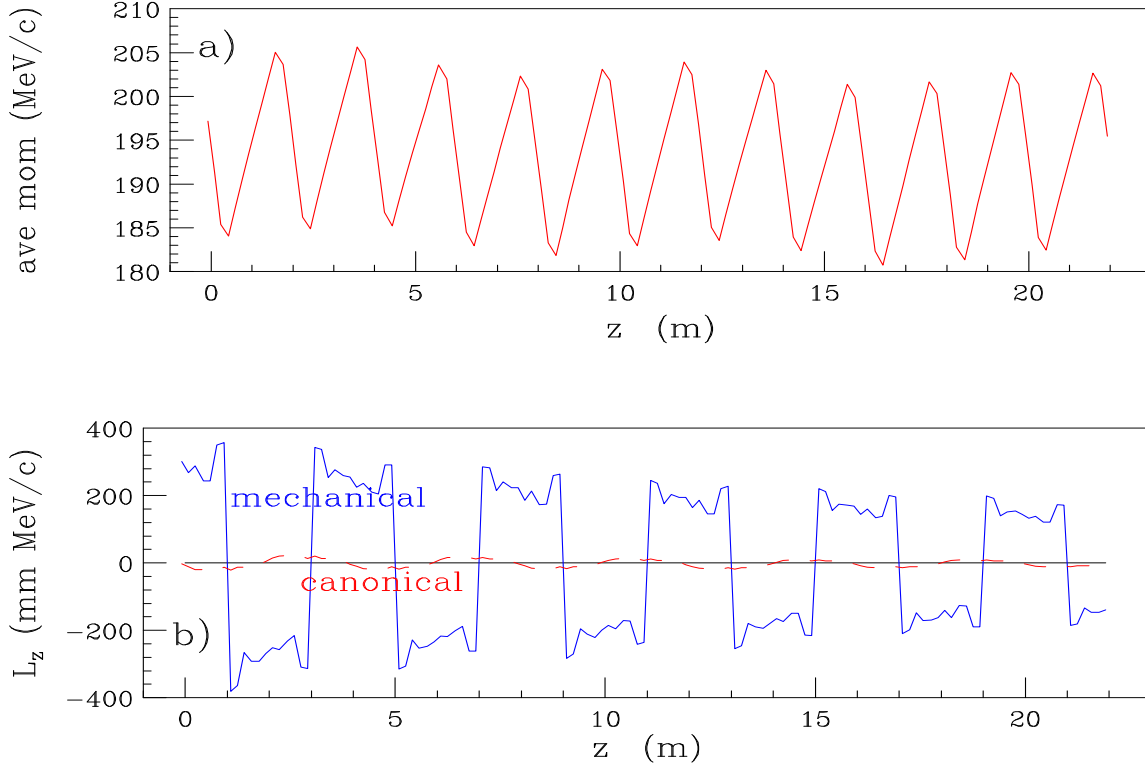,height=4.0in,width=6.0in}}
\caption[Average momentum  \textit{vs.} z  ]{a) Average momentum  \textit{vs.} z; b) Average
angular momentum: mechanical (solid curve) and canonical (dashed curve), \textit{vs.} z.}
 \label{simu8}
 \end{figure*}

Figure~\ref{simu8}(a) shows the average momentum of the beam as a function of distance along
the channel. The momentum drops as the beam crosses the liquid hydrogen absorbers. The gradient
and phase
 of the rf cavities have been adjusted so that the reacceleration given to
the reference particle 
equals the mean energy loss. This causes the average momentum of the beam to
remain in a
narrow band around 195~MeV/$c.$ Figure~\ref{simu8}(b) shows the mechanical and canonical
angular
momenta as a function of distance along the channel. The mechanical angular
momentum shows
the rotational motion of the beam around the axial solenoidal field. It
periodically reverses sign
when the solenoids alternate direction. The canonical angular momentum is
defined such that it
removes the axial field dependence \cite{aac_paper}. Without the
absorbers, the
beam would have a constant (0) value for the canonical angular momentum.
However, the
presence of absorbers causes the canonical angular momentum to grow and would
lead to severe
emittance growth by the end of a long channel. This growth is stopped by
alternating the direction
of the solenoid field, as shown in Fig.~\ref{simu8}(b) Simulations have shown that 2~m is a
reasonable
(half)  period for the field, since the net growth in canonical angular
momentum is small. In
addition  synchrobetatron resonances are avoided since the periodicity of the field forces the average
betatron wavelength
to be 2~m, whereas the synchrotron oscillation wavelength seen in the simulations for this
arrangement is $\approx 14$~m.

\begin{figure*}[bht!]
\centerline{\epsfig{file=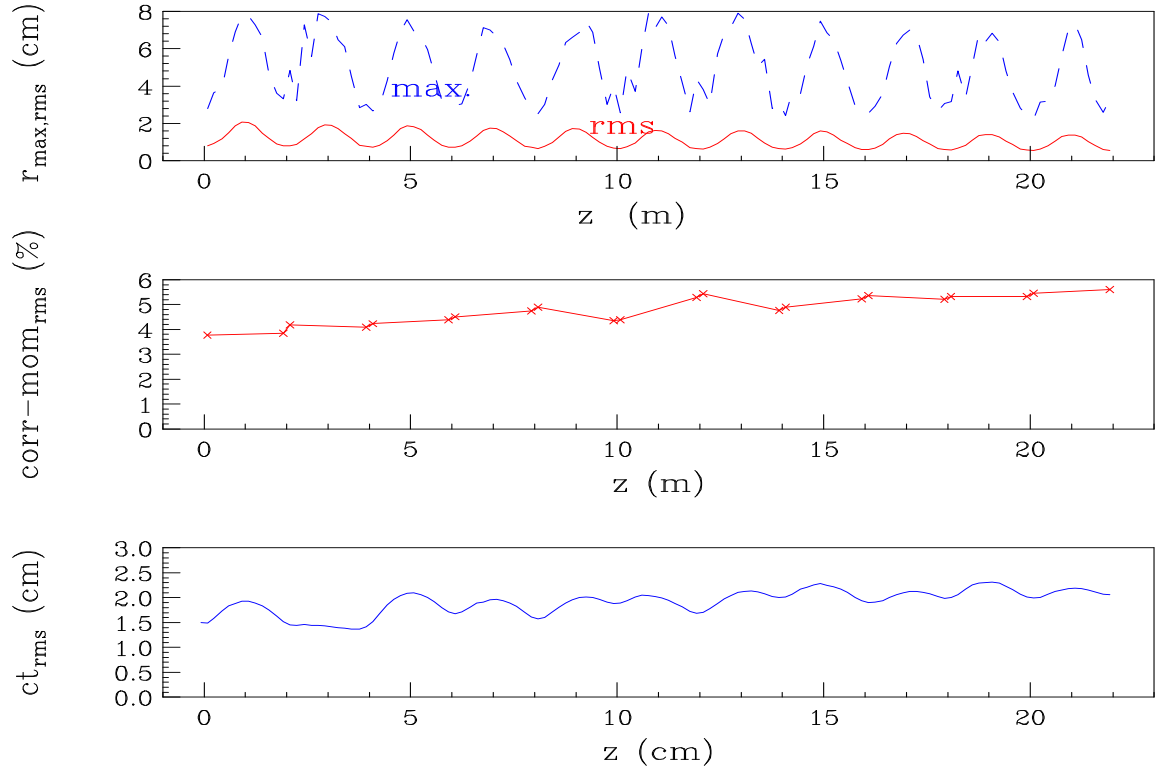,height=4.0in,width=6.0in}}
\caption[rms and maximum beam radii \vs~z ]{a) rms and maximum beam radii, b) rms corrected
momentum, c) rms bunch length; all \vs~z.}
 \label{simu6}
 \end{figure*}

Figure~\ref{simu6}(a) shows the rms and maximum radius of any particle in the beam
distribution as a
function of distance along the channel. The rms radius shows that most of
the beam is confined
to within 2~cm of the axis. The peak rms radius decreases towards the end of
the channel as a
result of the cooling. The maximum particle radius is about 8~cm, which
determines the radius of
the windows required in the rf cavities.  
 Figure~\ref{simu6}(b) shows the rms momentum
spread corrected for
the correlation between $p_z$ and transverse amplitude imposed on the initial
particle distribution.
The momentum spread grows as a function of distance since the alternating
solenoid system only
cools the transverse emittance.
 Figure~\ref{simu6}(c) shows the rms bunch length s a
function of distance
along the channel. Again this grows with distance since this channel does
not cool longitudinally.

\begin{figure*}[bht!]
\centerline{\epsfig{file=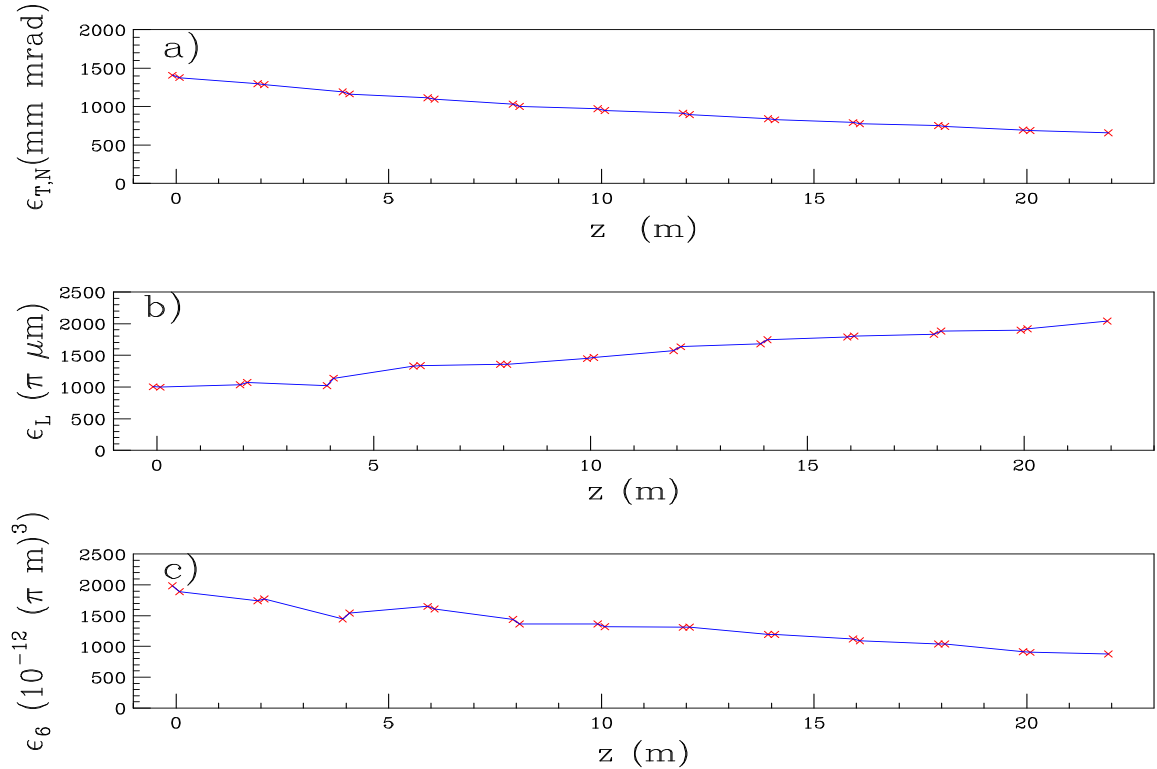,height=4.0in,width=6.0in}}
\caption[Emittance \vs~z  ]{Emittance \vs~z: a) transverse emittance; b) longitudinal emittance; and
c) 6-D emittance.}
 \label{coolingeg}
 \end{figure*}

Figure~\ref{coolingeg}(a) shows the decrease in transverse normalized emittance as a
function of distance along
the channel. The system provides cooling by a factor of $\approx $~2 in both the $x$ and $y$ transverse
phase
spaces. From the changing slope of the curve we note that the rate of cooling is
dropping. This sets $\approx$~22~m as
the maximum useful length for this type of system. It must be followed by a
longitudinal
emittance exchange region to reduce the momentum spread and bunch length
approximately
back to their starting values. Figure~\ref{coolingeg}(b) shows the increase in longitudinal
normalized emittance
in the channel due to the increase in momentum spread and bunch length. Finally,
Fig.~\ref{coolingeg}(c) shows the decrease in the 6-D normalized emittance as a function
of distance along the
channel. There is a net decrease in 6-D emittance by a factor of
$\approx$~2 in the channel. Table~\ref{cooltab} gives the initial and final beam parameters.

\begin{table*}[thb]
\caption{Initial and final beam parameters in a 15 T  transverse cooling section.}
\label{cooltab}
\begin{tabular}{llccc}
    & & initial & final & final/initial\\
\hline
Particles tracked    &        &  1000 & 980 & 0.98 \\
Reference momentum            &  MeV/$c$  &186 & 186 & 1.0\\  
Transverse Emittance & $\pi$ mm-mrad & 1400 & 600 & 0.43 \\
Longitudinal Emittance & $\pi$ mm-mrad    & 1100 & 2300 & 2.09 \\
6-D emittance~$\times 10^{-12}$& ($\pi$~m-rad)$^3$  & 2000 & 800 & 0.40 \\
rms beam radius in hydrogen  & cm            & 0.8 & 0.55 & 0.69 \\
rms beam radius in linac  & cm            & 2.0 & 1.4 & 0.70 \\
max beam radius in linac  & cm            & 7.0 & 7.0 & 1.0 \\
rms bunch length  & cm        & 1.5 & 2.2 & 1.5 \\ 
max bunch full width  & cm        & 13 & 19 & 1.5 \\ 
rms $\Delta p/p$              & \%        & 3.8  & 5.6 & 1.5 \\
\end{tabular} 
\end{table*}

 This simulation has been confirmed, with minor differences, by double precision GEANT \cite{paul} and PARMELA \cite{parmela} codes.

\subsection{31~T solenoid transverse cooling example}

As in the preceding example, the lattice consists of 11 identical 2~m long cells with the direction of the
fields in the solenoids alternating from one cell to the next. The maximum solenoidal field is higher
(31~T) than in the previous example, but the bore is smaller (8~cm), and the liquid hydrogen
absorber has a smaller diameter (6~cm). Between the 31~T solenoids there are 1.3~m long matching
sections with an inside diameter of 32~cm, superimposed on a 36~MeV/m reacceleration linac
operating at 805~MHz.

Table~\ref{cooltab2} gives the initial and final parameters for
 the 31~T example, together with the required emittances for a Higgs factory. In setting these
requirements a dilution of $20\%$ during acceleration  is assumed in each of the three emittances.

\begin{table*}[thb]
\caption{Initial and final beam parameters in a 31~T transverse cooling section.}
\label{cooltab2}
\begin{tabular}{llcccc}
    & & initial & final & final/initial& required\\
\hline
Particles tracked    &               & 4000&3984 &0.99&   \\
Reference momentum       &    MeV/$c$ & 186 & 186 &  \\
Transverse Emittance & $\pi$ mm-mrad & 460 & 240  & 0.52 & 240 \\
Longitudinal Emittance & $\pi$ mm-mrad    & 850 & 1600 & 1.9& \\
6-D emittance~$\times 10^{-12}$ &($\pi$~m-rad)$^3$  & 150 & 95 & 0.63&98 \\
rms beam radius in hydrogen  & cm            & 0.44 & 0.33 & 0.75& \\
rms beam radius in linac  & cm            & .4 & 1.1 & 0.80& \\
max beam radius in linac  & cm            & 6.0 & 6.0 & 1.0& \\
rms bunch length  & cm        & 1.5 & 1.8 & 1.2& \\ 
max bunch full width  & cm        & 11 & 19 & 1.7& \\ 
rms $\Delta p/p $             & \%        & 3.5  & 5.0 & 1.4& \\
\end{tabular} 
\end{table*}

\subsection{Bent solenoid emittance exchange example}

We have been considering using a system that exchanges longitudinal and transverse emittance by exploiting dispersion in a large acceptance channel, with a low-Z wedge absorber in the region of dispersion.

In a bent solenoid, in the absence of any dipole field, there is a drift perpendicular to the bend plane
of the center of the Larmor circular orbit, which is proportional to the particle's momentum \cite{toroid}. In our example we have introduced a uniform dipole field
over the bend to cancel this drift exactly for particles with the reference momentum. Particles with
momenta differing from the reference momentum then spread out spatially, giving the required
dispersion (0.4 m). The dispersion is removed, and the momentum spread reduced, by introducing
liquid hydrogen wedges \cite{neuffer_wedges}. The hydrogen wedges would be
contained by thin beryllium or aluminum foils, but these were not included in this simulation. 

\begin{figure*}[tbh!]
\centerline{\epsfig{file=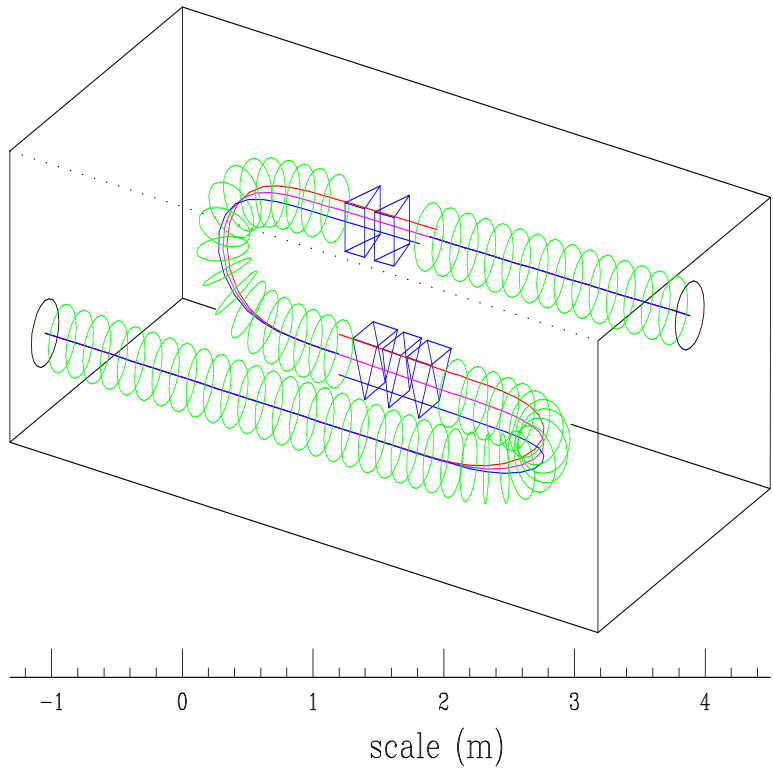,height=4.in,width=4.in}}
\vspace{0.5cm}
\caption{Representation of a bent solenoid longitudinal emittance exchange section}
 \label{cell_exch}
 \end{figure*}

After one bend and one set of wedges, the beam is asymmetric in cross section. Symmetry is restored
by a following bend and wedge system rotated by 90 degrees with respect to the first.
Figure~\ref{cell_exch} shows a representation of the two bends and wedges. The total solenoid
length was 8.5 m. The beam tube outside diameter is 20~cm, and the minimum bend radii is 34~cm. 

\begin{figure*}[bht!]
\centerline{\epsfig{file=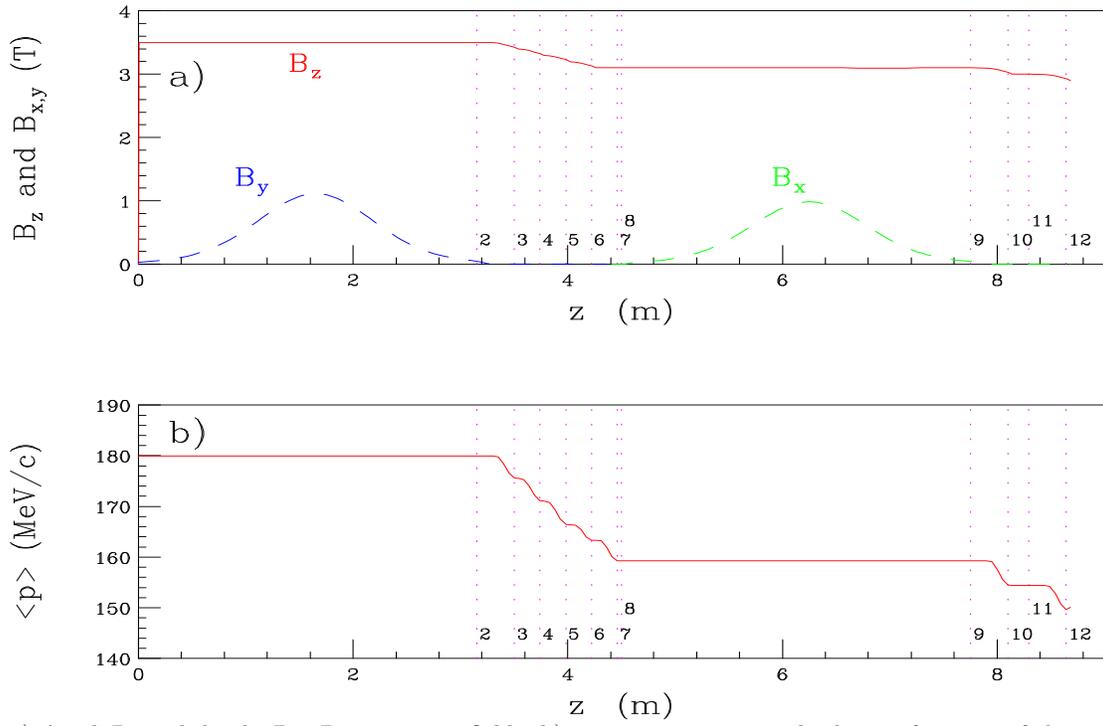,height=3.75in,width=5.75in}}
\caption[Axial $B_z$ and dipole $B_y$, $B_x$, magnetic fields   ]{a) Axial $B_z$ and dipole
$B_y$, $B_x$, magnetic fields; b)
average momentum; both as a function of the position along the cell.}
 \label{simu4}
 \end{figure*}

Figure~\ref{simu4}(a) shows the magnetic fields (B$_z$, B$_y$, and B$_x$) as a function of the
position along the cell. The solenoid bend curvature is exactly that given by the trajectory of a
reference particle (equal in momentum to the average momenta given in Fig.~\ref{simu4}(b)) 
in the given transverse fields. The actual shape of the bend turns out to be very important.
Discontinuities in the bend radius can excite perturbations
which increase the transverse emittance.  We have shown, for example,
that the transverse emittance growth in a bent solenoid depends on
discontinuities of the bend radius as a function of distance, and
its first and second derivatives, the size and tilt of the solenoidal
coils, auxiliary fields and the 6-D phase space of the beam. Thus optimization is not straightforward.
One solution to this problem is to have long, adiabatic bends. However, this adds undesirable length
to the emittance exchange section.
We are studying options with coupling sections to tight bends roughly
half a Larmor length long,  which seems to minimize transverse
emittance growth while also minimizing the length of the section.
Due to similar problems, the length and longitudinal distribution of
the wedge material has also been found to affect emittance growth. For example, 
the growth can be minimized when the vector sum of the Larmor
phases at the absorber elements is small or zero.

\begin{figure*}[htb!]
\centerline{\epsfig{file=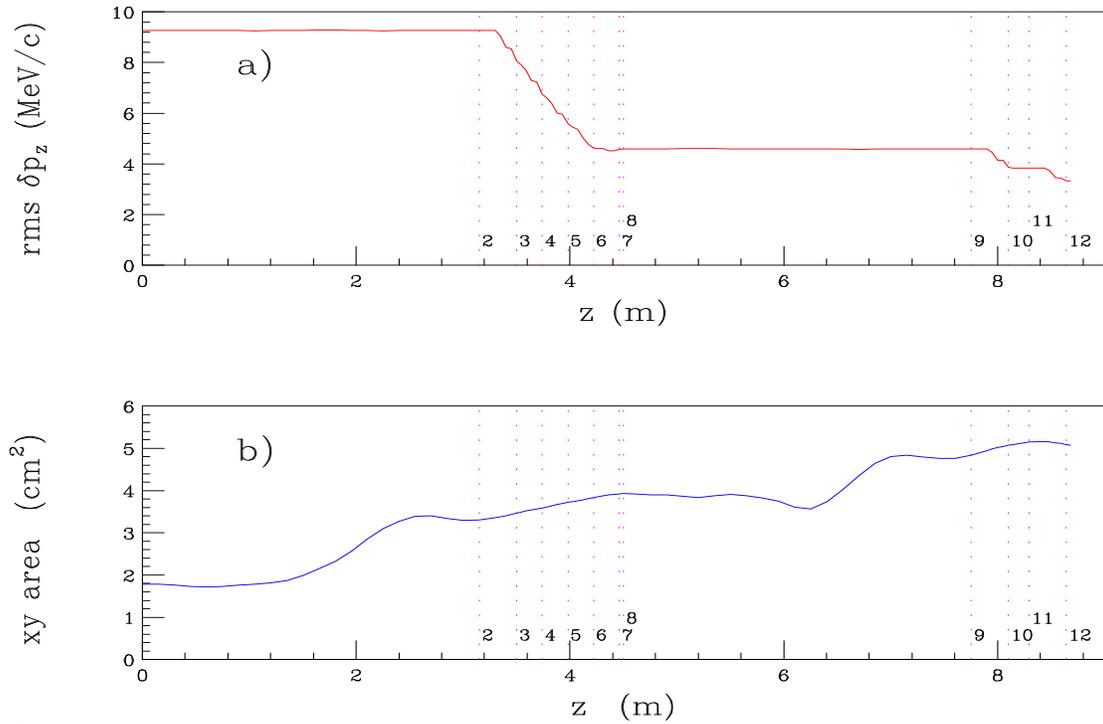,height=3.75in,width=5.75in}}
\caption[rms longitudinal $\delta p_z$ with respect to the reference momentum as a function of z]{a)
rms longitudinal $\delta p_z$ with respect to the reference momentum and b) transverse beam area,
both as a function of z. }
 \label{simu2}
 \end{figure*}

The simulations were performed using the program ICOOL.  The maximum beam
radius is 10~cm. Transmission was 100\%.
Figure~\ref{simu2}(a) shows the rms longitudinal momentum spread relative to the reference
momentum as a function of the position along the cell. The fractional spread  decreases from an
initial value of approximately 5\%, to a final value of approximately
2.2\%. At the same time, since this is an emittance exchange, the transverse beam area grows, as
shown in Fig.~\ref{simu2}(b). One notes that the area increases not only in the regions of bends
(region 1 and 8), but also in the regions of wedges (2-6 and 9-11). This is probably due to failures
in matching that have yet to be understood.

\begin{figure*}[thb!]
\centerline{\epsfig{file=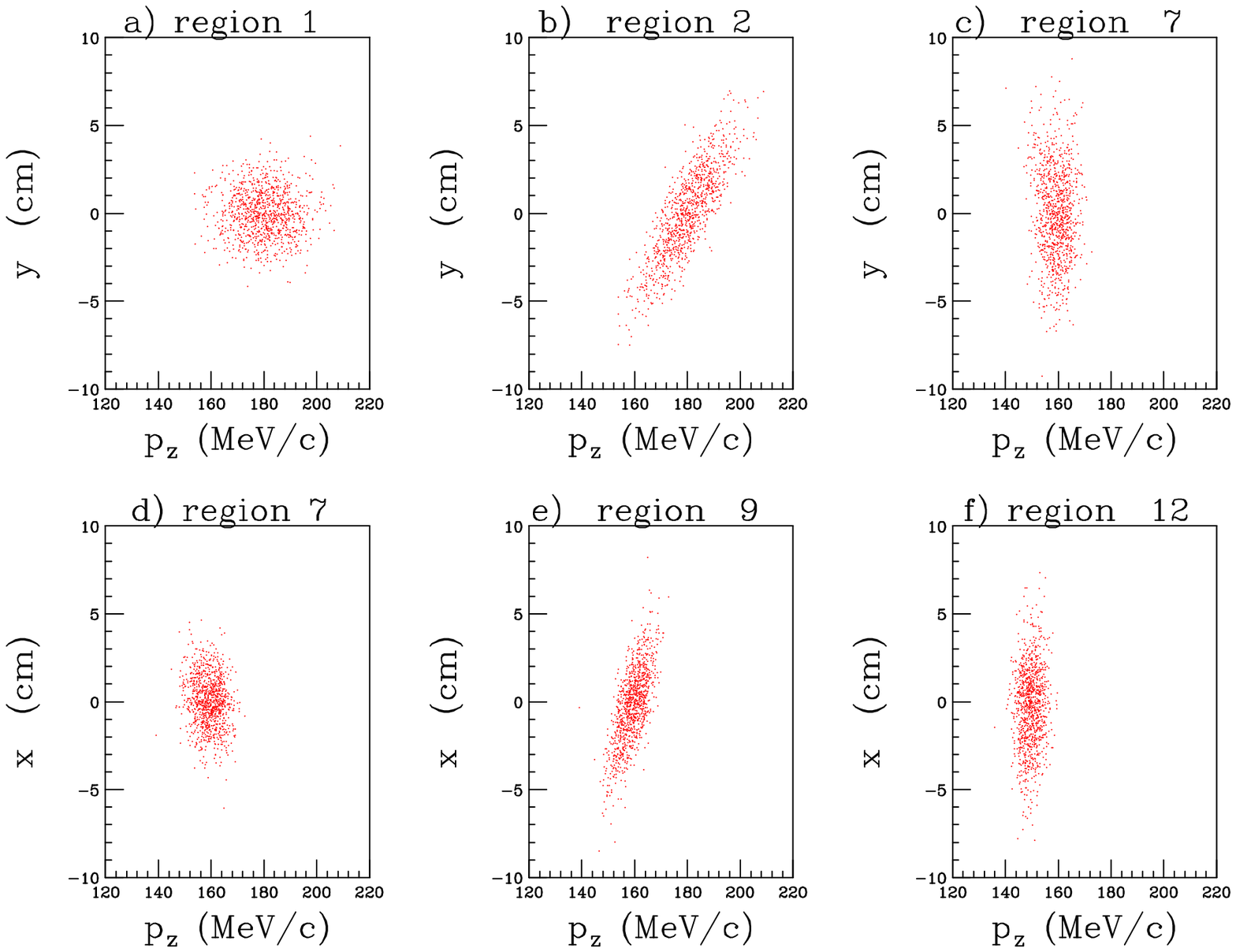,height=5.5in,width=5.5in}}
\caption[$y$ \vs~$p_z$ plots and $x$ \vs~$p_z$ plots before and after a wedge ]{$y$ \vs~$p_z$ plots:
a) at the start, b) after the first bend, c) after the first set of wedges. $x$ \vs~$p_z$ plots:  d) after the
first wedges, e) after the second bend, and f) at the end of the emittance exchange section, following
the second set of wedges.}
 \label{xfig5}
 \end{figure*}

Figure~\ref{xfig5} shows scatter plots of the transverse particle positions against their momenta. The
dispersion is clearly observed in Fig.~\ref{xfig5}(b) (after the first bend) and in Fig.~\ref{xfig5}(e)
(after the second). It is seen to be removed, with a corresponding decrease in momentum spread, in
Fig.~\ref{xfig5}(c) (after the first set of wedges) and Fig.~\ref{xfig5}(f) (after the second set of
wedges).

\begin{figure*}[htb!]
\centerline{
\epsfig{file=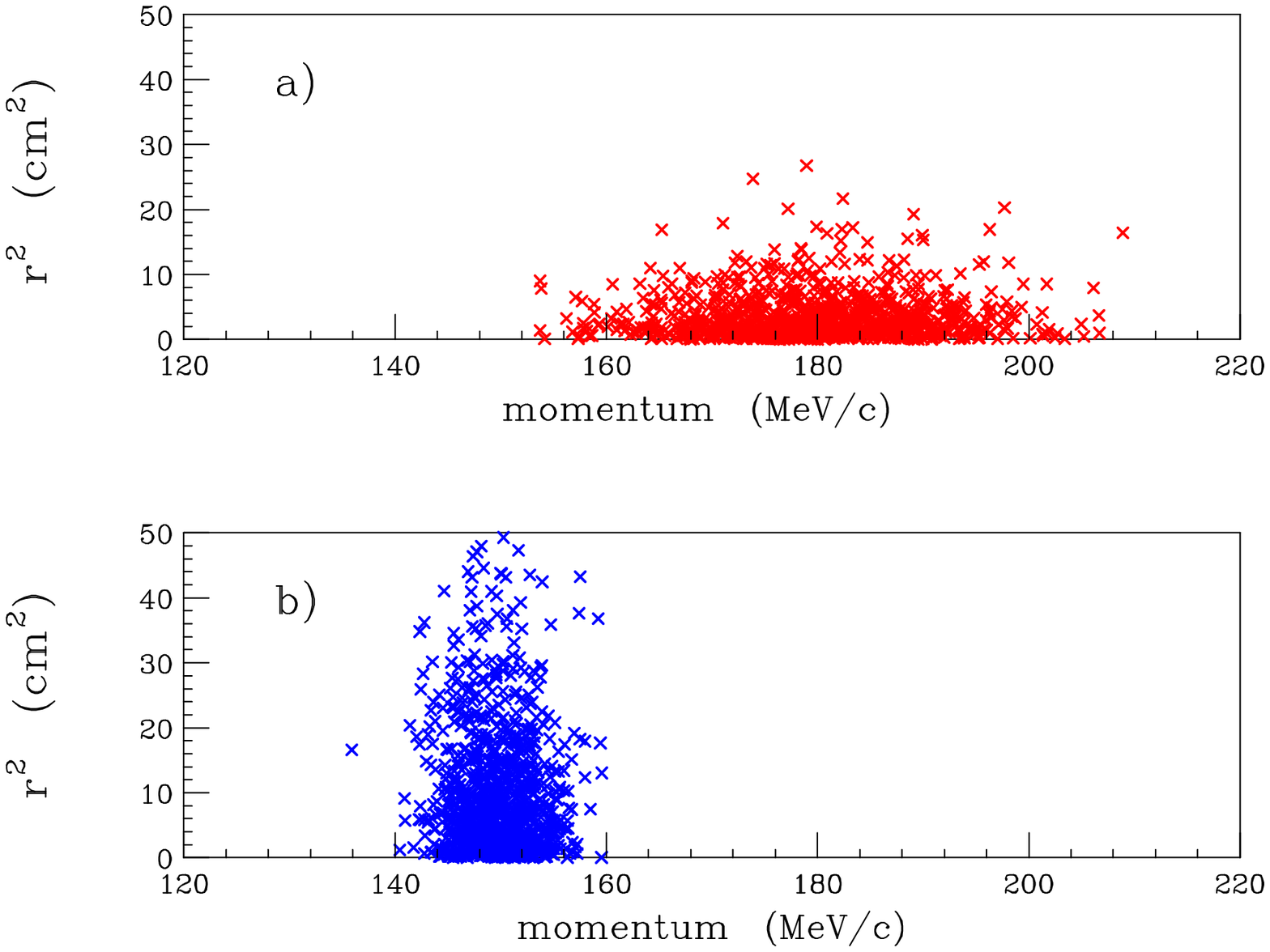,height=2.5in,width=3.5in}
}
\caption[Scatterplot of squared radii \vs~$p_z$]{Scatterplot of squared radii \vs~longitudinal
momentum: a) at the start, and b) at the end of the emittance exchange section.}
 \label{xfig1}
 \end{figure*}

Figure~\ref{xfig1} shows a scatterplot of the square of the particle radii vs.\ their longitudinal
momenta, (a) at the start, and (b) at the end of the emittance exchange section. The decrease in
momentum spread and rise in beam area are clearly evident. 

The initial and final beam parameters are given in table~\ref{xtab}. Although this example demonstrate a factor of $\approx 3$ reduction in the longitudinal momentum spread, there is a $37\,\%$ increase in the 5-D phase space. The simulations must be extended to include rf so that the 6-D emittance can be studied and the emittance
exchange section can be optimized. 

\begin{table*}[bth!]
\caption{Initial and final beam parameters in a longitudinal emittance exchange section.}
\label{xtab}
\begin{tabular}{llccc}
    & & initial & final & final/initial\\
\hline
Longitudinal Momentum spread & MeV/$c$        & 9.26 & 3.35 & 0.36 \\
Ave. Momentum         & MeV/$c$        & 180 & 150 & 0.83 \\
Transverse size       & cm            & 1.33 & 2.26 & 1.70 \\
Transverse Momentum spread  & MeV/$c$         & 6.84 & 7.84 & 1.15 \\ 
Transverse Emittance & $\pi$ mm-mrad & 870 & 1694 & 1.95 \\
Emit$_{\textrm{trans}}^2$ $\times$ $\Delta$p$_{\textrm{long}}$& ($\pi$ m-mrad)$^2$ MeV/$c$ & 7.0 & 9.6 & 1.37
\\
\end{tabular} 
\end{table*}

Emittance exchange in solid LiH wedges, with ideal dispersion and matching, has also been
successfully simulated using SIMUCOOL \cite{dave}. Dispersion generation by weak focusing
spectrometers \cite{balbekov} and dipoles with solenoids \cite{wan} have also been studied.

\subsection{RF for the cooling systems}

The losses in the longitudinal momentum of the muon beam from the cooling media have to be
restored using rf acceleration sections. 
These rf structures are embedded in solenoidal fields that 
reverse direction within each section.
% and vary in strength
%from 0 to 5~T.
In the two transverse cooling examples above, the rf frequency is 805~MHz and  the peak gradient
is 36~MeV/m. The magnetic fields that extend over the cavities vary from 0 to 10~T, reversing in
the center. It should be pointed out that in the earlier stages, the bunches are longer, and lower
frequencies will be required.
%\begin{center}
\begin{figure*}[thb!]
\centering
\epsfig{file=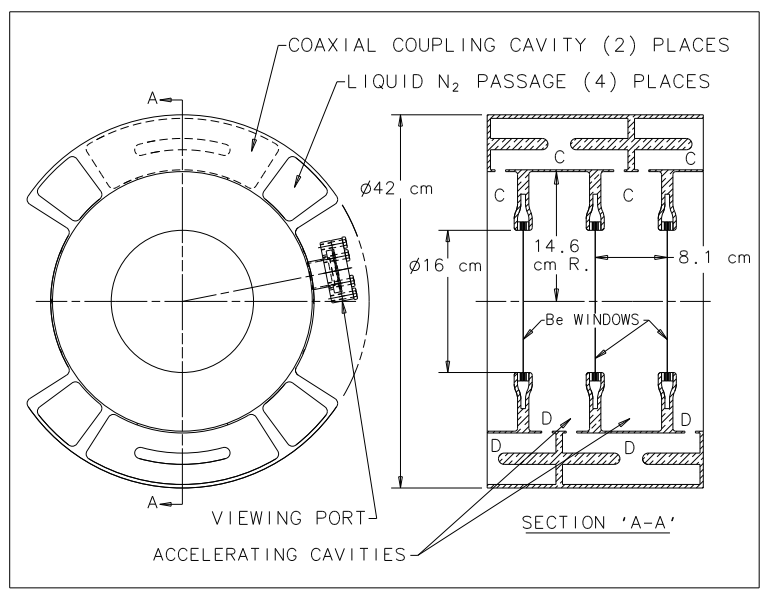,height=3.5in,angle=-90}
\vspace{-2.5cm}
\caption[Two full cell sections plus two half cell sections   ]{Two full cell sections plus two half cell
sections
of the interleaved $\pi$/2 mode accelerating cavities. The
volumes labeled \textbf{C} are powered separately from the volumes labeled \textbf{D}.}
\label{rf_inter}
\end{figure*}
%\end{center}

In order to realize maximum accelerating gradients within the acceleration
cavities, we take advantage of the penetrating properties of a muon
beam by placing thin windows between each rf cell, thereby creating an
accelerating structure closely approximating the classic pill-box cavity. This permits operating
conditions in which the axial accelerating field is equal to the maximum wall field and gives a high
shunt impedance. 

The windows in the 15~T example are
16~cm diameter, $125~\mu$m thick Be foils. In the 31~T case, they are 10~cm diameter and
$50~\mu$m thick.

\begin{figure*}[thb!]
\centering
\epsfig{file=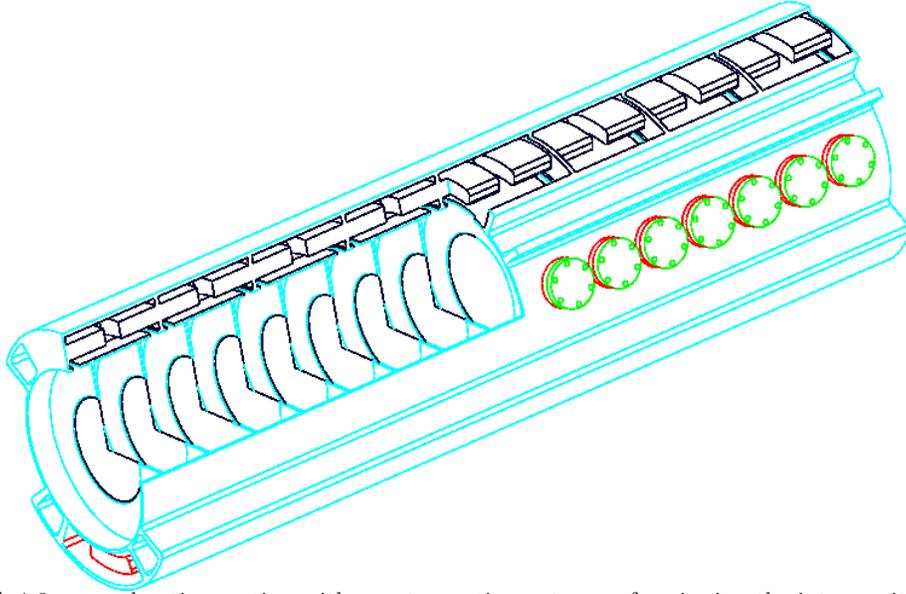,height=12cm,angle=-90}
\caption[A 1.3~m  acceleration section  ]{A 1.3~m  acceleration section with quarter section
cut away for viewing the inter-cavity windows.}
\label{rf_section}
\end{figure*}

 For these rf structures, we will use an interleaved cavity design in which
two parts are independently powered (Fig.~\ref{rf_inter}). 
The mode of the system will be referred to
as $\pi$/2 interleaved.
Each section supports a standing wave $\pi$ mode, with each acceleration
cell $\pi$/2 long, giving a good transit time factor.
To reduce the peak rf power requirements (by a factor of 2), we are considering operating the cells
at liquid nitrogen temperatures. 
%This also facilitates maximizing the shunt impedance. 
\begin{table*}[htb!]
\centering{\caption{Characteristics of the rf system.}
\begin{tabular}{lc}
%\multicolumn{2}{c}{RF Cavity Parameters} \\ 
%\noalign{\vspace{2pt}}
%\hline
%\noalign{\vspace{2pt}}
% &    \\
 RF frequency [MHz] & 805    \\
 Cavity Length [cm] & 8.1  \\
 Cavity Inner Radius [cm] & 14.6   \\
 Cavity Outer Radius [cm] & 21   \\
 $Q/1000$  & 2 $\times$ 20   \\
 Peak Axial Gradient [MV/m] & 36  \\
 Shunt Impedance  [M$\Omega$/m] & $2 \times 44$  \\
 $Zt^2$  [M$\Omega$/m] & $2 \times 36$ \\
 Fill Time (3 $\tau$)  [$\mu$s] & $2 \times 12$   \\
 RF Peak Power [MW/m] & $\frac{1}{2} \times 29$  \\
 Ave. Power (15Hz) [KW/m] & 5.3   \\
 Be window aperture [cm] & 16 (10 for 31 T case) \\
 Be window thickness [$\mu$m] & 127 (50 for 31~T case) \\
\end{tabular}
\label{rf_table}
}
\end{table*}

The characteristics of the rf systems currently being studied are summarized 
in Table~\ref{rf_table}. Figure~\ref {rf_section} shows a full 1.3~m section with interleaved
cavities.  
Each cell is 8.1~cm in length and the 1.3~m section consists of 16~cells.  

\subsection{The liquid lithium lens}
The final cooling element ultimately determines the luminosity of the
collider.  In order to obtain smaller transverse emittance as the muon
beam travels down the cooling channel, the focusing strength must increase, i.e.
the $\beta_{\perp}$'s must
decrease.  A current within a conductor produces an active
lens absorber, which can maintain the beam at small $\beta_{\perp}$
throughout
an extended absorber length, while simultaneously  attenuating
the beam momentum.  An active lens absorber, such as a
lithium lens, may prove to be the most efficient cooling element
for the final stages.

\begin{figure*}[hth!]
\epsfysize=4.5in
\vspace{-1.0cm}
\centerline{
\epsffile{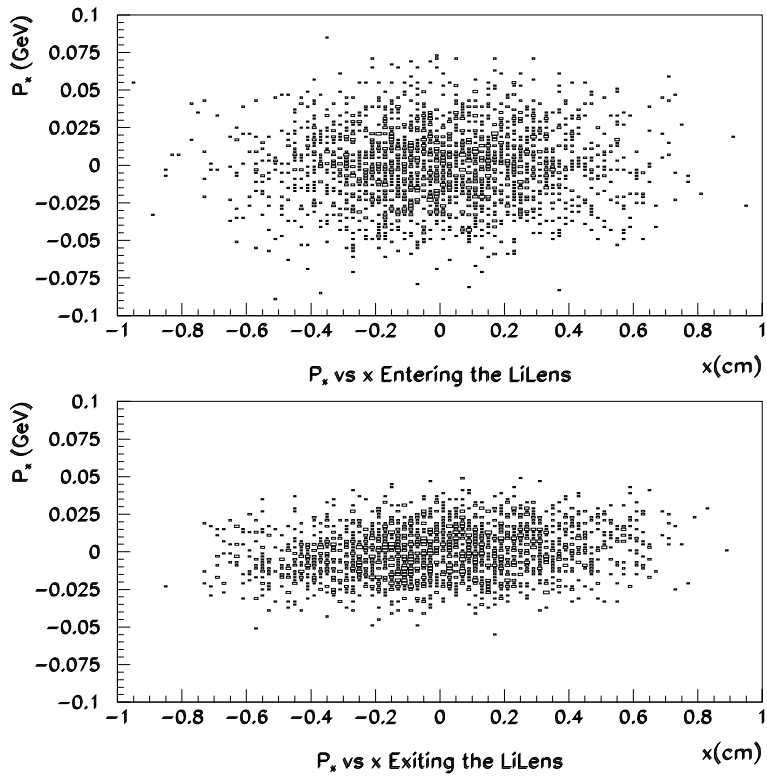}
}
\caption{$x$-$p_x$ phase space distribution at the beginning and at the
end of the absorber described in the text.}
\label{lilesim}
\end{figure*}

The cooling power of a Li lens is illustrated
in Figure~\ref{lilesim}, where the $x$ \vs~$p_x$ phase space distributions at
the beginning and at the end of the absorber are shown.  This example
corresponds to a 1~m long lens, with 1~cm radius, and a surface field of
 10~T.  The beam momentum entering the lens was 267~MeV/$c,$ with Gaussian
transverse spatial and momentum distributions:
$\sigma_x = \sigma_y = 2.89$~mm,
$\sigma_{p_x}=\sigma_{p_y} = 26.7$~MeV/$c,$ and a normalized emittance of
$\epsilon_{x,N} = 710$~mm-mrad.  The normalized emittance at the end of the
absorber was $\epsilon_{x,N} = 450$~mm-mrad (cooling factor
$\sim 1.57$), and the final beam momentum was 159~MeV/$c.$
The results were obtained using a detailed GEANT simulation of a single
stage. 

An alternative cooling scheme under study uses a series of Li
 lenses. The lens parameters would have to vary to match the changing beam emittance along the
section and in addition, acceleration of the beam between the lenses has to be included.

Lithium lenses have been used with high reliability
as focusing elements at FNAL and CERN \cite{reviewtev,ref26a,ref26b}.  Although these lenses
have
many similar properties to those required for ionization cooling, there
are
some very crucial differences which will require significant advances in
 lens technology: ionization cooling requires longer lenses
($\sim$ 1 meter), higher fields ($\sim 10~T$), and higher operation
rates (15~Hz).  The last requirement calls for operating the lenses with
 lithium in the liquid phase.

A liquid Li lens consists of a small diameter rod-like chamber filled
with
liquid Li through which a large current is drawn. 
  
 The azimuthal magnetic field
focuses the beam to give the minimum achievable emittance
$\epsilon_{x,N} \approx C \beta_{\perp}$ where the constant $C$ depends on 
the properties of the material, for example, $C_{Li}=79$ mm-mrad/cm.
The focusing term can be written as
$\beta_{\perp} \sim 0.08$[cm]$ \sqrt{p/J}$
with $p$ is the muon momentum in MeV/$c,$ and $J$
is the current density in MA/cm$^2$. Increasing $J$ is obviously desirable.
 Decreasing $p$ can also be useful. However, below about 250 MeV/$c$
the slope of ${dE\over dx}(E)$ tends to increase the longitudinal emittance.
The requirement for the highest current density causes 
large ohmic power deposition. The current density will be
limited by the maximum tolerable deposited energy, which
will produce instantaneous heating, expansion, and pressure effects.
Understanding these effects is part of the ongoing liquid Li lens R\&D.

The structural design of the lithium lens is determined by how the
pressure
pulse and heat deposition are handled.  We assume that the Li will be
flowing rapidly under high pressure, confined by electrical
insulators radially and by fairly thick Be windows longitudinally.
Operation at 15~Hz for long periods poses severe challenges.  Shock,
fatigue  and other failure modes are being evaluated, in addition
to studies of material compatibilities, corrosion and degradation to
insure
safe operation over long periods.  It seems that the
minimum required radius of the lens may be the most important parameter
to
determine,
since mechanical problems rise while losses decrease as a function of
radius.

Transferring the beam from one lens to another, with linacs to
reaccelerate and provide longitudinal focusing,
is also a challenging problem,  because of the multiple scattering
introduced
in the windows, straggling and the large divergence of the beams.
We are in the process of evaluating a number of designs for this
transfer channel, using detailed tracking simulations that include solenoids, quadrupoles and other
focusing elements together with Li lenses.

A group from BINP has designed, and is constructing, a 15~cm long liquid lithium lens prototype that will eventually be tested at FNAL.  It is planned to extend this R\&D program to design, construct, and test longer lenses.
 The design of two lenses, whose behavior will be tested at first on a
bench
and then with muon beams at the Ionization Cooling Demonstration Facility, will
then follow \cite{steveproposal}.

\subsection{Ionization cooling experimental R\&D}
An R\&D program has been proposed to 
design and prototype the critical sections of a muon ionization
cooling channel. The goal of this experimental R\&D program 
is to develop the muon ionization cooling hardware to the point
where a complete
ionization cooling channel can be confidently designed for the First Muon
Collider. Details can be found in Fermilab proposal 
P904 \cite{steveproposal}. A summary of the R\&D program can be found in ref.~\cite{steve_sum}.
% Following is a brief description of the main elements of the experimental program.

The proposed R\&D program consists of:
\begin{itemize}
\item Developing an appropriate rf re-acceleration structure. 
It is proposed to construct a 3-cell prototype rf cavity with 
thin beryllium windows, which will be tested at high power and within a high-field solenoid.
\item Prototyping initially a 2~m section, and eventually a 10~m section, 
of an alternating solenoid 
transverse cooling stage. It is proposed to test the performance of 
these sections in a muon beam of the appropriate momentum.
\item Prototyping an emittance exchange (wedge) section and measuring its 
performance in a muon beam of the appropriate momentum.
\item Prototyping and bench testing 
$\sim 1$~m long liquid lithium lenses, and 
developing lenses with the highest achievable surface fields, and hence 
the maximum radial focusing.
\item Prototyping a lithium lens--rf--lens system and measuring its 
performance in a muon beam of the appropriate momentum.
\item Developing, prototyping, and testing a hybrid lithium lens/wedge 
cooling system.
\end{itemize}

\begin{figure*}[thb!]
%\noindent
\begin{minipage}{.85\linewidth} % fig  
\centering\epsfig{figure=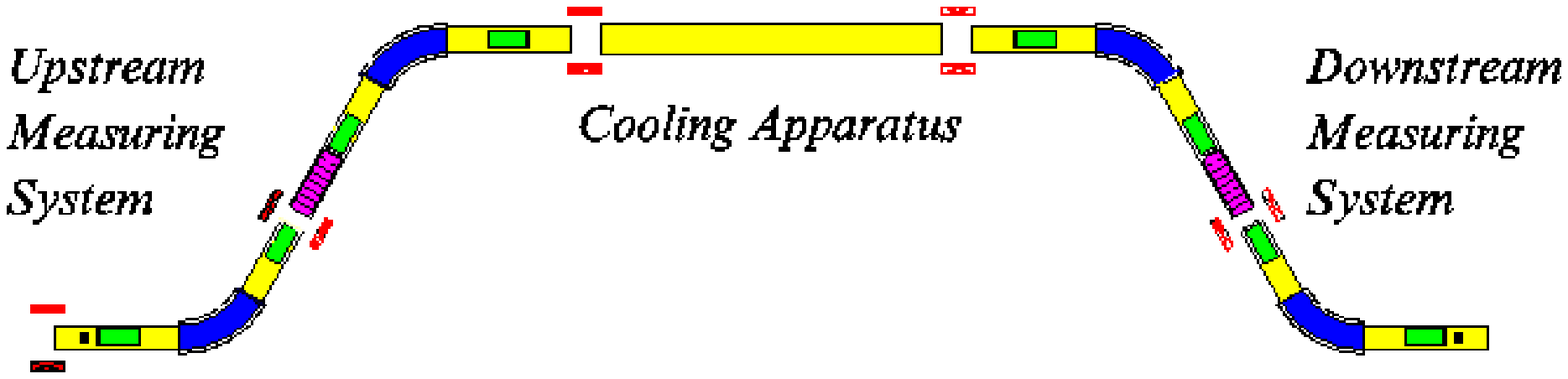,width=0.85\linewidth}
\end{minipage}\hfill
\begin{minipage}{.85\linewidth} % fig 
\centering\epsfig{figure=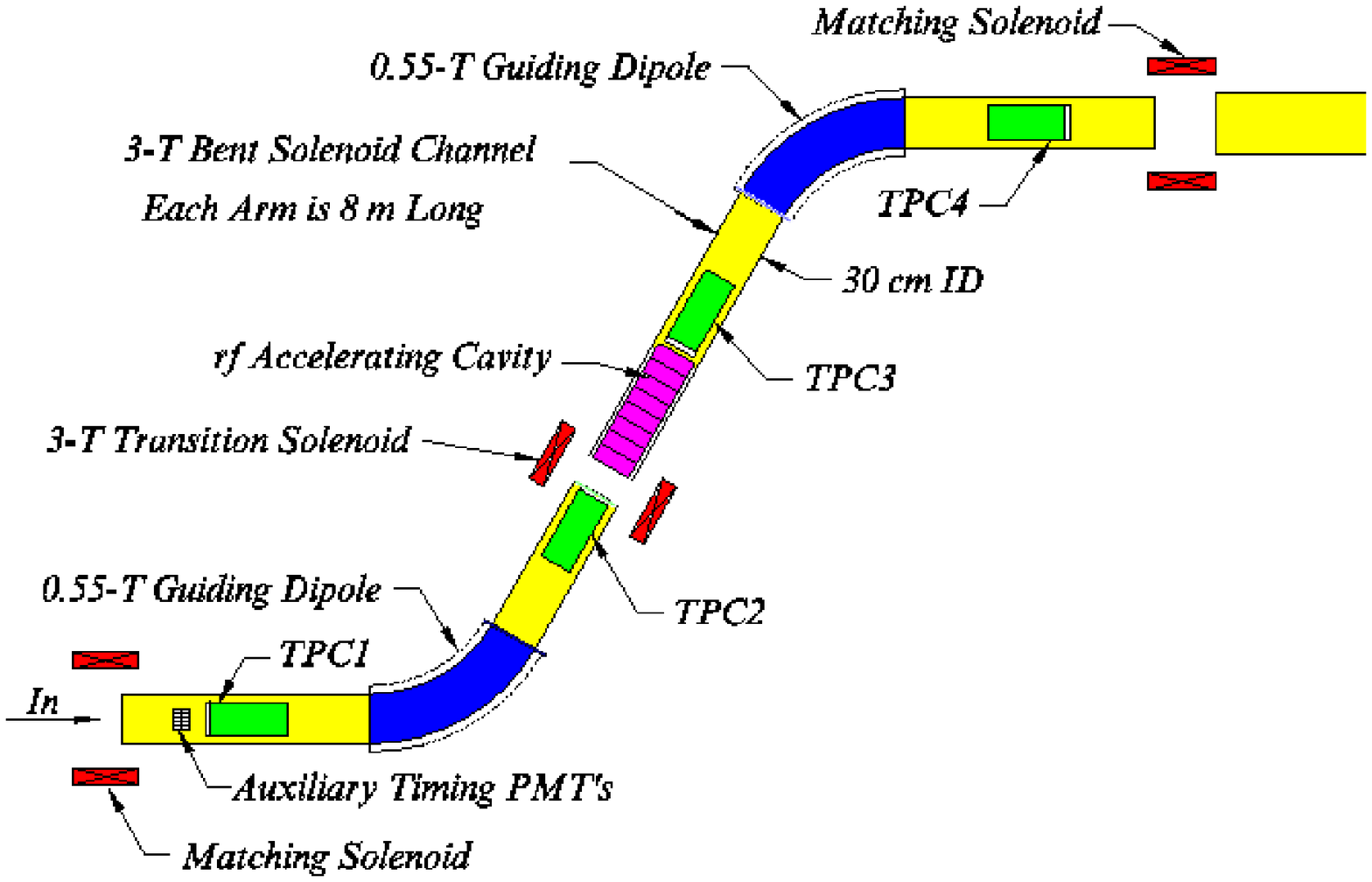,width=0.85\linewidth}
\end{minipage}
\caption{Schematic of the cooling test apparatus arrangement.}
\label{cooling_expt}
\end{figure*}

The measurements that are needed to demonstrate the cooling capability and 
optimize the design of the alternating solenoid, wedge, 
and lithium lens cooling stages will 
require the construction and operation of an ionization cooling test facility. 
This facility will need
\begin{enumerate} 
\item a muon beam with a central momentum that can be 
chosen in the range $100$-$300$~MeV/$c,$ 
\item an experimental area that can 
accommodate a cooling and instrumentation setup of initially $\sim 30$~m 
in length, and eventually up to $\sim 50$~m in length, and 
\item instrumentation to precisely measure the positions of the incoming 
and outgoing particles in 6-D phase space and confirm that they 
are muons.
\end{enumerate} In the initial design shown in Fig.~\ref{cooling_expt}, 
the instrumentation consists of 
identical measuring systems before and after the cooling apparatus \cite{lumcdpre}. 
Each measuring system consists of 
(a) an upstream time measuring device to determine the arrival time of the 
particles to one quarter of an rf cycle ($\sim\pm300$~ps), 
(b) an upstream momentum spectrometer in 
which the track trajectories are measured by low pressure TPC's on either 
side of a bent solenoid, 
(c) an accelerating rf cavity to change the 
particles momentum by an amount that depends on its arrival time, 
(d) a downstream momentum spectrometer, which is identical to the upstream 
spectrometer, and together with the rf cavity and the upstream spectrometer 
forms a precise time measurement system with a precision of a few ps. The 
measuring systems are 8~m long, and are contained within a high-field 
solenoidal channel 
to keep the beam particles within the acceptance of the cooling apparatus.

It is proposed to accomplish this ionization cooling R\&D program 
in a period 
of about 6~years. At the end of this period we believe that it will be 
possible to assess the feasibility and cost of constructing an ionization 
cooling channel for the First Muon Collider, and if it proves feasible, 
begin a detailed design 
of the complete cooling channel. 

%%%% TH LINES BELOW ARE GOOD FOR TWOCOLUMN FORMAT
%\begin{figure*}[thb!]
%\noindent
%\begin{minipage}{.85\linewidth} % fig 
%\begin{minipae}{2.05\linewidth} % fig 
%\centering\epsfig{figure=muonexp7a_c.eps,width=2.05\linewidth}
%\end{minipage}\hfill
%\begin{minipage}{2.05\linewidth} % fig 
%\centering\epsfig{figure=muonexp7_c.eps,width=2.05\linewidth}
%\end{minipage}
%\caption{Schematic of the cooling test apparatus arrangement.}
%\label{cooling_expt}
%\end{figure*}
%\input{prst6_new.tex}
\section{ACCELERATION}

\subsection{Introduction}

Following cooling and initial bunch compression, the beams must be rapidly 
accelerated. 
In this section some of the options in accelerator design
will be described and examples of acceleration scenarios presented.

Separate acceleration scenarios are given here for a low momentum spread ~100 GeV 
First Muon Collider (Higgs factory), and for a high luminosity 3 TeV collider. Ideally, though more difficult, this accelerator designed for the low energy machine should be extendable to the $\approx 250$~GeV beam energy and from
there to the $\approx 2$~TeV beam energy needed for a very high-energy collider.

While acceleration of muons to high energy is clearly possible,
an optimal and cost-effective acceleration complex is needed.
In the scenarios described below, a low-frequency linac would
take the beam from the end of cooling to an energy of $\approx$1 GeV followed by
recirculating-linac systems to take the beam to 50-70 GeV.
The multi-TeV energy regime can be reached through a series
of very rapid cycling synchrotrons. 
Variations on the acceleration model and potential difficulties are
discussed, including the use of Fixed Field Alternating Gradient (FFAG) accelerators in place of, or together with, the recirculating-linacs. Finally, topics for further study and research are described.

\subsection{Accelerator options}

 The acceleration time is limited by muon decay ($\tau_\mu =2.2 \mu~s$ at rest)
and requires that:
$$
eV_{rf}'\ >>\ 0.16 \textrm{MeV/m}
$$
where $eV_{rf}'$ is the acceleration rate. An acceleration rate whose value of 0.16~MeV/m is low for a linac, but very high for a conventional synchrotron.

At the lowest energies
($<$ 700 MeV), the momentum spread and beam sizes are so large that only a linac
is feasible, and
acceleration to full energy in a 
single-pass linac would be good, but it would be very expensive.

 Thus, following the initial linac, some form of recirculating acceleration is preferred. A synchrotron would be  possible, in principle, but the acceleration must occur so rapidly that
conventional magnet ramping is unlikely to be practical.  Two alternative multi pass
methods are being considered: recirculating linac accelerators similar to those
used at TJNAF and Fixed Field Alternating Gradient (FFAG) accelerators.

In a recirculating linac accelerator, the beam is circulated through
the same linac for several passes, with separate, energy matched, fixed-field
return paths for each pass.  
Each return path is
optically independent and can be separately designed.  In the
initial lattice design for the muon recirculating linac accelerator,
the return arcs are similar  alternating gradient (AG) systems
with the same dipole layout, but with differing quadrupole strengths
to allow separate tuning and chronicity matching in each arc.
Multiple aperture superconducting magnets have also
been designed %\cite{sk} 
which would reduce the diameter of the 
recirculating linac, lowering muon loss from decay
and possibly being more economical (see Fig~7.12 \cite{ref6a}).
In either case, both the linac and return transports must accommodate large
transverse emittances (rms) of $\approx 300~\pi$~mm-mrad.
Strong focusing is required not only to keep apertures down,
but also to minimize orbit deviations due to the large
momentum spreads, which in the initial stages 
of acceleration, can be as large as 10\%~rms.

More recently, an adaptation of the FFAG
(Fixed-Field Alternating Gradient) accelerator concept has been proposed for
$\mu+$ $\mu-$ acceleration \cite{fmillsffag,kingffag}.
In this variation, return transports are designed with a very large (factor of 5-10)
energy acceptance, so that separate energy turns can pass through the same
fixed-field elements.
More acceleration turns are possible than with a recirculating linac accelerator which reduces the rf requirements; but the orbits and
focusing properties are now energy-dependent.  Such FFAG configurations
require strong superconducting magnets 
with large apertures to accommodate
the energy-dependent spread in closed orbits.  The extra cost
associated with increased magnet apertures
must be evaluated against potential savings in
the number of magnets and reduced rf/turn requirements.

For the higher energy stages, the muon life time is greater and the needed rate of energy increase is less. Thus, above a few hundred GeV,
rapid accelerating synchrotrons become possible.
In a  rapid accelerating synchrotron, the beam is also multi-pass accelerated
through an rf system,
but the beam returns in a single arc, as the magnetic field
is ramped to match the increase in beam energy.
As above, more
acceleration turns are possible than with a recirculating  accelerator, but we now have a single moderate aperture return
transport. But rapid cycling synchrotrons have higher
power costs and the technical challenges associated with the
rapid acceleration time needed to conserve muons.

Thus a complete system would likely
include an initial linac followed by a sequence of recirculating linac accelerators and/or Fixed Field Alternating Gradient machines.
Depending on the final collider energy, one or more
rapid acceleration synchrotrons would follow.
Each system increases the
beam energy by a factor of 5-10.

\subsection{Scenario examples}

Several scenarios have been discussed earlier \cite{dn1,dn2}, see for instance the parameters (table~\ref{dntable}) used in a simulation of longitudinal motion discussed below. The ones given here are more recent, and more detailed, but they should not be 
taken to be definitive. They are examples that were derived to probe the design problems and to show that solutions should be possible.

\subsubsection{Acceleration for Higgs collider}

Table~\ref{acceleration1} gives an example of a sequence of
accelerators for a 100~GeV Higgs Factory, {\it i.e.} a machine with very low momentum spread (0.003\%, see table~\ref{sum}) and relatively large rms transverse emittance ($\approx~300~\pi$~mm-mrad). 

Following initial linacs, recirculating accelerators are used. 
The number of arcs in each recirculating accelerator is about 10.
In this example, conventional fixed field 2~T magnets are used, but the effective ramp frequencies that would be needed if pulsed magnets were used are given for reference.

\begin{table*}[htb!]
 \caption{Accelerator parameters for a Higgs Factory (100~GeV)} 
\label{acceleration1}

\begin{tabular}{llccccc|c}
Acc. type    &           &linac &linac &recirc&recirc&recirc& sums\\
Magnet type &           &      &      &warm  &warm  &warm  &     \\
rf type     &           &Cu    &Cu    &Cu    &Cu    &Cu    &     \\
\hline %    &           &      &      &      &      &      &     \\
$E^{\textrm{init}}$  &   (GeV)   &  0.10&  0.20&  0.70&    2 &    7 &      \\
$E^{\textrm{final}}$ &   (GeV)   &  0.20&  0.70&    2 &    7 &   50 &      \\
\hline %    &           &      &      &      &      &      &      \\
Circ.        &  (km)     &  0.04&  0.07&  0.07&  0.19&  1.74&  2.11\\
Turns       &           &    1 &    1 &    8 &   10 &   11 &      \\
Loss        &  (\%)     &  2.31&  3.98&  7.27&  7.91& 13.94& 31.06\\
Decay heat  &   (W/m)     &  0.89&  1.98& 11.04& 12.99& 12.44&      \\
%\hline %    &           &      &      &      &      &      &      \\
%$B_{pulse}$   &    (T)    &      &      &      &      &      &      \\
$B_{\textrm{fixed}}$   &  (T)    &      &      &    2 &    2 &    2 &      \\
Ramp freq.   & (kHz)     &  &  &  281 & 79.83&  8.00&      \\
\hline %    &           &      &      &      &      &      &      \\
Disp        &    (m)      &      &      &    1 &  1.50&    3 &      \\
$\beta_{\textrm{max}}$     &    (m)      &  0.83&  1.42&  3.00&  5.31& 21.08&      \\
%alpha      &   \%      &      &      &      &      &      &      \\
$\sigma_z^{\textrm{init}}$   &    (cm)   &  2.71&  2.22&  1.42&  1.64&  0.90&      \\
$\Delta p/p^{\textrm{init}}$   &    (\%)   &  3.58&  2.80&  1.64&  0.56&  0.32&      \\
$\sigma_y$        &   (cm)      &  1.09&  1.14&  1.01&  0.85&  0.94&      \\
$\sigma_x$         &   (cm)      &      &      &  1.93&  1.19&  1.34&      \\
Pipe full height&   (cm)      & 10.92& 11.42& 10.14&  8.49&  9.39&      \\
Pipe full width&   (cm)      & 10.92& 11.42& 19.28& 11.94& 13.35&      \\
\hline %    &           &      &      &      &      &      &      \\
rf freq.     &   (MHz)   &  200 &  200 &  200 &  200 &  400 &      \\
%lambda/sig 1&           & 55.30& 67.49&  106 & 91.46& 83.77&      \\
Acc/turn    &   (GeV)   &  0.20&  0.40&  0.17&  0.50&    4 &      \\
Acc time    & ($\mu s$) &      &      &    1 &    6 &   62 &      \\
$\eta$         &  (\%)     &  5.15&  5.36&  6.36&  2.84&  6.92&      \\
Acc. Grad.    &   (MV/m)  &    8 &    8 &    8 &   10 &   10 &      \\
Synch. rot's &           &  0.62&  0.63&  0.62&  3.92& 23.16&      \\
Cavity rad.  &   (cm)    & 54.37& 54.88& 54.88& 60.47& 38.26&      \\
Beam time   &   (ms)    &  0.00&  0.00&  0.00&  0.01&  0.06&      \\
rf time     &   (ms)    &  0.17&  0.17&  0.17&  0.18&  0.13&      \\
Tot. peak rf &   (GW)    &  0.05&  0.10&  0.05&  0.26&  4.71&  5.17\\
Ave. rf power&    (MW)     &  0.14&  0.24&  0.13&  0.68&  9.54& 10.73\\
rf wall     &    (MW)     &  0.64&  1.16&  0.46&  2.42& 28.06& 32.75\\
%    &           &      &      &      &      &      &      \\
\end{tabular}
\end{table*}

In this example, all the accelerating cavities are room temperature copper structures, and the accelerating gradients are modest ($<10$~MeV/m). Nevertheless, the acceleration is rapid enough that the total losses from decay are only 30\%.
The heating from these decays is also modest ($\approx 10$~W/m) because of the small number of turns and relatively low energy. Since no superconducting magnets or rf are used in this example, this heating should cause no problem.

In this machine, the transverse emittances are large and strong focusing is thus required, but the maximum momentum spread is moderate (up to 1.37 rms in the first recirculator) and is thus not likely to be a problem.

If the same machine is to also run at a high luminosity, with larger momentum spread, then
although the six dimensional emittance is the same as in the Higgs collider discussed above, the transverse emittance is smaller ($\approx 90 \pi$~mm-mrad~rms ), and the longitudinal emittance larger (by about a factor of 4) and the
 momentum spread in the first recirculating accelerator would be nearly 6\%~rms; or about 50\% full width. This is a very large momentum spread that could only be accepted in FFAG like lattices as discussed below.

Similarly, if the same acceleration is to be usable as the front end for a 250 + 250 GeV or higher energy machine then the transverse emittance will again be less, the longitudinal emittance even larger, and the problem of very large momentum spread will 
be worse. Clearly, although not absolutely needed for a Higgs Factory, it is desirable to solve this problem even in that First Collider.

A separate parameter set for the high luminosity 50 + 50~GeV collider and a 250 + 250~GeV collider could have been presented, but their parameters are very similar to those of the front end of the 3~TeV machine given below, and are thus omitted here.

\subsubsection{Acceleration for 3~TeV collider}

For a high energy machine, the muon accelerators are physically the largest component and are also probably the most expensive.
More work is needed on its design. Table~\ref{acceleration3} gives an early example of a possible sequence of
accelerators for a 3~TeV collider.

Linacs are used up to 700 MeV, followed by recirculating linac accelerators.
In the first of these, because of the very large longitudinal emittance, the momentum spread as the beam enters the first recirculating linac accelerator is 8.5\% rms, which is very large.
The lattice must have very strong focusing, small dispersion and large aperture. If this is not possible, higher energy linacs or lower frequency rf could relieve the requirement.
\begin{figure*}[thb!]
\begin{center}
\includegraphics[width=4in]{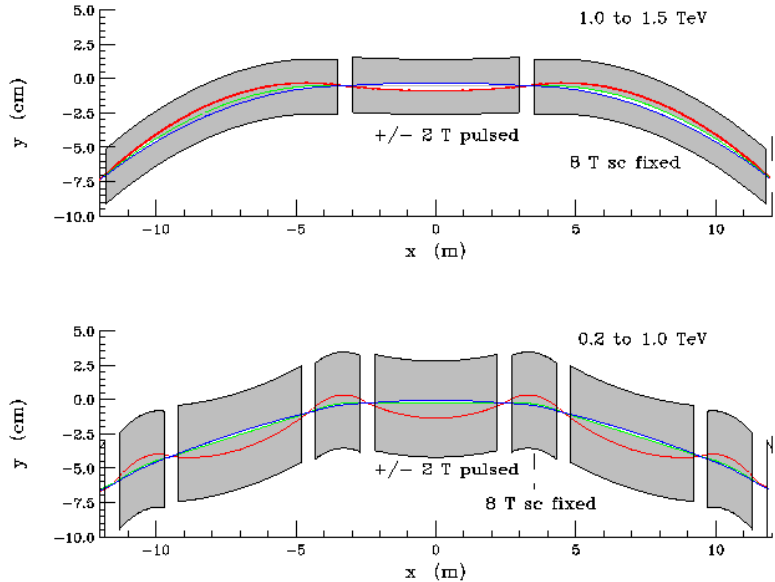} 
\end{center}
\caption[Schematic of hybrid superconducting-pulse magnet   ]{Schematic of hybrid superconducting-pulse magnet accelerator ring }
 \label{hybrid}
 \end{figure*}

For the final three stages, pulsed magnet synchrotrons \cite{summersshu} are used. In the 200~GeV ring, all the magnets in the ring are pulsed, but in the
last two rings a superconducting-pulsed hybrid solution is used. In these cases, if only pulsed magnets were used, then the power 
consumed would be too high, and because only low pulsed fields could be used,  the circumferences would also be very large. It is thus proposed to use rings with alternating warm
pulsed magnets and superconducting fixed  magnets \cite{summershybrid} (see figure~\ref{hybrid}). The fixed magnets are superconducting at 8 T; the pulsed magnets are warm with fields that swing from -~2~T to +~2~T. The effective ramp frequency is given 
in the table. 
Both of these rings are in the same tunnel, with the fraction of magnet length pulsed (vs. fixed) being different (73\% and 43\%).

In all the final three rings
superconducting rf is employed to minimize the peak power requirements and to obtain high wall to beam efficiency and thus keep the wall power consumption reasonable. In the final two rings the frequency and cavity 
designs are chosen to be the same as that in the TESLA \cite{tesla} proposal. 

%\squeezetable
{
\begin{table*}[tbh!]
 \caption{Parameters of Acceleration for 3~TeV Collider} 
\label{acceleration3}
\begin{tabular}{ll|ccccccc|c}
%\hline
Acc. type    &           &linac &recirc&recirc&recirc&synch &synch &synch & sums\\
Magnet type &           &      &warm  &warm  &warm  &warm  &hybrid&hybrid&     \\
rf type     &           &Cu    &Cu    &Cu    &SC Nb &SC Nb &SC Nb &SC Nb &     \\
\hline %    &           &      &      &      &      &      &      &      &     \\
$E^{\textrm{init}}$  &   (GeV)   &  0.10&  0.70&    2 &    7 &   50 &  200 & 1000 &      \\
$E^{\textrm{final}}$ &   (GeV)   &  0.70&    2 &    7 &   50 &  200 & 1000 & 1500 &      \\
\hline %    &           &      &      &      &      &      &      &      &      \\
Circ.        &  (km)     &  0.07&  0.12&  0.26&  1.74&  4.65& 11.30& 11.36& 29.52\\
Turns       &           &    2 &    8 &   10 &   11 &   15 &   27 &   17 &      \\
Loss        &  (\%)     &  6.11& 12.28& 10.84& 13.94& 10.68& 10.07&  2.65& 50.58\\
Decay heat  &   (W/m)     &  3.67& 15.02& 16.89& 15.91& 19.44& 30.97& 18.09&      \\
\hline %    &           &      &      &      &      &      &      &      &      \\
$B_{\textrm{pulse}}$   &    (T)    &      &      &      &      &    2 &    2 &    2 &      \\
$B_{\textrm{fixed}}$   &  (T)    &      &  0.70&  1.20&    2 &     &    8 &    8 &      \\
frac pulsed &     \%    &      &      &      &      &      &   73 &   43 &      \\
Ramp freq.   & (kHz)     & &  162 & 57.34&  8.00&  2.15&  0.50&  0.79&      \\
\hline %    &           &      &      &      &      &      &      &      &      \\
Disp.        &    (m)      &      &  0.40&  0.60&  0.80&    1 &    2 &    4 &      \\
$\beta_{\textrm{max}}$     &    (m)      &  0.89&  3.97&  8.75& 36.29& 52.20&  108 &  120 &      \\
Mom. compactn&   \%      &      &    1 & -0.25& -0.50& -0.50& -0.50&   -1 &      \\
$\sigma_z^{\textrm{init}}$   &    (cm)   & 16.34&  8.53&  5.29&  3.57&  1.59&  0.96&  0.78&      \\
$\Delta p/p^{\textrm{init}}$   &    (\%)   & 19.27&  8.49&  5.41&  2.47&  0.82&  0.35&  0.09&      \\
$\sigma_y$        &   (cm)      &  0.45&  0.45&  0.42&  0.48&  0.22&  0.16&  0.08&      \\
$\sigma_x$        &   (cm)      &      &  3.40&  3.25&  1.98&  0.82&  0.71&  0.36&      \\
Pipe full height&   (cm)      &  4.46&  4.52&  4.22&  4.77&  2.20&  1.62&  0.78&      \\
Pipe full width&   (cm)      &  4.46& 33.95& 32.49& 19.79&  8.20&  7.06&  3.62&      \\
\hline %    &           &      &      &      &      &      &      &      &      \\
rf Freq     &   (MHz)   &  200 &  100 &  200 &  200 &  800 & 1300 & 1300 &      \\
Acc./turn    &   (GeV)   &  0.40&  0.17&  0.50&    4 &   10 &   30 &   30 &      \\
Acc. time    & ($\mu s$) &      &    3 &    8 &   62 &  232 & 1004 &  631 &      \\
$\eta$         &  (\%)     &  3.82&  0.96&  1.97&  1.11& 10.15& 14.37& 12.92&      \\
Acc. Grad.    &   (MV/m)  &    8 &    8 &   10 &   10 &   15 &   25 &   25 &      \\
Synch. rot's &           &  0.81&  0.76&  1.02&  5.82& 19.14& 54.29& 31.30&      \\
Cavity rad.  &   (cm)    & 54.88&  110 & 60.47& 76.52& 19.13& 11.77& 11.77&      \\
rf time     &   (ms)    &  0.04&  0.12&  0.05&  0.56&  0.40&  1.25&  0.96&      \\
Tot. peak rf &   (GW)    &  0.21&  0.14&  0.59&  1.31&  1.06&  1.16&  1.04&  5.51\\
Ave. rf power&    (MW)     &  0.14&  0.25&  0.45& 11.04&  6.32& 21.91& 15.07& 55.18\\
rf wall     &    (MW)     &  0.64&  0.88&  1.62& 32.47& 18.59& 44.72& 30.76&  130 \\
%    &           &      &      &      &      &      &      &      &      \\
\end{tabular}
\end{table*}
}

%%%%%%%%%%%%%%%%%%%%%%%%%%%%%%%%%%%%%%%%%%%
\subsection{Design issues}

\subsubsection{Recirculating linac accelerator lattice issues}

Beam transport R\&D for recirculating linac accelerators follows the model of the Thomas Jefferson National Accelerator Facility (TJNAF).  The layout is a racetrack with linacs in the
straight sections and multi-pass return arcs.  At the ends of the
linacs the multi-pass beam lines are recombined.  A pulse magnet
at each separation/recombination point is used to guide the beam
into the energy matched return arc.  Some initial lattice design
concepts for recirculating linac accelerators are being developed.  The basic return arc
unit would be a FODO lattice, but with the quadrupole strengths
varied in order to perturb the arc dispersion function and obtain
nearly isochronous motion around the arcs.  The arcs are
dispersion matched by setting the arc phase advance to a multiple
of 2$\pi$.  Arc designs based upon the flexible momentum
compaction (FMC) module can also be used.

In the special case of the very low momentum spread Higgs factory, the transverse emittances are very large ($\approx 300~\pi$~mm-mrad {rms), and will require strong focusing in the lattices. Momentum acceptance in the rings is, in this case, not a problem.
But the longitudinal phase space of the muons in the other machines is much larger and requires, at low energies, either long bunches, or large momentum spreads. The requirement of high accelerating gradients argues for high frequencies, and thus short bunches. One therefore 
needs accelerators with large momentum acceptances. 

In the 3 TeV example above, the acceptance at injection into the first recirculating accelerator is 8.5\% rms.
This is very large by conventional standards, but far less than that in the FFAG lattices being studied \cite{fmillsffag}. Thus the early return arcs of such a recirculating linac accelerator would have to have very strong focusing, and be FFAG-like. Of course,
 if a true FFAG accelerator were to be used for its avoidance of the switchyards and multi-aperture magnets, then the 
specified momentum spread would certainly not be a problem.

Permanent, ferric or superferric ($\approx$2~T), or high field
superconducting magnets could be used for recirculating linac accelerators.  The lower field
magnets may be economic for initial turns, while high field
magnets minimize particle travel times, and therefore decay
losses.  Designs for multi-aperture superconducting magnets
suitable for recirculating linac accelerators have been developed \cite{ref6a}, and
superconducting magnets with as many as 18 apertures with
0.7--7~T fields have been designed. %\cite{sk}  
   A variety of magnet
configurations can be developed; cost/performance optimization
will be needed in developing a final choice.

\subsubsection{RF peak power requirements}

Because of the need for rapid acceleration, the peak rf powers are high, and the resulting numbers of power sources large. For the linacs and early recirculating accelerators, the powers are high because of the high gradients and low frequencies needed
to accelerate the long bunches. At these frequencies ($\approx$200~MHz), currently available sources (triodes and tetrodes) have relatively low maximum output powers and are expensive. Low temperature operation of the cavities, and superconducting or 
conventional SLED \cite{sled} systems, which would reduce the peak power requirements, are being considered. 

Study of the example suggests that in
the first two recirculators (up to 7~GeV) there is no hope for the rf to keep up with the beam loading. The cavity can only be filled in a suitable filling time (twice the time constant in these examples), and the rf voltage allowed to sag as the beam makes its multiple passes. If excessive sensitivity to beam current is to be avoided, then the stored energy must be large compared to that used, which is somewhat inefficient.

In the final recirculating accelerator, continuous filling (cw) is just possible, but requires yet higher peak power ($\approx 5$~GW total at 400~MHz)
because of the high acceleration rate.
The use of superconducting cavities can reduce losses, and thus reduce this peak rf power source requirement, and was included in the above 3 TeV example. At this frequency (400~MHz), klystrons are available with greater power ($\approx 20$~MW) 
than that of the sources at the lower frequencies, but
a yet higher power klystron (50-100~MW) could probably be developed and would be desirable.

\subsubsection{Pulsed magnet systems}

A pulsed current 4~T magnet has been designed for acceleration to 250~GeV in 360 $\mu$s \cite{ew}, but 
efficiency favors use of ferric materials
in rapid acceleration magnets, although this would limit peak magnetic
fields to $\approx$2~T.  The average field can be increased by
interleaving magnets swinging from -2~T to 2~T with fixed field 8~T
superconducting magnets.  

Faster pulsing magnets would require special materials to
minimize energy losses from eddy currents.  Options include
silicon steel, metglass laminations, or finemet laminated tape or
powdered solid.  A 30$\mu\textrm{m}$ metglass lamination suitable for
several kHz cycling has been developed. A design of suitable pulsed magnets \cite{summershybrid}(see Fig.~\ref{pulsed}) 
 has been shown to have sufficiently low losses for this application. The magnets employ cables made of many fine insulated strands (litz cable) and the yokes are made of
 very thin (0.28~mm), 3\%~Si-Fe laminations, possibly of metglas \cite{metglas1,metglas2} for the higher rate cases.
Detailed designs must be developed and prototypes constructed and the
practical limits of recycling scenarios should be determined. 

\begin{figure*}[thb!]
\centerline{
\epsfig{file=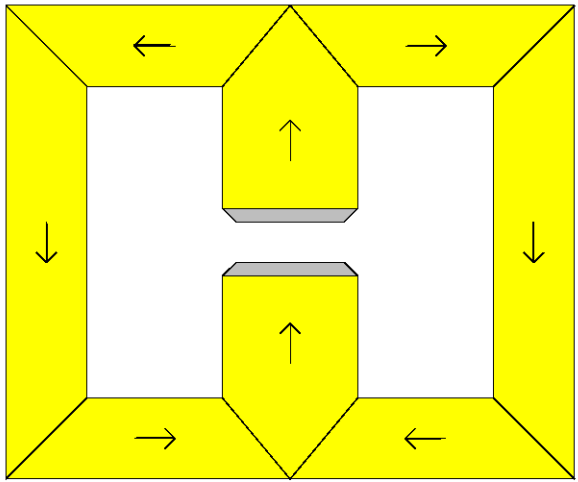,width=3.5in}}
\caption[A 2-D picture of an $H$ frame magnet lamination   ]{A 2-D picture of an $H$ frame magnet lamination with grain oriented $3\%$ Si-Fe steel.The arrows show both the magnetic field and the grain direction. }
 \label{pulsed}
 \end{figure*}

\subsubsection{Superconducting linacs}

While the gradients needed in the acceleration systems are not
excessive, they are larger than previous
experience at the lower frequencies.  The high peak power pulsed
operation poses power handling difficulties at lower energies and
high peak current presents collective effect (wakefield)
difficulties at higher energies.  Higher gradients and
efficiencies in all sections would improve performance.

The superconducting rf would operate in pulsed 
mode, matched to the
acceleration time of up to a few ms.  This pulse structure is
similar to the multibunch acceleration mode planned for TESLA (25~MV/m at 1300 ~MHz designs), and studies indicate that this design
could be adapted to $\mu^+-\mu^-$ acceleration.  At lower
frequencies, structures such as the CERN 350~MHz superconducting rf cavities
could be used.  These cavities have been tested in pulsed mode
operation, and tests indicate that pulsed
acceleration fields $>$ 10~MV/m are possible \cite{ic}.

The high single bunch intensities required for high intensities
imply large higher order mode losses and large wakefield effects
from the short, high intensity bunches.  Higher order mode (HOM) load designs adapted
from superconducting rf experience could be used.  HOM loads and wakefields are
expected to vary as $a^{-2}$ and $\lambda^{-2}$ and
$\sigma^{-1/2}$, where $a$ is the cavity aperture, $\lambda$ is
the acceleration wavelength and $\sigma$ is the bunch
length \cite{am,an}.  Calculations indicate that the wakefields would
limit bunch intensities to $\approx 2 \times 10^{12}$ with 1300~MHz superconducting rf in a recirculating linac accelerator scenario.  The longitudinal dynamics is
microtron-like or synchrotron-like and off-crest acceleration
enables compensation of the linear part of the wakefields, with
synchrotron-like phase stability \cite{dn1}.

\subsection{Simulations} 

\begin{table*}[tbh!]
 \caption{Parameters of Acceleration for 4~TeV Collider} 
\label{dntable}
\begin{tabular}{lccccc}
                       & Linac           & RLA1           & RLA2           & RCS1          & RCS2          \\
\hline
E (GeV)                & 0.1$\rightarrow$  1.5     & 1.5 $\rightarrow$ 10     & 10 $\rightarrow$ 70      & 70 $\rightarrow$ 250    & 250 $\rightarrow$ 2000  \\
f$_{rf}$ (MHz)         & 30 $\rightarrow$ 100      & 200          & 400            & 800           & 1300          \\ 
N$_{turns}$             & 1               & 9              & 11             & 33            & 45           \\
V$_{rf}$(GV/turn)      & 1.5             & 1.0            & 6              & 6.5           & 42            \\
C$_{turn}$(km)         & 0.3             & 0.16           & 1.1            & 2.0           & 11.5          \\
Beam time (ms)             & 0.0013          & 0.005          & 0.04           & 0.22          & 1.73          \\
$\sigma_{z,beam}$(cm)  & 50 $\rightarrow$ 8        & 4 $\rightarrow$ 1.7      & 1.7 $\rightarrow$ 0.5    & 0.5 $\rightarrow$ 0.25      & 0.25 $\rightarrow$ 0.12      \\
$\sigma_{E,beam}$(GeV) & 0.005 $\rightarrow$ 0.033 & 0.067 $\rightarrow$ 0.16 & 0.16 $\rightarrow$ 0.58       & 0.58 $\rightarrow$ 1.14      & 1.14 $\rightarrow$ 2.3       \\
Loss (\%)   & 5  & 7   & 6  & 7  & 10          \\
\end{tabular}
\end{table*}

A study \cite{dn2} followed the longitudinal motion of particles through 
a similar sequence of recirculating accelerators (see table~\ref{dntable}). Cavities similar to those proposed for TESLA \cite{tesla} were assumed. Figure~\ref{snowmass7.8}
\begin{figure*}[thb!] % fig 6
\includegraphics[height=6.in,width=6.in]{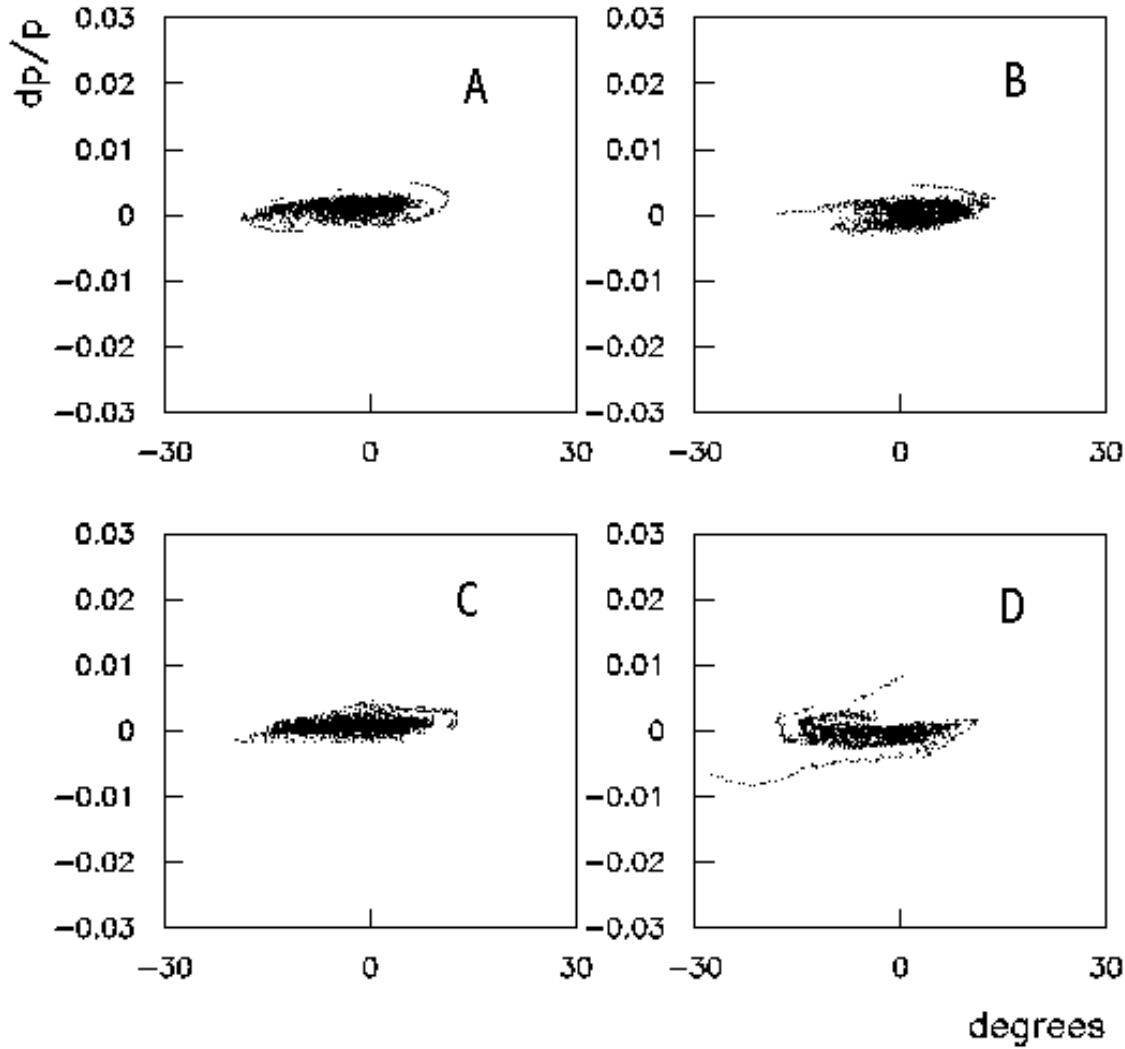}
\caption[Recirculating linac accelerator simulation results  with wakefields   ]{Recirculating linac accelerator simulation results  with wakefields, with beam accelerated
from 200 to $2000\,{\rm GeV}$ in a 10-turn recirculating linac accelerator. Longitudinal phase space plots for different bunch charges: A) very small number; B) $0.83\times 10^{12}$; C) $2.08\times 10^{12}$ and D) $4.17\times 10^{12}$  muons in a bunch.}
\label{snowmass7.8}
\end{figure*}
shows, after optimization of parameters, the final longitudinal phase space distributions corresponding to wakefields estimated for four different bunch charges: a) very small, b) $0.83\times 10^{12}$ muons, c) $2.08\times 10^{12}$ muons, and d) 
$4.17\times 10^{12}$ muons.

For the design beam charge of $2\times 10^{12}$ muons (approximately as for Fig.~\ref{snowmass7.8}(c)) the wakefield amplitude was estimated to be 2.5~MV/m, the accelerating phase was 35$^o$, and rf voltage depression 26\%. 
The simulation used an initial longitudinal phase space of 20~eV-s. It
 gave  negligible particle loss, a final longitudinal phase space of 21.6~eV-s, resulting in an increase of longitudinal emittance of only 8\%.

\subsection{Acceleration research needed}

As discussed above, possible acceleration configurations have
been developed, and critical longitudinal motion simulations have
been performed.  These calculations support the general
feasibility of acceleration of muons from cooling to collider
energies.  However the designs of acceleration systems have not
been fully detailed and much work would be needed to obtain a
buildable design.  Complete transport lattices for linacs and
return arcs have not yet been derived, and 6-D phase
space tracking of beams through the accelerators has not been
attempted.  Also the geometry of combining and separating
multi pass beams has not been worked out and optimized.

The rf requirements and systems have been specified at only the
rudimentary requirements level, and have not been developed to a
constructible level.  Optimal configurations and choices of
normal or superconducting rf must be developed, as well as more
optimal choices in acceleration frequencies.  The simple
wakefield models used in the initial simulations should be
expanded to obtain more realistic systems, and more precise
calculations of wakefield effects must be developed.

Rapid accelerating systems have only been outlined at the simplest
conceptual level.  Prototype magnet design and testing are needed
to test the limits of cycling rate and field strengths. 
Successful magnet concepts must then be specified in terms of
stable beam transport configurations, including focusing and
transport matching.  While beam is stored for only a few turns,
the individual bunch intensities are large enough that the
possibility of single bunch instabilities must be considered and
calculated.  The larger number of passes in a recirculating linac accelerator places greater
demands on the rf systems and higher order mode (HOM) loads, particularly for
superconducting systems.
\section{COLLIDER STORAGE RING}
\subsection{Introduction}
After one $\mu^+$ bunch and one $\mu^-$ bunch have been accelerated 
to collision energy, the two bunches
are injected into the collider ring, which is a fixed field storage ring.
Parameters for several possible collider storage rings are given 
in table~\ref{sum}. Collider ring lattices have been developed for two of
the collision energies in this table: 100~GeV and 3~TeV in the 
center of mass. 

Three operational modes are proposed in the above table for the 100~GeV
collider, each requiring  different machine optics. The following sections
discuss a 100~GeV collider lattice for two of the modes, the broad momentum
spread case ($\Delta p/p$ of $0.12\%$, rms)  and the narrow momentum spread 
case ($\Delta p/p$ of $0.003\%$), as well as a 3~TeV collider lattice.

\subsection{Collider Lattices}

\subsubsection{Design criteria}
Stringent criteria have been imposed on the collider lattice designs in order to attain
the specified luminosities.  The first and most difficult criterion to satisfy is
provision of an Interaction Region (IR) with extremely low
$\beta^*$ values at the collision point consistent with acceptable dynamic aperture.
The required $\beta^*$ values for the 100~GeV collider are
4~cm for the broad momentum spread case and 14~cm for the narrow momentum spread case.
For the 3~TeV machine, $\beta^*$ is only 3~mm.  These $\beta^*$ values
were tailored to match the longitudinal bunch lengths in order 
to avoid luminosity dilution from the hour-glass effect.  
Achieving this requirement in the 3~TeV lattice is complicated by
the high peak beta function values in the final focus quadrupoles
requiring 8-10~cm radial apertures.
The correspondingly weakened gradients
combined with the ultra-high energy make for a long final focus structure.
(In contrast, the lower energy and larger $\beta^*$ values
in the 100~GeV collider lead to an efficient, compact final focus telescope.)
Compounding the problem, particularly for the 3~TeV design, is 
the need to protect the superconducting coils
from the decay products of the muons.
Placing a tungsten shield between the vacuum
chamber and the coils can increase the radial aperture in
the 3~TeV quadrupoles by as much as 6~cm,
lowering  available gradients still further.
Final focus designs must also include collimators and background
sweep dipoles, and other provisions for protecting the magnets
and detectors from muon decay electrons.  Effective schemes have
been incorporated into the current lattices.
 
Another difficult constraint imposed on the lattice is that of isochronicity.
A high degree of isochronicity is required
in order to maintain the short bunch structure without
excessive rf voltage.
In the lattices presented here, control over the momentum compaction 
is achieved through
appropriate design of the arcs.

A final criterion especially important in
the lower energy colliders 
is that the ring circumference be
as small as feasible in order to minimize luminosity degradation through decay
of the muons. Achieving small circumference requires high fields in the bending 
magnets as well as a compact, high dipole packing fraction design.
To meet the small circumference demand, 8~T pole tip fields have been assumed 
for all superconducting magnets, with
the exception of the 3~TeV final focus quadrupoles, whose pole tips  are assumed
to be as high as 12~T.  In addition, design studies for still higher field dipoles
are in progress.

\subsubsection{rf system}
The rf requirements depend on the momentum compaction of the lattice
and on the parameters of the muon bunch. 
For the case of very low momentum spread,
synchrotron motion is negligible and the rf system is used solely to correct 
an energy spread generated through the
impedance of the machine. For the cases of higher momentum spreads,
there are two approaches. One is to make the momentum compaction zero
to high order through lattice design.
Then the synchrotron motion can be eliminated,
and the rf is again only needed to compensate the induced 
energy spread correction. Alternatively, if some
momentum compaction is retained, then a more powerful rf system is needed to maintain 
the specified short bunches.
In either case, rf quadrupoles will be required to generate BNS \cite{refbns,ref36} 
damping of the transverse head-tail instability.

\subsubsection{3~TeV CoM lattice}
The 3~TeV ring has a roughly racetrack design with two circular arcs
separated by an experimental insertion on one side, and a utility insertion 
for injection, extraction, and beam scraping on the other.
The experimental insertion includes the interaction region (IR)
followed by a local chromatic correction section (CCS) and a matching section. 
The chromatic 
correction section is  optimized to correct the ring's linear chromaticity,
which is mostly generated by the low beta quadrupoles in the IR. 
In designs of e$^+$e$^-$ colliders, it has been found that local chromatic correction
of the final focus is essential
\cite{chromatic,chromatica,chromaticb,chromaticc}, as was found to be the case here.
The 3~TeV IR and CCS are displayed in Fig.~\ref{3_tev_ir}.
The accompanying 3~TeV arc module in Fig~\ref{3_tev_arc}
is an example of a module which controls momentum compaction
(i.e. isochronicity) of the entire ring. 

\begin{figure*}[thb!]
\epsfxsize4.0in
\centerline{
\epsfig{figure=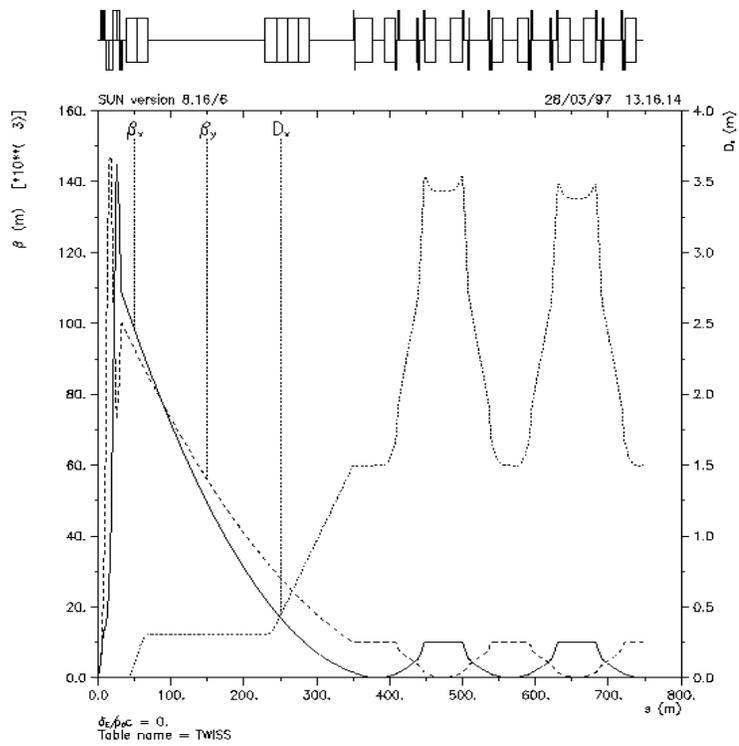,height=4.0in,width=4.5in,bbllx=30bp,
bblly=116bp,bburx=545bp,bbury=675bp,clip=}
}
\caption{Example~(a):
3~TeV IR and chromatic correction.}
\label{3_tev_ir}
\end{figure*}

\begin{figure*}[bth!]
\epsfxsize4.0in
\centerline{
\epsfig{figure=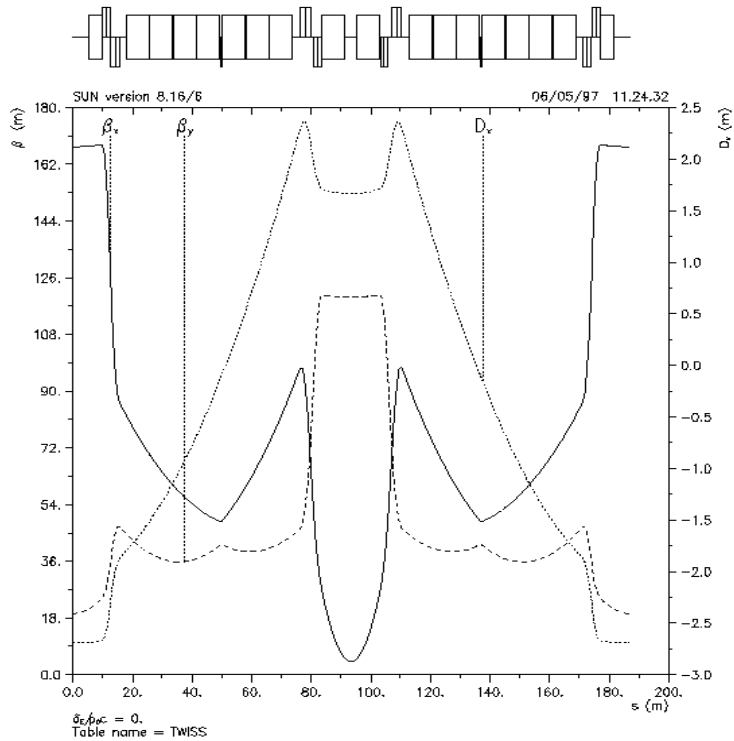,height=4.0in,width=4.5in,bbllx=30bp,
bblly=116bp,bburx=545bp,bbury=675bp,clip=}
}
\caption{Example~(a): 3~TeV arc module.}
\label{3_tev_arc}
\end{figure*}

\subsubsection{100~GeV CoM lattices}
For the 100~GeV CoM collider \cite{ref33}, two operating modes are contemplated:
a high luminosity case with broad momentum acceptance
to accommodate a beam with a $\Delta p/p$ of $\pm 0.12\%$ (rms), 
and one with a much narrower momentum acceptance
and lower luminosity for a beam with $\Delta p/p$ of $\pm 0.003\%$ (rms).
For the broad momentum acceptance case, $\beta^*$ must be 4~cm 
and for the narrow momentum acceptance case, 14~cm.
In either case, the bunch
length must be held comparable to the value of $\beta^{*}$.
The 100~GeV ring geometry is highly compact and
more complicated than a racetrack, but the lattice
has regions with the same functions as those of the 3~TeV ring.
 
Two independent 100~GeV lattice designs have evolved; these are described below
in separate sections and denoted Example (a) and Example (b), respectively.
The first design described is a lattice which has two optics modes.
In the high luminosity mode, the $\beta^*$ value is 4~cm with a
transverse and
momentum aperture sufficient to accept a normalized beam emittance
of $90\pi$ (rms) and a $\Delta p/p$ of $\pm 0.12\%$ (rms).
The second, lower luminosity mode has a $\beta^*$ value of 14~cm 
with a very large transverse acceptance,
but small, approximately monochromatic, momentum acceptance.

The second 100~GeV lattice described 
is another collider design 
with a 4~cm $\beta^*$ optics mode.
Although the number of magnets differ between the two lattices,
 the most important optics difference between the two 
is in the modules used in the arcs.  

\subsubsection{100~GeV CoM. Example~(a)}

The need for different collision modes in the 100~GeV machine
led to an Interaction Region design with two optics modes: 
one with broad
momentum acceptance ($\Delta p/p$ of $0.12\%$, rms) and a collision $\beta^*$ of 4~cm,
and the other basically monochromatic ($\Delta p/p$ of $0.003\%$, rms) and a
larger collision $\beta^*$ of 14~cm.  
The first lattice design, denoted Example~(a), shown in 
Fig.~\ref{50_gev_4cm} and Fig.~\ref{50_gev_14cm}, has a total circumference of 
about $350\,{\rm m}$ with  
arc modules accounting for only about a quarter of the ring circumference.

\begin{figure*}[thb!]
\epsfxsize4.0in
\centerline{
\epsfig{figure=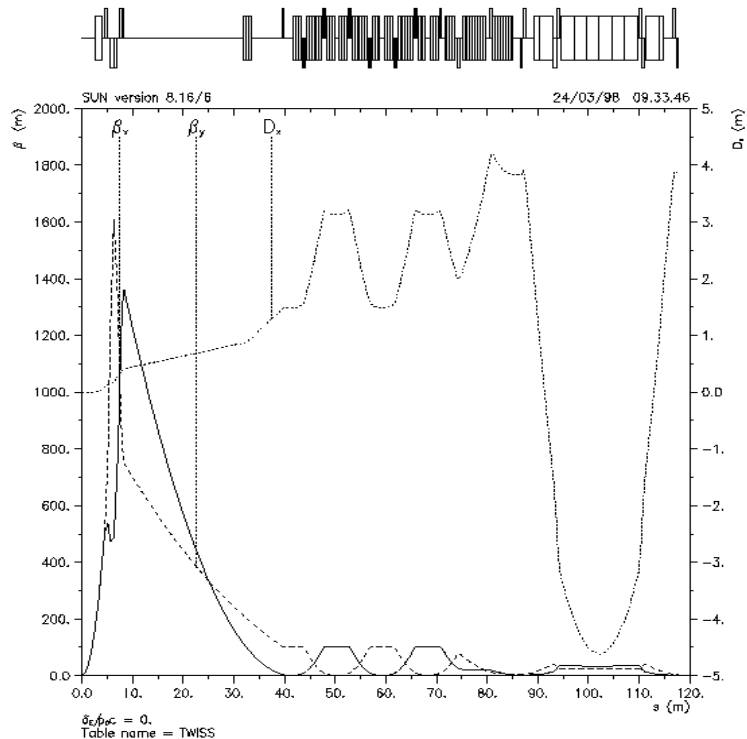,height=4.0in,width=4.5in,bbllx=30bp,
bblly=116bp,bburx=545bp,bbury=675bp,clip=}
}
\caption[Example~(a):  4~cm $\beta^*$ ]{Example~(a):  4~cm $\beta^*$ Mode showing
half of the IR, local chromatic correction, and one of three arc modules.}
\label{50_gev_4cm}
\end{figure*}

\begin{figure*}[thb!]
\epsfxsize4.0in
\centerline{
\epsfig{figure=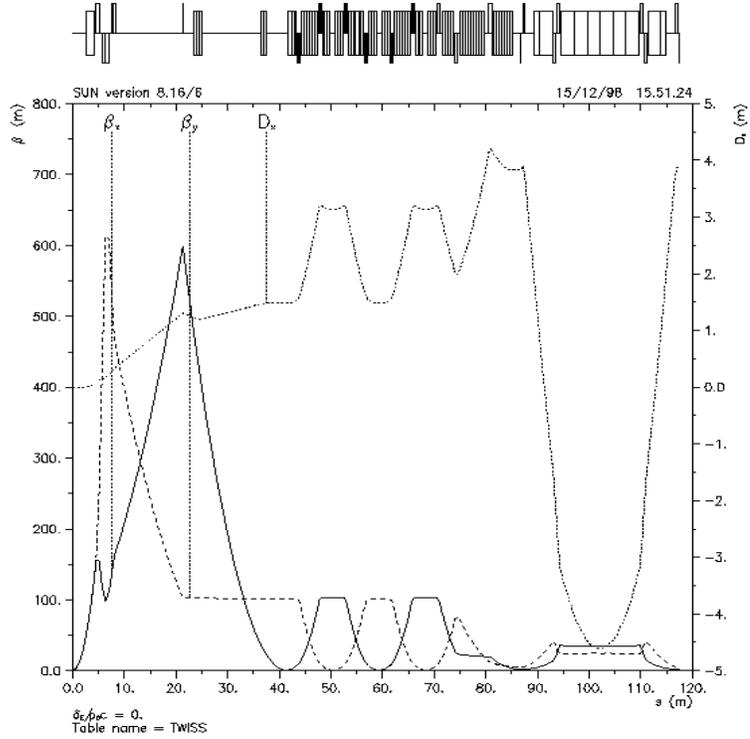,height=4.0in,width=4.5in,bbllx=30bp,
bblly=116bp,bburx=545bp,bbury=675bp,clip=}
}
\caption[Example~(a): 14~cm $\beta^*$  ]{Example~(a): 14~cm $\beta^*$ Mode showing
half of the IR, local chromatic correction, and one of three arc modules.}
\label{50_gev_14cm}
\end{figure*}

The low beta function values at the IP
are mainly produced by three strong superconducting quadrupoles
in the Final Focus Telescope (FFT) with pole tip fields
of 8~T.  The full interaction region is symmetric under 
reflection about the interaction point (IP).
Because of significant, large angle backgrounds from muon decay,
a background sweep dipole is included in the final focus telescope
and placed near the IP to protect the detector and the low $\beta$ 
quadrupoles \cite{carnik96}.
It was found that this sweep dipole, 2.5~m long with an 8~T field,
provides sufficient background suppression.
The first quadrupole is located 5~m away from the interaction point, 
and the beta functions reach a maximum value of $1.5\,{\rm km}$ 
in the final focus telescope, when the maxima of the beta functions 
in both planes are equalized. For this maximum beta value, 
the quadrupole apertures must be at least 11~cm in
radius to accommodate 5$~\sigma$ of a $90~\pi$~mm-mrad, 
50~GeV muon beam (normalized rms emittance) plus a 2 to 3~cm thick
tungsten liner \cite{scraping}.  The natural chromaticity
of this interaction region is about $-60.$

Local chromatic correction of the muon collider interaction region is
required to achieve broad momentum acceptance.
The basic approach developed by Brown \cite{chromatica} and others \cite{donald},
is implemented in the Chromatic Correction Region (CC). The CC contains
two pairs of sextupoles, one pair for each transverse
plane, all located at locations with high dispersion. 
The sextupoles of each pair are located at positions of equal, high beta value
in the plane (horizontal or vertical) whose chromaticity is to be corrected,
and very low beta waist in the other plane. Moreover, the two sextupoles of each pair
are separated by a betatron phase advance of near $\pi$, 
and each sextupole has a phase separation of $(2n+1){\pi\over 2}$ from the IP,
where $n$ is an integer.
The result of this arrangement is that
the geometric aberrations of each sextupole is canceled by its 
companion while the chromaticity corrections add.

The sextupoles of each pair are centered 
about a minimum in the opposite plane ($\beta_{min}<1m$), which
provides chromatic correction
with minimal cross correlation between the planes.
A further advantage to locating the opposite plane's
minimum at the center of the sextupole, is that this point is
${\pi\over 2}$ away from, or ``out of phase" with, the source of chromatic effects
in the final focus quadrupoles; i.e the plane not being chromatically corrected is
treated like the IP  in terms of phase to eliminate a second order 
chromatic aberration generated by an ``opposite-plane''
sextupole.

In this lattice example, the CC 
(Fig.~\ref{50_gev_ccs}) was optimized to be as short as possible.
The $\beta_{\textrm{max}}$
is only $100~\textrm{m}$ and the $\beta_{\textrm{min}}=0.7~\textrm{m}$, giving
a $\beta_{\textrm{ratio}}$ between planes of
about 150, so the dynamic aperture is not compromised
by a large amplitude dependent tuneshift.

This large beta ratio, combined with the opposite plane phasing,
allows the sextupoles for the opposite planes to be interleaved,
without significantly increasing the nonlinearity of the lattice.
In fact, interleaving improved lattice performance compared to that of
a non-interleaved correction scheme, due to a shortening
of the chromatic correction section, which 
lowers its chromaticity contribution \cite{wan1}.
The use of somewhat shallower beta minima with less variation in beta through the
sextupoles was made soften the chromatic aberrations, although this caused a 
slight violation of the exact $\pi$ phase advance separation between sextupole partners.
The retention of an exact $\pi$ phase advance difference between sextupoles 
was found to be less important to the dynamic aperture than elimination of
minima with $\beta_{\textrm{min}}<0.5~\textrm{m}$.

\begin{figure*}[thb!]
\epsfxsize4.0in
\centerline{
\epsfig{figure=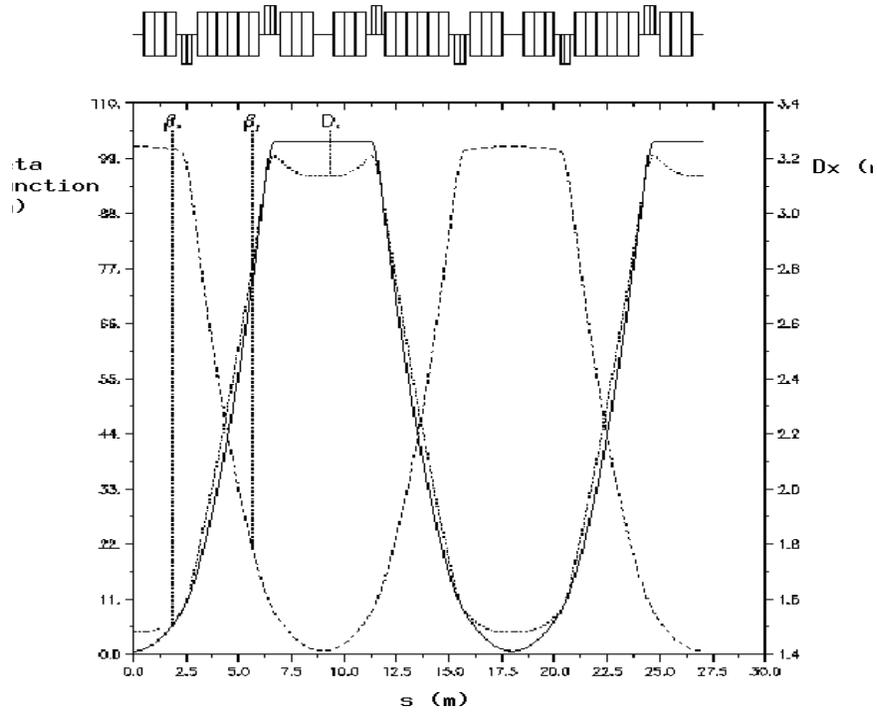,height=4.0in,width=4.5in,bbllx=60bp,
bblly=110bp,bburx=545bp,bbury=685bp,clip=}
}
\caption{Example~(a): The chromatic correction module.}
\label{50_gev_ccs}
\end{figure*}

The total momentum compaction contributions of the IR, CC, and matching sections
is about $0.04$.  The total length of these parts is $173\,{\rm m}$, while that
of the the momentum compaction correcting arc is 93~m.
From these numbers, it follows that this arc must have a negative momentum
compaction of about $-0.09$ in order to offset the positive contributions
from the rest of the ring.

The arc module is shown in Fig.~\ref{50_gev_arc}. It has the small beta 
functions characteristic
of FODO cells, yet a large, almost separate, variability
in the momentum compaction of the module which is a characteristic
associated with the flexible momentum compaction module \cite{ref32,ref32a}.
The small beta functions are achieved through the
use of a doublet focusing structure which produces
a low beta simultaneously in both planes.
At the dual minima, a strong focusing quadrupole
is placed to control the derivative of dispersion with
little impact on the beta functions.  
Negative values of momentum compaction as low as
$\alpha=-0.13$ have been achieved, and $\gamma_t=2~i$, has been achieved
with modest values of the beta function.

This arc module was able to generate the needed negative
momentum compaction with beta functions of $40\,{\rm m}$ or less.

\begin{figure*}[thb!]
\epsfxsize4.0in
\centerline{
\epsfig{figure=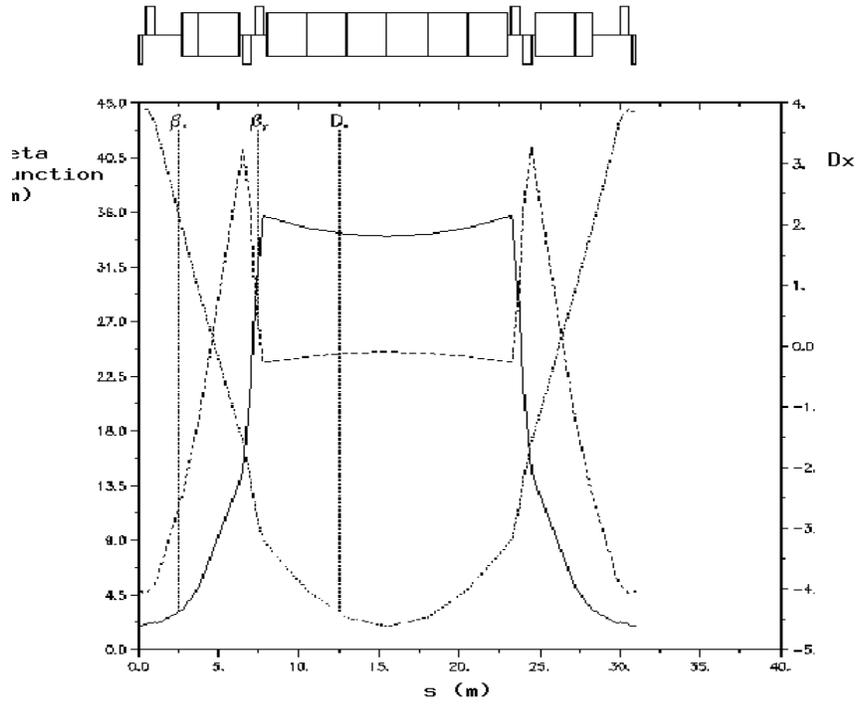,height=4.0in,width=4.5in,bbllx=48bp,
bblly=115bp,bburx=545bp,bbury=695bp,clip=}
}
\caption{Example~(a): A flexible momentum compaction arc module.}
\label{50_gev_arc}
\end{figure*}

\begin{figure*}[tbh!]
\epsfxsize4.0in
\centerline{
\epsfig{figure=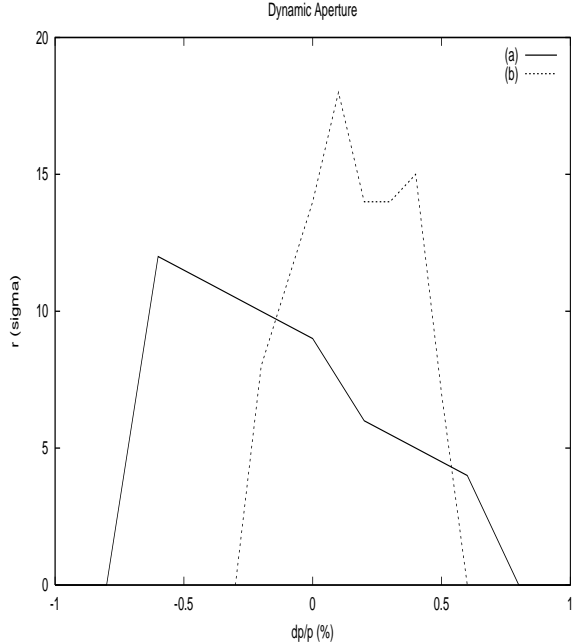,height=3.0in,width=3.5in,angle=-90}
}
\caption[Example~(a): A preliminary dynamic aperture ]{Example~(a): A preliminary dynamic aperture for the 4~cm
$\beta^*$ mode where $\sigma$ (rms) = 82$\mu m$ (solid line) and the
14~cm $\beta^*$ mode where $\sigma$ (rms) = 281$\mu m$ (dashed line).}
\label{50_gev_ap}
\end{figure*}

A very preliminary calculation of the dynamic aperture \cite{wan1} without optimization 
of the lattice or inclusion of errors and end effects
is given in Fig.~\ref{50_gev_ap}.  One would expect that simply turning
off the chromatic correction sextupoles in the 4~cm $\beta^*$ mode would  
result in a linear lattice with a large transverse aperture.
With only linear elements, the 4~cm $\beta^*$ optics was found to be strongly
nonlinear with limited on-momentum dynamic acceptance.

A normal form analysis
using COSY INFINITY \cite{cosy} showed that the variation of tune shift with amplitude was large, which was the source
of the strong nonlinearity in the seemingly linear lattice.
To locate the source of this nonlinearity, a lattice 
consisting of the original IR
and arcs only (no CC), was studied. Numerical
studies confirmed  similar dynamic aperture and
variation of tune shift with amplitude.
This ruled out the possibility that the dynamic aperture was limited by
the low beta points in the local chromatic correction section and points 
to the IR as the source of the
nonlinearity. These findings were also verified \cite{ohnuma} using a Runge-Kutta integrator to track through the IR and a linear matrix for the rest of the
lattice. Further analytical study using perturbation theory showed that the first order
contribution to the tune shift with amplitude is proportional to $\gamma^2_{x,y}$ and
$\gamma_x \gamma_y$, which are large in this IR. These terms come from the nonlinear
terms of $p_x/p_0$ and $p_y/p_0$, which, to  first order, equal the angular
divergence of a
particle. As a demonstration, a comparison to the LHC low beta IR was done. Taking into
account only the drift from the IP to the first quadrupole, the horizontal detuning
at 10$\sigma$ of the present IR ($\beta^{*}$ $=$ 4~cm) is 0.01, whereas the detuning of
the entire LHC lattice is below 1E-4. This also explains the fact that the on-momentum
aperture of the wide momentum spread mode remains roughly constant 
despite various versions and correction attempts.                                              

It was therefore concluded and later shown that the dynamic aperture of the more
relaxed $\beta^*$ of 14~cm would not have the same strong nonlinearities
due to the reduced angular terms.
In fact, the variation of tune shift with amplitude was less by an order of magnitude;
hence the large transverse acceptance shown in Fig.~\ref{50_gev_ap} (dashed line).

\subsubsection{100~GeV CoM. Example~(b)}
%\mbox{~~}\\[-0.95in]
The second lattice design, Example~(b), 
is shown in Fig.~\ref{f1} starting from the IP.
The 1.5~m background clearing dipole is 2.5 m away from the IP and is 
followed by the triplet
quadrupoles with the focusing quadrupole in the center.  
The interaction region (IR) stops at 
about 24~m from the IP.  Because of the small low betatron
functions in both transverse planes, the betatron functions at the 
final focusing triplets increase to $\sim 1550$~m.  The natural chromaticities,
of order $\sim -40$, are high, requiring local correction.
Due to the size limitation of the collider ring, it appears that we have room
for only two pairs of interleaved sextupoles on each side of the IP, each
pair correcting chromaticity in one transverse plane.
The correction section on each side of the IP spans
a distance of roughly 61.3~m.
\begin{figure*}[bht!]
\centering{\epsfig{figure=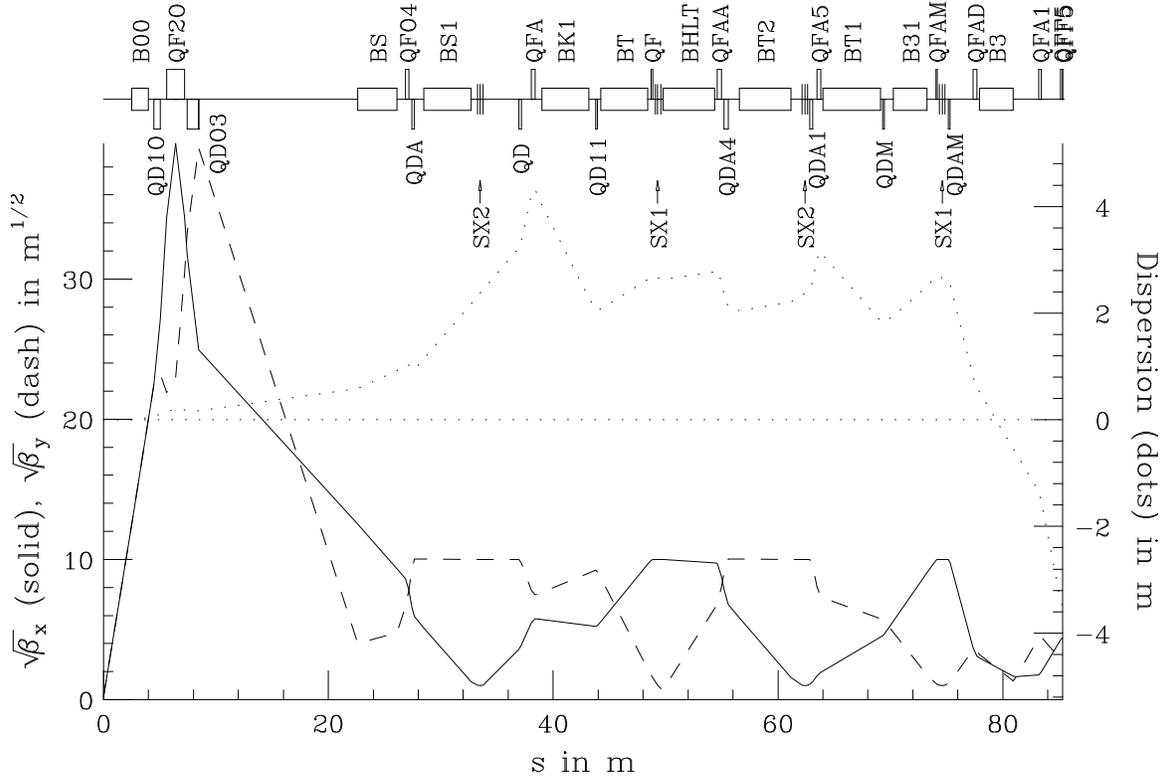,width=6.0in,bbllx=73bp,bblly=342bp,bburx=530bp,bbury=658bp,clip=}}
\caption[Example (b): Lattice structure of the IR including 
local chromaticity corrections]{Example (b): Lattice structure of the IR including 
local chromaticity corrections.
The maximum and minimum $\beta_x$ are 1571.74 and 0.040~m,
the maximum and minimum $\beta_y$ are 1550.94 and 0.040~m, while
the maximum and minimum dispersions are  4.31 and $-3.50$~m.
The natural horizontal and vertical chromaticities are $-41.46$ and $-39.90$,
giving a transition gamma of $\gamma_t=5.52$.  
The total module length is 85.32~m
with a total bend angle of 1.307~rad.}
\label{f1}
\end{figure*}

The SX1's are the two horizontal
correction sextupoles.  They should be placed at positions with the
same betatron functions and dispersion function, and
separated horizontally and vertically by phase advances $\Delta\psi_x$
and $\Delta\psi_y=\pi$ 
so that their nonlinear effect
will be confined in the region between the two sextupoles.  Their horizontal
phase advances should also be  
integral numbers of $\pi$ from the triplet focusing F-quadrupole
so that the chromaticity compensation for that quadrupole will be most 
efficient \cite{donald}.  
The SX2's are the
two vertical correction sextupoles which should be 
placed similarly at designated
locations.
In general, it will be difficult to satisfy all the requirements mentioned;
especially in this situation, luminosity arguments limit the lattice size.
For this lattice, the Twiss properties at the centers of 
the four correction sextupoles are 
listed in table~\ref{t1}, where all the figures given by the lattice code
are displayed.
An attempt was made to satisfy all the requirements at the expense of
having $\Delta\psi_y/(2\pi)=0.60$ instead of 0.50 for the SX1's.
This trade-off is explained below.

\begin{table*}[bht!]
\vskip -0.05in
\caption{Twiss properties of the IR correction sextupoles.}
\label{t1}
\begin{tabular}{ccccddc}
%\vspace{-0.25in}\mbox{~~~~}&&&&&&\\
&Distance  & \multicolumn{2}{c}{~~~Phase Advances} & 
\multicolumn{2}{c}{~~~~~Betatron Functions (m)} & Dispersion \\
&(m) &$\psi_x/(2\pi)$ &$\psi_y/(2\pi)$ &~~~$\beta_x$ &~~~$\beta_y$ &(m)\rule[-0.08in]{0in}{-0.08in} \\
\tableline
SX2 & 33.5061 & 0.48826 & 0.74953 &   1.00000 &  100.00012 & 2.37647\\ 
SX2 & 62.3942 & 0.98707 & 1.24953 &   1.00000 &  100.00009 & 2.37651\\  
SX1 & 49.3327 & 0.74892 & 0.87703 & 100.00023 &    1.00000 & 2.66039\\
SX1 & 74.6074 & 1.24892 & 1.47987 &  99.99967 &    0.99992 & 2.65817\\   
\end{tabular}
\end{table*}

The second order effects of the sextupoles
contribute to the amplitude dependent tune spreads, which, if too large,
can encompass
resonances leading to dynamical aperture limitation.  For example, in
this lattice,
\begin{eqnarray}
\nu_x\!&=8.126337 - ~100~\epsilon_x - \,4140~\epsilon_y,
\nonumber \\
\nu_y\!&=6.239988 - 4140~\epsilon_x - ~54.6~\epsilon_y,
\label{amp-tune}
\end{eqnarray}
where $\epsilon_x$ and $\epsilon_y$ are the horizontal and vertical
unnormalized emittances in $\pi$m.
In order to
eliminate these tune spreads due to the sextupole nonlinearity, 
the sufficient conditions are \cite{nonlinear}:
\begin{displaymath}
\sum_k\frac{S_ke^{i\psi_{xk}}}{\sin\pi\nu_x}=0\,,~~~~
\sum_k\frac{S_ke^{i3\psi_{xk}}}{\sin3\pi\nu_x}=0\,,~~~~
\end{displaymath}
\begin{equation}
\sum_k\frac{\bar S_ke^{i\psi_{xk}}}{\sin\pi\nu_x}=0\,,~~~~
\sum_k\frac{\bar S_ke^{i\psi_{\pm k}}}{\sin\pi\nu_\pm}=0\,,
\label{restriction}
\end{equation}
where for the $k$th thin normal sextupole with strength 
$S_{Nk}=
~\raisebox{-1.3ex}{$\stackrel
{\textstyle\lim}{\scriptstyle\ell\rightarrow0}$}~
%\stackk{\lim}{\scriptstyle\ell\rightarrow0}
[B''\ell/(B\rho)]_k$,
\begin{equation}
S_k=\left[S_{_N}\beta_x^{3/2}\right]_k\,,
\quad
\bar S_k=\left[S_{_N}\beta_x^{1/2}\beta_y\right]_k\,,
\end{equation}
$\psi_{\pm k}=(2\psi_y\pm\psi_x)_k$, 
and $\nu_{\pm k}=(2\nu_y\pm\nu_x)_k$.
The 5 requirements come about because there are 5 first order resonances
driven by the sextupoles when the residual tunes of the ring satisfy
$[3\nu_x]=0$, $[\nu_\pm]=0$ and two $[\nu_x]=0$.
The nominal tunes shown in Eq.~(\ref{amp-tune}) are far from these
resonances.  Therefore, the sines in the 
denominators of Eq.~(\ref{restriction})
can be omitted in this discussion.  
Since the strengths of SX1 and SX2 are similar, we have
$S_{\rm SX2}\ll\bar S_{\rm SX1}\ll\bar S_{\rm SX2}\ll S_{\rm SX1}$.
In fact, they are roughly in the ratios of
$1:(\beta_{\rm max}/\beta_{\rm min})^{1/2}
:\beta_{\rm max}/\beta_{\rm min}:(\beta_{\rm max}/\beta_{\rm min})^{3/2}$,
which amount roughly to 1:10:100:1000 in this lattice.
In above, $\beta_{\rm max}$ represents either $\beta_x$ at the SX1's or
$\beta_y$ at the SX2's, and 
$\beta_{\rm min}$ represents either $\beta_y$ at the SX1's or
$\beta_x$ at the SX2's.
Thus, the first two restrictions in Eq.~(\ref{restriction}) are the most
important, implying that all  $\beta_{\rm max}$ and $\beta_{\rm min}$
for each pair of SX1's must be made equal  and 
$\Delta\psi_x=\pi$ between them must be strictly obeyed.
The third restriction is the next important one, for which $\bar S_{\rm SX2}$
must be made equal for each pair of SX2's and their horizontal phase
difference must equal $\pi$.
The only two parameters left are
$\Delta\psi_y$ 
between a pair of SX1's and  $\Delta\psi_y$ between a pair of 
SX2's.  They affect
the restrictions for the
$\nu_\pm$ resonances only,
where the effective sextupole strengths $\bar S_{\rm SX1}$
and $\bar S_{\rm SX2}$ are involved.  Thus if we allow one restriction
to be relaxed,
the relaxation of $\Delta\psi_y=\pi$ for the SX1's 
will be least harmful.

Flexible momentum compaction (FMC) modules \cite{ref32a} 
are used in the arc.  The momentum compaction of the arc has to be 
made negative in order
to cancel the positive momentum compaction of the IR, so that the whole ring
becomes quasi-isochronous. 
This is accomplished in three ways:
1) removing the central dipole of the usual FMC
module; 2) increasing the length of the first and last dipoles, and 3) 
increasing the negative dispersion at the entrance. 
 Two such modules will be 
required for half of the
collider ring, one of which is shown in Fig.~\ref{f2}.
To close the ring
geometrically, there will be a $\sim72.0$~m straight section between the two sets of FMC modules.  The total length of the collider ring is now only
$C=354.3$~m.  
This is a nice feature, since a small ring allows a larger number of 
collisions before the muons decay appreciably.  
Note that the IR and local
 correction sections take up 48.2\% of the whole ring.
The momentum compaction factor
of this ring is now $\alpha_0=-2.77\times10^{-4}$.
The rf voltage required to maintain a bunch with rms length 
$\sigma_{\rule[0.05in]{0in}{0.05in}\ell}$
and rms momentum spread $\sigma_{\rule[0.05in]{0in}{0.05in}\delta}$ is 
 $V_{\rm rf}=|\eta|EC^2\sigma_\delta^2/(2\pi h\sigma_\ell^2)$,
where $\eta$ is the slippage factor and $E$ the muon energy.
On the other hand, 
if the bucket height
is taken as $k$ times the rms momentum spread of the bunch, 
the rf harmonic is given by 
$h=C/(k\pi\sigma_{\rule[0.05in]{0in}{0.05in}\ell})$.  Thus, for 
$\sigma_{\rule[0.05in]{0in}{0.05in}\ell}=4$~cm and 
$\sigma_{\rule[0.05in]{0in}{0.05in}\delta}=0.0012$, this lattice requires 
an rf voltage
of $V_{\textrm {rf}}\approx 88k$~kV.  Since $\alpha_0$ is
negative already, its absolute value can be further lowered easily
if needed.  However,
we must make sure that the contributions from the higher order momentum
compaction are small in addition.
\begin{figure*}[bht!]
\centering{\psfig{figure=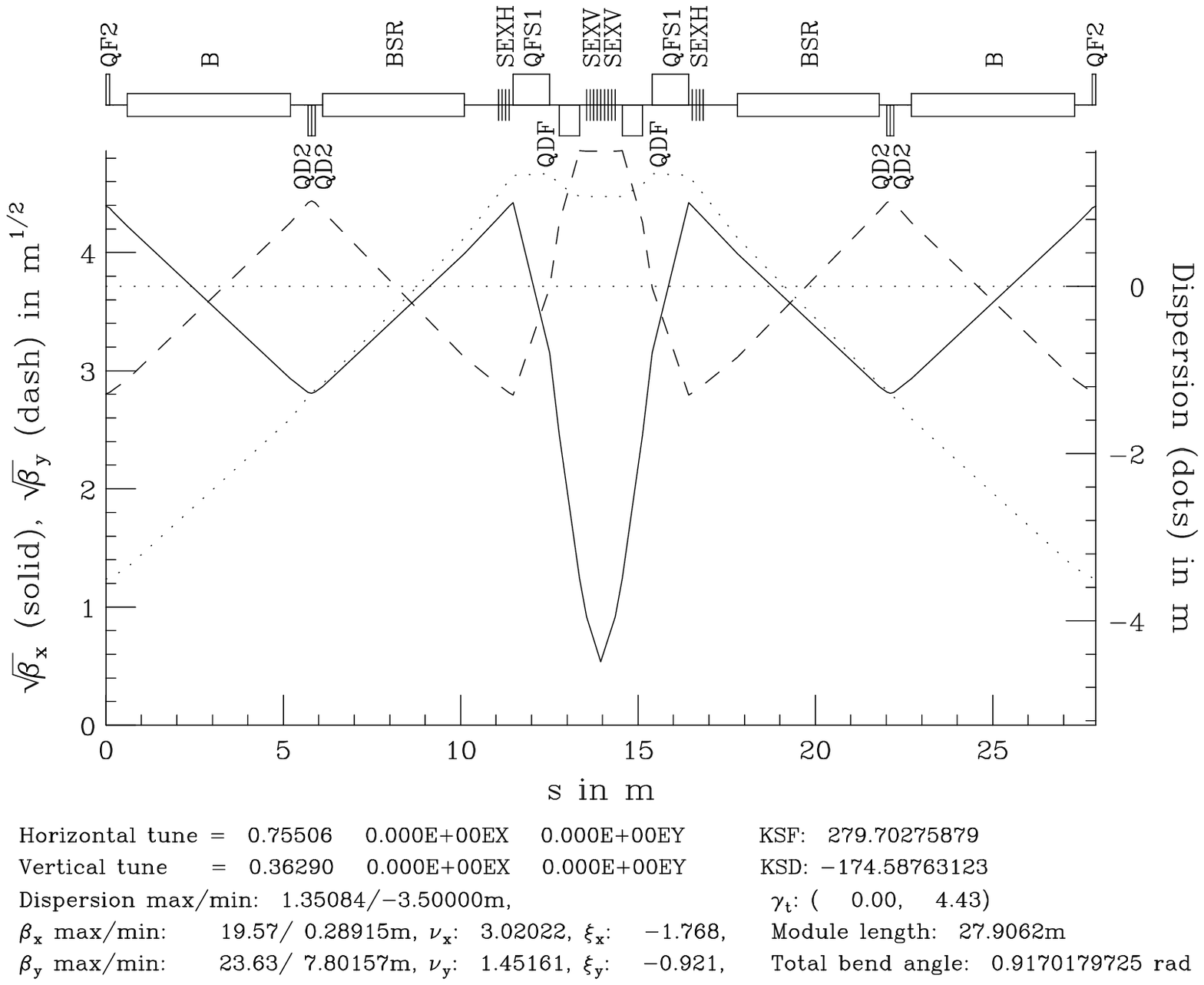,width=6.0in,bbllx=73bp,bblly=342bp,bburx=530bp,bbury=658bp,clip=}}
\caption[Lattice structure of the flexible momentum compaction module]{Example (b):
Lattice structure of the flexible momentum compaction module.
The maximum and minimum $\beta_x$ are 19.57 and 0.29~m,
the maximum and minimum $\beta_y$ are 23.63 and 7.80~m,
the maximum and minimum dispersions are  1.35 and $-3.50$~m.
The natural horizontal and vertical chromaticities are $-1.77$ and $-0.92$,
giving a transition gamma of $\gamma_t=i4.43$.  
The total module length is 27.91~m
with a total bend angle of 0.917~rad.}
\label{f2}
\end{figure*}

The dynamical aperture of the lattice is computed by tracking particles with 
the code COSY INFINITY \cite{cosy}.
Initially 16 particles with the same momentum offset and 
having vanishing $x'$ and $y'$ are placed uniformly on a circle in
the $x$-$y$ plane. 
The largest radius that provides survival of the 16 particles
in 1000 turns is defined here as the dynamical aperture at this momentum 
offset and is plotted in solid in Fig.~\ref{f3} in units of the rms radius of the beam.  
(At the 4~cm low beta IP, 
the beam has an rms radius of $82~\mu$m.)  As a reference,
the 7~$\sigma$ aperture
spanning $\pm6$~sigmas of momentum offset is also displayed as a
semi-ellipse in \textcolor{red}{dashdot}.
To maximize the aperture, first, the tunes must be chosen to avoid
parametric resonances.  The on-momentum 
amplitude dependent horizontal and vertical tunes
are given in Eq.~(\ref{amp-tune}).
With the designed rms
$\epsilon_x= \epsilon_y=0.169\times10^{-6}~\pi$m, 
the on-momentum tune variations are at most 0.0007. 
Second, the chromaticity variations with momentum must be as small as 
possible.  This is shown in Fig.~\ref{f3} (right hand side plot).  Note that there are no families 
of sextupoles to correct for the higher order chromaticities in this
small ring with only four FMC modules.
As the  momentum spread varies from $-1$ to 0.9\%,
$\nu_x$ varies from 8.16698 to 8.07459, and $\nu_y$ from 6.28305 to 6.22369 
for the center of the beam.

During aperture tracking we notice that
particle loss occurs mostly in the horizontal direction. 
We are convinced that the small momentum aperture is a result of the 
large dispersion swing in the lattice
from $+4.5$ to $-3.5$~m.
For example, $4.5$~m dispersion and 0.6\%
momentum offset translates into a 2.7~cm off-axis motion.  
The nonlinearity of the lattice will therefore diminish the dynamical
aperture.  A resonant strength study using, for example, swamp plots and
normalized-resonance-basis-coefficient analysis \cite{yan} actually reveals that 
this lattice and some of its variations are unusually nonlinear. 
Recently, we make a modification of the FMC arc modules which have
a smaller dispersion swing from $-2.6$ to $+2.0$~m only.  The IR  
has not been changed except for the matching to the arc modules.
 The aperture has been tracked with TEAPOT \cite{teapot} 
in the same way as COSY and is plotted as \textcolor{magenta}{dashes} 
in Fig.~\ref{f3} (left hand side plot).  We see that the momentum
aperture has widened appreciably.  The dynamical aperture near on-momentum, however, is one sigma less than the lattice presented here.  
Nevertheless, it is not clear that this decrease
is significant because all
tracking has been performed in steps of one sigma only.
However this type of aperture is still far from satisfactory, because
so far we have been studying
a bare lattice.  The aperture will be reduced 
when fringe fields, field errors, and misalignment
errors are included.

We suspect that the aperture for small momentum spread is limited by
the dramatic changes in betatron functions near the IP \cite{ohnuma}.  
These changes
are so large that Hill's equation would no longer be adequate and  the 
exact equation for beam transport must be used.  This equation brings in
nonlinearity and limits the aperture, which can easily be demonstrated
by turning off all the sextupoles.
In other words, although the momentum
aperture can be widened by suitable deployment of sextupoles, the
on-momentum dynamical aperture is determined by the triplet quadrupoles and
cannot be increased significantly by the sextupoles. 
Some drastic changes in the low beta design may be necessary.

\begin{figure*}[hbt!]
\dofigs{3.25in}{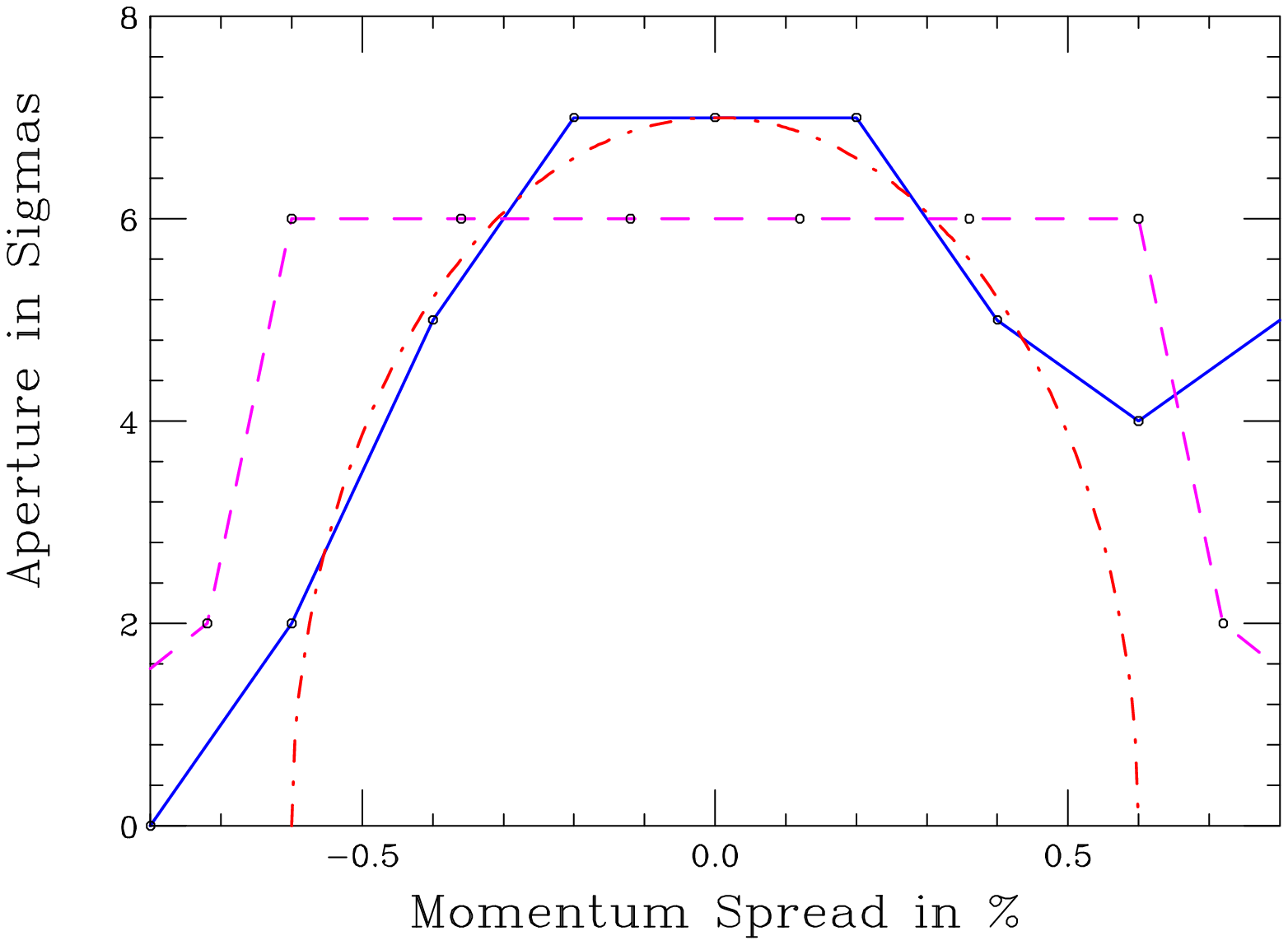}{3.25in}{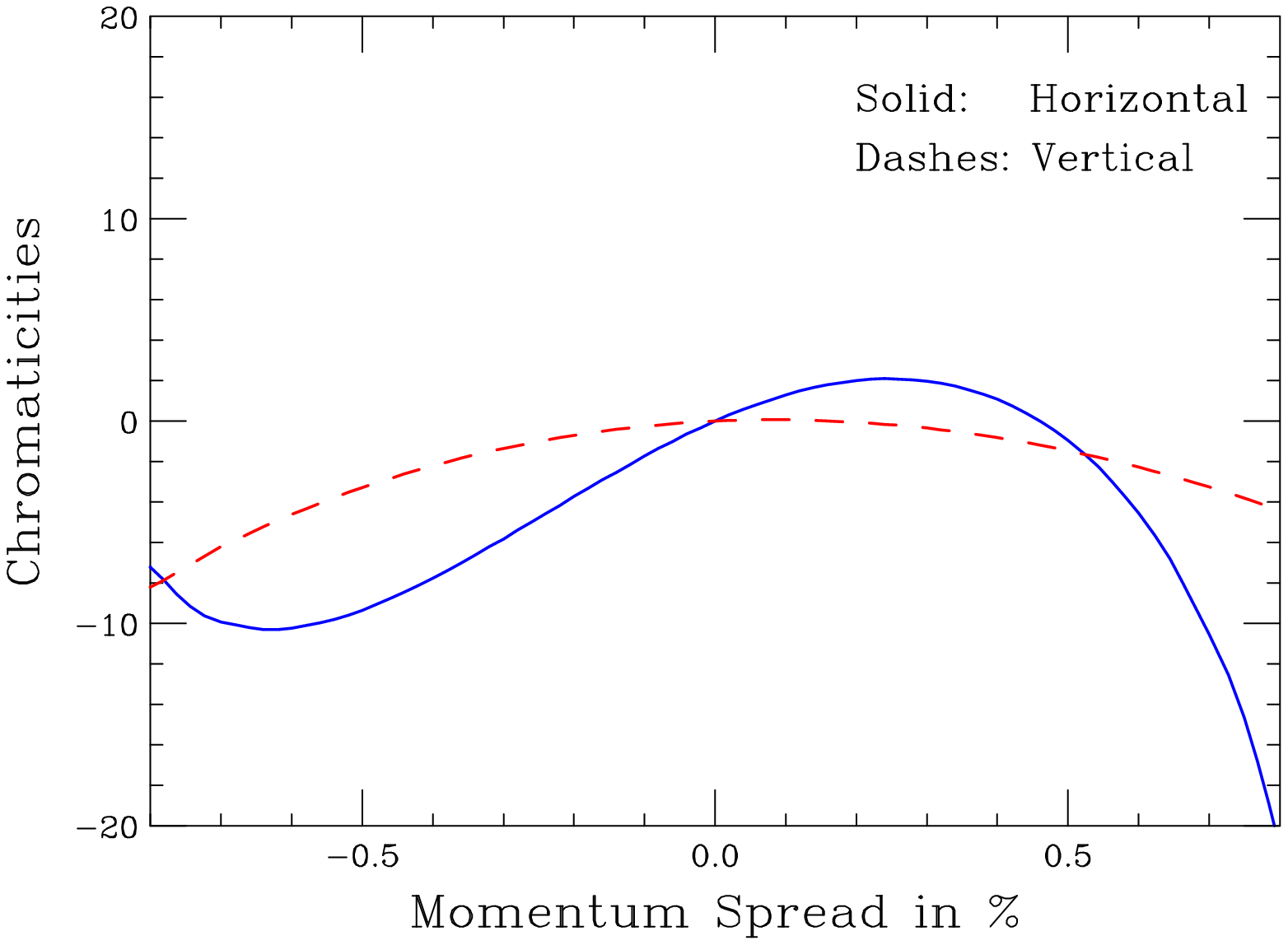}

%\small
\caption[Example~(b): Dynamical aperture and chromaticities {\it vs.}  momentum offset]
{Example~(b): Left hand side plot is dynamical aperture of the lattice {\it vs.} momentum offset. 
COSY calculation in \textcolor{blue}{solid}, $7~\sigma$ in \textcolor{red}{dot-dashes}, and 
TEAPOT calculation with modified FMC modules in \textcolor{magenta}{dashes}. 
Right hand side plot is  chromaticities {\it vs.} momentum offset.}
\label{f3}
%\vspace{-.5in}
\end{figure*}

\subsection{Scraping}

It has been shown \cite{ref42} that detector backgrounds
originating from beam halo
can exceed those from decays in the vicinity
of the interaction point (IP). Only with a dedicated beam cleaning system
far enough from the IP can one mitigate this problem \cite{scraping}.
Muons injected with large momentum errors or betatron oscillations will be lost 
within the first few turns. After that, with active scraping,
the beam halo generated through
beam-gas scattering, resonances and beam-beam 
interactions at the IP
reaches equilibrium and beam losses remain constant throughout the
rest of the cycle.

Two beam cleaning schemes have been designed \cite{scraping}, one for muon colliders at 
high energies, and one for those at low energies. 

The studies \cite{scraping} showed that no absorber, ordinary or
magnetized, will suffice for beam cleaning at 2~TeV;
in fact, the disturbed
muons are often lost in the IR, but a simple metal collimator was found to be satisfactory at 100 GeV.

\subsubsection{Scraping for high energy collider}

At high energies, a 3~m long electrostatic deflector (Fig.~\ref{scrap1}) separates 
muons with amplitudes larger than 3~$\sigma$ and deflects them into
a 3~m long Lambertson magnet, which extracts these downwards through a deflection
of 17~mrad. A vertical septum magnet is used in the vertical scraping section 
instead of the Lambertson to keep the direction of extracted beam down. 
The shaving process lasts for the first few turns.
To achieve practical distances and design apertures for the separator/Lambertson 
combinations,
$\beta$ functions must reach a kilometer in the 2~TeV case, but only 100~m at 50~GeV.
The complete system consists of a vertical scraping 
section and two horizontal ones for positive and negative momentum 
scraping (the design is symmetric about the center, so
scraping is identical for both $\mu^+$ and $\mu^-$). The system provides the  scraping power of a factor of 1000; that is, for every 1000 halo muons, one remains.

\subsubsection{Scraping for low energy collider}

At 50~GeV, 
collimating muon halos with a 5~m long 
steel absorber (Fig.~\ref{scrap2}) in a simple compact utility section
does an excellent job. Muons lose a significant fraction of their energy in 
such an absorber (8\% on average) and have broad angular and spatial distributions.
Almost all of these muons are then lost in the first 50-100~m downstream 
of the absorber with only 0.07\% of the scraped muons reaching the low $\beta$ quadrupoles in the IR,~\ie~a scraping power is 1500 in
this case, which is significantly better than with an earlier septum scraping system design \cite{scraping} similar to that developed for the high energy collider.

\begin{figure*}[thb!]
\centering{{\epsfig{figure=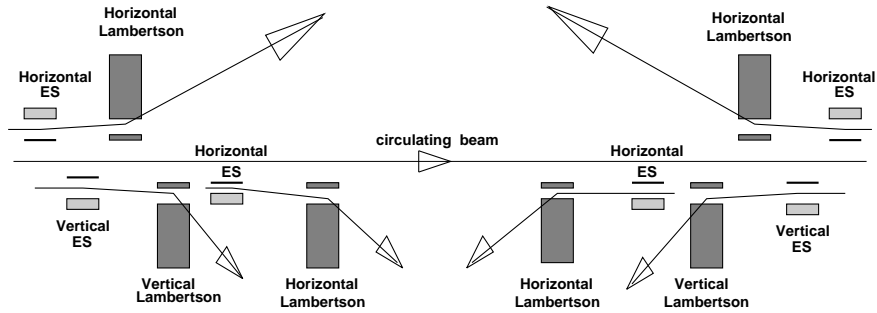,width=1.6in,angle=270}}}
\vspace{10pt}
\caption{Schematic view of a \mumu collider beam halo extraction.}
\label{scrap1}
\end{figure*}

\begin{figure*}[thb!]
\centering{{\epsfig{figure=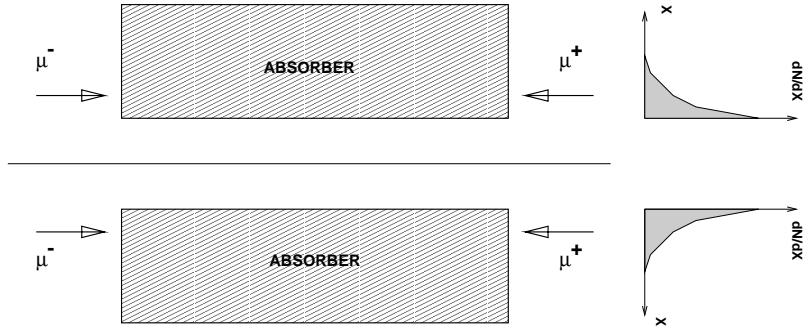,width=1.7in,angle=270}}}
\vspace{10pt}
\caption{Scraping muon beam halo with a 5~m steel absorber.}
\label{scrap2}
\end{figure*}

\subsection{Beam-beam tune shift}
Several studies have considered beam emittance growth due to the beam-beam tune shift and none 
have observed significant luminosity loss. For instance, a study \cite{furman}, using the high 
energy collider parameters (see table~\ref{sum}), in which particles were tracked 
assuming Gaussian beam field distributions, and no muon decay, showed a luminosity
 loss of only 4\%. With muon decay included, the loss contribution from beam-beam effects is even less. 
Another study \cite{chen} using a particle in cell approach with no assumptions about field symmetry obtained a similar result. 
Collisions between beams displaced by 10\% of their radius also gave little loss. 
But all these studies assumed an ideal lattice, and none considered whether small losses due to nonlinearities give rise to an  
unacceptable background.

\subsection{Impedance/wakefield considerations} 

A study \cite{ref35} has examined the resistive wall impedance longitudinal 
instabilities in rings at several energies. At the higher energies and larger 
momentum spreads, solutions were found with small but finite momentum 
compaction and moderate rf voltages. For
 the special case of the Higgs Factory, with its very low momentum spread, a 
solution was found with no synchrotron motion, but rf was provided to correct 
the first order impedance generated momentum spread. The remaining 
off-momentum tails which might generate background could be removed by a 
higher harmonic rf correction without affecting luminosity.
Solutions to the higher energy and larger momentum spread cases without synchrotron motion are also being considered.

Given the very slow or nonexistent synchrotron oscillations, the transverse beam breakup instability is significant. This instability can
be  stabilized using rf quadrupole \cite{ref36} induced BNS damping. 
For instance, the required tune shift with position in the bunch, calculated using 
the two particle model approximation \cite{ref38}, is only $1.58\times10^{-4}$ for the 3~TeV case 
using a 1~cm radius aluminum pipe. This stabilizes the resistive wall instability.  
However, this application of BNS damping to a quasi-isochronous ring, and other 
head-tail instabilities due to the  chromaticities $\xi$ and $\eta_1$, needs more study.

\subsection{Bending magnet design}

The dipole field assumed in the 100~GeV collider lattices described above was 8~T. This field can be obtained using $1.8^o$ niobium titanium (NbTi) \textit{cos theta} superconducting magnets similar to those developed for the LHC. The only complication is the n
eed for a tungsten shield between the beam and coils to shield the latter from beam decay heating.

The $\mu$'s decay within
the rings ($\mu^- \rightarrow\ e^-\overline{\nu_e}\nu_{\mu}$), producing
electrons whose mean energy is approximately $0.35$ that of the muons.   With no shielding, the average power deposited per unit length would be about 2 kW/m
in the 4 TeV machine, and 300 W/m in the 100 GeV Higgs factory.
Figure~\ref{shieldingnew} shows the power penetrating tungsten shields of different 
thickness \cite{ref6a,carnik96,scraping,shield96}. One sees that 3 cm in the low energy case, 
or 6~cm at high energy would reduce the power to below 10 W/m, which can reasonably 
be taken by superconducting magnets. 

\begin{figure*}[bht!]
\centerline{
\epsfig{file=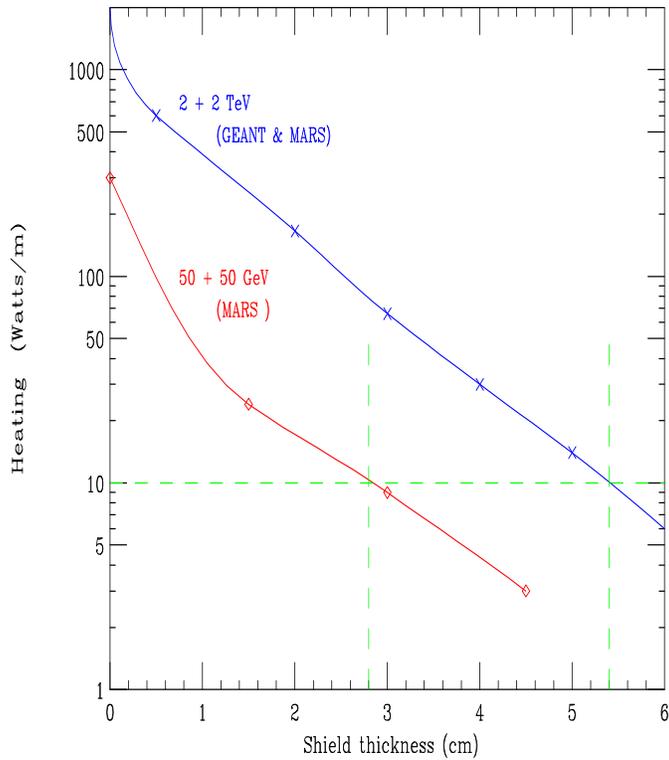,height=3.95in,width=3.45in}
} 
\caption{Power penetrating tungsten shields vs.\ their thickness for  a) 4~TeV, and b) 0.1~TeV, colliders. \label{shieldingnew}}
 \end{figure*}

Figure~\ref{costheta} shows the cross section of a baseline magnet suitable for the 100~GeV collider. 

\begin{figure*}[bht!]
\centering{
\epsfig{figure=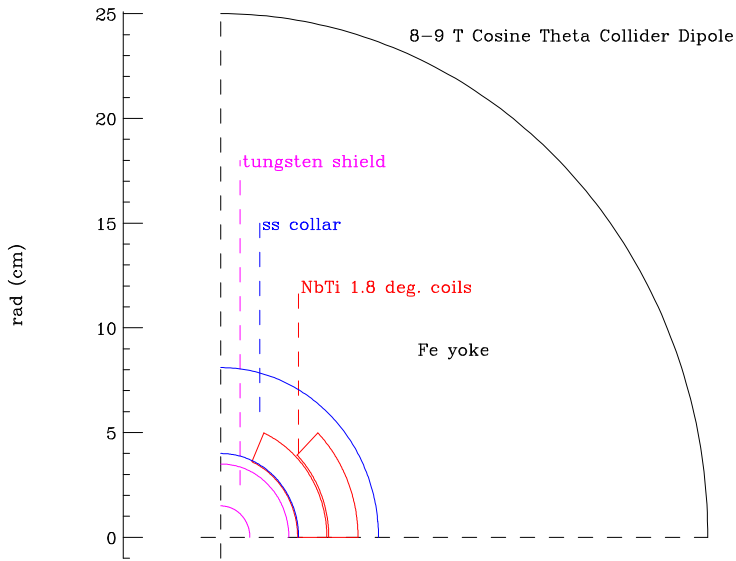, width=4in}
}
\caption{Cross Section of a baseline dipole magnet suitable for the 100~GeV collider.
 \label{costheta}}
 \end{figure*}

The quadrupoles could use warm iron
poles placed as close to the beam as practical. The coils could then be either 
superconducting or warm, placed at a greater distance from the beam and shielded from it by the poles. 

The collider ring could be made smaller, and the luminosity increased, if higher field dipoles were used. In the low energy case, the gain would not be great since less than half the circumference is devoted to the arcs. For this reason, and to avoid yet 
another technical challenge, higher field magnets are not part of the baseline design of a 100 GeV collider. But they would give a significant luminosity improvement for the higher energy colliders, and would be desirable there. There have been several studies of possible designs, three of which (two that are promising and one that appears not to work) are included below.

\subsubsection{Alternative racetrack Nb$_3$Sn dipole}

A higher field magnet based on Nb$_3$Sn conductor and racetrack coils is presently being designed. The Nb$_3$Sn conductor allows higher  fields and provides a large temperature margin 
over the operating temperature, but being brittle and sensitive to bending or other stress, presents a number of engineering challenges.

In this design, the stress levels in the conductor are reduced by the use of a rectangular coil block geometry and end support problems are reduced by keeping the coils flat.
In the more conventional \textit{cos theta} designs, the conductor is distributed around a cylinder and the forces add up towards the midplane; in addition, the ends, as they arc over the cylinder, are relatively hard to support.

The geometry of the cross section is shown in Fig.~\ref{g_mag}. It uses all 2-D flat racetrack 
coils. Each quadrant of the magnet aperture has two blocks of 
conductors. The block at 
the pole in the first quadrant has a return block in the second 
quadrant, similar to that in 
a conventional design. The height of this block is such that it 
completely clears the 
bore. In a conventional design, the second block, the midplane block, 
would also have a 
return block in the second quadrant. That would, however, require the 
conductor block to 
be lifted up in the ends to clear the bore and thus would lose the 
simple 2-D geometry. In 
the proposed design, the return block retains the 2-D coil geometry, 
as it is returned on the 
same side (see Fig.~\ref{g_mag}) and naturally clears the bore. Since the return block does not 
contribute to the field, this design uses 50\% more conductor. This, 
however, is a small 
penalty to pay for a few magnets where the performance and not the 
cost is a major issue. 
The field lines are also shown in Fig.~\ref{g_mag}.

Preliminary design parameters for two cases are given in table~\ref{magnetdesign}. The first 
case is one 
where the performance of the cable used is the same that 
is in the LBL D20  magnet, which created a central field of 13.5~T. The second case is the one 
where the cable is graded and two types of cable are used, and
it is assumed that a reported 
improvement in cable performance is realized.  
It is expected to produce a 
central field of 14.7~T when operated at $4.2~{}^{o}$K. 
%\begin{center}
\begin{table*}[tbh!]
\caption[Preliminary design parameters for a racetrack Nb$_3$Sn dipole]{Preliminary design parameters for a racetrack Nb$_3$Sn dipole with two different types of cable.}
\label{magnetdesign}
\begin{tabular}{ll}
\multicolumn {2}{l}{\emph{Case 1}:
 Same conductor as in LBL 13.5~T D20 magnet without grading} \\ \hline

Central field at quench & 13~T at $4.2~^{o}\!K$\\

Coil dimensions & $25~\textrm{mm} \times 70$~mm\\ 

Total number of racetrack coils in whole magnet & 6\\ 

Total number of blocks per quadrant in aperture & 2 (+1 outside the aperture)\\ 

Yoke outer radius &500~mm (same as in D20)\\

Field harmonics & a few parts in $10^{-5}$ at 10~mm\\

Midplane gap (midplane to coil) & 5~mm (coil to coil 10~mm)\\

Minimum coil height in the end & 45~mm (Note: coils are not lifted up.)\\
% &  \\
%\vspace{.1in}
\hline
\multicolumn {2}{l}{\emph{Case 2}: Newer conductor and graded}\\ 
\hline
Central field at quench & 14.7~T at $4.2~^{o}\!K$\\
Grading & 70~mm divided in two 35~mm layers \\
Overall current densities &  370~A/mm$^2$ and 600~A/mm$^2$\\
Peak fields & 16~T and 12.5~T\\
Copper current density & 1500~A/mm$^2$\\
Other features are the same as in \emph{Case 1} & \\
\end{tabular}
\end{table*}

\begin{figure*}[hbt!]
\centering{\psfig{figure=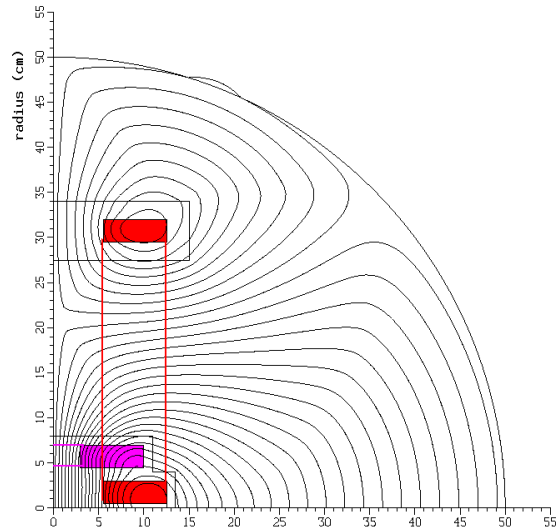,height=3.0in,clip=}}
\caption[Cross section of alternative high field race track coil dipole magnet     ]{Cross section of alternative high field ($\approx 15$~T) 
 race track coil dipole magnet with Nb$_3$Sn conductor.}
\label{g_mag}
\end{figure*}

\subsubsection{Alternative \textit{Cos Theta} Nb$_3$Sn dipole}

\begin{figure*}[hbt!]
\begin{center}
\includegraphics[width=4.5in,angle=90]{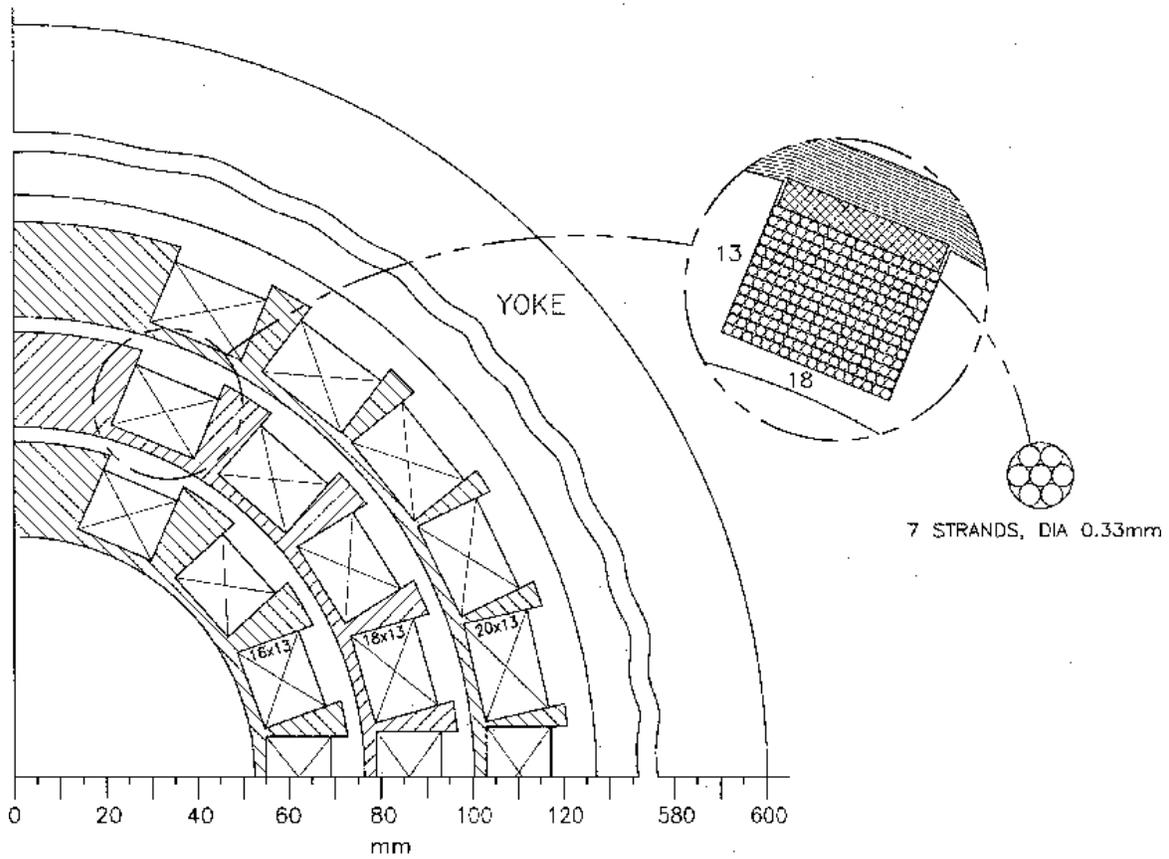}   
\end{center}
\caption{Cross section of alternative high field slot dipole made with 
Nb$_3$Sn conductor.}
\label{slotmag}
\end{figure*}
In this case the problem with the brittle and sensitive conductor is solved by winding the 
coil inside many separate slots cut in metal support cylinders. There is no build up of 
forces on the coil at the mid-plane. The slots continue around the ends, and 
thus solve the support problem there too.

Figure~\ref{slotmag} shows this alternative Nb$_3$Sn dipole \textit{cos theta} design. It is an extension of the 
concept used to build helical magnets \cite{willenhelical} for the polarized proton program at RHIC \cite{rhic}.
The magnet is wound with pre-reacted, kapton-insulated, B-stage impregnated, low current cable. 
The build up of forces is controlled by laying the cables in machined slots in a metal support 
cylinder. After winding, the openings of the slots are bridged by metal spacers and the coils pre-compressed 
inward by winding B-stage impregnated high tensile thread around the spacers. After curing, the outside of 
each coil assembly is machined prior to its insertion into an outer coil, or into the yoke. There are 3 layers. 
The inner bore is 55~mm radius, the outer coil radius approximately 118~mm, and the yoke inside radius is 127~mm. 
The maximum copper current density is 1300~A/mm$^2$.

Using the same material specifications as used in the above high field option, 
a central short sample field of 13.2 T was calculated. This is somewhat less 
than the block design discussed above, but could be improved by increasing the cable diameters to improve 
the currently rather poor (64\%) cable to cable-plus-insulator ratio.

\subsubsection{Study of C-magnet dipole}

\begin{figure*}[hbt!]
\centering{
\epsfig{figure=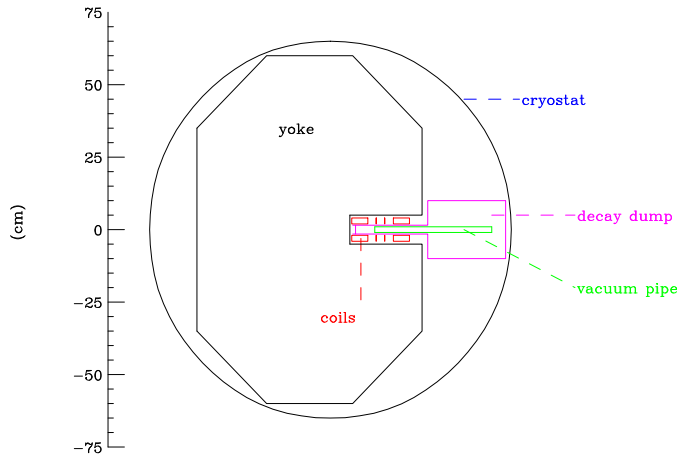,width=3.5in}
}
\caption{Cross section of an unsuccessful alternative high field C magnet with 
open mid-plane.}
\label{cmag}
%\end{center}
\end{figure*}

Figure~\ref{cmag} shows the cross section of a high field dipole magnet in which it was hoped to bring the coils closer to the beam pipe without suffering excessive heating from beam decay.  The coil design \cite{willencmag} appeared reasonable, but the required avoidance of coil heating was not achieved.

Decay electrons are generated at very small angles ($\approx \ 1/\gamma$) to the beam, and with an average energy about 1/3 of the beam.
Such electrons initially spiral inward (to the right in Fig.~\ref{cmag}) bent by the high dipole field. In the high energy case, these electrons also radiate a significant fraction of their energy as ($\approx$ 1~GeV) synchrotron gamma rays, some of which
 end up on the outside (to the left in Fig.~\ref{cmag}).
The concept was to use a very wide beam pipe, allow the electrons to exit between the coils, and be absorbed in an external cooled dump. Unfortunately a preliminary study found that a substantial fraction of the electrons did not reach the dump. They were
 bent back outward before reaching it by the return field of the magnet coils and the nature of the curved ring geometry. Such electrons were then trapped about the null in the vertical field and eventually hit the upper or lower face of the unshielded 
vacuum pipe. They showered, and deposited unacceptable levels of heat in the coils. 

Another idea called for collimators between each bending magnet that would catch such trapped electrons. This option has not been studied in detail, but the impedance consequences of such periodic collimators are expected to be unacceptable. 

Further study of such options might find a solution, but the use of a thick cylindrical heavy metal shield appears practical, adequate, and is thus the current baseline choice.

\subsection{Energy scale calibration}
 
 In order to scan the width of a Higgs boson of mass around 100~GeV, one needs
to measure the energy of the individual muon stores to an accuracy of a few
parts per million, since the width of a Higgs boson of that mass is expected to
be a few MeV. Assuming that muon bunches  can be produced with modest 
polarizations of $\approx 0.25,$ and that the polarization can be maintained
from turn to turn in the collider, it is possible to use the precession of the
polarization in the ring to measure accurately the average energy of the muons
 \cite{ref7}. The total energy of electrons produced by muon decay observed in
the calorimeter placed in the ring varies from turn to turn due to the $g-2$
precession of the muon spin, which is proportional to the Lorentz factor $\gamma$
of the muon beam. Figure~\ref{efig3} shows the result of a fit of the total
electron energy observed in a calorimeter to a functional form that includes 
muon decay and spin precession. Figure~\ref{efig4} shows the 
fractional error $\delta\gamma/\gamma$ obtained from a series of such fits
plotted against the fractional error of measurement in the total electron
energy that depends on the electron statistics. It has been  shown that precisions of a few parts per million in $\gamma$ are 
possible with modest electron statistics of $\approx 100,000$ detected.
It should be noted that there are $3.2\times 10^6$ decays per meter for a muon intensity of
$10^{12}$ muons.

\begin{figure*}[thb!]
\centering
\epsfxsize = 5in
\epsffile{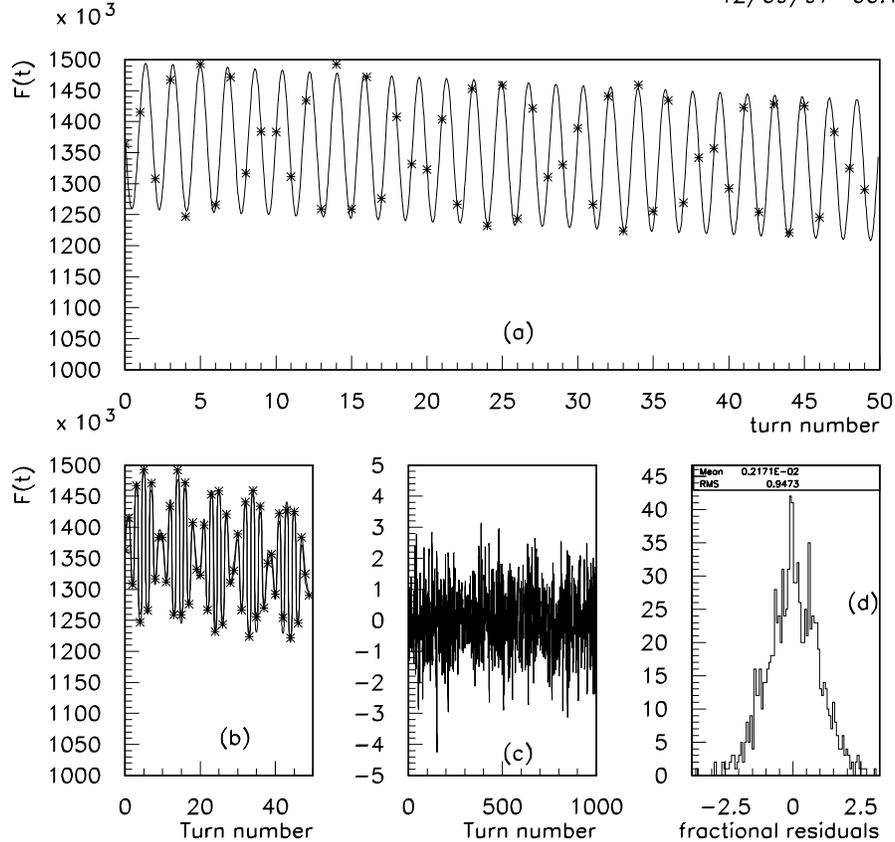}
\caption[ Energy detected in the calorimeter during the first 50 turns in a
50~GeV muon storage ring   ]{a) Energy detected in the calorimeter during the first 50~turns in a
50~GeV muon storage ring (points). An average polarization value of ${\hat P}=-0.26$ is
assumed and a fractional fluctuation of $5\times 10^{-3}$ per point. The curve is the
result of a MINUIT fit to the expected functional form. b) The
same fit, with the function being plotted only at integer turn
values. A beat is evident. c) Pulls  as a function of turn number. d) Histogram
of pulls. A pull is defined by (measured value-fitted value)/(error in
measured-fitted). }
\label{efig3}
\end{figure*}
\begin{figure*}[hbt!]
\centering
\epsfxsize = 5in
\epsffile{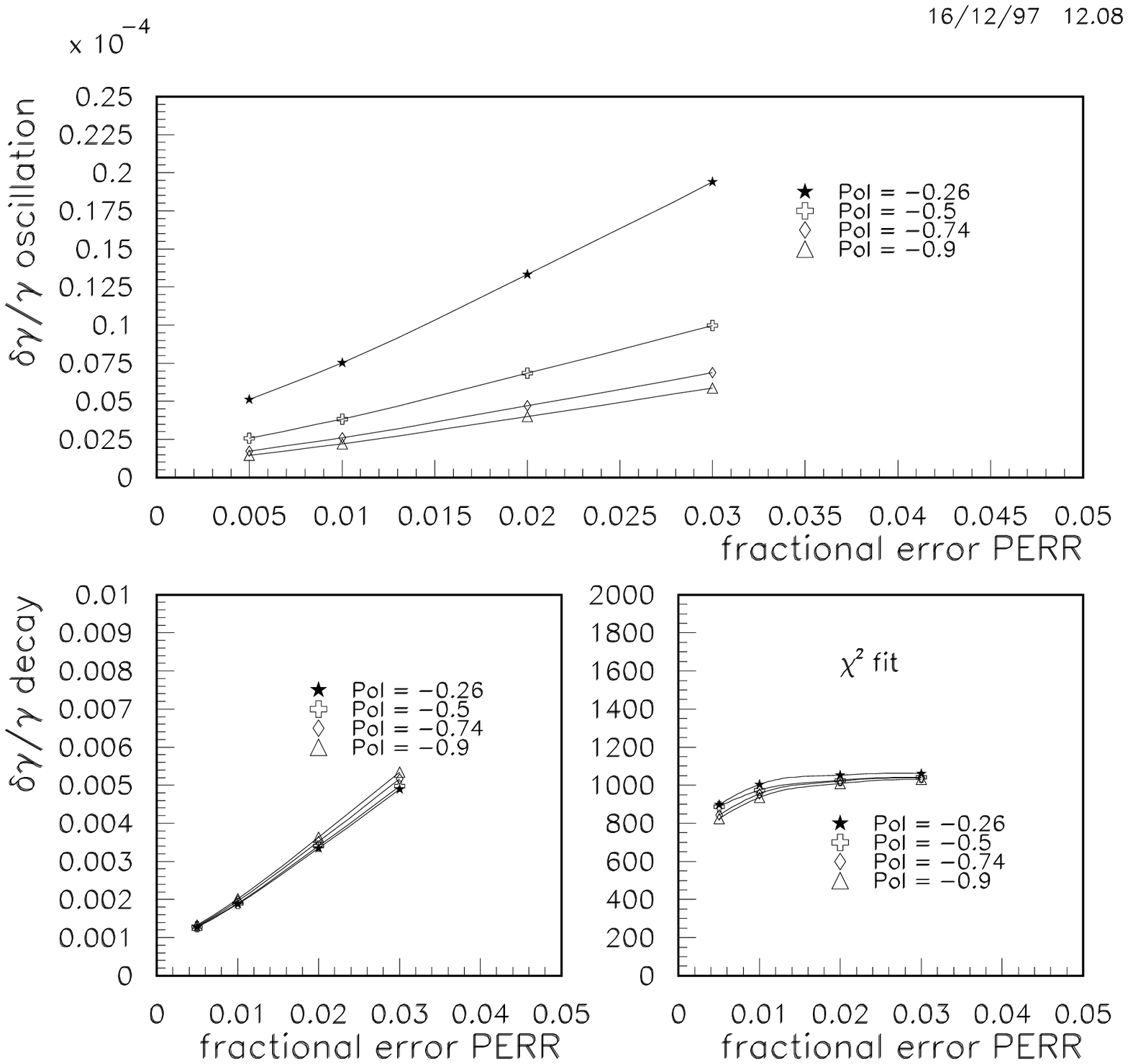}
\caption[Fractional error in $\delta\gamma/\gamma$   ]{a) Fractional error in $\delta\gamma/\gamma$ obtained from the
oscillations as a function of polarization $\hat P$ and the fractional error in
the measurements PERR. b) Fractional error in $\delta\gamma/\gamma$ obtained
from the decay term as a function of polarization $\hat P$ and the fractional
error in the measurements PERR. c) The total $\chi^2$ of the fits for 1000
degrees of freedom. PERR is the percentage measurement error on the total electron energy in the
calorimeter measuring the decay electrons. }
\label{efig4}
\end{figure*}

 Our current plans to measure the energy due to decay electrons entail an electromagnetic calorimeter that is
segmented both longitudinally and transversely and placed inside an enlarged beam
pipe in one of the straight sections in the collider ring. The length of the
straight section upstream of the calorimeter can be chosen to control the total number of decays  and hence the rate of energy deposition. The sensitive material can be
gaseous, since the energy resolution is controlled by decay fluctuations
rather than sampling error. In order to measure the total number of electrons
entering the calorimeter, we plan to include a calorimeter layer with little
absorber upstream of it as the first layer.

 This scheme will enable us to calibrate and correct the energy of individual
bunches of muons and permit us to measure the width  of a low mass Higgs boson.
\section{RADIATION AND BACKGROUNDS}

\subsection{Conventional radiation}
The proton source generates a 4~MW proton beam, which is comparable to the
proposed spallation source \cite{sns}. This is a very high power and will, as
in the spallation source, require great care in reducing unwanted particle
losses, as well as  careful machine shielding,  and target and beam dump
design. Initial studies of the target and capture solenoid region have been
performed with the MARS code, and preliminary specifications for shielding
determined, but more work is needed.

The cooling and accelerator chain is rather clean, since a relatively small
fraction of the muons decay, and their energies are low. Power deposited in the
accelerators is typically 10-30~W/m (see table~\ref{acceleration1} and table~\ref{acceleration3}).

If no muons are lost, then the only sources of radiation are the muon decays
yielding electrons and neutrinos. The neutrino radiation we discuss below. The
electrons shower in the collider beam pipe shields, depositing most of their
energy there and a relatively small amount in the magnet coils and yoke. Radioactivation levels, as calculated by MARS \cite{snowrad}, after five years of 4~TeV collider operation are given in table~\ref{beamradiation} for the cases immediately after turn off and 1 day after turn off. It is seen that the areas in the tunnel that are
outside the magnets are relatively free of radioactivation. Special procedures
will be needed when the shield pipe has to be opened, as for instance when a
magnet is changed. For the lower energy colliders, the  radioactivation levels
are proportionally less.

\begin{table}[htb!]
\caption[4~TeV (CoM) collider ring radioactivation levels ]{4~TeV (CoM)
collider ring radioactivation levels (mrem/hour) after turn off, for parameters
in table~\ref{sum}}
\label{beamradiation}
\begin{tabular}{lcc}
             & immediate   & after 1 day    \\
\hline
Inside face of shield     & 9000 & 4000  \\
Outside face of shield    &  200 & 170   \\
Outside of coils          & 30   &  14   \\
Outside of yoke           & 3    &  1.4  \\
\end{tabular}
\end{table}
If muons are lost either accidentally, by scraping, or deliberately after some
 number of turns, then the muons penetrate to considerable distances in the 
soil/rock (3.5~km at 2~TeV, 800~m at 250~GeV) and deposit their energy directly
 or through their interaction products. Figure~\ref{snowmass10.6} 
and Fig.~\ref{snowmass10.7} show the distribution of radiation levels, assuming 25\% of all muons (4 bunches of $2 \times 10^{12}$ at 15~Hz) are dumped into 
soil/rock with density 2.24~g/cm$^3$. 
 The outer contours correspond to the federal limits, reaching at maxima of 
18~m (2~TeV) and 14.5~m (250~GeV). To confine this radiation beneath the ground
one can deflect the extracted beams down by 4.5~mrad at 2~TeV and about 10~mrad
at $\leq$250~GeV.
If any water were present in the soil/rock, then the first two meters around
the tunnel and around the
aborted beam axis  would require insulation or drainage up to a distance of
2.5~km at 2~TeV or 550~m at 250~GeV.

\begin{figure*}[tbh!]
\centerline{\epsfig{file=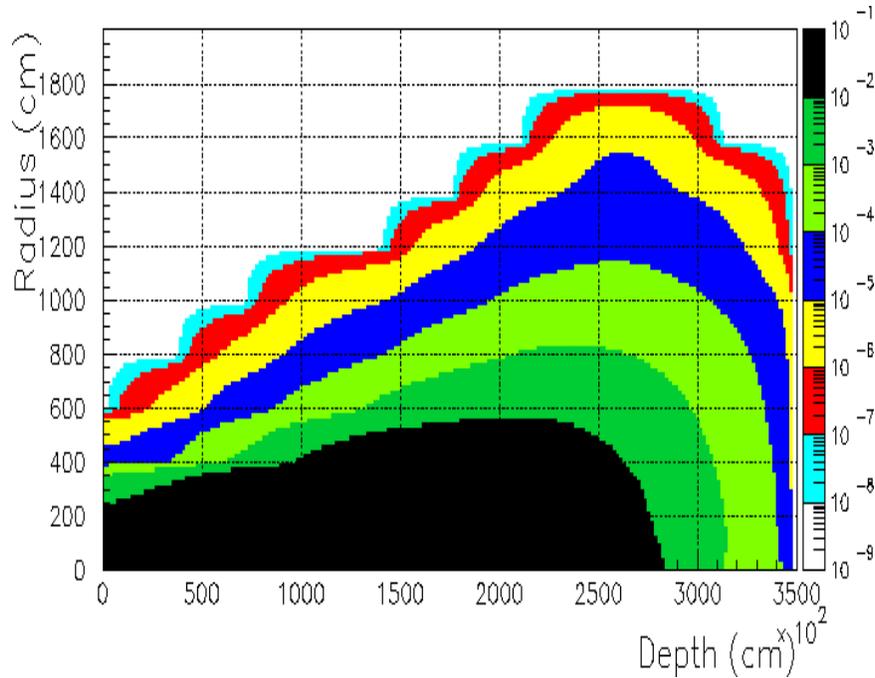,height=3.5in,width=4.5in}}
\caption[MARS isodose contours in the soil/rock for 2~TeV muons extracted at 
$3\times 10^{13}$ per second]
{Isodose contours in the soil/rock ($\rho$=2.24\,g/cm$^3$) for 2~TeV muons 
extracted at $3\times 10^{13}$ per second. Right scale is dose
rate in rem/s.} 
\label{snowmass10.6} 
\end{figure*}
\begin{figure*}[tbh!]
\centerline{\epsfig{file=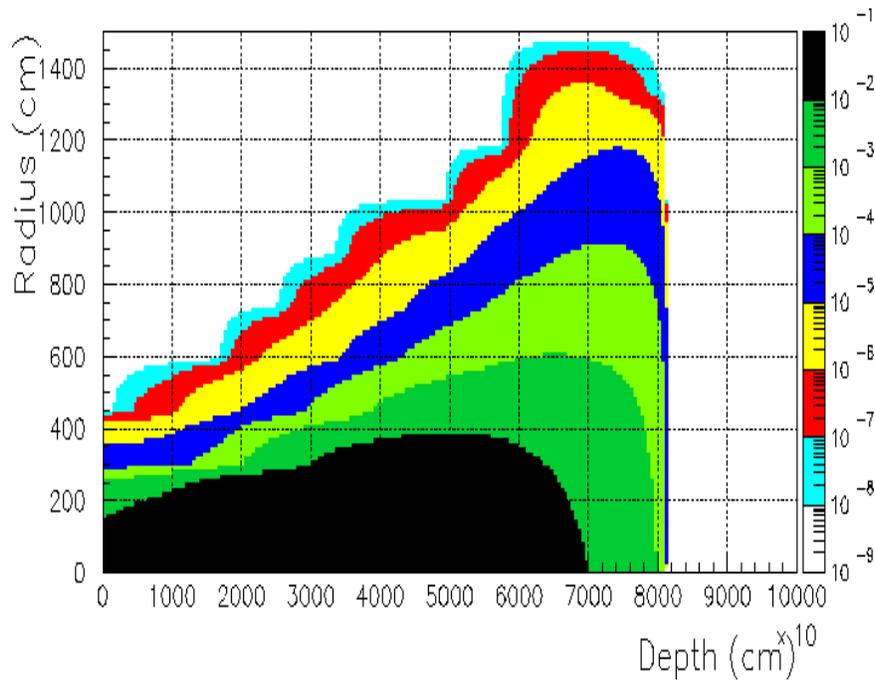,height=3.5in,width=4.5in}}
\caption[MARS isodose contours in the soil/rock for 250~GeV muons extracted at 
$3 \times 10^{13}$ per second]
{Isodose contours in the soil/rock ($\rho$=2.24\,g/cm$^3$) for 250~GeV muons extracted at $3 \times 10^{13}$ per second. Right scale is dose
rate in rem/s.} 
\label{snowmass10.7} 
\end{figure*}
\subsection{Neutrino induced radiation}
It has been shown \cite{ref8b,king_phd,ref41a,ref41b,ref41c} that the neutrinos created
in muon beam decays can generate excessive secondary radiation at large
distances from a muon collider (see Fig.~\ref{nurad}). The surface radiation
dose $D_B(Sv)$ in units of equivalent \cite{data_book} doses ($Sv$) over a time
$t(s)$, in the plane of a bending magnet of field B(T), in a circular collider
with beam energy $E(TeV)$, average bending field $<B(T)>$, at a depth $d(m)$
(assuming a spherical earth), with muon current (of each sign) of
$I_{\mu}\textrm{(muons/s/sign)}$ is given by:
\begin{equation}
D_B\ \approx\ 4.4\times 10^{-24}\ {I_\mu \ E^3\ t \over d}\ {<B> \over B}
\end{equation}
and the dose $D_S$ at a location on the surface, in line with a high
beta straight section of length $\ell(m),$ is:
\begin{equation}
D_S\ \approx\ 6.7\times 10^{-24}\ {I_\mu \ E^3\ t \over d}\ {\ell  <B>}.
\end{equation}
The equation for $D_S$ assumes that the average divergence angles satisfy the
condition: $\sigma_\theta << {1\over \gamma}$. This condition is not satisfied
in the straight sections approaching the IP, and these regions, despite their
length, do not contribute a significant dose. 

\begin{figure*}[tbh!]
\centerline{\epsfig{file=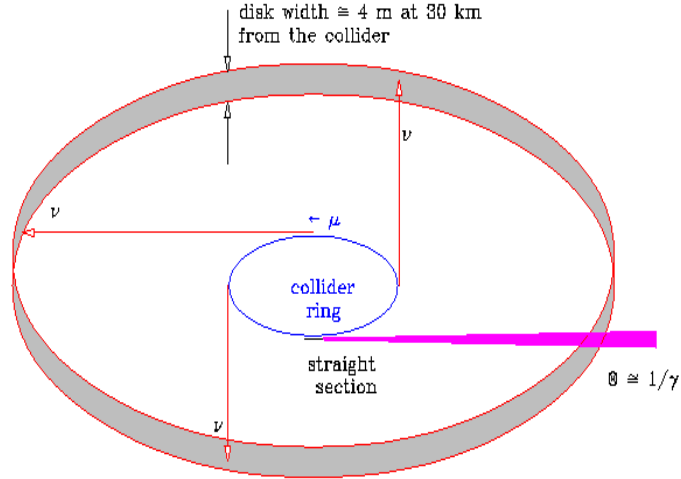,height=2.5in,width=3.5in}}
\caption[Neutrino radiation disk    ]{Neutrino radiation disk. For a 3~TeV CoM
collider the neutrino radiation width is $\approx $~4 m at a distance of 30~km.
A \textit{hot spot} produced by 0.1 m straight section in the ring contains
roughly twice the number of neutrinos on the disk on average, depending on the
details of the collider lattice. } 
\label{nurad} 
\end{figure*}
For the 3~TeV parameters given in table~\ref{sum} and muon currents 
$I_{\mu}=6\times 10^{20}\,\mu^-/yr,$ $<B>= 6~T,$ $B= 10~T$ and
$\textrm{depth}=500$~m, and taking the Federal limit on off-site radiation
dose/year, $D_{\textrm{Fed}},$ to be 1~mSv/year (100~mrem/year), the annual dose
$D_B$ (1 year is defined as $10^7$ s), in the plane of a bending dipole is,
\begin{equation}
D_B = 1.07\times 10^{-5}\  \textrm{Sv}\ \approx 1\% \ D_{\textrm{Fed}},
\end{equation}
and for a straight section of length 0.6 m is:
\begin{equation} 
D_S = 9.7\times 10^{-5}\ \textrm{Sv}\ \approx 10\% \ D_{\textrm{Fed}},
\end{equation}
which may be taken to be within a reasonable limit. The general trend of these  expressions has been verified by Monte Carlo simulations \cite{mokhovrada} using MARS. In particular, for the 3~TeV case the needed depth to stay within 
$1\%\ D_{\textrm{Fed}},$ is 300~m instead of 500~m. 

Special care will be required in the lattice design to assure that no field-
free region longer than 0.6~m  is present. This may sound difficult, but it may
be noted that the presence of a field of even 1~T, is enough
to reduce the dose to a level below the Federal limit. The application of such
a field over all rf and other components seems
possible \cite{mokhovrada}. 

For lower energy machines, the requirements rapidly get easier: a 0.5~TeV
machine at 100~m depth could have 25~m long sections, for the same surface
dose. For a 100~GeV machine the doses are negligible. 

For machines above 3~TeV, various strategies can be employed: 
\begin{itemize}
\item  The machines could be built at greater depths (mines many km deep are
common). 
\item  The vertical beam orbits in the machine could be varied so as
to spread the plane of radiation and thus reduce the peak doses. \item  The
specific locations in line with straight sections could be purchased and
restricted. 
\item  Straight sections could be shortened further by using
continuous combined function magnets. 
\item The machines could be built on an
island, but this could have difficulties associated with access to power and
other utilities. 
\end{itemize}

But for any large increase in energy, to 10~TeV for instance, some reduction
in muon beam flux probably will  be required. The resultant loss of luminosity
might be made up in a number of ways \cite{skrinskiultimate}:

\begin{itemize} 
\item  The beam-beam tune shift constraint could be avoided by
introducing a conducting medium (e.g. liquid lithium) at the interaction
point \cite{skrinskibeambeam}. 
\item  The focusing strength could be increased
by the use of plasma or other exotic focusing method. 
\item  Better cooling
could be developed. Optical stochastic cooling \cite{ref21}, for instance,
might reduce the emittances by many orders of magnitude, thus greatly reducing
the required beam currents. Indeed, such cooling would require lower currents
to function appropriately. 
\end{itemize}

Such options will need future study.
\subsection{Muon decay background}
 
With $4\times 10^{12}$ muons per bunch in a $2\: +\: 2$~TeV  collider ring there
are approximately $4\times 10^5$ muon decays per meter giving rise to high
energy electrons. These off-energy, off-axis electrons undergo bremsstrahlung
when they traverse magnetic fields. When they exit the beam pipe they interact and
produce electromagnetic showers and, to a lesser extent, hadrons and muons. Much
of this debris can be locally shielded, so the primary concern is muon decays
near the interaction point \cite{ref6a}. This is the background we discuss
in some detail below.

\begin{figure*}[tbh!]
\centerline{\epsfig{file=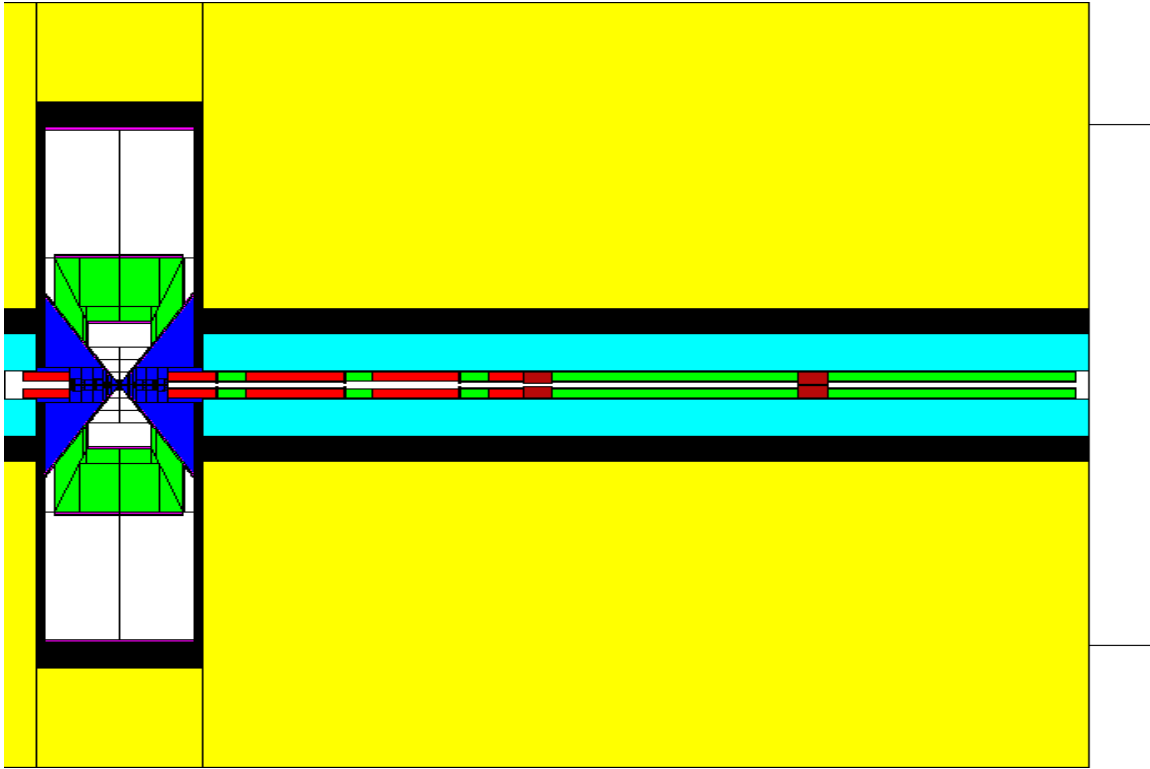,height=4.in,width=6.0in}}
%\vspace{0.5cm}
\caption[Region up to 130~m from the IP, of the 2 + 2~TeV interaction
region modeled in GEANT]{Region up to 130~m from the IP, of the 2 + 2~TeV interaction region modeled in GEANT. The triangular blue regions represent
tungsten shielding. On the right hand of the figure, the red areas represent
quadrupoles in the beam line. The areas around the IP represent the various
detector volumes used in the calculations of particle fluences. The detector
(white and green areas) is 10~m in diameter and 20~m long.}
\label{ip_fig}
\end{figure*}

Detailed Monte Carlo simulations of electromagnetic, hadronic and muon
components of 
the background \cite{ref6a,carnik96,ref42,shield96,mokhovrada,ref39,nikolai96}
have been performed using the MARS \cite{mars} and GEANT \cite{ref40} codes.
The most recent study \cite{ref39} has been done with GEANT. 
Figure~\ref{ip_fig} shows the final 130~meters of the 2 + 2~TeV detector
region in this study.
It includes the final four quadrupoles, dipoles and a solenoidal field
surrounding the detector.
This study:
   \begin{itemize}
\item
followed shower neutrons and photons down to 
$40\,$keV and electrons to $25\,$keV.
\item 
used a tungsten shield over the beam, extending outward to an angle of
20 degrees from the axis.
%extending in to within 14 cm of the interaction point.
\item
Inside this shield, the clear radius has a minimum, in the high energy
cases, at a distance from the IP of 1.1~m (80~cm for 50+50~GeV). At this
point, and in an expanding cone beyond it, the clear radius is maintained at
approximately 4~sigma of the beam size. 
  \item Between this minimum aperture point and the IP, the clear radius
follows an inverse cone, increasing as it approaches the IP, with an angle
a little greater than the 4~sigma of the beam divergence. These cones are 
designed so that the detector could not `see' any surface directly
illuminated by the initial decay electrons, whether in the 
forward or backward (albedo) direction (see figure~\ref{shield}).          
  \item
The resulting open space between the IP and the tip of the cone is
approximately 3~cm in the 4~TeV and 500~GeV CoM cases, and approximately 6~cm in the 100~GeV CoM case.
   \item
The inner surface of each shield is shaped into a series of collimating
steps and slopes to maximize the absorption of electron showers from
electrons at very small angles to the cone surface, thus reducing the
funnelling of low energy electrons down the pipes.
     \item
Further upstream, prior to the first quadrupole (from 2.5 to 4~m in the
Higgs case), an 8~T dipole, with collimators inside, is used to sweep decay
electrons before the final collimation.
   \end{itemize}

Note that there is currently an inconsistency, in the very low $\Delta p/p$ Higgs
Factory Case, between the short open space between shields (+/- 6 cm) and
the rms source length ($\sigma_{\textrm{source}}=1/\sqrt(2) \sigma_z$) of 10~cm. Some modifications to the parameters and shielding design will be
required for this case.

\begin{figure*}[tbh!]
\centerline{\epsfig{file=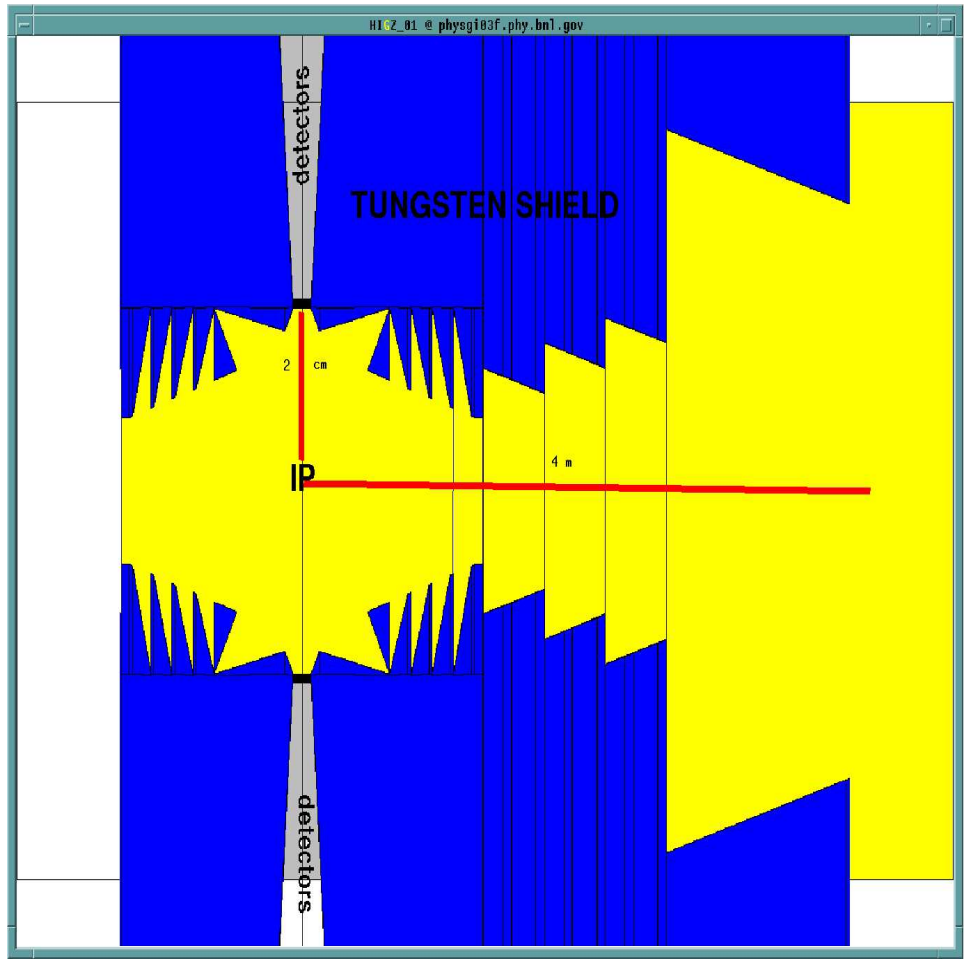,height=5.in,width=5.0in,clip=}}
\vspace{0.25cm}
\caption[Detail of the tungsten shielding designed for the 50 + 50 GeV
case.   ]{ Detail of the tungsten shielding designed for the 50 + 50 GeV
case. It is designed so that the detector is not connected by a straight
line with any surface hit by decay electrons in forward or backward
directions. The picture extends out to a radius of 6~cm and, on the right, to a
distance 4~m from the IP. The dipole from 2.5-4.0~m is not shown. }
\label{shield}
\end{figure*}

Every modern detector will have to be able to identify and reconstruct
secondary vertices such as those associated with b-quark decays. In order to estimate the viability of a vertex
detector we have to show that the occupancy of its elements is not higher than
about $1\%.$ Figure~\ref{occ} shows the occupancy as a function of radial
distance from the interaction point for the three CoM energies studied : 0.1,
 0.5 and 4~TeV. The occupancy was calculated for silicon pads of $300~\mu m \times 300~\mu m$, and assuming interaction probabilities of 0.003 and 0.0003 for low
energy photons and neutrons respectively. One can observe that the
total occupancy (left figure) is above one percent for small radii. Most of the hits is due  to conversions of photons.  The 
occupancy due to hits resulting from charged particles is below $1\%$ 
(right hand figure). One can lower the
occupancy at small radii by using smaller pixel sizes, as indicated in table~\ref{backgnd} below, as well as by using innovative detector
ideas as described in the next section.

Table~\ref{backgnd} gives the hit density for the Higgs factory from the
various sources and the occupancy of pixels of the given sizes; in each case
the number is given per bunch crossing. The hit density for the higher energy
machines is found to be somewhat lower due to the smaller decay
angles of the electrons. %
\begin{table*}[htb!]
\caption{Detector backgrounds from $\mu$ decay}
\label{backgnd}
%\centering \protect
\renewcommand{\arraystretch}{1,1}
\begin{tabular}{llcccc}
%\tableline
Radius             &$cm$   &      5   & 10  &20 & 100   \\
\hline
Photons hits     &$cm^{-2}$& 26 & 6.6 & 1.6 & 0.06   \\
Neutrons hits    &$cm^{-2}$ & 0.06 & 0.08 & 0.2 & 0.04     \\
Charged hits     &$cm^{-2}$ &8 & 1.2 & 0.2 & 0.01    \\
Total hits       &$cm^{-2}$ & 34 & 8 & 2 & 0.12   \\
\hline
Pixel size     & $\mu m \times \mu m $& $60\times 150$&$60\times 150$&$300\times
300$&$300\times 300$ \\
Total occupancy  & \%  & 0.6 & 0.14 & 0.4 & 0.02 \\
Occupancy charged & \% & 0.14& 0.02 & 0.04 & 0.002 \\
\end{tabular}

\end{table*}

\begin{figure*}[tbh!]
\centerline{\hbox to 4.in{\hfil      \hfil}\hfill
   \hbox to 4.in{\hfil       \hfil}}
\dofigs{3.5in}{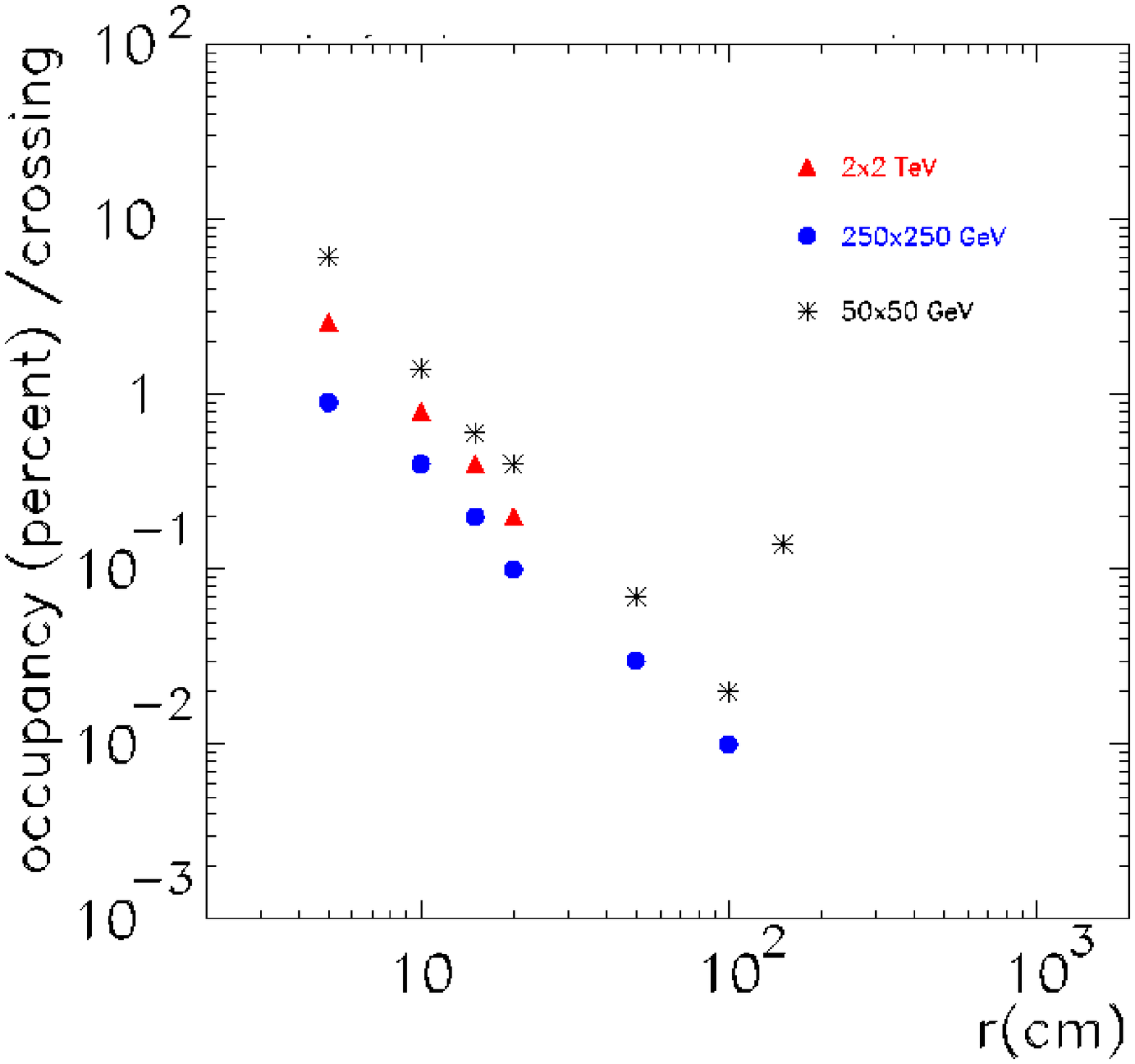}{3.5in}{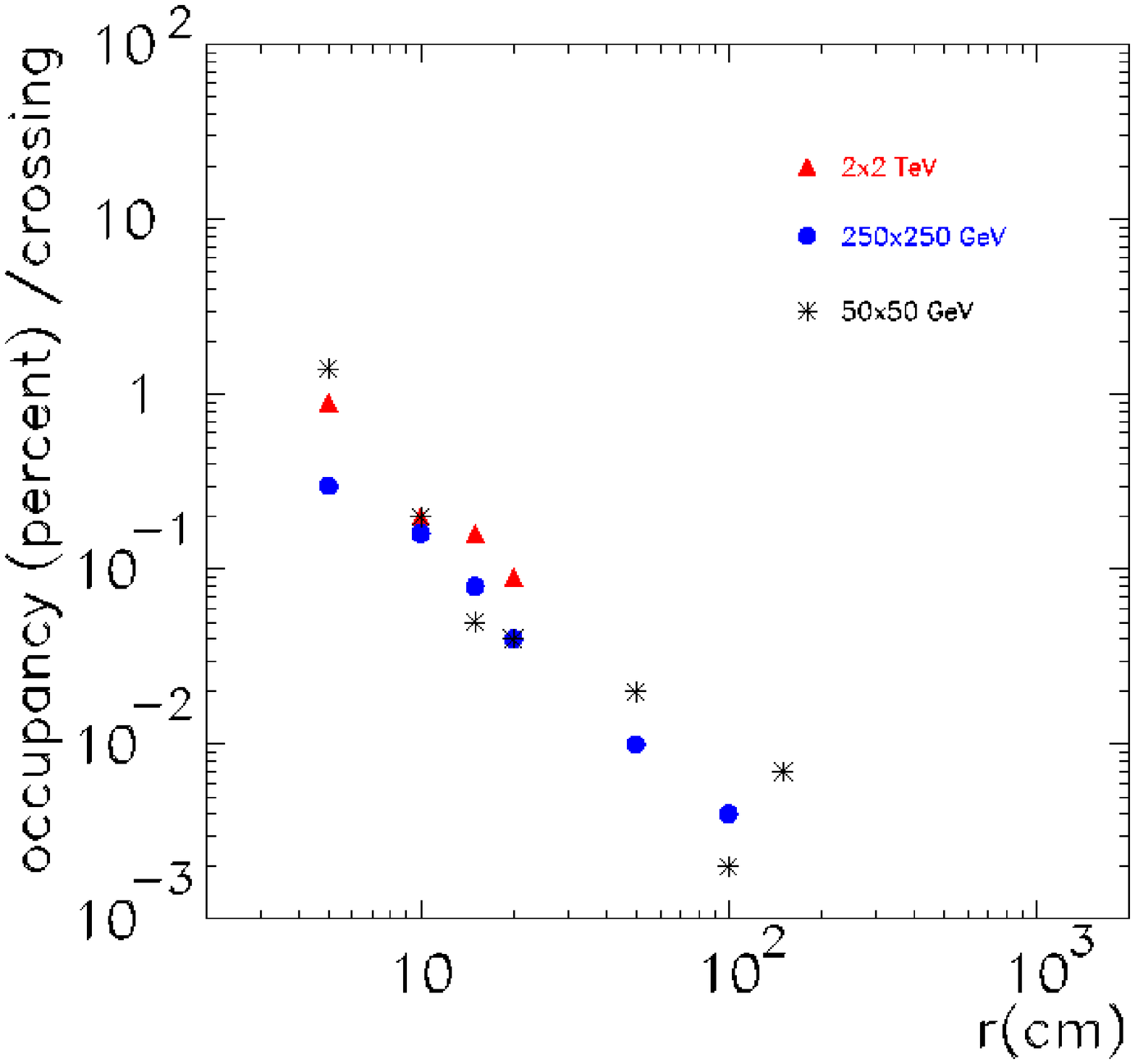}
\caption[Occupancy as a function of the radius]{Occupancy for $300~\mu\,m\times 300~\mu\,m$ silicon pads, as a function of the radius for the three energies studied.
Left figure shows the total occupancy and the right figure shows the occupancy
from hits resulting from charged particles.}
\label{occ}
\end{figure*}
The radiation damage by the neutrons on a silicon detector has also been
estimated. In the Higgs case, at 5~cm from the vertex, the number of hits from
neutrons above 100~keV is found to be $1.8\times 10^{13}$ per year.
This is significantly less than that  expected at the LHC which is now ordering
silicon detectors claimed to survive $5\times 10^{14}$ hits, 
approximately three times that assumed here. The damage for silicon detectors
in the higher energy machines is of the same order (see table~\ref{neutron}).

\begin{table*}[htb!]
\caption[Radiation damage by neutrons on silicon detectors   ]{Radiation damage
by neutrons on silicon detectors. The working assumptions are: 1000 turns,
15~Hz and 1~year=$10^7$~s. An acceptable number of hits per year is $1.5\times
10^{14}$. }
\label{neutron}
%\centering \protect
\renewcommand{\arraystretch}{1,1}
\begin{tabular}{ccccc}
%\tableline
CoM &$\mu$'s/bunch &neutrons/cm$^2$/crossing & Hits/year&Lifetime   \\
(TeV) & $(10^{12})$ & (above 100~KeV) & $(10^{13})$ & (years)\\
4 & 2 & 100 & 3 & 5 \\
0.5 & 4 & 50 & 3 & 5    \\
0.1 & 4 & 30 & 1.8 & 8   \\
\end{tabular}
\end{table*}

\subsection{Halo background}

Muon halo refers to those muons which are lost from the beam bunch as it
circulates around the collider ring. In conventional electron or proton
accelerators, beam particles which are lost away from the IP are of little
concern as they can be locally shielded. However, muons can traverse long
distances and therefore have the potential to generate background in a
detector. The magnitude of this background depends on a detailed knowledge of
the injected beam profile and a credible model for beam halo and beam losses.
More work is needed before these are well enough understood. Nevertheless, it
is clear that the beam will need careful preparation  before  injection into
the collider, and  the injection system will have to be precise and free of
ripple. 

The collimation system described in the previous subsection was designed to
scrape the beam both initially and during the 1000 turns, to assure that all
loss occurs at the scraper and not near the IP. That study indicated
suppressions  better than 10$^3$ of background in the detector \cite{scraping}. 

Beam loss must be limited as far as possible. Gas scattering has been
studied \cite{snowmass8.5.3} and shown to give a negligible contribution. The
effects of beam-beam scattering are under study and need
further work.  Momentum spread tails from uncorrected wakefield effects must
be controlled. Assuming that the total loss from all causes, after injection
and the first few turns is less than $10^{-4}$ in 1000~turns, (i.e. $10^{-7}$
per turn), then the number of background muons passing through the detector
should be less than 800 ($2\times 4\times 10^{12} \times 10^{-7}\times 10^{-3}$) per turn. This is a low density of tracks per cm$^2$ and should be acceptable, but lower losses or better scraping would be desirable.

\subsection{Pair production}
Coherent beam-beam electron pair production (beamstrahlung) has been shown \cite{chen,ginzburg} to be negligible, but the
incoherent pair production (i.e. \mumu  $\rightarrow$ \ee) in
the 4~TeV collider case is significant.
\begin{figure*}[bht!]
\dofigs{3.5in}{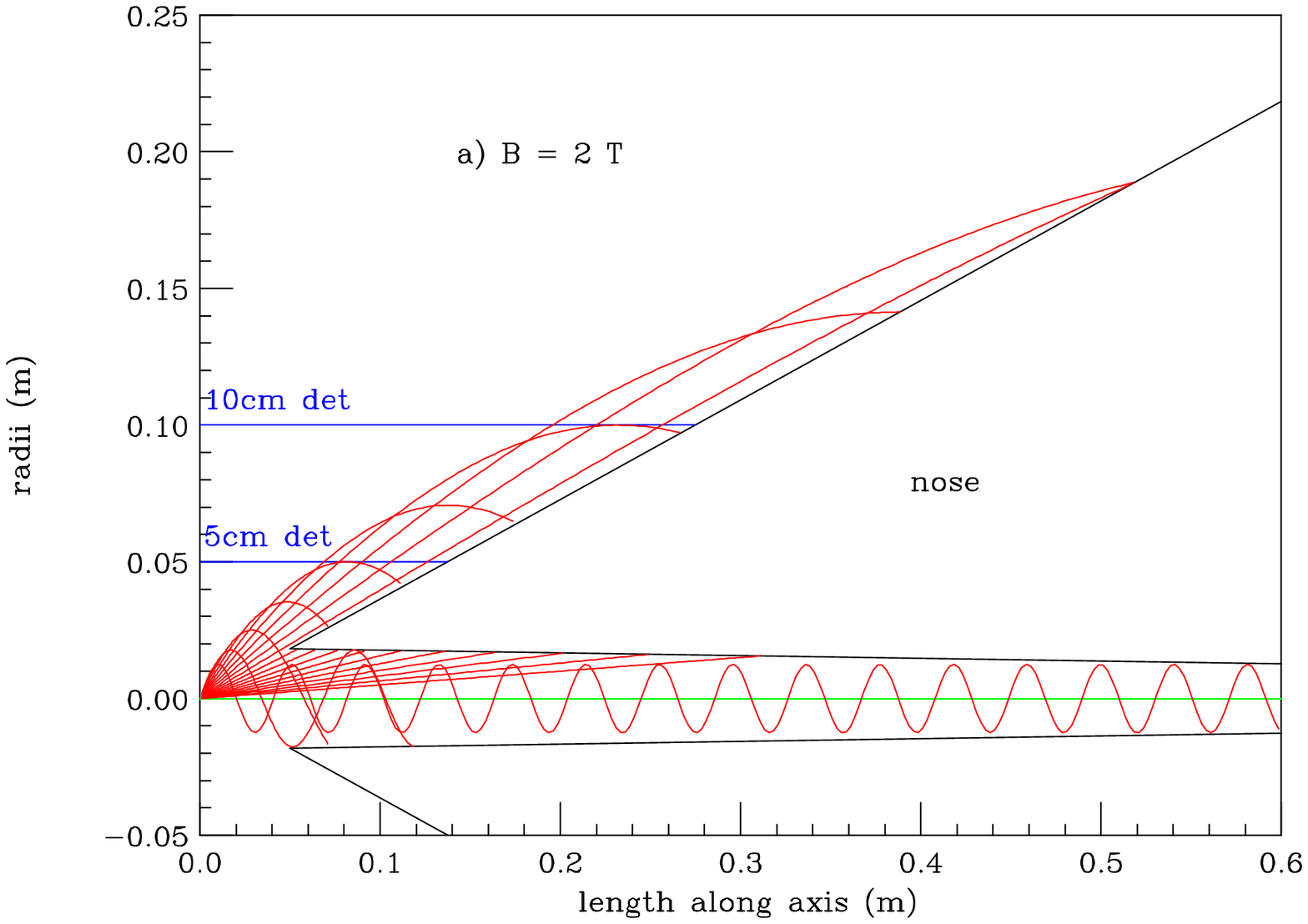}{3.5in}{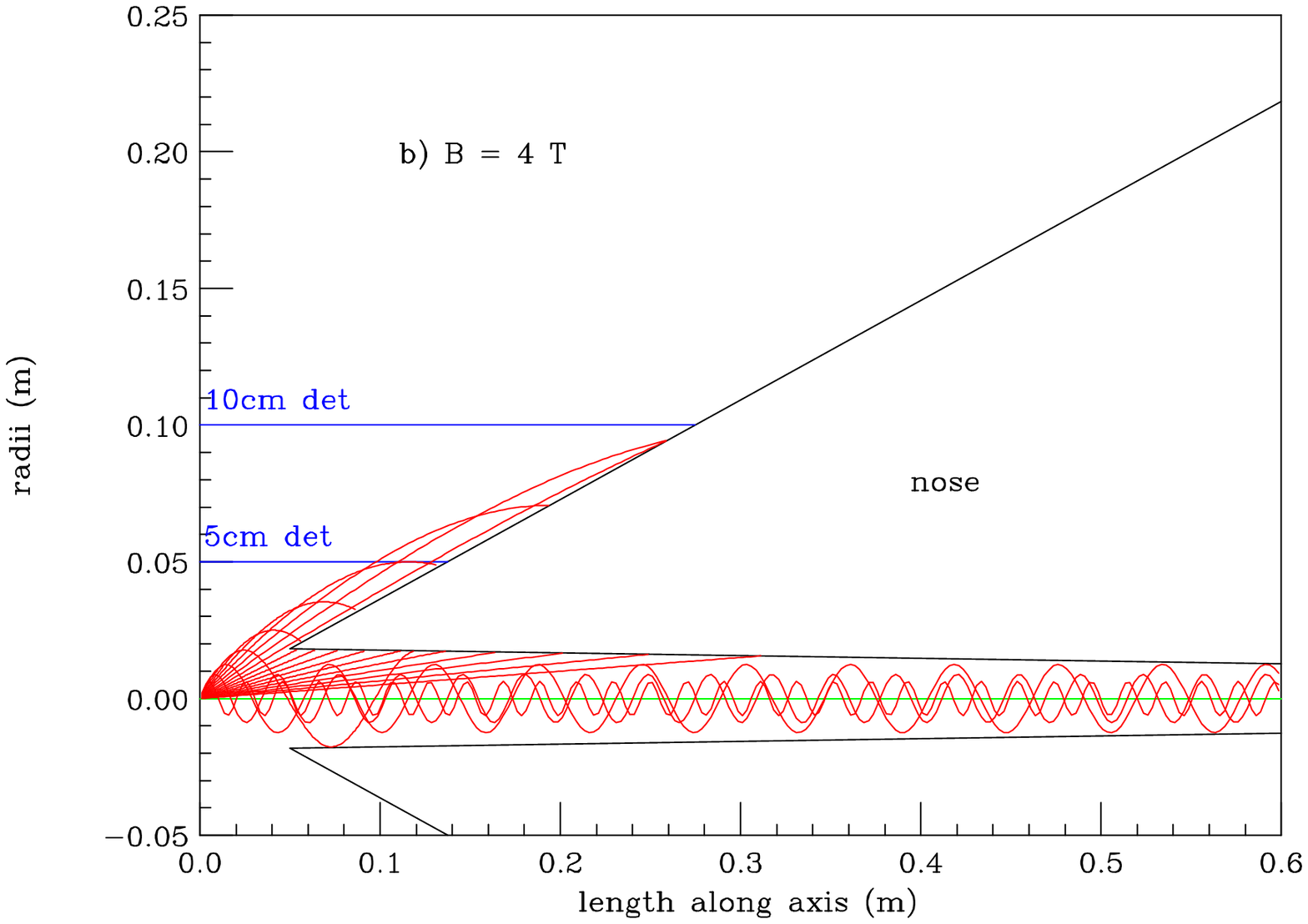}
\caption[Radius vs. length of electron pair tracks ]
{Radius vs. length of electron pair tracks for
initial momenta from 3.8 to 3000 MeV 
in geometric steps of $\sqrt{2};$ (a) for
a solenoid field of 2 T, (b) for 4 T.
\label{pair}}
\end{figure*}

The cross section is estimated to be  $10\,$mb \cite{ginzburg}, which would give
rise to a  background of  $\approx 3\times 10^4$  electron pairs per bunch
crossing.  The electrons at production do not have significant  transverse
momentum but the fields of the on-coming $3 \mu$m bunch can deflect them
towards the detector. A simple program was written to track electrons from
close to the axis  (the worst case) as they are deflected away from the bunch
center.  Once clear of the opposing bunch the tracks spiral under the 
influence of the experimental solenoid field. Figures~\ref{pair} shows the
radii vs. length of these electron tracks for initial momenta from 3.8 to
3000~MeV in geometric steps of $\sqrt{2}.$ Fig.~\ref{pair}(a) is for a solenoidal
field of 2~T and Fig.~\ref{pair}(b) for 4~T. In the  2~T case tracks with
initial energy below 30~MeV do not make it out  to a detector at 10~cm, while
those above 100~MeV have too small an  initial angle and remain within the
shield. Approximately 10\%~(3000 tracks) of  these are in this energy range and
pass through a detector at 10~cm. The track fluences at the ends of the detector
are less  than 10~tracks per cm$^2$ which should not present a serious problem.
At 5~cm, there are 4500~tracks giving a fluence of 30~per cm$^2$, which is also
probably acceptable. If  the detector solenoid field is raised to 4~T, then no
electrons reach 10~cm and the flux at 5~cm is reduced by a factor of 2. 

\subsection{Bethe-Heitler muons}

\begin{table*}[htb!]
\caption{Bethe-Heitler Muons }
\label{tab-bheitler}
%\centering \protect
\renewcommand{\arraystretch}{1,1}
\begin{tabular}{lccc}
%\tableline
CoM Collider Energy (TeV)  & 4 & 0.5  & 0.1  \\
Assumed source length (m) & 130 & 33 & 20 \\
\hline
$\mu\ (p_{muon }>1\ \textrm{GeV}/c)$ per electron & $5.4\times 10^{-4}$ & $8.3\times
10^{-5}$ & $9.6\times 10^{-6}$ \\
Beam $\mu$'s per bunch         &   $2\times 10^{12}$ &    $2\times 10^{12}$ &   
$4\times 10^{12}$ \\
Bethe-Heitler $\mu$'s per bunch crossing ($\times 10^3$)   & 28  &  17.5  &
6.1  \\
$<p_{muon}>$ initial (GeV) &     22      &   9.5    &
4.4  \\
\hline
$\mu$'s entering calorimeter     &     220      &     160      & 25    \\
$<p_{muon}>$  (GeV)  &    15.4    &   6.3   &
1.8  \\
$<E_{dep}>$   (GeV)                & 2.9 &    1.3 &   0.4 \\
Total $E_{dep}$  (GeV)          &  640    &    210 &    10    \\
$E_{dep}$ pedestal subtracted  (GeV)& 50    &  25  &   1    \\
Fluctuation in $E_{dep}$ (GeV)  &    55    &   15    &   1   \\
$E_{trans}$ pedestal subtracted  (GeV)  &  15    &   15          &   .5  \\
Fluctuation in $E_{trans}$  (GeV)  &   40      &  8     &  0.5  \\
\end{tabular}
\end{table*}
  
The GEANT/MARS studies \cite{ref6a,shield96,snowrad} also found a significant
flux of muons with quite high energies, from $\mu$ pair production in
electromagnetic showers (Bethe-Heitler).  Figures~\ref{fig-bheitler1}
and~\ref{fig-bheitler2} show the trajectories of typical muons from their
sources in the shielding around the beam pipe to the detector.
Figure~\ref{fig-bheitler1} is for a 4~TeV CoM collider, where the muons have
high energy  and long path lengths. A relatively long  (130~m) section of beam
pipe prior to the detector is shown. Figure~\ref{fig-bheitler2} is for the
100~GeV CoM collider for which, since the muons have rather short path lengths,
only a limited length of beam pipe is shown. Note that the scales are extremely
distorted: the 20$^o$ shielding cones on the right hand  of the figures
appear at steeper angles.
\begin{figure*}[bht!]
\centerline{\epsfig{file=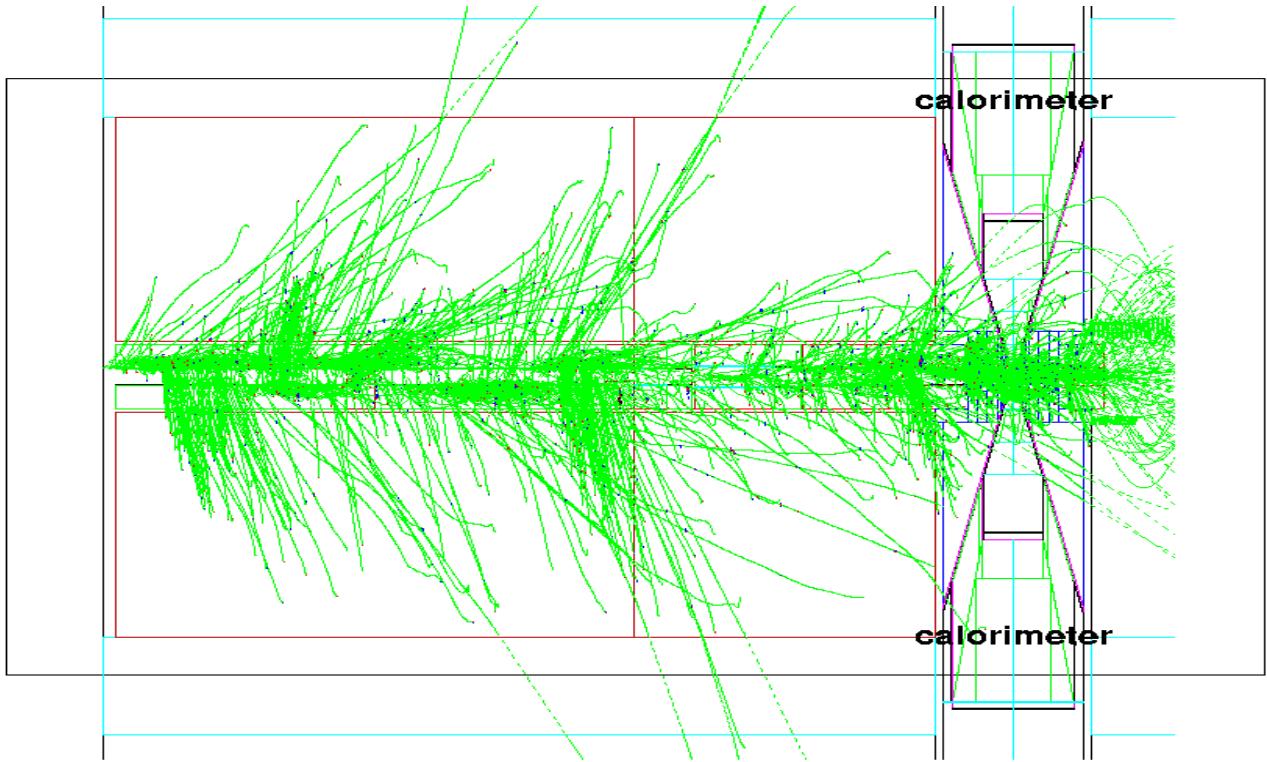,height=7.0in,width=4.0in,angle=-90}}
\vskip 0.5cm
\caption[Trajectories of Bethe-Heitler muons for a 4~TeV collider]{Trajectories of typical Bethe-Heitler muons from their source in the
shielding around the beam pipe to the detector for a 4~TeV CoM collider. As indicated in the text the scales are extremely distorted, the total horizontal length is $\approx 130$~m and the outer edge of the calorimeter is $\approx 4$~m.  Notice that $<1\%$ of the tracks end in the calorimeter (see
table~\ref{tab-bheitler}).}
\label{fig-bheitler1}
\end{figure*}

\begin{figure*}[hbt!]
\centerline{\epsfig{file=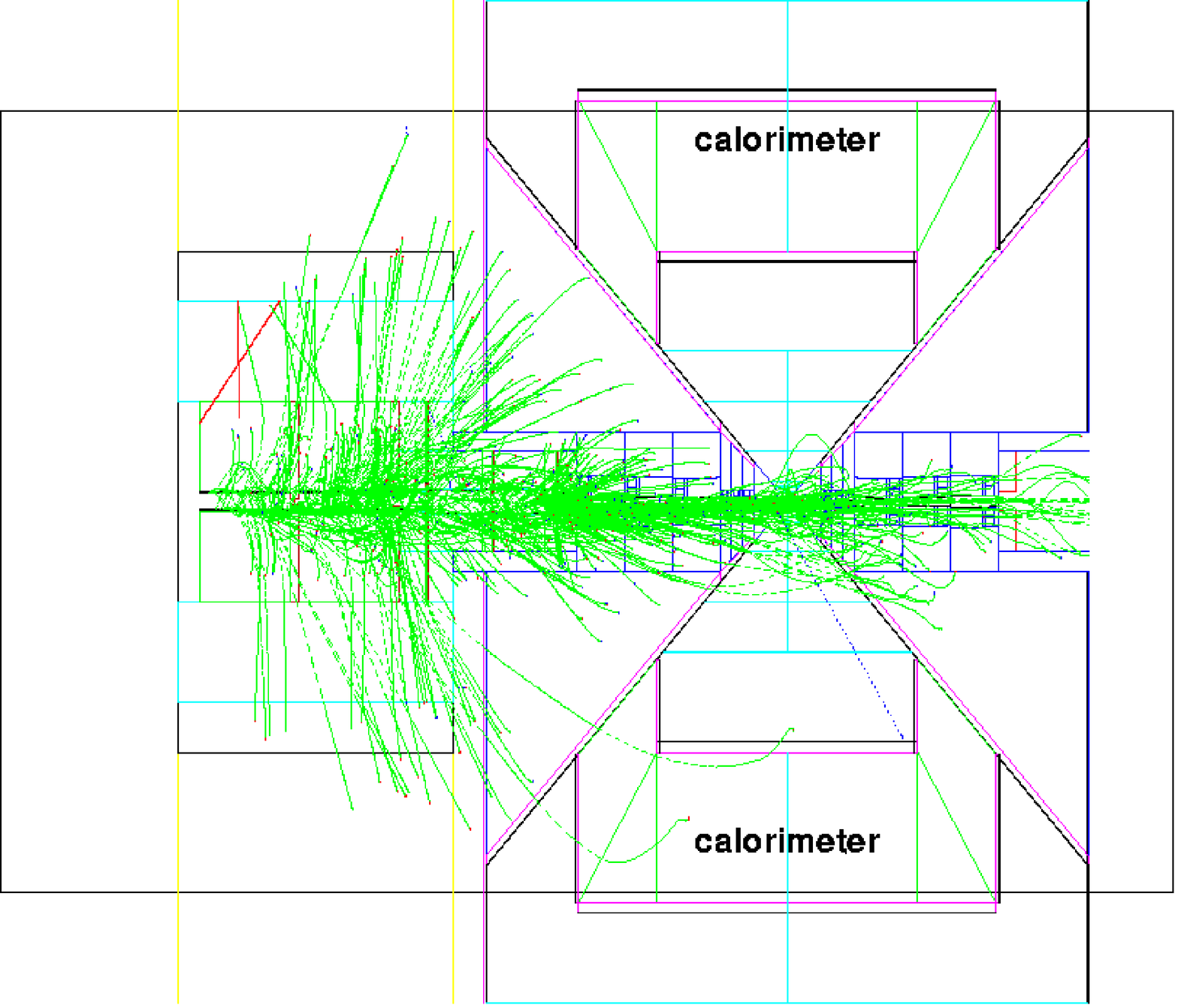,height=7.0in,width=4.0in,angle=-90}}
\vskip 0.5cm
\caption[Trajectories of Bethe-Heitler muons for a 100~GeV collider]{Trajectories of typical Bethe-Heitler muons from their source in
the shielding around the beam pipe to the detector for a 100~GeV CoM collider. As indicated in the text the scale is extremely distorted, the total horizontal length is $\approx 20$~m and the outer edge of the calorimeter is $\approx 4$~m. Notice that $<0.5\%$ of the tracks end in the calorimeter (see
table~\ref{tab-bheitler}).}
\label{fig-bheitler2}
\end{figure*}

The most serious effect appears to arise when these muons make deeply inelastic
interactions and deposit spikes of energy in the electromagnetic and
hadronic calorimeters. This is not serious in the Higgs case, for which the
fluxes
and cross sections are low, but at the higher collider energies they generate
significant fluctuations in global parameters, such as transverse energy and
missing transverse energy.

Table~\ref{tab-bheitler} gives some parameters of the muons for three different
machine energies. In the 4~TeV and 500~GeV CoM cases, massive lead
shielding outside the focus quadrupoles has been included. 

Figures~\ref{spikes1} and~\ref{spikes2} show energy deposition from 
Bethe-Heitler muons in a typical bunch crossing. These depositions are 
 plotted against the cosine
of the polar angle and azimuthal angle in the calorimeter for 4~TeV and for
500~GeV CoM, respectively. The massive lead shielding referred to above was not
included in this study. Right hand plots in  Figs.~\ref{spikes1}
and~\ref{spikes2} show the same distributions with a 1~ns timing cut. It is
seen that the timing cut, if it is possible, is effective in removing energy
spikes at small rapidity, but has little effect in the forward and backward
directions. The overall reduction in energy deposition is about a factor of
two.  

The energy spikes can cause at least three problems: 1) they affect the triggers and
event selections based on overall or transverse energy balance; 2) they can
generate false jets and 3) they can give errors in the energies of real jets.
After a pedestal subtraction, the effects on energy balances do not seem
serious. The generation of false jets can be eliminated by a longitudinal
energy distribution cut without introducing significant inefficiency. Energy
errors in real jets appear to be the most serious problem. They can be reduced
by the application of radial energy distribution cuts, but such cuts introduce
significant inefficiencies for lower energy jets. More study is needed.

Earlier studies \cite{nikolai96} with MARS, using less sophisticated
shielding, gave results qualitatively in agreement with those from GEANT.

\begin{figure*}[hbt!]
\centerline{\epsfig{file=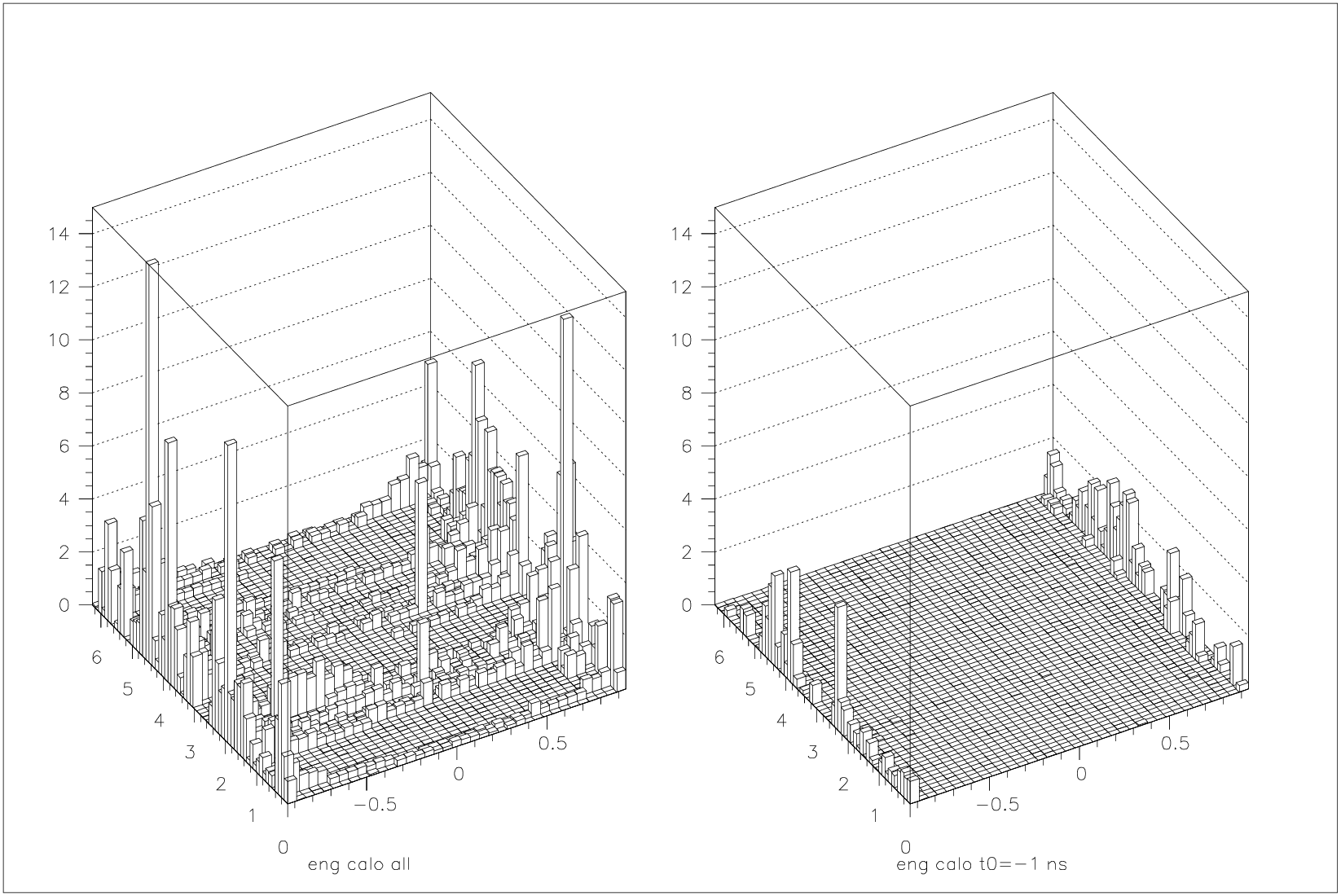,height=6.0in,width=4.0in,
angle=-90}}
\vskip 1cm
\caption[Energy deposition from Bethe-Heitler muons for a 4~TeV CoM collider]{Left hand plot shows the energy deposition from Bethe-Heitler muons
\textit{vs.}  the cosine of the polar angle and azimuthal angle in the
calorimeter for a 4~TeV CoM collider. Right hand plot shows the same
distributions with a 1~ns timing cut.}
\label{spikes1}
\end{figure*}

\begin{figure*}[hbt!]
\centerline{\epsfig{file=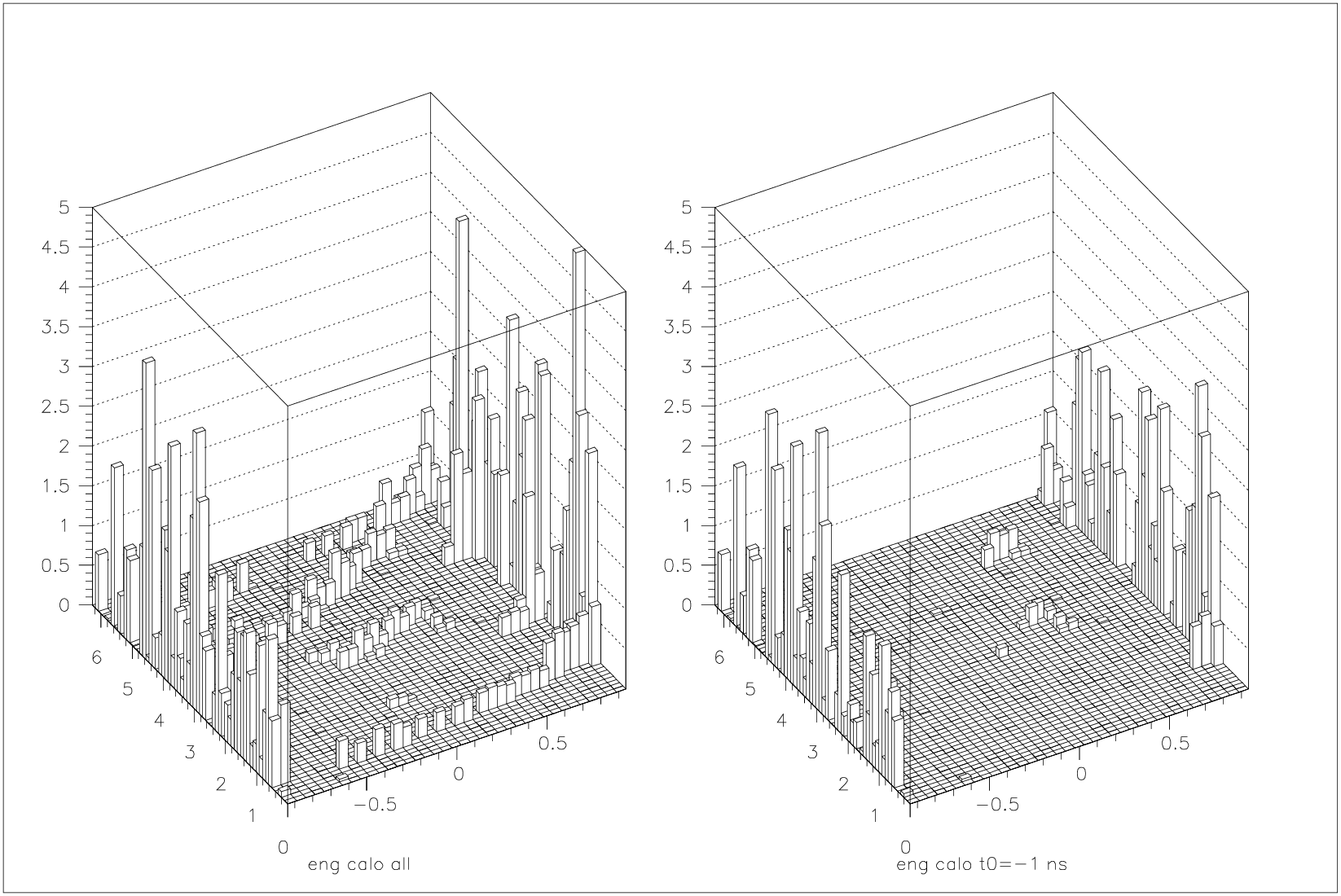,height=6.0in,width=4.0in,
angle=-90}}
\vskip 1cm
\caption[Energy deposition from Bethe-Heitler muons for a 0.5~TeV CoM collider  ]{Left
hand plot shows the energy deposition from Bethe-Heitler muons
\textit{vs.}  the cosine of the polar angle and azimuthal angle in the
calorimeter for a 0.5~TeV CoM collider. Right hand plot shows the same
distributions with a 1~ns timing cut.}
\label{spikes2}
\end{figure*}
\section{DETECTOR SCENARIOS }
\label{det}

 The background consists of neutral and charged particles. 
 For neutrons, the longitudinal and radial fluences were found to be
comparable. The photons (average energy about 1~MeV) show a clear
 radial source. The charged particles and the photons do not all point
back to the interaction point, but to the general vicinity of the
IP, namely to the region where the 20~degree tungsten shield becomes
thinner. The flux of secondary muons (Bethe-Heitler pairs) is mainly
longitudinal.  

We would expect this background to pepper the tracking volume with 
random hits and produce significant energy pedestals in the calorimeter 
cells. These effects are considered in more detail in the following sections. 
In general, in designing a strawman detector that must operate in a large 
background flux we will want to employ as many detector channels as is 
practical. A strawman muon collider detector design with a few times $10^6$ 
non-pixel channels would seem reasonable \cite{stevebackdet}. Over the last few years, development
of pixel detectors has resulted in a quantum jump in the number of electronic channels.  For example,
the SLD vertex detector \cite{sld} contains $300\times 10^6$~pixels, and similar 
numbers of pixels are planned for the LHC vertex detectors. Hence, a  strawman
muon collider vertex detector employing $10^8-10^9$ pixels  would seem
reasonable. 

\subsection{Silicon vertex detector schemes}
 
From table~\ref{neutron}, it can be seen that the
radiation damage to silicon detectors is acceptable in terms of the number of
hits per year and the resultant lifetime of the detector. This prompts \cite{stevebackdet} us to
consider the following options for silicon vertex detector design for the muon
collider:

\begin{figure*}[htb!]
\leavevmode
\centering
\epsfysize=5cm
\epsfbox{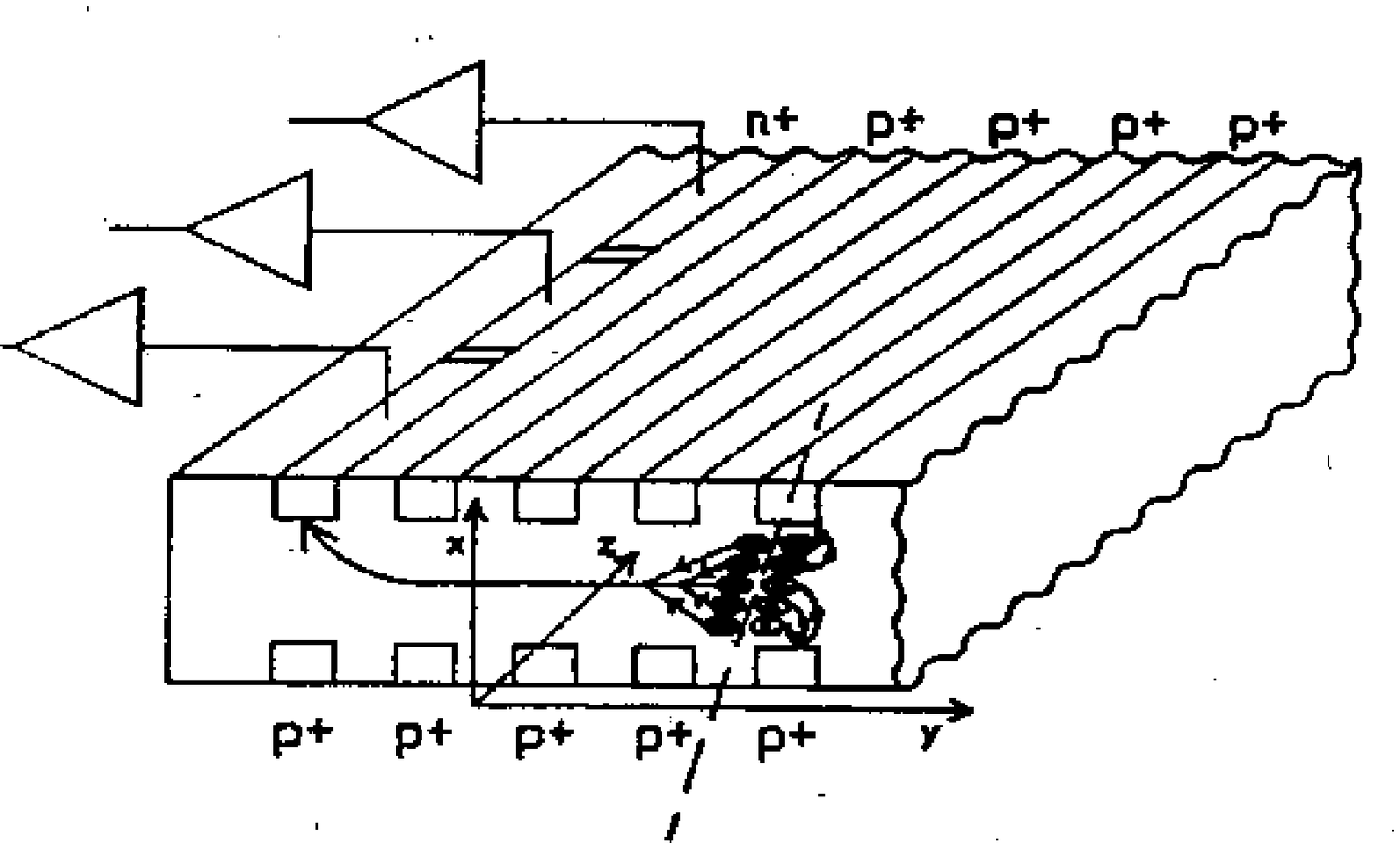}
\caption{Silicon drift vertex detector.}
\label{fig:silicon_drift}
\end{figure*}

\begin{figure*}[htb!]
\leavevmode
\centering
\epsfysize=8cm
\epsfbox{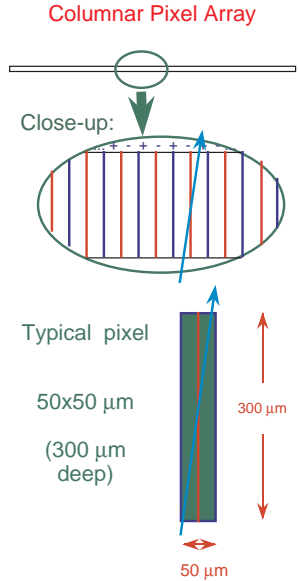}
\caption[Columnar pixel geometry]{Columnar pixel geometry. Courtesy of A.~Sill.}
\label{fig:columnar_pixels}
\end{figure*}

\begin{figure*}[bth!]
\leavevmode
\centering
\epsfxsize=7.5cm
\epsfbox{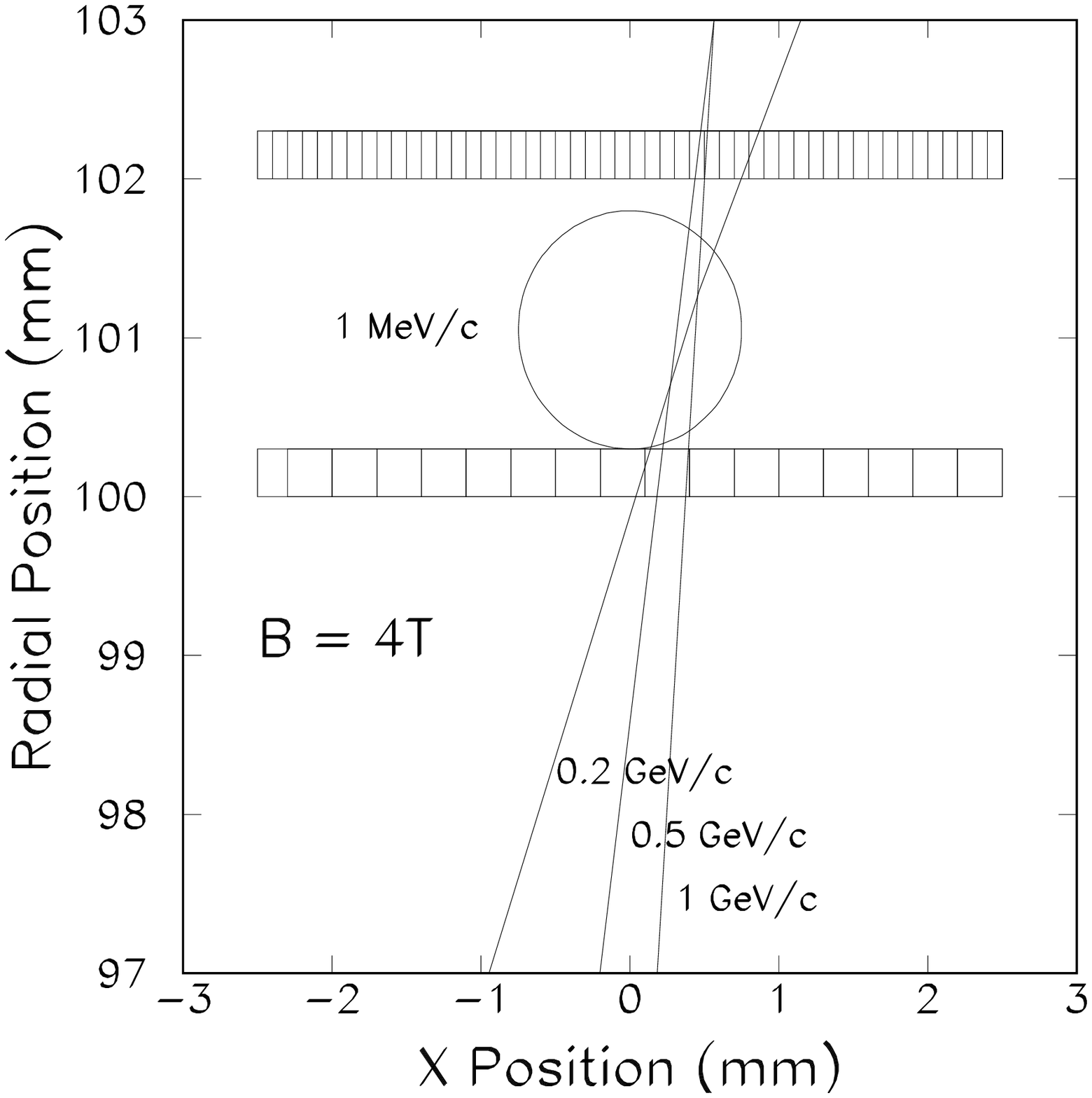}
\caption[Pixel micro-telescope geometry, showing trajectories of 0.2~GeV/$c,$ 
0.5~GeV/$c,$ and 1~GeV/$c$ tracks]{Pixel micro-telescope geometry~\cite{micro_telescope}, showing trajectories of 0.2~GeV/$c,$ 
0.5~GeV/$c,$ and 1~GeV/$c$ tracks coming from the IP and bending in a 4~T field.}
\label{fig:micro_telescope}
\end{figure*}

\begin{itemize}
\item Silicon drift detector. 
The idea, which is described in the muon collider feasibility study \cite{ref6a,Snowmass96}, 
is to exploit the time gap between bunch crossings by using the silicon drift detector technology \cite{silicon_drift} (see Fig.~\ref{fig:silicon_drift}). 
Using $50 \times 300$~$\mu$m$^2$ detectors it should be possible to obtain 
a resolution of a few microns in the drift direction. This would facilitate 
a very precise vertex detector, although questions of radiation hardness remain to be resolved for this option.
\item Columnar pixels \cite{parker}. The idea is to exploit the very well localized primary vertex position by 
using long thin tracking pixels that point at the IP and therefore 
record large ionization signals only for tracks coming from the IP 
(Fig.~\ref{fig:columnar_pixels}). 
For example, one can construct  $50 \times 50$~$\mu$m$^2$ pixels 
that are 300~$\mu$m deep. The pixels are produced using controlled 
feed-through-drilling technology to create a lattice of 
anodes and cathodes that extend through the 300~$\mu$m thick wafer.
\item Pixel micro-telescopes \cite{micro_telescope}. 
The idea is to replace a single pixel layer with two layers separated 
by a small distance, and read them out by taking the AND between 
appropriate pairs. The distance between the layers is optimized so that 
soft MeV tracks (which are associated with almost 80\% of the predicted 
background hits) produced in one layer curl up in the magnetic field 
before reaching the second layer. Thus, the pixel micro-telescope is blind 
to the soft background hits and also blind to tracks that do not come from 
the IP. In the example shown in Fig.~\ref{fig:micro_telescope} 
the top measurement layer has a 
finer granularity than the bottom confirmation layer. The corresponding 
rows in the two pixel layers can be read out with different clock speeds 
to maintain the correct correspondence at the input into the AND gate that 
registers valid hits in the telescope. If the readout rows are 
the ones parallel to the beam direction, then variable clock speeds 
can be used to maintain the correct accepted direction with respect to the IP.
\end{itemize}
\noindent
The challenge of a high 
background environment is clearly fruitful ground for new ideas. The 
above considerations suggest that, 
provided silicon detectors can be used in the inner tracking 
volume, it should be possible to construct a vertex detector able to 
tag secondary vertices from short lived particles at a muon collider. 
Detailed simulations are currently underway  to  establish this more concretely.

\begin{figure*}[hbt!]
\leavevmode
\centering
\epsfysize=10cm
\epsfbox{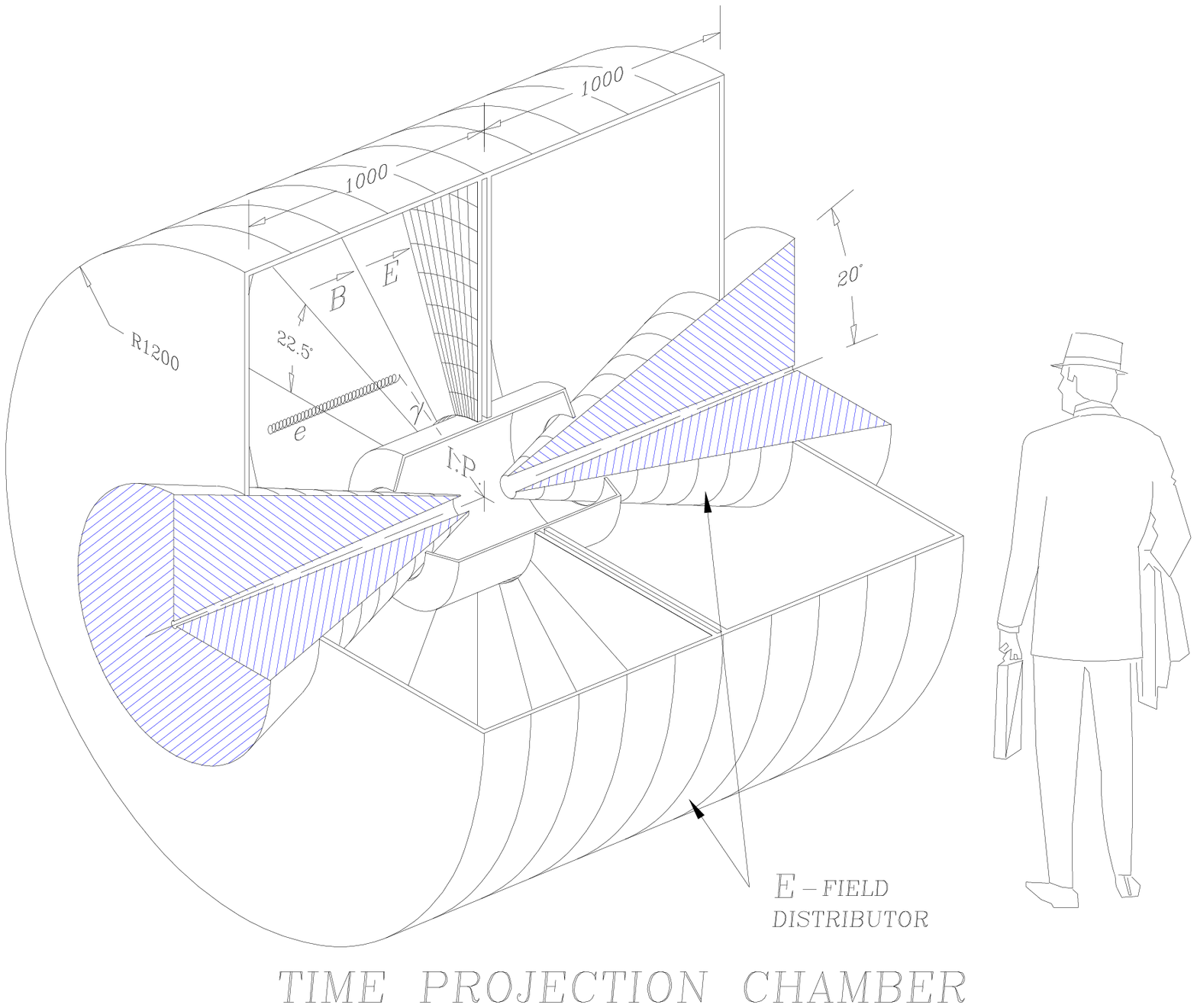}
\caption{Outer tracker TPC.}
\label{fig:tpc}
\end{figure*}

\subsection{Outer tracking schemes}

 The predicted 
background fluxes  for a Higgs factory detector at a radius of 50~cm are 200~photons/cm$^2,$ 
350~neutrons/cm$^2,$ and 0.08~charged tracks per cm$^2.$ 
The neutron flux is therefore about the same as the flux in the inner 
tracking volume, whereas the photon and charged particle fluxes are 
significantly less than those predicted at smaller radii. There are two 
alternative tracking strategies under consideration:
\begin{itemize}
\item Low field, large tracking volume drift chamber option. 
This option, which is described in the muon collider book \cite{snowmass9.4.2}, 
uses a TPC to 
exploit the 20~$\mu$s time between bunch crossings. This option is viable 
for the very high energy muon collider ($1.5\times 1.5$~TeV).
The large neutron 
flux necessitates choosing a gas that does not contain hydrogen. A 
mixture of 90\%~neon plus 10\%~CF$_4$ gives a drift velocity of 
9.4~cm/$\mu$s, which is close to that required to match the bunch crossing time. High-$p_T$ tracks from 
the IP embedded in the predicted background flux 
have been simulated for the TPC shown in 
Fig.~\ref{fig:tpc}. The simulation includes ionization, drift and 
diffusion of the electrons in the gas, multiplication, and other details 
of the detection process. The majority of the background hits arises from 
low energy Compton recoils yielding very low energy electrons that have 
a radius of curvature of less than 1~mm in the 2~T field. Their 
projection on the readout plane covers not more than one readout pitch 
(0.3 $\times$ 0.4~cm$^2$). These background electrons, together with the 
nuclear recoils from neutron scatters, yield large pulses 
that can be removed by cutting on the maximum acceptable pulse height. 
The simulation predicts that with an average background flux of 
100~photons/cm$^2$, reasonable pulse height cuts remove only 1\% of the 
effective TPC volume, and yield tracks of high quality.
However, it was realized 
that positive ion build-up may be a problem with the design shown 
in Fig.~\ref{fig:tpc}. If this problem 
can be overcome, the design shown in the figure yields a simulated momentum 
resolution of about 1.2\% for tracks with p$_T = 50$~GeV/$c.$
\item High field, compact silicon tracker option. 
An alternative strategy is to make a compact tracker by using silicon in 
a high field (for example, 4~T). 
\begin{figure*}[htb!]
\leavevmode
\centering
\epsfysize=3cm
\epsfbox{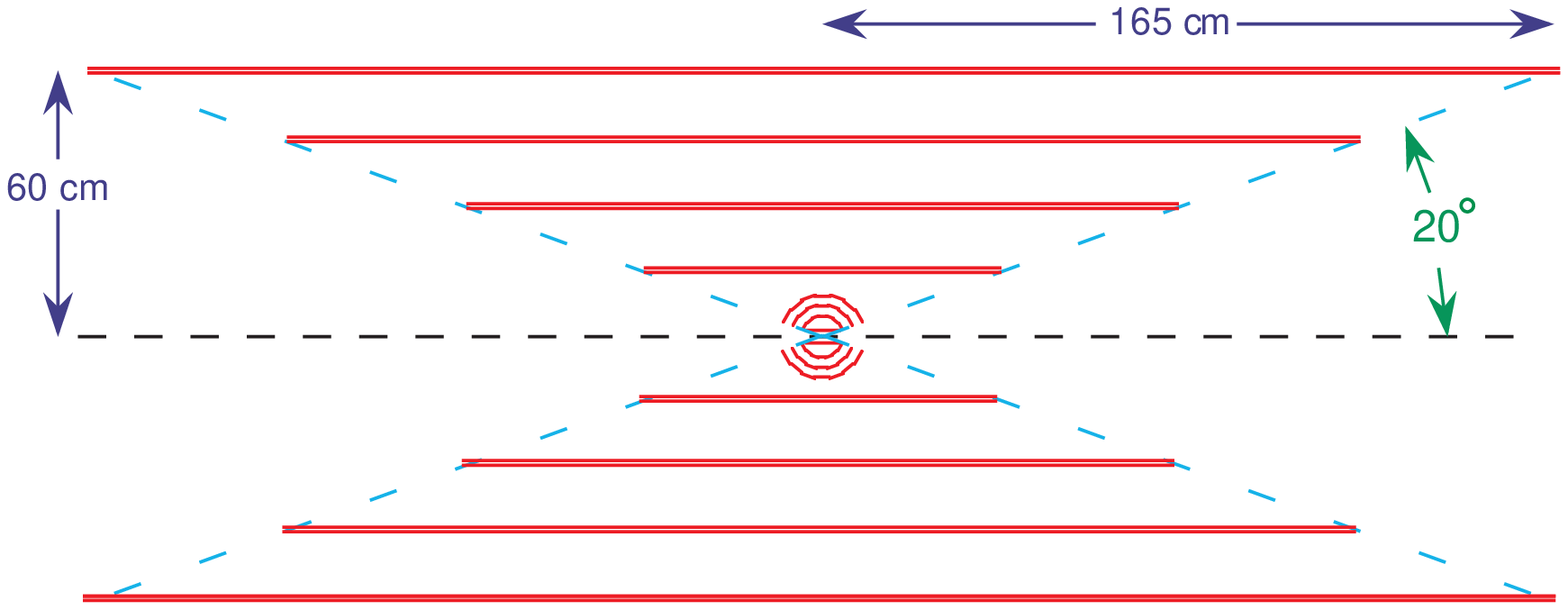}
\caption{Compact tracker geometry in a 4~T field.}
\label{fig:alans_tracker}
\end{figure*}
As an example, consider the geometry 
shown in Fig.~\ref{fig:alans_tracker} in which a 4-layer pixel vertex detector 
is embedded in a 4-layer small angle stereo cylindrical silicon microstrip 
detector with a $50 \times 300$~$\mu$m$^2$ resolution. 
We take the inner layer of the vertex detector 
to consist of a cylinder of $50 \times 300$~$\mu$m$^2$ pixels, and the outer 
3 vertex layers to consist of spherical shells of $50 \times 50$~$\mu$m$^2$ 
columnar pixels or pixel micro-telescopes. The system is assumed to correspond 
to 15\% of a radiation length at $90^\circ$. Using a parametric calculation 
of the momentum resolution, including multiple scattering, we obtain 
$\sigma_p/p^2 = 10^{-4}~(10^{-2})~$(GeV/$c)^{-1}$ for p = 100~GeV/$c$~(1~GeV/$c).$ 
\end{itemize}
\noindent
Both the low field and high field tracking solutions look 
interesting and should be pursued with more complete simulations. 

\subsection{Electromagnetic calorimeter schemes}

Background particles entering the electromagnetic calorimeter  
are expected to give rise to significant 
energy pedestals in the calorimeter cells. Consider a 4~m long calorimeter 
that is 25~radiation lengths deep, has an inner 
radius of 120~cm, and is constructed from $2 \times 2$~cm$^2$ cells. 
This gives a total of $10^5$ electromagnetic calorimeter towers. 
The GEANT background 
calculation predicts that each cell sees on average $n_{\gamma} = 4$ 
background photons per crossing with a mean energy $E_{\gamma} = 1-2$~MeV. 
If an electromagnetic shower 
occupies 9 cells, then the mean background pedestal will be about 
70~MeV. This pedestal can be subtracted from the measured energies. 
The precision of the resulting electron and photon energy measurements 
will depend on the fluctuations in the mean background energy per cell. 
For an electromagnetic shower occupying 9 cells, the fluctuations in the pedestals are predicted to be about 10~MeV. This takes into account the fluctuations in the number and the energies of the background photons. 
\subsection{Hadronic calorimeter schemes}

None of the energy generated by background photons in the electromagnetic 
calorimeter is expected to penetrate into the hadronic calorimeter.
GEANT calculations show that the total kinetic energy deposited by neutrons
 in the calorimeter is of the order of 140~TeV with an average energy of 30~MeV
per neutron.
In order to estimate what fraction of the kinetic energy of the neutrons
will be visible, we should consider the materials involved.
For this simulation we have presumed an equal mix by volume of liquid argon
(as active medium) and copper (as absorber).
At 30~MeV we expect only a small fraction of the neutrons to knock off protons
and only about 10\% of the proton ionization to be visible in the liquid.
Presuming a hadronic calorimeter with $10^4$ towers, with the material composition
described above, the average energy read in the liquid argon will be of 
the order of 10~MeV per tower with a fluctuation of 5~MeV.
In summary, a $50 \times 50$~GeV collider with $4\times 10^{12}$ muons per bunch, the photons and neutrons are expected to generate pedestals of 800 and 100~GeV
respectively. 
The estimates for pedestal fluctuations are at or below the 
level of the expected electronic noise. Therefore we believe that the subtraction of these pedestals would present little problem both for the electromagnetic and the hadronic calorimeters. The presence of the high neutron background should be taken into account in choosing materials for calorimetry. Liquid argon seems a natural choice for the electromagnetic calorimeter.

The energy deposited by the Bethe-Heitler muons in the calorimeter is given in
table~\ref{tab-bheitler} as a function of the center of mass energy of the
collider. For low center of mass energies, such as the Higgs factory, the Bethe-Heitler muons are not a problem, since there are fewer of them and they leave
less energy by catastrophic bremsstrahlung in the calorimeter. For the higher
energy option (4~TeV in the CoM or higher), one should explore ways
to correct for the energy deposition in the calorimeter, such as pattern 
recognition of the muon tracks or by using timing information.

\subsection{Muon detector schemes}

The predicted background flux is expected to be relatively modest 
beyond a radius of 3~m in the vicinity of the muon detector. Several possible 
technologies for muon detectors at a muon collider were discussed during 
Snowmass\cite{Snowmass96}:
\begin{itemize}
\item Cathode strip chambers. 
The idea, which is described in the muon collider book \cite{snowmass9.4.2}, 
is to use MWPCs with 
segmented cathodes and a short (35~ns) drift time to provide prompt signals 
for triggering. The precision of the coordinate measurements would be expected 
to be of order 50~$\mu$m $\times$ a few mm. 
\item Threshold Cherenkov counter.
The idea is to use a gas Cherenkov radiator to exploit the 
directionality of Cherenkov radiation in order to select high-$p_T$ muons 
coming from the IP. The device would also give excellent timing resolution 
(of order 2~ns). 
\item Long drift jet chamber with pad readout \cite{newatac} (Fig.~\ref{fig:muzaffer}). 
Drift time provides the z~coordinate, and pad readout provides 
the r-$\phi$~coordinates. Directionality at the trigger level is provided 
by the pattern of pad hits within a limited time window. The drift field 
is provided by cathode strips on grooved G-10 plates. Using 90\% argon plus 
10\% CF$_4$ and a maximum drift distance of 50~cm, the maximum drift time 
is 5~$\mu$s.
\end{itemize}
\noindent

\begin{figure*}[htb!]
\leavevmode
\centering
\epsfxsize=9.5cm
\epsfysize=8.0cm
\epsfbox{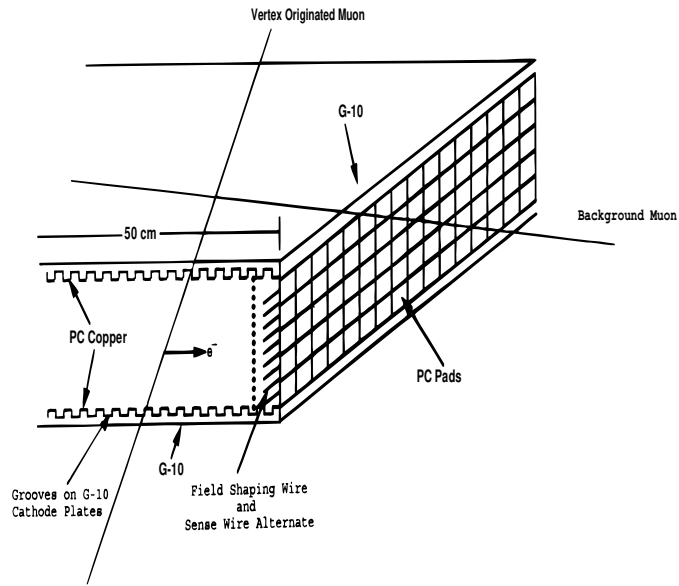}
\caption{Long drift jet chamber with pad readout for muon detection 
at a muon collider. }
\label{fig:muzaffer}
\end{figure*}

At high energy in the CoM, the channel $\mu^+\mu^- \rightarrow \mu^+\mu^- + \textrm{Higgs boson}$ becomes particularly attractive to study using the muon collider, if the forward going muons from the interaction can be detected \cite{newraja}. The method provides a capability to search for any missing neutral state such as the Higgs boson via the missing mass technique. We are investigating methods to improve our forward muon detection capability.
\section{CONCLUSIONS}

Unlike protons, muons are point-like but, unlike electrons, they emit
relatively little synchrotron radiation and therefore can be accelerated and collided in rings. 

Another advantage resulting from the low synchrotron radiation is
the lack of beamstrahlung and the possibility of very small collision energy
spreads. A beam energy spread of $\Delta$E/E of  0.003\%  
 is considered feasible for a 100~GeV machine. It has been shown that by observing spin precession, the absolute energy could
be determined to a small fraction of this width. These features become 
important in conjunction  with the large s-channel Higgs production ($\mu^+\mu^-\rightarrow h$, 
43000~times larger than for $e^+e^-\rightarrow h$), allowing precision
measurements of the Higgs mass, width and branching ratios. A higher energy muon collider can also distinguish the nearly degenerate heavy
Higgs bosons $H^0$ and $A^0$ of the minimal supersymmetric extension of the
standard model, since these states can also be produced in the $s$ channel.
We have also examined the ability of the muon collider to study
techni-resonances, do a high luminosity study of $Z$ boson physics, scan the
$W$ and $t\bar{t}$ thresholds to make precision mass measurements as well as
SUSY and strongly interacting W boson physics. The high luminosity proton driver
and  the cold low energy muons permit the study of rare kaon and muon decays.
Muon storage rings will permit low-systematics  studies of neutrino oscillations
for a wide range of mixing angle and $\delta m^2$ phase space with
hitherto unattainable sensitivity.

Such machines are clearly desirable. The issues are:
\begin{itemize} 
\item { whether they can be built and physics done with them}
\item { what they will cost}.
\end{itemize}
 Much progress has
been made in addressing the first question and the answer, so far, appears to
be yes. It is too early to address the second. 

We have studied
machines with CoM energies of 0.1, 0.4 and 3~TeV, defined
parameters and simulated many of their components. Most recent work has been done on
the 0.1~TeV \textit{First Muon Collider}, the energy taken to be representative
of the actual mass of a Higgs particle. A summary of progress and challenges follows:
\paragraph{Proton  driver}
The specification of the proton driver for the three machines is assumed the
same: $10^{14}$  protons/pulse at an energy above 16~GeV and 1-2~ns \textit{rms} bunch
lengths. There have been three studies of how to achieve these parameters. The most
conservative, at 30~GeV, is a generic design. Upgrades of the FNAL (at 16~GeV)
and BNL (at 24~GeV) accelerators have also been studied. Despite the very short bunch
requirement, each study has concluded that the specification is attainable.
Experiments are planned to confirm some aspects of these designs.

\paragraph{Pion production and capture}
Pion production has been taken from the best models available, but an
experiment (BNL-E910) that has taken data, and is being analyzed, will refine these
models. The assumed 20~T capture solenoid will require state-of-the-art technology. Capture, decay and phase rotation have been simulated, and
have achieved the specified production of 0.3~muons per initial proton. The
most serious remaining issues for this part of the machine are:
\begin{enumerate}
\item { The nature and material of the target:}
The baseline assumption is that a liquid metal jet will be used, but the
effects of shock heating by the beam, and of the eddy currents induced in the
liquid as it enters the solenoid, are not yet fully understood.  
\item { The
maximum rf field in the phase rotation:} For the short pulses used, the current
assumptions would be reasonably conservative under normal operating conditions,
but the effects of the massive radiation from the nearby target are not known.
\end{enumerate}
Both these questions can be answered in a target experiment planned to start within the next two years at the BNL AGS. 

Polarization of the muon beams represents a significant physics
advantage and is an important feature of a muon collider. Polarized muon beams are possible. Muons are produced with $100\%$ polarization
in the rest frame of the pion, but they travel in all directions. By accepting
the forward going muons, it is easy to obtain $25\%$ polarization in
either beam easily. The amount of polarization can be increased with an
accompanying price in luminosity.
\paragraph{Cooling}
The required ionization cooling is the most difficult and least understood
element in any of the muon colliders studied.
 Ionization cooling is a phenomenon that occurs whenever there is energy
loss in a strong focusing environment. 

 But achieving the
nearly $10^6$ reduction required is a challenge. Cooling over a wide range has
been simulated using lithium lenses and ideal (linear matrix) matching and
acceleration.  Examples of limited sections of solenoid lattices with
realistic accelerating fields have now been simulated, but the specification
and simulation of a complete system has not yet been done. Much theoretical
work remains: space  charge and wakefields must be included; lattices at the
start and end of the cooling sequences must be designed; lattices including
liquid lithium lenses must be studied, and the sections must be
matched together and simulated as a full sequence. The tools for this work
are nearly ready, and this project should be completed within two years.

Technically, one of the most challenging aspects of the cooling system appears
to be: 
\begin{itemize}
\item { High gradient rf} (e.g. 36~MV/m at 805~MHz) operating in strong
(5-10~T) magnetic field, with beryllium foils between the cavities. 
\end{itemize}
An experiment is planned that will test such a cavity, in the required fields,
in about two years time. On an approximately six year time scale, a \textit{Cooling Test Facility} is being proposed that could test ten meter lengths of different
cooling systems. If they are required, then an urgent need is to develop:
\begin{itemize}
\item { Lithium Lenses:} (e.g.\ 2~cm diameter, 70~cm long, liquid lithium
lenses with 10~T surface fields and a repetition rate of 15~Hz).
\end{itemize}
The use of 31~T solenoids could avoid their need, at least in the low energy \textit{First Muon Collider}, which would ease the urgency of this rather long term R\&D, but both options would require long-term R\&D. Meanwhile a short lithium lens
is under construction at BINP (Novosibirsk, Russia).

\paragraph{Acceleration}

The acceleration system is probably the least controversial, although
possibly the most expensive, part of a muon collider. Preliminary parameters
have been specified for acceleration sequences for a 100~GeV and a 3~TeV
machine, but they need refinement. In the low energy case, a linac is followed
by three recirculating or FFAG accelerators. In the high energy accelerator, the
recirculating or FFAG accelerators are followed by three fast ramping synchrotrons
employing alternating pulsed and superconducting magnets. The parameters do
not appear to be extreme, and it does not appear as if serious problems are
likely. 

\paragraph{Collider}

The collider lattices are challenging because of the requirement of  very low
beta functions at the interaction point, high single bunch intensities, and short bunch lengths.
 However, the fact that all muons will decay after about 800~turns means that
slowly developing instabilities  are not a problem. Feasibility lattices have
been generated for a 4~TeV case, and more detailed designs for 100~GeV machines are been studied. In the latter case, but still without errors, $5\sigma$~acceptances in
both transverse and longitudinal phase space have been achieved in tracking
studies. Beam scraping schemes have been designed for both the low energy
(collimators) and high energy (septum extractors) cases.

The short bunch length and longitudinal stability problems are avoided if the rings, as
specified, are sufficiently isochronous, but some rf is needed to remove the
impedance generated momentum spread. Transverse instabilities (beam breakup)
should be controlled by rf BNS damping.

The heating of collider ring superconducting magnets by electrons from muon
decay can be controlled by thick tungsten shields, and this technique also
shields the space surrounding the magnets from the induced radioactivity on the
inside of the shield wall. A conceptual design of magnets for the low energy
machine has been defined.

Although much work is yet to be done (inclusion of errors, higher order
correction, magnet design, rf design, etc), the collider ring does not appear
likely to present a serious problem.

\paragraph{Neutrino radiation and detector background}

Neutrino radiation, which rises as the cube of the energy, is not 
serious for machines with center of mass energies below about 1.5~TeV. It is
thus not significant for the First Muon Collider; but above  2~TeV, it
sets a constraint on the muon current and makes it harder to achieve desired
luminosities. However, advances in cooling and correction of tune shifts may
still allow a machine at 10~TeV with substantial luminosity ($>\ 10^{35}\
\textrm{cm}^{-2}\textrm{s}^{-1}$).

Background in the detector was at first expected to be a very serious
problem, but after much work, shielding systems have evolved that limit most
charged hadron, electron, gamma and neutron backgrounds to levels that are
acceptable. Muon background, in the higher energy machines, is a
special problem that can cause serious fluctuations in calorimeter
measurements. It has been shown that fast timing and segmentation can help
suppress this background, and preliminary studies of its effects on a physics
experiment are encouraging. The studies are ongoing.

\paragraph{Detector scenarios}
We have considered several options for the experimental detector components for
 various CoM energy colliders. Much work needs to be done to optimize the physics reach at each energy by feeding back the results of detailed simulations of backgrounds and signal to the detector design. Only then will the feasibility of doing physics with a muon collider be fully explored. 
\paragraph{Summary}
Much progress has been made since Snowmass, but much still needs to be done. A
time scale of two years should allow completion of simulation studies and the
experimental testing of crucial technical challenges. Prototype
construction and testing will take another 4-6~years. The
construction of a \textit{First Muon Collider} by about 2010 does seems possible.
\section{ACKNOWLEDGMENTS}
This research was supported by the U.S. Department of Energy under Contracts No.
DE-ACO2-98CH10886, DE-AC02-76CH03000 and DE-AC03-76SF00098.
%\bibliography{report}      % temp delete
%\bibliographystyle{prsty}  % temp delete
%\end{document}             % temp delete

%***********************************************************************
% include here the new  file pr_11.bbl
% next comment out \bibliography{report} and \bibliographystyle{prsty}
% next move \end{document} to the end of the file
%***********************************************************************

%\begin{thebibliography}{100}
%\input prst00_new.bbl

\end{document}